\documentclass[british,english]{book}
\usepackage[totalwidth=385pt,totalheight=584pt]{geometry}
\usepackage{amsmath,amsthm,amssymb,epsf,graphics,graphicx,mathrsfs,yfonts,eufrak
,color,dsfont,bbm,esint}
\usepackage[latin9]{inputenc}
\usepackage[toc]{appendix}
\usepackage{psfrag}
\usepackage{amssymb}
\usepackage[unicode=true,
 bookmarks=false,
 breaklinks=true,pdfborder={0 0 1},backref=false,colorlinks=true]
 {hyperref}
 
 \usepackage[font=small,labelfont=bf]{caption}

\usepackage{tikz}
\usetikzlibrary{chains}

\usetikzlibrary{plotmarks,calc,decorations, decorations.pathmorphing, patterns}
\usetikzlibrary{arrows}

\tikzstyle dynkin node=[very thick,shape=circle,draw,inner sep=0pt,minimum size=3mm]
\tikzstyle dynkin line=[very thick]
\tikzstyle inverse line=[gray,line width=1.46pt,line cap=round, dash pattern=on 0pt off 2\pgflinewidth]
\tikzstyle red phase=[red,decoration={snake,amplitude=0.1mm,segment length=1.6mm},decorate]
\tikzstyle blue phase=[blue,decoration={snake,amplitude=0.1mm,segment length=0.9mm},decorate]
\tikzstyle green phase=[green,decoration={snake,amplitude=0.1mm,segment length=0.9mm},decorate]
\tikzstyle brown phase=[brown,decoration={snake,amplitude=0.1mm,segment length=0.9mm},decorate]
\newcommand{\boundellipse}[3]
{(#1) ellipse (#2 and #3)
}

\tikzstyle arrow=[thick,rounded corners=18pt,-latex]
\tikzstyle box=[draw,rounded corners,outer sep=4pt]
\tikzstyle B node=[outer sep=0pt]
\tikzstyle Q node=[inner sep=1pt,outer sep=0pt]

\graphicspath{ {./images/} }
\definecolor{MyDarkBlue}{rgb}{0.2,0.2,0.6}
\definecolor{MyDarkRed}{rgb}{0.7,0.1,0.2}
\hypersetup{
 pdftitle={ODE/IM correspondence in Toda field theories and fermionic basis in sin(h)-Gordon model},
 pdfauthor={Stefano Negro},
 citecolor=MyDarkRed,
 linkcolor=MyDarkBlue,
 urlcolor=MyDarkBlue
}

\usepackage{fbb}
\usepackage{tocloft}
\newcommand{\bz}{\overline{z}}
\newcommand{\bp}{\overline{\partial}}

\newcommand{\tr}{\textit{tr}\;}

\newcommand\blank[1]{}

\renewcommand{\hat}{\widehat}
\newcommand\eq{\begin{equation}}
\newcommand\en{\end{equation}}
\newcommand\bea{\begin{eqnarray}}
\newcommand\eea{\end{eqnarray}}
\newcommand\nn{\nonumber}
\newcommand\ba{\(\begin{array}}
\newcommand\ea{\end{array}\)}
\newcommand{\resection}[1]{\setcounter{equation}{0}\section{#1}}

\renewcommand{\theequation}{\thesection.\arabic{equation}}

\newcommand\bzero{\boldsymbol{0}}
\newcommand\balpha{\boldsymbol{\alpha}}
\newcommand\bbeta{\boldsymbol{\beta}}
\newcommand\bgamma{\boldsymbol{\gamma}}

\newcommand\boeta{\boldsymbol{\eta}}
\newcommand\bepsilon{\boldsymbol{\epsilon}}

\newcommand\bkappa{\boldsymbol{\kappa}}
\newcommand\bpi{\boldsymbol{\pi}}

\newcommand\bmu{\boldsymbol{\mu}}
\newcommand\bnu{\boldsymbol{\nu}}

\newcommand\blambda{\boldsymbol{\lambda}}
\newcommand\btheta{\boldsymbol{\theta}}

\newcommand\bPsi{\boldsymbol{\Psi}}

\newcommand\bPi{\boldsymbol{\Pi}}

\newcommand\bLambda{\boldsymbol{\Lambda}}

\newcommand\bv{\mathbf{v}}

\newcommand\Rth{{\mathbbm R}}

\newcommand\mg{\mathfrak g}
\newcommand\mL{\mathcal L}

\newtheorem*{Mconj}{Main conjecture}
\newtheorem*{Def}{Definition}
\newtheorem*{LiouvArn}{Liouville-Arnold Theorem}
\newtheorem*{Conj}{Conjecture}

\newcommand{\mychapter}[2]{
    \setcounter{chapter}{#1}
    \setcounter{section}{0}
    \chapter*{#2}
    \addcontentsline{toc}{chapter}{#2}
}
\newcommand{\mynewchapter}[2]{
    \setcounter{chapter}{#1}
    \setcounter{section}{0}
    \subsubsection*{#2}
    \addcontentsline{toc}{chapter}{#2}
}
\newcommand{\mysection}[2]{
    \setcounter{chapter}{#1}
    \setcounter{section}{0}
    \subsubsection*{#2}
    \addcontentsline{toc}{section}{#2}
}

\begin{document}

\begin{titlepage}
\begin{center}
\textsc{\huge 
ODE/IM correspondence in Toda field theories and fermionic basis in sin(h)-Gordon model}\\[2cm]
{\LARGE Stefano Negro}\\[1cm]
{\footnotesize{}Dipartimento di Fisica,}\\
{\footnotesize{}Universit\'{a} di Torino,}\\
{\footnotesize{}via Pietro Giuria 1, 10125, Torino, Italy}\\[0.3cm]
\href{mailto:steff.negro@gmail.com}{\ttfamily steff.negro@gmail.com}\\[3cm]

ABSTRACT
\end{center}
This article is the author's PhD thesis as it has been submitted in June 2014 to the Universit\`{a} degli Studi di Torino, with minor additions and revisions. Some of the results contained in this thesis, in particular concerning the second part, have been previously published in \cite{Dore_Fald_Negr_Tate_12,Negr_Smir_13_1,Negr_Smir_13_2,Negr_14}.\\

The first part of this work consists of a study of the ODE/IM correspondence for simply-laced affine Toda field theories. It is a first step towards a full generalisation of the results of S. Lukyanov and A. Zamolodchikov \cite{Luky_AZam_10} to a general affine Lie-Ka\v{c}-Moody algebra $\hat{\mathfrak g}$. In order to achieve our goal, we investigate the structure of evaluation representations of $\hat{\mathfrak g}$ and show how their tensor products are related by what we call \emph{projected isomorphisms}. These isomorphisms are used to construct a set of quadratic functional relations, called $\psi$-system, for the solutions to complex differential equations associated to $\hat{\mathfrak g}$. Finally, from the $\psi$-system we derive a set of Bethe Ansatz equations satisfied by the eigenvalues of some particular boundary problem for the above mentioned differential equations. This algebro-differential setting was brought to its general and mathematically rigorous form, for the massless case, by D. Masoero, A. Raimondo and D. Valeri in 2015 \cite{Maso_Raim_Vale_15_1,Maso_Raim_Vale_15_2}.

The second part of this work deals with the study of one-point functions in sine- and sinh-Gordon models. The approach to the computation of these quantities follows a powerful method, which we call \emph{fermionic basis}, developed by H. Boos, M. Jimbo, T. Miwa, F. Smirnov and Y. Takeyama for the XXZ, quantum Liouville and quantum sine-Gordon models \cite{Boos_Jimb_Miwa_Smir_Take_07,Boos_Jimb_Miwa_Smir_Take_09,Jimb_Miwa_Smir_09,Boos_Jimb_Miwa_Smir_10,Jimb_Miwa_Smir_11_1,Jimb_Miwa_Smir_11_2}. We show how the determinant formula for one-point functions obtained there can be generalised to the sinh-Gordon model. In doing so we give an interpretation of the fermionic basis in terms of certain symmetries of the system. This new perspective will also allow us to solve trivially the reflection relations introduced by V. Fateev, D. Fradkin, S. Lukyanov, A. Zamolodchikov and Al. Zamolodchikov in \cite{Fate_Frad_Luky_AZam_AlZa_99}. We then provide analytical and numerical results supporting our finding.\\

\end{titlepage}
\newpage{}

\frontmatter
\begin{titlepage}

\begin{center}

\textsc{\huge \textbf{Universit\`a degli studi di Torino}}\\[1cm]
\includegraphics[width=0.24\textwidth]{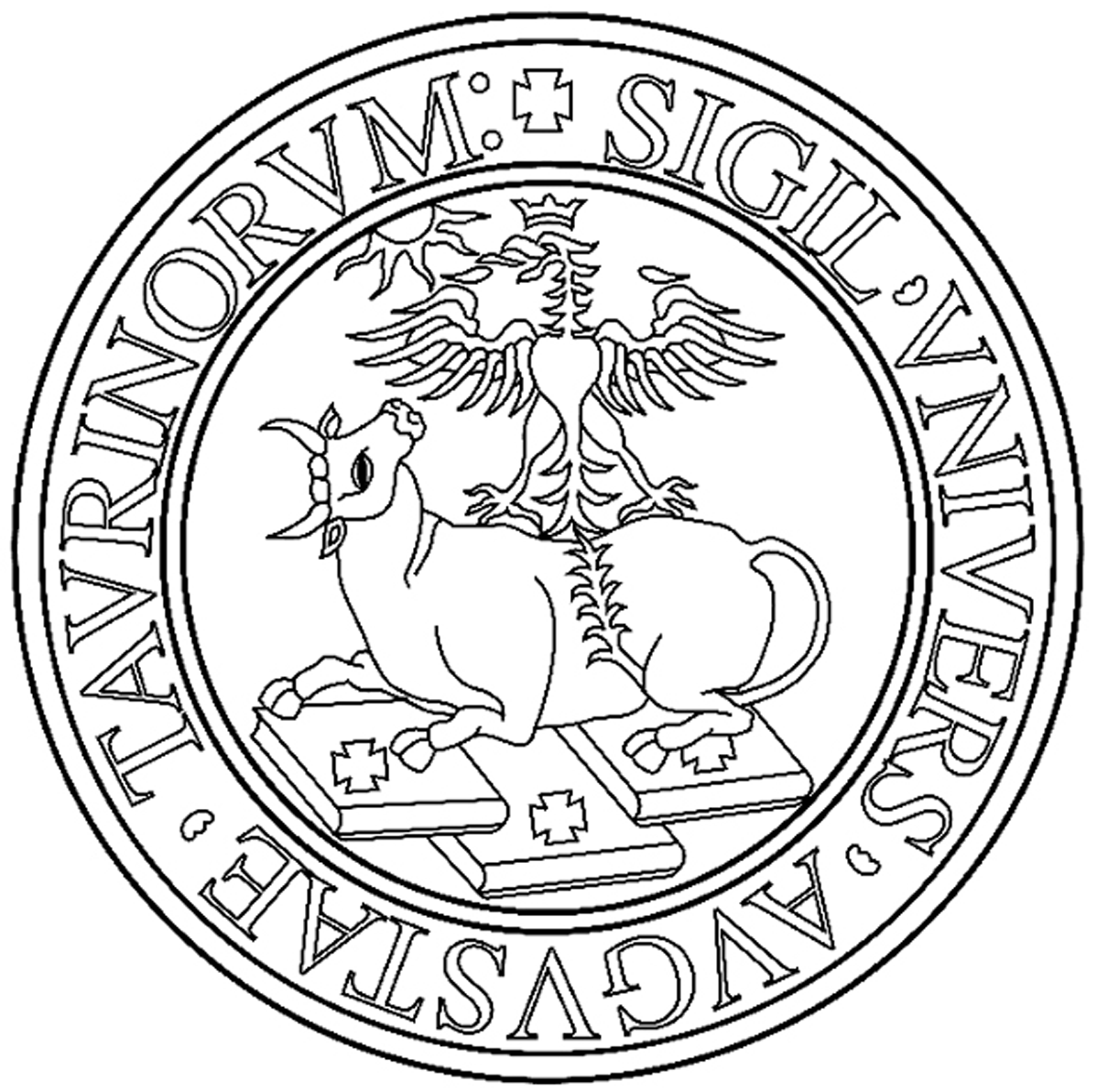}\\[1cm]
\textsc{\Large Scuola di dottorato in}\\
\textsc{\Large ``Scienze della natura e tecnologie 
innovative"}\\
\textsc{\Large Indirizzo in}\\
\textsc{\Large ``Fisica e astrofisica"}\\[0.7cm]
\textsc{\large Dipartimento di Fisica Teorica}\\[1.5cm]


\huge \textbf{ODE/IM correspondence in Toda field theories and fermionic basis 
in sin(h)-Gordon model}\\[0.7cm]
\normalsize\textit{\large PhD thesis}\\
\normalsize\textit{\large June 2014}\\[1cm]

\textsl{\Large Author}\\[0.4cm]
{\LARGE Stefano Negro}\\[2cm]

\begin{minipage}{0.4\textwidth}
\begin{flushleft}
\textsl{\large Supervisor}\\[0.3cm]
{\Large Dott. Roberto Tateo}
\end{flushleft}
\end{minipage}
\begin{minipage}{0.4\textwidth}
\begin{flushright}
\textsl{\large Co-Supervisor}\\
\Large Prof. Fedor Smirnov
\end{flushright}

\end{minipage}

\end{center}
\end{titlepage}

\tableofcontents

\newpage
\pagestyle{plain}

\mynewchapter{0}{Foreword}

This thesis collects the work I have done during the three years and a half of my PhD along with my advisors Roberto Tateo and Fedor A. Smirnov. It comprises two subjects, the ODE/IM correspondence and the fermionic basis formalism, which, at first sight and in many of their aspects, are rather disconnected. For this reason they will be presented here in two separate parts, each one as self-contained as possible. The first one will deal with the study of the ODE/IM correspondence for the $2D$ Toda Field Theories, while the second will be devoted to the development of the fermionic basis formalism in the quantum sin(h)-Gordon model, a particular case of Toda Field Theory. As diverse as these two topics might appear, there is an important contact point between them, beyond the trivial fact that both of them deal with integrable field theories. This fact, which is the main motivation behind the twofold nature of my work, will be outlined in the last part, dedicated to conclusions and perspectives.\\

Before delving into the hardcore matter of this thesis allow me to present a ``bird's eye" view on the developments of the integrability from its birth up to the mid nineties of the last century.

\mynewchapter{0}{A (not so) short history of integrability}

The history of Integrable Systems is as old as that of Classical Mechanics and the two were, for for the greatest part of 18th century, more or less coinciding. Following the formulation of Isaac Newton's laws of motion \cite{Newt1726}, eminent mathematicians and physicists such as Daniel Bernoulli, Alexis Clairaut, Jean-Baptiste d'Alembert, Leonhard Euler and Joseph-Louis Lagrange devoted many works to the problem of finding exact solutions to Newton's equations. In particular, between the 1750s and the 1780s, Lagrange and Euler managed to reformulate these equations into a form which has been known since as \emph{Lagrangian mechanics} \cite{Lagr_1788}, whose central element is the \emph{Lagrangian function} $L$ which summarises the dynamics of the system in question; the Equations of Motion of Newton are then rewritten as \emph{Euler-Lagrange equations}:

\begin{equation*}
\frac{d}{d t} \frac{\partial L}{\partial \dot{\mathbf q}} = \frac{\partial L}{\partial \mathbf q} \;, \qquad L = T - V \; : \begin{array}{c c c}\mathbbm R^n \times \mathbbm R^n & \longrightarrow & \mathbbm R \\ \rotatebox[origin=c]{90}{$\in$} & & \rotatebox[origin=c]{90}{$\in$} \\ (\mathbf q, \dot{\mathbf q}) & \longmapsto & L(\mathbf q, \dot{\mathbf q})  \end{array} \;,
\end{equation*}
where $T$ is the kinetic energy of the $n$-dimensional system, $V$ its potential energy while $\mathbf q$ and $\dot{\mathbf q}$ are the generalised coordinates and velocities. It became quickly evident, although the reason was not completely clear at that point, that only a handful of ``nice" systems, such as the Kepler problem which Newton solved himself, allowed closed form solutions; these systems were called \emph{soluble} or \emph{integrable}.

It is at this moment that the histories of Integrable Systems and Classical Mechanics begin to separate: Clairaut, Lagrange, Pierre-Simon Laplace and, later, Sim\'eon Denis Poisson and Carl Friedrich Gauss began directing part of their efforts towards the development of methods to obtain approximate solutions to problems of celestial mechanics; these works would lead, ultimately, to the creation of the \emph{perturbation theory}.

On the other hand, the work of Lagrange and Euler was continued in the 19th century by Johann Friedrich Pfaff and Augustin-Louis Cauchy and peaked with the development of a general method of integrating the equation of dynamics, introduced by Sir William Rowan Hamilton \cite{Hami_1834,Hami_1835} and Carl Gustav Jacob Jacobi \cite{Jaco_1837_1,Jaco_1837_2} in the 1830s. Hamilton introduced a formalism analogous to that of Euler-Lagrange in order to describe the wave nature of optical systems which were beyond the reach of Lagrangian mechanics; Jacobi then imported Hamilton's ideas in mechanics and showed their relations with the method of Euler and Lagrange, eventually arriving at the so-called \emph{Hamilton-Jacobi} formalism. The starting point of this is the description of the dynamics of a system in the following \emph{canonical form}:

\begin{equation*}
	\left\lbrace \begin{array}{l}
	\dot{\mathbf q} = \frac{\partial H}{\partial \mathbf p} \\ \\
	\dot{\mathbf p} = -\frac{\partial H}{\partial \mathbf q}
	\end{array}\right. \; ; \qquad \mathbf p = \frac{\partial L}{\partial \dot{\mathbf q}} \; , \quad H = \mathbf p\dot{\mathbf q} - L\vert_{\mathbf p,\mathbf q}\; ,
\end{equation*}
where $\mathbf p$ is the \emph{generalised momentum} relative to the \emph{generalised coordinate} $\mathbf q$ and the \emph{Hamiltonian function} $H$ corresponds to the total energy of the system. These equations can be represented in terms of the \emph{Poisson bracket} $\{\cdot,\cdot\}$, introduced by Poisson in 1809 \cite{Pois_1809}:

\begin{equation*}
	\left\lbrace \begin{array}{l}
	\dot{\mathbf q} = \{q,H\} \\ \\
	\dot{\mathbf p} = \{p,H\}
	\end{array}\right. \; , \qquad \{A,B\}\doteq \sum_{j=1}^n \frac{\partial A}{\partial q_i}\frac{\partial B}{\partial p_i} - \frac{\partial A}{\partial p_i}\frac{\partial B}{\partial q_i}\; .
\end{equation*}
The essence of the Hamilton-Jacobi method lies then in the existence of a \emph{canonical transformation} of variables $S$:
\begin{equation*}
	\left\lbrace \begin{array}{l}
	\mathbf p = \frac{\partial S}{\partial \mathbf q} \\ \\
	\mathbf Q = \frac{\partial S}{\partial \mathbf P}
	\end{array}\right. \;; \qquad S \; : \begin{array}{c c c}\mathbbm R^n \times \mathbbm R^n & \longrightarrow & \mathbbm R \\ \rotatebox[origin=c]{90}{$\in$} & & \rotatebox[origin=c]{90}{$\in$} \\ (\mathbf P, \mathbf q) & \longmapsto & S(\mathbf P, \mathbf q)  \end{array} \;,
\end{equation*}
such that in the transformed canonical equations

\begin{equation*}
	\dot{\mathbf Q} = \frac{\partial K}{\partial \mathbf P} \; , \qquad \dot{\mathbf P} = - \frac{\partial K}{\partial \mathbf Q} \; ,
\end{equation*}
the new Hamiltonian function $K = H\vert_{\mathbf Q,\mathbf P}$ does not depend on $Q$. If this is the case, then the canonical equations are immediately integrated

\begin{equation*}
	\mathbf P(t) = \mathbf P_0 \; , \qquad \mathbf Q(t) = \mathbf Q_0 + t\frac{\partial K}{\partial P}\Big\vert_{P_0} \; , 
\end{equation*}
and the problem of integrating a system reduces to a search for a generating function $S(\mathbf P,\mathbf q)$ satisfying the non-linear \emph{Hamilton-Jacobi} equation

\begin{equation*}
	H(\frac{\partial S}{\partial\mathbf q},\mathbf q) = K(\mathbf P) \; .
\end{equation*}
The functions $\mathbf P$ are called \emph{first integrals of motion} and are constant in time; moreover they are in \emph{involution}, meaning that their Poisson bracket vanishes identically:

\begin{equation*}
	\lbrace P_i,P_j\rbrace = 0 \; .
\end{equation*}

With this formalism available, in 1855, Jacques Edmond \'Emile Bour \cite{Bour_1855} and Joseph Liouville \cite{Liou_1855} formulated the first consistent definition of integrability:

\begin{LiouvArn}
	If in an Hamiltonian system with $n$ degrees of freedom there exist $n$ first integrals of motion that are independent and in involution, then it is possible to find a canonical transformation to canonical coordinates such that the Hamiltonian depends only on the integrals of motion and the canonical equations can be solved explicitly by quadratures.
	
These canonical coordinates are called \emph{action-angle} coordinates and the systems for which a set of such coordinates exist are known as \emph{completely integrable in the Liouville sense} (although they're often referred to simply as ``integrable").
\end{LiouvArn}

Liouville's definition of integrable Hamiltonian systems naturally covered many known examples, like the already mentioned Kepler problem, harmonic oscillators solvable by trigonometric functions, the rigid bodies (also known as \emph{spinning tops}) of Euler-Poinsot \cite{Poin_1834} and Lagrange \cite{Lagr_1788} type and the geodesic motion on an ellipsoid introduced by Jacobi \cite{Jaco_1839}.

Soon after the Liouville-Arnold theorem had been stated, Carl Gottfried Neumann discovered a new integrable Hamiltonian system, solvable by means of hyperelliptic functions \cite{Neum_1859}. With this model as a prototype, a series of integrable systems, all more or less related to hyperelliptic functions, was discovered and solved by authors such as Gustav Robert Kirchhoff \cite{Kirc_1870}, Rudolf Friedrich Alfred Clebsch \cite{Cleb_1871}, Vladimir Andreevich Steklov \cite{Stek_1893}, Heinrich Martin Weber \cite{Webe_1878} and Sofia Vasilyevna Kovalevskaya \cite{Kova_1889,Kova_1890}. However, the excitement for this series of discoveries was seriously blunted by a result attained by Jules Henri Poincar\'e: following the road traced by its predecessors in the field of approximate solutions to mechanical problems, he proved in 1890 \cite{Poin_1890} that, in general, there exists no analytic integral of motion which can be represented in the form of a convergent power series in a small parameter. This fact effectively appointed to the integrable systems the status of exceptions among the Hamiltonian ones and the interest in the search for new systems and in the analysis of the structure of integrability almost vanished for the next 70 years\footnote{It is worthy of note the fact that, although it crushed the high hopes that were placed on the concept of integrability, the result of Poincar\'e opened the door to a completely new area in the study of dynamical systems which would ultimately lead to the birth of the Theory of Chaos.}.

What has been briefly sketched above has been the main stream of studies on integrable system in the 18th and 19th century. Although many important milestones were set along this path, the direct origin of the breakthrough that, in the seventies of the 20th century, brought back to the fore the studies on integrability has to be searched in a rather different area of physics: the study of solitary waves. It has been thanks to a Scottish engineer, John Scott Russel, that these studies were born; in 1834 he observed the formation and the propagation of a solitary wave in a canal which he describes as a singular and beautiful phenomenon \cite{Russ_1834}:

\begin{quote}
\textit{``I was observing the motion of a boat which was rapidly drawn along a narrow channel by a pair of horses, when the boat suddenly stopped - not so the mass of water in the channel which it had put in motion; it accumulated round the prow of the vessel in a state of violent agitation, then suddenly leaving it behind, rolled forward with great velocity, assuming the form of a large solitary elevation, a rounded, smooth and well-defined heap of water, which continued its course along the channel apparently without change of form or diminution of speed. I followed it on horseback, and overtook it still rolling on at a rate of some eight or nine miles an hour, preserving its original figure some thirty feet long and a foot to a foot and a half in height. Its height gradually diminished, and after a chase of one or two miles I lost it in the windings of the channel. Such, in the month of August 1834, was my first chance interview with that singular and beautiful phenomenon which I have called the Wave of Translation."}
\end{quote}

The report of Russel's observation invoked a controversy in the scientific community, since the described phenomenon could not be explained by means of the existing theories of hydrodynamics. In particular Sir George Biddell Airy and Sir George Gabriel Stokes were quite suspicious about this finding and the search for an explanation went on for 40 years without a satisfying result until 1871, when Joseph Valentin Boussinesq \cite{Bous_1871} and later Lord Rayleigh \cite{Rayl_1876} published the first theoretical treatment of the solitary waves. In a subsequent work, Boussinesq introduced a peculiar nonlinear partial differential equation in connection with his hydrodynamic studies\footnote{We write the equation in its modern expression; the equation introduced by Boussinesq is in a different but equivalent form.} \cite{Bous_1877}:

\begin{equation*}
	\partial_t u(t,x) + 6 u(t,x)\partial_x u(t,x)+\partial_x^3 u(t,x) = 0 \; ,
\end{equation*}
where $u(t,x)$ represent the height of the wave at the time $t$ and position $x$.
This equation was rediscovered 18 years later by Diederik Johannes Korteweg and Gustav de Vries \cite{Kort_deVr_1895} and has since then taken the name of \emph{Korteweg-de Vries} equation, in short \emph{KdV}. Its distinctive feature is that, contrary to those appearing in the study of dynamics, it possesses an infinite number of degrees of freedom, in other words it is a field equation, like the equation describing the vibrations of a continuous medium or Maxwell's equations that were derived some years before \cite{Maxw_1861}. However the KdV equation is nonlinear, meaning that the familiar linear superposition principle does not apply to its solutions; even so, the equation admits simple solitary wave solutions which had been found rather easily:

\begin{equation*}
	u(t,x) \equiv u(x-c t) = \frac{c}{2} \textrm{sech}^2\left[\frac{\sqrt{c}}{2}(x-x_0 - c t)\right] \; ;
\end{equation*}
they describe a wave which maintains its shape as it travels towards the right with phase speed $c$, precisely the type of phenomenon witnessed by Russell. As interesting as the questions concerning these peculiar phenomena were, the theory of solitary waves did not go far beyond gaining some insight in the KdV equation, in part also due to the fact that the mathematical tools available then where not mature enough to efficiently handle their nonlinear nature.

So at the dawn of the 20th century both the studies on integrability and on solitary waves went in a sort of hibernation as physicists became more and more interested in the questions on the very structure of time and space posed by the works of Ernst Waldfried Mach, Henri Poincar\'e, Hendrik Antoon Lorentz and Albert Einstein on the one side and the baffling properties of the atomic world that the experiments of Heinrich Rudolf Hertz, Pieter Zeeman, Sir Joseph John Thomson, Ernest Rutherford and many others were disclosing on the other. The whole scientific community delved with excitement into the study of these new fascinating topics, producing a flurry of discoveries which peaked with the formulation of the theories of the Special and General Relativity \cite{Eins_1905,Eins_15,Eins_17} and the birth of the Quantum Mechanics \cite{Heis_25,Schr_26}. In the 40 years that followed these results, the outstanding efforts of a plethora of eminent scientists, too many to be cited in these lines without turning this introduction into a meaningless list, accumulated, opening perspectives unthinkable even at the beginning of the century and bringing about the establishment of the Standard Model and of modern General Relativity in the late 60s and 70s, and in the same years, to the appearance of the first studies on String Theory.

During these restless years, while the attention of the mainstream physics was captured by these topics, a smaller community of scientists moved from the results obtained in the late 19th century in the field of thermodynamics and statistical mechanics, especially from the works of James Clerk Maxwell, Ludwig Edward Boltzmann and Josiah Willard Gibbs, towards the investigation of the properties of materials and of collective behaviours, such as phase transitions. The framework provided by Gibbs \cite{Gibb_1902}, being extremely general and powerful, allowed for the immediate incorporation of the ideas of quantum mechanics and during the first half of the century, this new avenue, which in many respects is complementary to the above mentioned studies on quantum mechanics, would steadily annex various topics which were considered as separate, such as crystallography, elasticity, magnetism and many others, becoming one of the biggest branches of physics, known today as \emph{Condensed Matter Physics}. The first studies pointed towards a microscopical description of the electric and magnetic properties of materials, much in the spirit of Gibbs' statistical mechanics, and brought about the appearance, in 1900, of Paul Karl Ludwig Drude's model of electrical conductivity, later remodelled by Arnold Johannes Wilhelm Sommerfeld, and that of the so-called spin models, the prime examples being the \emph{Heisenberg model}, named after Werner Karl Heisenberg, and the \emph{Ising model}, first studied by Wilhelm Lenz and Ernst Ising in 1920\footnote{In the same year, Ising obtained the solution to the model in $1$ dimension, showing that it admits no phase transition, meaning it didn't reproduce the behaviour of magnetic material of possessing two distinct phases: ferromagnetic and paramagnetic. He wrongly assumed that this was the case for any dimension.}. These last two models are of particular relevance for the history of integrability, as they were the first examples of interacting many-body models to be solved exactly; quite remarkably, first came the solution to the quantum isotropic Heisenberg model, also called \emph{spin $1/2$ XXX chain}, which Hans Albrecht Bethe obtained in 1931 \cite{Beth_31} by means of a method which was to be named \emph{Bethe Ansatz}. This method, which would prove pivotal in the context of quantum integrability, allowed Bethe to explicitly obtain the eigenfunctions\footnote{The question of completeness of the eigenvalue spectrum obtained by Bethe was addressed and solved in the 1977 by Donald G. Babbitt and Lawrence E. Thomas \cite{Babb_Thom_77}.} of the Heisenberg model Hamiltonian, revealing the presence of particle-like excitations, called \emph{Magnons} or \emph{spin-waves}, which scatter in pairs, a peculiarity which would appear in any quantum integrable model. Later, in 1944, Lars Onsager obtained the free energy of the Ising model in $2$ dimensions and vanishing magnetic field \cite{Onsa_44}, effectively solving it.

These models were successful in the application of quantum mechanics to condensed matter problems and showed how macroscopic properties of material emerged from their atomic structures. However many particular phenomena, most notably the \emph{Kondo effect} and \emph{superconductivity}, which were discovered in the early 20th century, still could not be explained satisfyingly. After the second World War, the recognition of the relevance of collective behaviours in solids, like the aforementioned magnons, brought condensed matter physicists to import in their studies ideas from the still young Quantum Field Theory. In order to give a theoretical explanation to the phenomenon of superconductivity, Vitaly Lazarevich Ginzburg and Lev Davidovich Landau interpreted the transition to the superconductive regime as a second-order phase transition. In 1950 \cite{Ginz_Land_50} they introduced a mean-field theory, called the \emph{Ginzburg-Landau model}, which describes second order phase transitions as the results of a spontaneous breakdown of symmetry and introduces the notion of \emph{order parameter} distinguishing between phases. This gave birth to a new method for the study of phase transitions and collective behaviours, called \emph{Statistical Field Theory}. The following twenty years saw a steady increase in the importance of the concepts of phase transition and spontaneous symmetry breaking, not only in the context of Condensed Matter but also in that of Quantum Field Theory and, more generally, in the modern approach to physics. The studies on this subject would bring about the introduction of the fundamental concepts of \emph{scaling} and \emph{critical exponents}, thanks to Leo Philip Kadanoff, Benjamin Widom and Michael Ellis Fisher, that were collected and unified in the 1972 by Kenneth Geddes Wilson under the formalism of the \emph{Renormalization Group} \cite{Wils_72}.

By then the time was ripe for integrability to step back onto the stage; a fertile substratum of results and ideas had been laid during the first seventy years of the century and the very concept of integrable system was slowly, almost silently, resurfacing as more and more models of statistical nature, such as the $6$-vertex model studied by Elliott Hershel Lieb \cite{Lieb_67_1,Lieb_67_2,Lieb_67_3,Lieb_67_4}, showed to be exactly solvable\footnote{For an excellent review on exactly solvable models see Baxter's famous book \cite{Baxt_82}}. At the same time, scientists involved in the study of the Standard Model were beginning to search for methods complementary to the Perturbative Theory, in order to study those regions of the parameter space lying outside the reach of perturbative expansions and, hopefully, to give an answer to puzzling questions such as the problem of quark confinement. The only missing thing was a spark to light the match and this came in 1965, when Martin David Kruskal and Norman Zabusky published a work on the numerical solutions of the KdV equation \cite{Krus_Zabu_65}. Motivated by the work on the \emph{Fermi-Pasta-Ulam} problem, carried on in the fifties by Enrico Fermi, John Pasta and Stanis\l{}aw Marcin Ulam (see \cite{Zabu_63} for a review of the problem), they discovered that the KdV equation admits solutions in which many solitary waves coexist; these waves displayed a remarkable stability and their non-linear interaction, unexpectedly, didn't disrupt their identity, rather they conserved the same velocity and shape as before the collision, behaving like particles, and, for this reason, they received the name of \emph{solitons}. This observation led Clifford Spear Gardner, John Morgan Greene, Kruskal and Robert Miura to the introduction, just two years after the publication of Zabusky and Kruskal results, of a new powerful method to exactly solve the KdV equation. In their seminal paper \cite{Gard_Gree_Krus_Miur_67} they show the existence of a linearising transformation, which maps the initial value $u(0,x)$ of the KdV problem to the spectral and scattering data of the Schr\"odinger operator $-\frac{d^2}{dx^2} - u(0,x)$; the nonlinear evolution yielding $u(t,x)$ is then transformed in a linear evolution for that data which is readily solved and the solution $u(t,x)$ to the KdV problem is finally obtained via the inverse map, the so-called \emph{Inverse Scattering Transform}\footnote{The procedure of reconstructing the potential of a Schr\"odinger equation from the scattering data was developed in the 50s by Isra\"\i l Moyseyovich Gel'fand, Boris Levitan \cite{Gelf_Levi_51}, Vladimir Alexandrovich Marchenko \cite{Marc_50} and Mark Grigorievich Krein \cite{Krei_51} and involves a linear integral equation known as \emph{Gel'fand-Levitan-Marchenko equation}.}. In the same paper they also show how the eigenvalues of the Schr\"odinger operator are first integrals (that is, conserved quantities) of the KdV equation, thus proving the existence of infinite conservation laws hidden under the nonlinear equation. Finally they explicitly presented multi-soliton solutions, showing that their interactions amount to a series of independent pair scatterings: exactly the same behaviour displayed by the magnons in the Heisenberg XXX chain.

This surprising connection between the KdV equation and the Schr\"odinger operator was readily reformulated in a more general setting by Peter David Lax \cite{Lax_68}. He considered a self-adjoint differential operator $L(t)$, depending on a parameter $t$ through a function $u(t,x)$, which acts on the space of $L^2(\mathbbm R)$ square integrable functions on the real line and its eigenvalue equation

\begin{equation*}
	L(t)\psi(x,t;\lambda) = \lambda \psi(x,t;\lambda) \; .
\end{equation*}
If the spectrum $\{\lambda\}$ does not depend on time then it can be shown that a unitary matrix $U(t)$ exists such that

\begin{equation*}
	L(t)U(t) = U(t)L(0) \; , \quad U^\ast(t)U(t) = U(t)U^\ast(t) = \mathbbm I \; ,
\end{equation*}
that is, $L(t)$ and $L(0)$ are unitarily equivalent. Differentiating this relation with respect to $t$ and multiplying on the right by $U^\ast(t)$, he obtained

\begin{equation*}
	\frac{d}{dt}L(t) = B(t)L(t) - L(t)B(t) = [B(t),L(t)] \; , \quad B(t) \doteq \frac{d U(t)}{d t} U^\ast(t) \; ,
\end{equation*}
which is known as the \emph{Lax equation}, while $L$ and $B$ are called a \emph{Lax pair}. The operator $B(t)$ is skew-adjoint $B^\ast(t)=-B(t)$ (as follows directly from its definition) and controls the time evolution of the system:

\begin{equation*}
	B(t)\psi(x,t;\lambda) = \partial_t\psi(x,t;\lambda) \; .
\end{equation*}
The Lax equation describes the infinitesimal \emph{isospectral flow} for the operator $L$ which corresponds to an evolution equation for the field $u(t,x)$. Asking that this relation reproduces a particular (non-)linear partial differential equation for $u(t,x)$ fixes the form of $B(t)$. For example, it is a matter of simple algebraic calculation to check that the Lax equation for the following linear problem

\begin{equation*}
	\left\{\begin{array}{l}
		L(t)\psi(x,t;\lambda) = \left[-\frac{d^2}{dx^2} - u(t,x)\right]\psi(x,t;\lambda) = \lambda\psi(x,t;\lambda)\\
		\\
		B(t)\psi(x,t;\lambda) = \left[-4\frac{d^3}{dx} -3\left(u\frac{d}{dx}+\frac{d}{dx}u\right)\right]\psi(x,t;\lambda) = \partial_t\psi(x,t;\lambda)
	\end{array}\right. \; ,
\end{equation*}
reproduces the Korteweg-de Vries equation. More generally one might chose $B(t) = \sum_{i=1}^N a_{2i-1}(x,t) \frac{d^{2i-1}}{dx^{2i-1}}$ and obtain a whole series of partial differential equations corresponding to different isospectral flows of the operator $L$.

Inspired by these discoveries, in 1971, Ludvig Dmitrievich Faddeev and Vladimir Evgen'evich Zakharov officially reintroduced the concept of integrability \cite{Fadd_Zakh_71}, showing how the KdV equation may be viewed as an infinite-dimensional integrable Hamiltonian system, where the spectral and scattering data play the r\^ole of action-angle variables, the Inverse Scattering Transform being the (inverse of the) action-angle canonical transformation and the infinite set of conserved quantities corresponding to an infinite set of Poisson commuting Hamiltonians. Just one year later, Aleksei Shabat and Zakharov \cite{Shab_Zakh_72} successfully extended the Lax formalism to the case of $L$ and $B$ being matrix differential operators and managed to solve the nonlinear Schr\"odinger equation, originally introduced in \cite{Chia_Garm_Town_64}. Subsequently Mark Ablowitz, David Kaup, Alan Newell and Harvey Segur applied the method of Shabat and Zakharov to the \emph{sine-Gordon equation} \cite{Ablo_Kaup_Newe_Segu_73} and then structured it into the so-called \emph{AKNS formalism}, an extremely general method which allows to obtain, given any suitable linear matricial eigenvalue problem, a non-linear partial differential equation. A further generalisation of the Lax formalism to systems with two spatial dimensions, as the KP equation introduced by Boris Borisovich Kadomtsev and Vladimir Iosifovich Petviashvili in 1970 \cite{Kado_Petv_70}, was performed again by Shabat and Zakharov in 1974 \cite{Shab_Zakh_74}. For future development, the most important achievement of these two works was the substitution of the Lax scheme with the more general \emph{Zero Curvature Condition}: the nonlinear equation is now obtained from the commutativity condition of a pair of covariant derivatives

\begin{equation*}
	\left\{\begin{array}{l}
		\nabla_x \doteq \partial_x + U(x,t;\lambda)\\ \\
		\nabla_t \doteq \partial_t + V(x,t;\lambda)
	\end{array}\right. \; , \qquad [\nabla_x,\nabla_t]=0 \; \Leftrightarrow \ \textrm{Nonlinear equation} \; ,
\end{equation*}
where $U$ and $V$ are matricial functions of the spectral parameter $\lambda$ and depend on the coordinates through the field $u(x,t)$ of the nonlinear equation. The r\^oles of the spectral problem and of the evolution equation of Lax approach are now played by a system of two auxiliary linear equations:

\begin{equation*}
	\nabla_x \Psi(x,t;\lambda) = 0 \; , \quad \nabla_t \Psi(x,t;\lambda) = 0 \; ,
\end{equation*}
where the spectral parameter can enter in a nonlinear fashion. The scattering data of the problem is then obtained from the large $L$ limit of the \emph{holonomy} $T$, also known as \emph{monodromy matrix} $M$:

\begin{equation*}
	T(L;\lambda) \doteq\  \stackrel{\leftarrow}{\exp}\left\{\int_{-L}^L U(x,t;\lambda) dx\right\} \; ; \qquad M(\lambda) \doteq \lim_{L\rightarrow\infty} T(L;\lambda) \; ,
\end{equation*}
where $\stackrel{\leftarrow}{\exp}$ stands for the path ordered exponential. These articles were to be followed in 1979 by an important work \cite{Shab_Zakh_79} where the authors devised a procedure, known as \emph{Zakharov-Shabat construction}, allowing to build consistent Lax pairs (or better the covariant derivatives) giving rise to integrable systems.

While these new methods were being developed and studied, Rodney Baxter was working on the extension of the results obtained by Lieb for the 6-vertex model, starting from the ideas of Hendrik Anthony Kramers, Gregory Hugh Wannier \cite{Kram_Wann_41} and Lars Onsager \cite{Onsa_44} on the concept of \emph{Transfer Matrix}. In the remarkable papers \cite{Baxt_72_1,Baxt_72_2} he managed not only to obtain the partition function of the 8-vertex model, but, while doing so, he also introduced the \emph{Yang-Baxter equation}\footnote{This equation bears the names of both Baxter and Chen-Ning Yang who independently discovered it \cite{Yang_68} some years before.} and the related concepts of \emph{$R$ matrix} and \emph{$Q$ operators}; what's more, he discovered an unexpected link between classical two dimensional lattice models and one dimensional quantum spin chains. He showed how the transfer matrix for the 8-vertex model and the Hamiltonian of the quantum XYZ model, that is the completely anisotropic Heisenberg model, are connected by a simple formula; this allowed him to use the results for the former to derive the ground-state energy of the latter. In the subsequent papers \cite{Baxt_73_1,Baxt_73_2,Baxt_73_3}, by means of a highly non-trivial generalisation of the Bethe Ansatz, he completed the work, obtaining the eigenvectors and eigenvalues of the XYZ model Hamiltonian and, consequently, of the 8-vertex model transfer matrix, thus solving them completely. The outstanding results of Baxter and the powerful method he devised were then collected by Ludvig Faddeev, Evgeny Sklyanin and Leon Takhtajan and nicely incorporated, together with the algebraic interpretation of the inverse scattering provided by Mark Adler \cite{Adle_79} and Bertram Kostant \cite{Kost_79}, into a very elegant and general framework: the \emph{Quantum Inverse Scattering}, often also called \emph{Algebraic Bethe Ansatz} \cite{Fadd_Skly_Takh_79}.

During the seventies, another different approach to the integrable systems was proposed: the \emph{Thermodynamic Bethe Ansatz}. This method was introduced by Roger Dashen, Shang-Keng Ma and Herbert Bernstein \cite{Dash_Keng_Bern_69} and Cheng-Ning Yang and Cheng-Ping Yang \cite{Yang_Yang_69} and rests on the concept of Bethe Ansatz and on the idea of factorisable $S$-matrix. By then it was clear that the possibility of expressing multi-particle (or multi-soliton, multi-magnon etc...) interactions in terms of pairwise scatterings was a common feature of all the integrable systems; in other words the $n$-bodies $S$-matrix of an integrable system, which contains the information on the interactions, can be always factorised in a product of $2$-bodies $S$-matrices. Using this fact and the relation between Quantum Field Theories and Statistical Models, the above mentioned authors devised a method to obtain the thermodynamics quantities of a model in terms of the corresponding Quantum Field Theory's $S$-matrix. The concept of factorisable interaction bore also a renewed interest in the concept of \emph{bootstrap}, introduced some years before, which proposed the possibility of building the $S$-matrix starting from basic physical assumptions on its structure and on the spectrum of the theory; although it was quickly abandoned in the general Quantum Field Theory, due to lack of sufficient imposable constraints to obtain a consistent theory, the factorisability of the interaction typical of integrable models granted the chance of arriving at the $S$-matrix starting uniquely from its spectrum (and imposing some unitarity and analyticity constraint). With this method Alexander and Alexei Zamolodchikov derived the $S$-matrices of the quantum sine-Gordon \cite{AZam_77} and of the nonlinear $O(n)\ \sigma$-models \cite{AZam_AlZa_78}; in the following years many authors used this method to derive the $S$ matrices of various models.

These two lines of research are at the basis of the modern approach to the Quantum Integrable Systems along with a theory which saw the light in the early eighties thanks to Alexander A. Belavin, Alexander M. Polyakov and Alexander B. Zamolodchikov: the \emph{Conformal Field Theory}. Although the relevance of the conformal symmetry for string theory was pointed out by Polyakov \cite{Poly_81}, the fuel for this discovery came, again, from the study of statistical models, in particular of the so-called \emph{critical phenomena}. As early as 1970 \cite{Poly_70}, Polyakov had shown how the correlation functions of a statistical model at a critical point are invariant under conformal transformations which, for a generic dimension $d>2$, form a finite group of dimension $\frac{1}{2}(d+1)(d+2)$; as a consequence, the conformal invariance can say relatively little about the model, just slightly more than what rotational and translational invariance alone can. However, when considering the $d=2$ case, an infinity of transformations appear: these correspond to all the possible \emph{analytic mappings} of the complex plane; the algebra related to this infinite group is called \emph{Virasoro algebra}. Although not all of these transformations are globally well-defined, they are nonetheless locally conformal and it is perfectly natural to assume that a local field theory would be sensitive to local symmetries. Starting from this, Belavin, Polyakov and Zamolodchikov, in their fundamental 1984 work \cite{Bela_Poly_AZam_84}, combined the representation theory of the Virasoro algebra, developed shortly before by Viktor G. Kac \cite{Kac_79}, Boris L. Feigin and Dimitry B. Fuchs \cite{Feig_Fuch_83}, with the idea of the existence of a local operator algebra, showing how to build completely solvable conformal field theories: the \emph{minimal models}. Following this initial step, an intense activity at the boundary between mathematical physics and statistical mechanics followed and gave a physical meaning to the minimal models, identifying them with various two dimensional statistical systems at their critical point. The construction of Belavin, Polyakov and Zamolodchikov relies heavily on the assumption that the product of local quantum operators (such as fields) can always be expressed as a linear combination of local operators and that this expression satisfies the associativity principle, in other words that the local operators form an \emph{associative algebra}: this is an expression of the bootstrap approach hinted at above. As we remarked then, a successful application of the bootstrap program is hopeless unless the model possesses enough symmetries or, which is the same, the number of local fields is finite, which is precisely the case for the minimal conformal field theories.

The studies on two dimensional conformal field theory grew very rapidly both in number and in variety of approaches and goals, which ranged from applications to string theory, to incorporation of additional structures (as fields with higher-spin or fractional statistics, Lie algebra symmetries, Superalgebras etc...), to analysis of perturbations of the conformal models. This last direction is of particular importance as, in 1987, Alexander B. Zamolodchikov published a work \cite{AZam_87} in which he manages to build explicitly the integrals of motion of some particular perturbations of minimal models arguing that these might be integrable, a fact that he proved two years later in \cite{AZam_89}. This important work unveiled a contact point between the approaches to quantum integrability named above; on the one hand it allowed the application of factorised $S$-matrix and Thermodynamic Bethe Ansatz methods in the context of perturbed Conformal Field Theories, a route followed by many authors, among whom we recall Alexei B. Zamolodchikov \cite{AlZa_90}, Timothy R. Klassen and Ezer Melzer \cite{Klas_Melz_90}, Philippe Christe and Marcio J. Martins \cite{Chri_Mart_90} and Giuseppe Mussardo \cite{Chri_Muss_90}. On the other hand the research on Algebraic Bethe Ansatz had led towards the discovery of deformations of Lie algebras, first observed by Peter P. Kulish and Yu N. Reshetikhin \cite{Kuli_Resh_83}, which were then formalised and given the name of \emph{Quantum Groups} by Vladimir G. Drinfeld \cite{Drin_85} and Michio Jimbo \cite{Jimb_86}; the connection between these mathematical structures and the integrable perturbations of Conformal Field Theory was unveiled shortly after, thanks to the works of Reshetikhin and Fedor A. Smirnov \cite{Resh_Smir_90}, Denis Bernard and Andr\'e LeClair \cite{Bern_LeCl_91} and many others. However, the real contact point between Algebraic Bethe Ansatz and Conformal Field Theories was discovered some years later when Vladimir B. Bazhanov, Sergei L. Lukyanov and Alexander B. Zamolodchikov in a remarkable series of papers \cite{Bazh_Luky_AZam_96,Bazh_Luky_AZam_97,Bazh_Luky_AZam_99} presented the construction of the Conformal Field Theory analogues of Baxter $T$ and $Q$ operators and of the Yang-Baxter equation, effectively implementing the Quantum Inverse Scattering Method for the Conformal Field Theory; they also showed how the $T$ operator's eigenvalues satisfy a set of functional equations equivalent to the Thermodynamic Bethe Ansatz, which allows the determination of the spectrum of the theory.

And we've finally arrived at the end of the nineties of the last century, a decade marked by the ground breaking discovery of Juan Maldacena \cite{Mald_98}: the \emph{AdS/CFT correspondence}. This new, exciting finding ``opened the floodgates" (to borrow the words of Polyakov) and stimulated an impressive amount of work in the last fifteen years, especially in the field of integrability, due to the AdS part of the correspondence dealing with a string theory which, in most of its formulations, is an integrable model. Thus the range of possible applications of the study of integrable models widened considerably and, correspondingly, the methods of analysis evolved rapidly and multiplied, and today constitute one of the most active areas of mathematical physics.\\

This historical sketch has no pretension of being exhaustive or comprehensive; many fundamental discoveries and results were not addressed and many important figures were not introduced, one reason being the limited amount of space-time at disposal for this task. There is also another reason, more subtle and important: I wished to present the historical development of integrability in the most linear and clear way possible. This representation is intrinsically artificial as the evolution of ideas in science and, more generally, in human knowledge is highly non-linear and non-local: it is better represented as an evolving network where each node, representing an idea, a concept, a discovery, is potentially linked to all the other ones. This fact is more evident when looking at ancient times, when human thought was not really compartmentalised, but it is more than true in our days too: I think the reader is familiar with those ``breakthrough" ideas that suddenly put in contact previously disjointed areas of human thought which then begin to talk and exchange ideas, often becoming more than the mere sum of the two. The attempt to reduce this complex structure, retaining its completeness, to a series of nested currents which flow linearly - safe from the occasional merging, crossing and divergence - is hopeless. However it is possible to choose a scale suitable to the description of a particular subject and then to trace some directions, some larger flows inside this network which in first approximation give an idea of how said subject has evolved in time. Isolating these currents and disentangling them from the whole comport the risk, inherent to the subjective point of view of the writer, of cutting away important branches.\\ My goal was extracting a limpid historical portrait of the integrability, my wish is having been successful in this without being too much of a clumsy gardener.

\mynewchapter{0}{Acknowledgements}

First of all I wish to thank my supervisors: Roberto Tateo and Fedor Smirnov. It is almost to be taken for granted that a student acknowledge its supervisors for their support and their help in directing its scientific growth, however the gratitude I want to express is not reducible to a formulaic set phrase. Both Roberto and Fedor were able to convey the flow of my work in fruitful directions with intelligence and sensibility without putting dams on its way but letting it move according to my personal affinities, teaching me how to spot obstacles and how to avoid them; they recognised my qualities and my weak spots, helping me to hone the former and strengthen the latter; they always had respect for my way of working, giving me the time I needed to learn and understand new things and, most importantly, their care was not directed solely to the output of my work, but mainly to me as a physicist \emph{in fieri}.\\
I'm grateful to Patrick Dorey for his participation to the project I worked on (and am still working on!) with Roberto and for the help during this last year: this thesis wouldn't have been finished without him.\\
I also wish to thank Clare Dunning for her precious suggestions.

The research of these years was founded by the \emph{Universit\`a degli Studi di Torino}, by the \emph{Universit\`a Italo-Francese} grant ``Vinci" and by the People Programme (Marie Curie Actions) of the European Union's Seventh Framework Programme FP7/2007-2013/ under REA Grant Agreement No 317089 (GATIS).

Then I want to express my gratitude to all the people that have orbited and/or are orbiting around me; Ana ``the boss" Serra, Falcions, Vittino, Vadacchi, Palba, Cristiano, Marghe, The Scott and the dead PhD who shared with me the open-space office a.k.a. ``The Green Mile", succeeding in the deed of making it a really nice place to be; Emanuele, Sasha and Tresa with whom I endured the arrows of the evil parisians and spent some very nice and not so very parisian evenings; Istvan and Rouven, office-mates, burger-mates and beer-mates: Durham is indeed a fun place with you guys around! Thanks to to ``Infinito Love" for the time he spent with me.\\
I surely forgot to mention someone, anyway the probability that they will actually read these lines is infinitesimal, so, please, forgive me.

Afgano Rosso, I'm grateful of having met you: amongst all these countless people, you've got eyes with a view.

Finally a huge thank you goes to my extended family: Marta (almost a sister, almost a companion, still the woman of my life), Carol (kindred souls never lose themselves), Martina (my other half, my mirror), Stefania and Felice without whom, really, all this won't exist.\\

Thank you!

\vspace{9cm}
\begin{flushright}
Toda joia, Toda beleza.
\end{flushright}

\mainmatter
\pagestyle{myheadings}
\part{The ODE/IM Correspondence in Toda Field Theories}
\mychapter{0}{Introduction}
\pagestyle{headings}
In a broad sense, the ODE/IM correspondence can be described as a recently found 
link between 2D \emph{Integrable Models} and the theory of \emph{Ordinary 
Differential Equations}. More precisely this link is founded on the formal 
equivalence of some functional relations appearing, on one hand, in the study of 
spectral characteristics for ODE (and, more generally, linear problems) and, on 
the other hand, in the analysis of the integrals of motion spectra for the 
Integrable Models\footnote{The acronym IM can be, in fact, understood as 
standing for both \emph{Integrable Models} and \emph{Integrals of Motion}. Some 
authors, as Lukyanov \cite{Luky_13}, prefer this last interpretation since still 
no indication for an extension of the correspondence beyond the relation hinted at
here has been found.}.

The first instance of this correspondence has been worked out in 1999 by P. Dorey 
and R. Tateo \cite{Dore_Tate_99}; they observed how the functional relations 
(Y-systems, Q-T relations) emerging from the Thermodynamic Bethe Ansatz (TBA for 
short) analysis of certain 2D perturbed CFT minimal models\footnote{These are 
perturbations of the CFT of $\mathbbm Z_h$ parafermions by the thermal operator 
of conformal dimension $\Delta = \overline{\Delta} = 2/(h+2)$, which result in an 
integrable massive quantum field theory associated with the $\mathfrak a_{h-1}$ Lie 
algebra \cite{Kobe_Swie_79}.} coincide with the exact quantisation condition for 1D 
anharmonic oscillator (a result due to A. Voros in 
\cite{Voro_92,Voro_94,Voro_98}). Shortly after, this observation was proved and 
generalised by V. Bazhanov, S. Lukyanov and A. Zamolodchikov in 
\cite{Bazh_Luky_AZam_01}. Since these initial results, the ODE/IM correspondence 
has been used in various branches of physics, such as condensed matter 
\cite{Grit_Altm_Deml_Polk_06}, PT-symmetric quantum mechanics 
\cite{Dore_Dunn_Tate_01}, boundary CFT \cite{Luky_Vitc_AZam_04} and non-compact 
sigma models \cite{Tesc_07}.

The question whether this correspondence extends to massive integrable models has been lingering for more than ten years, until S. Lukyanov and A. Zamolodchikov, building on the results of D. Gaiotto, G. Moore and A. Neitzke \cite{Gaio_Moor_Neit_09,Gaio_Moor_Neit_10} on superconformal field theories, managed to establish, in 
the work \cite{Luky_AZam_10}, an ODE/IM correspondence for the quantum sine- and sinh-Gordon model. In virtue of their simplicity, these two models have always been a first step toward the understanding and establishment of methods and techniques of investigation in 2D integrable QFT. In fact the results of S. Lukyanov and A. Zamolodchikov have been recently extended to the Tziz\'eica-Bullough-Dodd \cite{Dore_Fald_Negr_Tate_12} model and constitute the primary background for the analysis carried on in the first part of this thesis, whose aim is to establish an ODE/IM correspondence for the entire family of affine Toda QFT \cite{Negr_Tate_14}.

In this part we intend to present the fundamental aspect of the ODE/IM 
correspondence and work out the particular case of Toda QFT. In order to do so, 
we will first give, in chapter \ref{chap:ODE/IM}, an elementary introduction to the ODE/IM in its simplest incarnation, while in chapter \ref{chap:Toda} we will introduce the Toda 
Field Theories (ToFT) and their more interesting aspects. Chapter \ref{chap:repthe} is devoted to an analysis of the relation between the representation theory for Lie algebras and the so-called $\psi$-system, a set of relations which is of fundamental importance for the correspondence. Finally, in chapter 
\ref{chap:ODE/IM_Toda}, we will explicitly present the construction of the Bethe Ansatz Equations from the linear problem associated to the ToFT based on the Lie algebras $\mathfrak a_r$ and $\mathfrak d_r$ and will specialise the framework to the simple cases $\mathfrak a_3\cong\mathfrak d_3$ and $\mathfrak d_4$.

\chapter[The ODE/IM Correspondence]{The ODE/IM Correspondence: an Introduction}
\label{chap:ODE/IM}
\renewcommand{\chaptermark}[1]{ \markboth{#1}{} }
\renewcommand{\sectionmark}[1]{ \markright{#1}{} }
\markboth{Chapter 1 - The ODE/IM Correspondence: an Introduction}{}

The aim of this chapter is to introduce the main ideas behind the ODE/IM 
correspondence in a simple and self-contained fashion; in order to do so we will 
focus on the simplest example, namely the connection between certain 
second-order ordinary differential equations and integrable models associated 
with the Lie algebra $\mathfrak{su}(2)\equiv\mathfrak a_1$.

We begin by introducing, in section \ref{sec:int_mod}, the \emph{six-vertex} 
model and its quantum counterpart, the $XXZ$ model whose continuum limit result 
in CFT associated with Lie algebra of the $\mathfrak a$ series. We will define the \emph{fundamental integrability objects}, that is the $T$- and $Q$-functions, and display the functional relations they satisfy: the $T$-$Q$ relation, the fusion hierarchy and the quantum wronskian relation. These last two sets of functional equations are known in the literature with the name, respectively, of $T$- and $Q$-systems \cite{Kuni_Naka_Suzu_11}. From the fusion hierarchy we will then derive the $Y$-system and "resum" it, in the case in which the hierarchy truncates, to a non-linear integral equation, known as Thermodynamic Bethe Ansatz equation \cite{AlZa_90}. Then, in section \ref{sec:ODE}, we will present the other side of the correspondence, that is the Ordinary Differential Equations, their spectral characteristics and the functional relation between them. We will first show how, defining the eigenvalue problem for an ODE on the complex plane, the WKB solutions, in the vicinity of irregular singularities, will display the so-called \emph{Stokes phenomenon}. After explaining this property of solutions to ODEs, we will exploit it in order to obtain a set of functional relations for the \emph{spectral determinants}, that is functions encoding the eigenvalues of specific boundary problems. These relations, which gives the exact spectrum of an eigenvalue problem associated to the starting ODE \cite{Voro_94,Voro_98}, coincide exactly with the $T$-$Q$ relations obtained in the section \ref{sec:int_mod}, thus letting us interpret the eigenvalues as Bethe roots for a certain integrable model. Finally we will completely glue together the two sides of the correspondence together in section \ref{sec:corresp}.

This introduction follow very closely the review \cite{Dore_Dunn_Tate_07}; the figures used in this chapter are also taken from the said review.

\resection{Integrable models}
\label{sec:int_mod}

\subsection{The six-vertex model}
\label{subsec:six-vert}

\begin{figure}[t]
\centering
\includegraphics[scale=1]{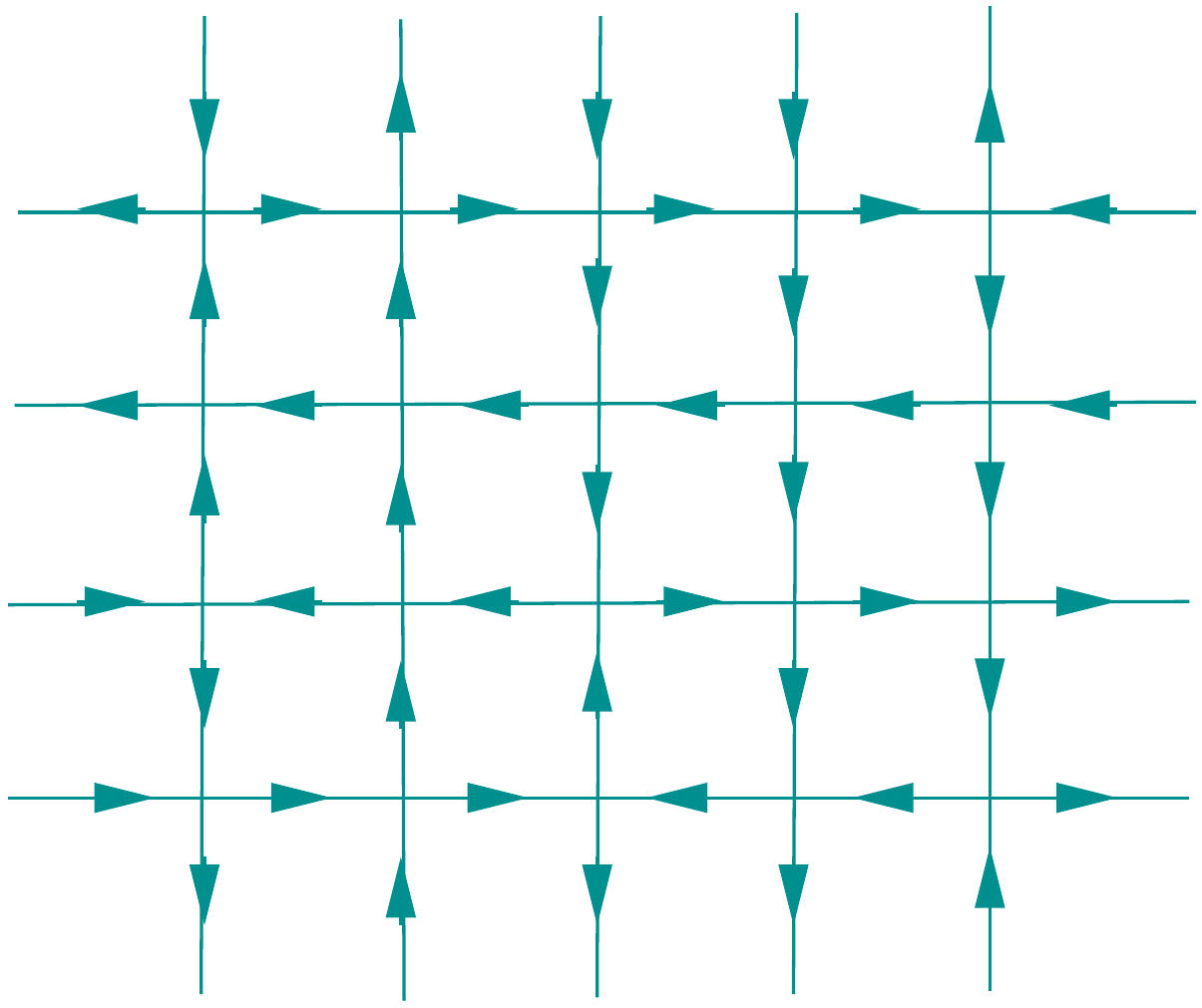}
\caption{A possible configutation of the spins for the six-vertex model.}
\label{pic:sixvert_conf}
\end{figure}

Let us consider an $N \times N'$ bidimensional square lattice with $N/2$ and 
$N'/2$ even\footnote{This is assumed in order to avoid some signs; ultimately 
the thermodynamic limit $N,N' \rightarrow \infty$ will be taken. While sometimes 
the parity of $N$ and $N'$ can be relevant, this is not our case.} and periodic 
boundary conditions imposed. On each link of the lattice we place a spin 
$\frac{1}{2}$ variable, whose values we represent as ``directions" on the link. 
In Figure \ref{pic:sixvert_conf} a possible configuration of the model is depicted.

We agree that arrows pointing up or right stand for spin valued $+\frac{1}{2}$ 
and left or down for $-\frac{1}{2}$. We allow only configurations preserving the 
``flux of arrows" through each node of the lattice, meaning that, around each 
vertex, the four spins can assume six different configurations. We assign to 
each of these a local Boltzmann weight, asking for an overall $\mathbbm Z_2$ 
symmetry (that is, the weights are invariant under a simultaneous reversal of all 
the spins). we are left with three parameters, which reduce to two, since the 
overall normalisation factors out from all physical quantities:

\begin{align}
	& W \left[
		\begin{array}{l l l}
			 & \uparrow & 
			\\
			\rightarrow & & \rightarrow
			\\
			 & \uparrow & 
		\end{array}\right] =
	W \left[
		\begin{array}{l l l}
			 & \downarrow & 
			\\
			\leftarrow & & \leftarrow
			\\
			 & \downarrow & 
		\end{array}\right] = a(\nu,\eta) = \sin (\eta + \mathbbm i\nu) \; ;
	\\
	& W \left[
		\begin{array}{l l l}
			 & \downarrow & 
			\\
			\rightarrow & & \rightarrow
			\\
			 & \downarrow & 
		\end{array}\right] =
	W \left[
		\begin{array}{l l l}
			 & \uparrow & 
			\\
			\leftarrow & & \leftarrow
			\\
			 & \uparrow & 
		\end{array}\right] = b(\nu,\eta) = \sin(\eta - \mathbbm i\nu) \; ;
	\\
	& W \left[
		\begin{array}{l l l}
			 & \uparrow & 
			\\
			\rightarrow & & \leftarrow
			\\
			 & \downarrow & 
		\end{array}\right] =
	W \left[
		\begin{array}{l l l}
			 & \downarrow & 
			\\
			\leftarrow & & \rightarrow
			\\
			 & \uparrow & 
		\end{array}\right] = c(\eta,\nu) = \sin(2 \eta) \; ;
\end{align}
The parameters $\nu$ and $\eta$ are usually called the \emph{spectral parameter} 
and the \emph{anisotropy}, respectively. In what follows we will consider $\eta$ 
as a fixed parameter, dropping its explicit dependence in the formulae.

The partition function, encoding the properties of the system, is defined in the 
usual way as the weighted sum of all the configurations:

\eq
	Z = \sum_{\lbrace\sigma\rbrace} \prod_{\langle i, j \rangle} W\left[
		\begin{array}{l l l}
			 & \cdot & 
			\\
			\cdot & & \cdot
			\\
			 & \cdot & 
		\end{array}\right] = e^{-\frac{F}{T}}
\en
where $T$ is the temperature and $F$ is the Helmholtz free energy.

A neat way to compute the partition function is to define the \emph{transfer 
matrix} $\mathbbm T$, which basically is the sum of configurations along a single 
line of the lattice. Introducing the multi-indices notation $\boldsymbol{\alpha} 
\doteq (\alpha_1,\alpha_2,\ldots,\alpha_N)$, we can write

\eq
	\mathbbm T_{\boldsymbol{\alpha}}^{\boldsymbol{\alpha'}}(\nu) = 
\sum_{\lbrace\beta_i\rbrace} W \left[
		\begin{array}{l l l}
			\! \! \! \! & \! \! \!\alpha'_1 & \! \! \! \! \! \! \!
			\\
			\! \! \beta_1 \! \!&\! \! \! &\! \! \! \! \beta_2 \! \! 
\!
			\\
			\! \! \! \! &\! \! \! \alpha_1 & \! \! \! \! \! \! \!
		\end{array}\right]\!\!(\nu)\:W \left[
		\begin{array}{l l l}
			\! \! \! \! & \! \! \!\alpha'_2 & \! \! \! \! \! \! \!
			\\
			\! \! \beta_2 \! \!&\! \! \! &\! \! \! \! \beta_3 \! \! 
\!
			\\
			\! \! \! \! &\! \! \! \alpha_2 & \! \! \! \! \! \! \!
		\end{array}\right]\!\!(\nu)\: \cdots \:W \left[
		\begin{array}{l l l}
			\! \! \! \! & \! \! \!\alpha'_N & \! \! \! \! \! \! \!
			\\
			\! \! \beta_N \! \!&\! \! \! &\! \! \! \! \beta_1 \! \! 
\!
			\\
			\! \! \! \! &\! \! \! \alpha_N & \! \! \! \! \! \! \!
		\end{array}\right]\!\!(\nu)\:
\en

It is clear that the partition function is obtained simply by tracing over the 
$N'$-fold product of the transfer matrix:

\eq
	Z = \textrm{Tr}\left[\mathbbm T^{N'}\right]
\en
and, supposing we know how to diagonalise $\mathbbm T$

\eq
	\mathbbm 
T_{\boldsymbol{\alpha}}^{\boldsymbol{\alpha'}}\Psi_{\boldsymbol{\alpha'}}^{(j)} 
= t_j \Psi_{\boldsymbol{\alpha}}^{(j)},
\en
the thermodynamic quantities can be expressed entirely in terms of the 
eigenvalues $\left\lbrace t_j \right\rbrace$. For example the free energy 
density in the limit $N'\rightarrow \infty$ can be expressed as

\eq
	f \doteq \frac{F}{N N'} = -\frac{T}{N N'}\log Z = -\frac{T}{N 
N'}\log\textrm{Tr}\left[\mathbbm T^{N'}\right] \sim -\frac{T}{N}\log t_0
\label{eq:freeen}
\en
where we made the hypothesis $t_0>t_1>\ldots$.

This is indeed a nice way to reformulate the problem, however, $\mathbbm T$ is a 
$2^N\times2^N$ matrix and its diagonalisation is impossible to tackle 
head-front; we need to exploit some properties in order to simplify the task. 
The standard technique to carry on this job is known as \emph{Bethe ansatz} 
which, broadly, amounts to two steps:

\begin{itemize}
	\item make an educated guess for the generic form of the eigenvectors of 
$\mathbbm T$, depending on a certain number $n$ of supplementary parameters 
$\left\lbrace \nu_i \right\rbrace_{i=1}^n$, called \emph{roots};
	\item impose the physical constraints (such as the boundary conditions) 
to obtain a certain set of equations for the roots, called the \emph{Bethe 
ansatz equations} (BAEs).
\end{itemize}

To each set of roots solving the BAEs corresponds an eigenvector of $\mathbbm T$ 
and letting also $n$ vary, we obtain the totality of them (at least in the 
$N\rightarrow \infty$ limit). The justification to this method basically sits on 
the property of integrable model's $S$ matrices of being factorisable in terms 
of two-body $S$ matrices only; this, as a consequence, means that the 
wave-functions are superpositions of two-particle wave-functions, which are easy 
to write down. The interested reader may find more details in \cite{Kore_Bogo_Izer_93}.

Working out the two points above for the six-vertex model, we obtain the 
following BAEs:

\eq
	\prod_{j=1}^n \frac{\sinh(\nu_j-\nu_k +2 \mathbbm i\eta)}{\sinh(\nu_j-\nu_k -2 \mathbbm i
\eta)} = - \left[ \frac{a(\nu_k,\eta)}{b(\nu_k,\eta)} \right]^N \quad ; \qquad k 
= 1,\ldots,n
\en
which is a set of $n$ equations in $n$ unknowns. There exist a discrete set of 
solutions for each fixed $n$ and to each of these correspond an eigenvector 
$\vert \Psi (\nu\vert\lbrace \nu_j \rbrace)\rangle$ of $\mathbbm T$ with 
eigenvalue

\eq
	t(\nu\vert\lbrace \nu_j \rbrace) = [a(\nu,\eta)]^N \prod_{j=1}^n 
g(\nu_j-\nu) + [b(\nu,\eta)]^N \prod_{j=1}^n g(\nu-\nu_j)
\en
where $g(\nu)\doteq a(\nu-\mathbbm i\eta,\eta)/b(\nu-\mathbbm i\eta,\eta) = -\sin(2\eta + \mathbbm i
\nu)/\sin(\mathbbm i\nu)$. In order to fix the solutions corresponding to a defined 
eigenvector one has to impose some supplementary conditions; in particular for 
the ground state one asks the roots to be $n=N/2$, distinct, real and 
symmetrically placed about the origin in the interval $[-\eta,\eta]$ (we are 
considering the parameter region $0<2\eta<\pi$).

The periodic boundary conditions were fundamental for the application of this 
method, since they are the constraint which has to be imposed in order to obtain 
the BAEs. Actually one can impose slightly more general conditions without 
spoiling the integrability; let us introduce a twist by modifying the 
local Boltzmann weights on one single column of the lattice, say the 
$N^{\textrm{th}}$:

\eq
	W \left[
		\begin{array}{l l l}
			\! \! \! \! & \! \! \!\alpha'_N & \! \! \! \! \! \! \!
			\\
			\! \! \beta_N \! \!&\! \! \! &\! \! \! \! \beta_1 \! \! 
\!
			\\
			\! \! \! \! &\! \! \! \alpha_N & \! \! \! \! \! \! \!
		\end{array}\right]\!\!(\nu)
	\longrightarrow
	e^{2 \mathbbm i\phi \beta_1} W \left[
		\begin{array}{l l l}
			\! \! \! \! & \! \! \!\alpha'_N & \! \! \! \! \! \! \!
			\\
			\! \! \beta_N \! \!&\! \! \! &\! \! \! \! \beta_1 \! \! 
\!
			\\
			\! \! \! \! &\! \! \! \alpha_N & \! \! \! \! \! \! \!
		\end{array}\right]\!\!(\nu)
\en
with $\beta_1$ taking the values $\pm 1/2$.

Repeating the calculations for the Bethe ansatz one ends with a more general set 
of BAEs:

\eq
	\prod_{j=1}^n\frac{\sinh(\nu_j-\nu_k+2 \mathbbm i\eta)}{\sinh(\nu_j-\nu_k-2 \mathbbm i
\eta)}=-e^{-2 \mathbbm i\phi}\left[\frac{a(\nu_k,\eta)}{b(\nu_k,\eta)} \right]^N
\label{eq:twistBAE}
\en
and a more general transfer matrix $\mathbbm T(\nu,\phi)$ with eigenvalues

\eq
	t(\nu,\phi\vert\lbrace\nu_i\rbrace) = e^{-\mathbbm i\phi}[a(\nu,\eta)]^N 
\prod_{j=1}^n g(\nu_j-\nu) + e^{\mathbbm i\phi}[b(\nu,\eta)]^N \prod_{j=1}^n 
g(\nu-\nu_j)
\label{eq:twistTeigen}
\en

\subsection{The XXZ model}
\label{subsec:XXZ}

As it is known, there exists a connection between classical 
$(D+1)$-dimensional models and quantum $D$-dimensional ones. In our particular 
case, the six-vertex model is related to the one-dimensional XXZ spin chain, 
also known as \emph{Heisenberg magnet}. This is a system, defined on a 
one-dimensional lattice, with spin $\frac{1}{2}$ variables sitting on the 
lattice sites and interacting only with their nearest neighbours. The 
Hamiltonian of said model is

\eq
	H_{\textrm{XXZ}} = -\frac{1}{2} 
\sum_{j=1}^{N}\left(\sigma_j^x\sigma_{j+1}^x + \sigma_j^y\sigma_{j+1}^y - cos 
2\eta\,\sigma_j^z\sigma_{j+1}^z\right)
\label{eq:hamXXZ}
\en
where $\sigma_j^{\alpha}$ represents a Pauli matrix acting non-trivially only on the $j$-th 
site of the lattice:

\eq
	\sigma_j^x = \left(	\begin{array}{cc}
					0 & 1 \\
					1 & 0
				\end{array}\right)_j \; , \quad
	\sigma_j^y = \left(	\begin{array}{cc}
					0 & -\mathbbm i\\
					\mathbbm i& 0
				\end{array}\right)_j \; , \quad
	\sigma_j^z = \left(	\begin{array}{cc}
					1 & 0 \\
					0 & -1
				\end{array}\right)_j \; .
\en

One must obviously clarify the kind of boundary conditions in 
$H_{\textrm{XXZ}}$; in order to reproduce the twist we introduced in the 
six-vertex model, we set

\eq
	\sigma_{N+1}^z = \sigma_1^z \quad , \qquad \sigma_{N+1}^{\pm} = e^{\pm 2 
\mathbbm i\phi}\sigma_1^{\pm}
\en
where we introduced the linear combinations

\eq
	\sigma_j^{\pm} \doteq \sigma_j^x \pm \mathbbm i\sigma_j^y \; .
\en
which are sometimes referred to as \emph{annihilation/creation operators}. The 
reason for this nomenclature reside in the fact that the reference state chosen as starting point for the application of the Bethe Ansatz is the completely aligned state $\vert \uparrow_1, \uparrow_2, \ldots, \uparrow_N\rangle = \bigotimes_{j=1}^N \binom{1}{0}_j$ (or, equivalently, the state $\vert \downarrow_1, \downarrow_2, \ldots, \downarrow_N\rangle = \bigotimes_{j=1}^N \binom{0}{1}_j$); the operators then 
simply act as

\eq
	\sigma_j^+\vert \uparrow_1, \uparrow_2, \ldots, \uparrow_N\rangle = 0 
\qquad \sigma_j^-\vert \uparrow_1, \uparrow_2, \ldots, \uparrow_N\rangle = \vert 
\uparrow_1, \uparrow_2, \ldots, \downarrow_j, \ldots, \uparrow_N\rangle \; ,
\nonumber
\en
while $\sigma_j^z$ act diagonally with eigenvalue $+1$ if the $j$-th spin points 
upward or $-1$ conversely.

From these definitions the relation that runs between the 
XXZ spin chain and the six-vertex model is not clear at all; in fact the connection was first based, 
\cite{Suth_67,Lieb_67_2,Lieb_67_3,Lieb_67_4}, on the identity of the BAEs and the fact that the 
six-vertex model's transfer matrix eigenvalues coincided with those of 
$H_{\textrm{XXZ}}$, previously studied in \cite{Yang_Yang_66_1,Yang_Yang_66_2}. Subsequently 
Baxter \cite{Baxt_72_1,Baxt_72_2} showed that the six-vertex transfer matrix $\mathbbm 
T(\nu)$ and the Hamiltonian (\ref{eq:hamXXZ}) are effectively connected by the 
relation

\eq
	H_{\textrm{XXZ}} = -\frac{1}{2}\cos2\eta \;\mathbbm I -\mathbbm i\left.\sin2\eta 
\frac{d}{d\nu}\log\mathbbm T(\nu)\right\vert_{\nu=-\mathbbm i\eta}
\en
with $\mathbbm I$ being the identity matrix (more precisely the $N$-fold tensor 
product of $2\times2$ identity matrices).

Since the Hamiltonian (\ref{eq:hamXXZ}) commutes, for all values of $\phi$, with 
the total spin operator $S^z = \frac{1}{2}\sum_{i=1}^N \sigma_i^z$, the spectrum 
of the model splits in disjoint sectors, labeled by an integer number $m=0$, 
representing the number of down spins in the states. In the six-vertex model, 
these sectors are mapped into sets of solutions of BAEs of fixed $n=N/2-m$. 
Obviously the ground state $\vert \uparrow_1, \uparrow_2, \ldots, 
\uparrow_N\rangle$ lies in the $m=0$ sector and is thus mapped in the six-vertex 
ground state, the solution of the BAE with $n=N/2$, as expected.

\subsection{The T-Q relation}
\label{subsec:TQrel}

A particularly elegant reformulation of the BAEs exists, elaborated by 
R. J. Baxter, which allows to encode all the elaborated structure of the Bethe 
Ansatz in a simple relation between two functions. As a first step, let us remark 
that the transfer matrices, taken at different values of the spectral parameter 
$\nu$, commute

\eq
	\left[\mathbbm T(\nu) , \mathbbm T(\nu')\right] = 0 \quad , \qquad \forall 
\nu, \nu' \; .
\en
The standard proof of this property can be found, for example, in \cite{Kore_Bogo_Izer_93}. This means that, when diagonalising $\mathbbm T(\nu)$, the 
dependence on the spectral parameter is inherited by the eigenvalues only and 
one can focus on them as functions of $\nu$. Moreover, from the explicit form of 
Boltzmann weights, one can show that these functions are entire and periodic in 
$\nu$ with period $\mathbbm i\pi$.

Now, let us suppose that, for each eigenvalue $t(\nu)$ (we drop the index $j$), 
there exists a second function $q(\nu)$ also entire (and, at least for the 
ground state, $\mathbbm i\pi$-periodic) and satisfying the relation:

\eq
	t(\nu)q(\nu) = e^{-\mathbbm i\phi}\left[a(\nu,\eta)\right]^N q(\nu+2 \mathbbm i\eta) + 
e^{\mathbbm i\phi} \left[b(\nu,\eta)\right]^N q(\nu-2 \mathbbm i\eta)\; .
\label{eq:TQrelation}
\en
In the following we will refer to this as the \emph{T-Q relation}, although this 
name officially belongs to the corresponding relation between the matrices 
$\mathbbm T(\nu)$ and $\mathbbm Q(\nu)$, from which the (\ref{eq:TQrelation}) is 
obtained when projecting on an eigenvector.

It is quite remarkable that such a simple relation can encode the whole Bethe 
Ansatz structure, but, in fact, the equation (\ref{eq:TQrelation}), together 
with the request of entirety of $t(\nu)$ and $q(\nu)$ are equivalent to the 
BAEs\footnote{It is worth remarking that, actually, there's no need to pass 
through the Bethe Ansatz machinery to obtain the eigenvalues of $\mathbbm T(\nu)$ 
(and thus ``solve" the model); nor it is needed to explicitly build the 
eigenvectors. In fact this method can be applied also to models which cannot be 
dealt with through the Bethe Ansatz, such as the eight-vertex model, that Baxter solved 
in \cite{Baxt_72_1,Baxt_72_2}} as we are going to show for the case 
where $q(\nu)$ is $\mathbbm i\pi$-periodic. Suppose that the function $q(\nu)$ has zeroes 
$\lbrace \nu_i\rbrace_{i=1}^n$; given its entirety and the periodicity, we can 
write it as a product over its zeroes, that is, up to an overall normalisation, 
which is irrelevant to our purpose:

\eq
	q(\nu) = \prod_{j=1}^n \sinh(\nu_j - \nu)
\label{eq:qprodrep}
\en
If we now evaluate the T-Q relation at $\nu=\nu_i$ we see that, given the fact 
that $t(\nu)$ is entire and thus non-singular at $\nu_i$, the left-hand side 
vanishes, leaving us with

\eq
	\frac{q(\nu_i-2 \mathbbm i\eta)}{q(\nu_i + 2 \mathbbm i\eta)} = -e^{-2 \mathbbm i
\phi}\left[\frac{a(\nu_i,\eta)}{b(\nu_i,\eta)}\right]^N
\en
which, using the product representation of the function $q(\nu)$, becomes

\eq
	\prod_{j=1}^n\frac{\sinh(\nu_i-\nu_l+2 \mathbbm i\eta)}{\sinh(\nu_i-\nu_l-2 \mathbbm i
\eta)} = -e^{-2 \mathbbm i\phi}\left[\frac{a(\nu_i,\eta)}{b(\nu_i,\eta)}\right]^N \; .
\en
These are precisely the BAEs (\ref{eq:twistBAE}), with the zeroes $\lbrace 
\nu_i\rbrace_{i=1}^n$ playing the r\^ole of roots. In a similar fashion one can 
obtain, from the T-Q relation, the expression (\ref{eq:twistTeigen}) for the 
eigenvalues $t(\nu)$.

\subsection{The quantum Wronskian}
\label{subsec:wronski}

Let us now derive another important functional relation; the starting point is 
the following identity, consequence of the invariance, under simultaneous 
reversal of all the spins, possessed by the Boltzmann weights

\eq
	t_0(\nu,\phi) = t_0(\nu,-\phi) \equiv t_0(\nu,\vert\phi\vert) \; .
\en
From this it follows that, redefining the function $q(\nu)$ as

\eq
	\hat q_0(\nu,\phi) \doteq e^{-\nu \frac{\phi}{2 \eta}} q_0(\nu,\phi) 
\; ,
\en
the T-Q relation can be recast into the following two relations

\begin{align}
	t_0(\nu,\vert\phi\vert) \hat q_0(\nu,\phi) &= 
\left[a(\nu,\eta)\right]^N\hat q_0(\nu+2 \mathbbm i\eta,\phi) + \left[b(\nu,\eta) 
\right]^N\hat q_0(\nu-2 \mathbbm i\eta,\phi) \; ;	\label{eq:tildeq1}
\\	\nonumber
\\
	t_0(\nu,\vert\phi\vert) \hat q_0(\nu,-\phi) &= 
\left[a(\nu,\eta)\right]^N\hat q_0(\nu+2 \mathbbm i\eta,-\phi) + \left[b(\nu,\eta) 
\right]^N\hat q_0(\nu-2 \mathbbm i\eta,-\phi) \; .	\label{eq:tildeq2}
\end{align}
These equations, together with the periodicity of $t_0$ and the quasi-periodicity 
of $\hat q$, implies that $\hat q_0(\nu,\phi)$ and $\hat q_0(\nu,-\phi)$ are two 
independent ``Bloch-wave solutions" of the following functional equation

\eq
	t_0(\nu,\vert\phi\vert)\hat q(\nu) = 	\left[a(\nu,\eta)\right]^N\hat 
q_0(\nu+2 \mathbbm i\eta) + \left[b(\nu,\eta) \right]^N\hat q_0(\nu-2 \mathbbm i\eta) \; ,
\en
which is a finite-difference analogue of a second order differential equation. 
It is thus natural to consider their Wronskian or something equivalent; to this 
end we consider the two equations (\ref{eq:tildeq1}), (\ref{eq:tildeq2}), 
multiplied by $\hat q_0(\nu,-\phi)$ and $\hat q_0(\nu,\phi)$ respectively and 
subtracted, obtaining

\eq
	\left[a(\nu,\eta)\right]^N \Delta(\nu+\mathbbm i\eta) - 
\left[b(\nu,\eta)\right]^N \Delta(\nu-\mathbbm i\eta) = 0 \; ,
\en
where we defined

\eq
	\Delta(\nu) \doteq \hat q_0(\nu+\mathbbm i\eta,-\phi)\hat q_0(\nu-\mathbbm i
\eta,\phi) - \hat q_0(\nu+\mathbbm i\eta,\phi)\hat q_0(\nu-\mathbbm i\eta,-\phi) \; .
\label{eq:deltaeq}
\en
Now, from the same definition of $\hat q_0(\nu,\phi)$ and from the 
representation (\ref{eq:qprodrep}), we see that $\Delta(\nu)$ and, consequently, 
the function $\mathcal W(\nu) \doteq \Delta(\nu)/\sinh^N(\nu)$ are periodic 
with period $2 \pi \mathbbm i$; however, the definitions of $a(\nu,\eta)$ and 
$b(\nu,\eta)$, together with (\ref{eq:deltaeq}) and the fact that $n$ is even, 
tell us that $\mathcal W(\nu)$ is also periodic with period $2 \mathbbm i \eta$. So, for 
$\eta/\pi$ irrational, $\mathcal W(\nu)$ has to be a constant and, by 
continuity, it must be so for all values of $\eta$.

Evaluating $\mathcal W(\nu)$ at $\nu \rightarrow \infty$ we obtain the 
\emph{quantum Wronskian relation}

\begin{align}
	\Delta(\nu) &= e^{-\mathbbm i\phi}q_0(\nu+\mathbbm i \eta,\phi)q_0(\nu-\mathbbm i\eta,-\phi) - 
e^{\mathbbm i\phi}q_0(\nu+\mathbbm i \eta,-\phi)q_0(\nu-\mathbbm i\eta,\phi) =	\nonumber
\\	\label{eq:wronsky}
\\ &= - 2 \mathbbm i \sin\phi \sinh^N(\nu) \; ,	\nonumber
\end{align}
which was first discussed in \cite{Bazh_Luky_AZam_97}\footnote{It has to be 
remarked that, in deriving this relation, we made various assumptions: $\phi\neq 
0$, $N/2$ even and we also derived the functional relation for the ground-state 
eigenvalue only. A more general treatment can be found, for example, in 
\cite{Pron_Stro_99} and \cite{Korf_05_1,Korf_05_2}}.

\subsection{The fusion hierarchy}
\label{subsec:fusion}

Another important functional relation can be derived from (\ref{eq:tildeq1}) and 
(\ref{eq:tildeq2}); if we multiply them, respectively, by $\hat 
q_0(\nu-2\mathbbm i\eta,-\phi)$ and $\hat q_0(\nu-2\mathbbm i\eta,-\phi)$, their difference gives 
us

\begin{align}
	t_0(\nu,\vert\phi\vert) &= -\left[a(\nu,\eta)\right]^N \frac{\hat 
q_0(\nu+2\mathbbm i\eta,-\phi)\hat q_0(\nu-2\mathbbm i\eta,\phi)-\hat q_0(\nu+2\mathbbm i\eta,\phi)\hat 
q_0(\nu-2\mathbbm i\eta,-\phi)}{\Delta(\nu-\mathbbm i\eta)} =	\nonumber
\\
\\	&= \frac{\mathbbm i}{2\sin\phi}\Big(\hat q_0(\nu+2\mathbbm i\eta,-\phi)\hat 
q_0(\nu-2\mathbbm i\eta,\phi)-\hat q_0(\nu+2\mathbbm i\eta,\phi)\hat q_0(\nu-2\mathbbm i\eta,-\phi)\Big)	
\nonumber
\end{align}
which is reminiscent the quantum Wronskian relation; in fact, this identity and 
the (\ref{eq:wronsky}) are two elements of a whole hierarchy we can build. Let us 
define the following column vector

\eq
	\vec q^{\;(k)} \doteq \frac{1}{\sqrt{-2 \mathbbm i \sin\phi}} \Big(e^{-\mathbbm i 
\frac{k}{2}\phi}q_0(\nu-\mathbbm i k\tilde{\eta},\phi),e^{\mathbbm i \frac{k}{2}\phi}q_0(\nu-\mathbbm i 
k\tilde{\eta},-\phi)\Big)^T \; ,
\en
where $\tilde{\eta} = -\eta+\pi/2$, and the determinants

\eq
	\mathcal W[k,k'](\nu) \doteq \det (\vec q^{\;(k)},\vec q^{\;(k')}) 
\; .
\en
Then, if we set

\eq
	t^{(k/2)}(\nu) \doteq \mathcal W[k+1,-k-1](\nu) \quad , \qquad 
\forall k =-1,0,1,\ldots \; ,
\en
we immediately see that

\eq
	t^{(-1/2)}(\nu) = 0 \; ,\qquad  t^{(0)}(\nu) = 
\left[\mathbbm i\cosh(\nu)\right]^N \  , \quad  t^{(1/2)}(\nu) = t_0(\nu) \; .
\en

Now, by exploiting the Pl\"ucker-type relation

\eq
	\det(\vec a_0 , \vec a_1)\det(\vec b_0, \vec b_1) = \det(\vec b_0, \vec 
a_1)\det(\vec a_0, \vec b_1) + \det(\vec b_1,\vec a_0)\det(\vec b_0,\vec a_0)
\en
and the easily demonstrated property

\eq
	\mathcal W[k+a,-k-a](\nu) = \mathcal W[k,-k](\nu-\mathbbm i a \tilde{\eta}) \; ,
\en
we can show that the following sets of bilinear functional relations hold

\begin{align}
	&t^{(m)}(\nu-\mathbbm i\tilde{\eta}) \, t^{(m)}(\nu + \mathbbm i \tilde{\eta}) =	
\nonumber
\\
&=t^{(0)}\big(\nu-\mathbbm i(2m+1)\tilde{\eta}\big) \, t^{(0)}\big(\nu + 
\mathbbm i(2m+1)\tilde{\eta}\big)+t^{(m-1/2)}(\nu) \, t^{(m+1/2)}(\nu) 	\nonumber
\\	\label{eq:fushier}
\\
&t^{(1/2)}(\nu) \, t^{(m)}(\nu - \mathbbm i(2m+1)\tilde{\eta}) = 	\nonumber
\\
&=t^{(0)}(\nu-\mathbbm i\tilde{\eta}) \, t^{(m+1/2)}(\nu - 
2\mathbbm im\tilde{\eta})+t^{(0)}(\nu+\mathbbm i\tilde{\eta}) \, t^{(m-1/2)}\big(\nu - 
\mathbbm i(2m+2)\tilde{\eta}\big) \; .	\nonumber
\end{align}
These are called \emph{fusion hierarchies}, since they can also be obtained from 
a process of ``fusion" of the transfer matrix $\mathbbm T$, without having to 
introduce the auxiliary function $q(\nu)$ (more on this can be found in 
\cite{Bazh_Luky_AZam_96,Kuli_Skly_82,Kiri_Resh_86,Kiri_Resh_87,Klum_Pear_92,
Kuni_Naka_Suzu_94,Kuni_Saka_Suzu_98}).

In general, the fusion hierarchies (\ref{eq:fushier}) are an infinite set of 
functional relations. However, let us take $\eta/\pi$ to be a rational number and 
see what happens. The particular case we wish to mention is $\eta = 
\frac{\pi}{2} \frac{M}{M+1}$ (that is $\tilde{\eta} = \frac{\pi}{2M+2}$) with 
$2M \in \mathbbm Z^+$ and $\phi = \tilde{\eta}= \frac{\pi}{2M+2}$\footnote{This 
particular choice is a matter of convenience: the truncation of the fusion 
hierarchies holds also for generic values of $\phi$, given $\eta/\pi$ is 
rational.}; due to the $\mathbbm i \pi$-periodicity of the function $q(\nu)$ it is easy to 
see that

\eq
	t^{(M+1/2)}(\nu) = 0
\en
and, by comparing the form of $\vec q^{\;(2M+1)}$ and $\vec q^{\;(-2M-1)}$ with 
that of $\vec q^{\;(1)}$ and $\vec q^{\;(-1)}$,

\eq
	t^{(M)}(\nu) = t^{(0)}(\nu) \; .
\en
The infinite hierarchy has been \emph{truncated} to a finite set of functional 
relations, known in the literature as a \emph{T-system}. It follows quite 
immediate from the truncation properties above and the (\ref{eq:fushier}), that 
the following symmetry exists

\eq
	t^{(m)}(\nu) = t^{(M-m)}(\nu) \quad , \qquad m= 0,\frac{1}{2}, \ldots , 
\frac{M}{2} \; .
\en

The phenomenon of truncation is very important, since it grants us a closed set of 
functional relations which, when garnished with suitable analyticity properties, 
can be recast into integral equations, as it is shown in (\ref{subsec:NLIE}) in the simpler case of the continuum limit, allowing the model to be solved.

\subsection{The continuum limit}
\label{subsec:continuum}

The six-vertex model happens to lie, in the whole region $0< \eta < \pi/2$, at 
the phase transition of the eight-vertex model, thus, as we have seen, if we let 
the number of sites of the lattice $N\rightarrow \infty$ while the lattice 
spacing $d\rightarrow 0$ in such a way that the size of the system $L = N d = 
\textrm{const.}$, then the system should be oblivious of short distance features and show 
universal behaviour. In fact the logarithm of the ground-state eigenvalue of the 
transfer matrix behaves as

\eq
	\log t_0(N) \underset{N\rightarrow\infty}{\sim} -\frac{f}{T} N + 
\frac{\pi c_{\textrm{eff}}}{6N} + \ldots  \  .
\en
Here, after the expected term $- f N/T$, with $f$ being the free energy per site 
(\ref{eq:freeen}), we have a term which depends algebraically on the system 
size: a clear consequence of the scaling symmetry which is characteristic of 
second-order phase transitions. Defining the rescaled free energy as

\eq
	F \doteq -\log t_0(L) - \frac{f}{T} L \; ,
\en
we see that all the terms in $\log t_0(L)$ give vanishing contribution for 
$d\rightarrow 0$ aside from the first two, which means

\eq
	F(L) = -\frac{\pi c_{\textrm{eff}}}{6L} \; .
\en
This is the expected behaviour of the free energy of a conformal field theory 
(CFT) living on an infinite cylinder of circumference $L$. The effective central 
charge $c_{\textrm{eff}}$ coincides with the standard Virasoro central charge $c$ 
in the case of unitary theories; the six-vertex model and the $XXZ$ spin chain 
are both described by a CFT with central charge $c^{6V}=1$ and effective central 
charge

\eq
	c_{\textrm{eff}}^{6V} = 1-\frac{6 \phi^2}{\pi(\pi-2\eta)} \leq 1 \; .
\label{eq:6Vceff}
\en

The eigenvector $\Psi^{(0)}$ corresponding to the eigenvalue $t_0$ becomes the 
CFT ground state, the remaining states, often called \emph{excited states}, are 
assigned an energy the same way as done to extract the free energy, that is 
rescaling the logarithm of the corresponding eigenvalue of the transfer matrix. 
For periodic boundary conditions the energies are 
\cite{Alca_Barb_Batc_88,Alca_Barb_Batc_87,Card_86,Woyn_87,Karo_88}

\eq
	F_{\vert\lbrace m_i,m_i'\rbrace;k,k'\rangle} = \xi \left[-\frac{\pi 
c_{\textrm{eff}}}{6L}+\frac{2\pi}{L} \left(x_{k,k'}+\sum_i(m+m')\right)\right] 
\; ,
\label{eq:CFTenergies}
\en
where $k,k' \in \mathbbm Z \;$, $m_i,m_i' \in \mathbbm Z^+ \, , \; \forall i$ and 
$x_{k,k'} = k^2 x + k'^2/4x$ with $x=(\pi-2 \eta)/2 \pi$. The parameter $\xi$ is 
the \emph{velocity of light} and is a model-dependent quantity; in the 
particular case of the six-vertex model is $\xi^{6V}=1$ while for the XXZ chain, 
with the notation used here, is $\xi^{XXZ}= \pi \sin 2\eta/2 \eta$, but it can 
be rescaled to $1$ multiplying $H_{\textrm{XXZ}}$ by an overall factor. The 
states associated with energies (\ref{eq:CFTenergies}) where $m=m'=0$ are the 
primary fields, while states with nonzero values of $m$ and $m'$ are the 
descendants. A similar structure emerges also in the twisted case 
\cite{Alca_Barb_Batc_87}.

The thermodynamic limit can also be applied, with given care, to the T-Q and BAE 
structure introduced above and it happens to simplify their forms. In order to 
show this, first we use the following new variables

\eq
	E_i'= e^{2\nu_i} \quad , \qquad \omega = -e^{-2 \mathbbm i \eta} = e^{2 \mathbbm i 
\tilde{\eta}} \; ,
\en
to rewrite the BAE

\eq
	\prod_{\ell = 1}^n 
\left(\frac{E_{\ell}'-\omega^2E_j'}{E_{\ell}'-\omega^{-2}E_j'}\right) = 
-\omega^{2n-N}e^{-2 \mathbbm i \phi}\left(\frac{1+\omega 
E_j'}{1+\omega^{-1}E_j'}\right)^N \quad , \qquad j=1,\ldots,n \; .
\en
Let us continue, for simplicity, to concentrate on the ground state for which all 
the $\nu_i$ lie on the real axis and $n=N/2$; this means the factor 
$\omega^{2n-N}$ disappears from the BAE and the $E_j' \in \mathbbm R^+ \;, \ 
\forall j$. When we take the $n \rightarrow \infty$ limit, the number of roots 
diverges, while the BAE for the $\nu_i$ lying furthest left and right along the 
real axis somewhat simplify, at least for $\eta>\pi/4$, allowing to retrace a 
scaling behaviour. In fact the left edge root $\nu_{\textrm{min}}$ behaves as 
\cite{Klum_Batc_Pear_91}

\eq
	\nu_{\textrm{min}} \underset{N\rightarrow\infty}{\sim} 
-\frac{2\eta}{\pi}\log N \quad \Rightarrow \quad E_{\textrm{min}}' 
\underset{N\rightarrow\infty}{\sim} E_{\textrm{min}} N^{-\frac{4\eta}{\pi}} \; ,
\en
while the right edge root behaves the same way, given the symmetry

\eq
	q_0(-\nu,\phi)=q_0(\nu,-\phi) \quad \Rightarrow \quad \nu_i(\phi) = 
-\nu_{\frac{N}{2}+1-i}(-\phi)
\en
Thus we can renormalise the BAE by substituting each $E_i'$ with 
$N^{-4\eta/\pi}E_i$ and send $N\rightarrow\infty$ while keeping the $E_i$ 
finite. The result is\footnote{A similar result can be found also for 
$\eta\leq\pi/4$. In that case, however, the product has to be regulated in order to 
grant its convergence.}

\eq
	\prod_{\ell=1}^\infty \left(\frac{E_{\ell} - \omega^2 
E_j}{E_{\ell}-\omega^{-2}E_j}\right) = -e^{-2 \mathbbm i \phi} \; .
\label{eq:BAEcont}
\en

Applying these considerations to the functions $q_0(\nu)$ and $t^{(n)}(\nu)$ one 
finds

\eq
	q_0(\nu) \rightarrow q_0(E) \doteq 
\lim_{N\rightarrow\infty}\left[e^{N\frac{\nu}{2}}q_0(\nu)\right]_{\nu=\frac{1}{2
}\log(E N^{-\frac{4\eta}{\pi}})} = 
\prod_{\ell=1}^\infty\left(1-\frac{E}{E_{\ell}}\right)
\en
and that the T-Q relation simplifies to

\eq
	t_0(E)q_0(E) = e^{\mathbbm i\phi}q_0(\omega^2E)+e^{-\mathbbm i\phi}q_0(\omega^{-2}E) \; ,
\label{eq:continuumTQ}
\en
while the fusion relations become

\begin{align}
	t^{(m)}(\omega^{-1}E)t^{(m)}(\omega E) &= 1 + 
t^{(m-\frac{1}{2})}(E)t^{(m+\frac{1}{2})}(E)	\nonumber
\\	\label{eq:contfushier}
\\	t^{(\frac{1}{2})}(E)t^{(m)}(\omega^{2m+1} E) &= 
t^{(m+\frac{1}{2})}(\omega^{2m}E) + t^{(m-\frac{1}{2})}(\omega^{2(m+1)}E) \; .	
\nonumber
\end{align}

As we have seen for $\eta/\pi \in \mathbbm Q$ the hierarchy truncates; in our new 
variables this happens when $\omega$ is a root of unity. For $\eta = \pi 
M/(2M+2)$, with $2M \in \mathbbm Z^+$, and $\phi = \pi/(2M+2)$ the truncated 
hierarchy equations can be neatly written in the following form

\eq
	t^{(\frac{m}{2})}(\omega^{-1}E)t^{(\frac{m}{2})}(\omega E) = 1+ 
\prod_{j=1}^{h-1} \Big(t^{(\frac{j}{2})}(E)\Big)^{G_{j,m}} \  , \quad 
m=1,2,\ldots,h-1 \; ,
\label{eq:TSystem}
\en
where $h=2M$, $\omega = e^{\pi \mathbbm i/(M+1)}$ and $G_{a,b}$ is the incidence matrix 
of the Dynkin diagram associated with the Lie algebra $\mathfrak a_{h-1}$.

\subsection{Non-linear integral equations from truncated hierarchies}
\label{subsec:NLIE}

Let us show how to transform the truncated fusion hierarchy (\ref{eq:TSystem}) into a set of Non-Linear Integral Equations (NLIE): we concentrate on the continuum limit for its simplicity, however the procedure remains more or less the same for finite $N$.

Since the following argument applies for any simple-laced Lie algebra, we keep the treatment general and set $r= h-1$, denoting the rank of the algebra, and $T_a(E) = t^{(a/2)}(E)$ so that the $T$-system associated to a simply-laced Dynkin diagram with incidence matrix $G_{a,b} = G_{b,a}$ reads

\eq
	T_a(\omega^{-1}E)T_a(\omega E) = 1 + \prod_{b=1}^r\left(T_b(E)\right)^{G_{a,b}} \; , \quad a=1,\ldots ,r \; .
\label{eq:genTsystem}
\en

Now, let us introduce the $Y$-functions as

\eq
	Y_a(E) \doteq \prod_{b=1}^r\left(T_b(E)\right)^{G_{a,b}} \; ,
\en
so that, substituting in (\ref{eq:genTsystem}), raising to power $G_{b,a}$ and taking a product over $a$, one gets

\eq
	Y_b(\theta + \mathbbm i\frac{\pi}{r+1})Y_b(\theta - \mathbbm i\frac{\pi}{r+1}) = \prod_{a=1}^r \left(1+Y_a(\theta)\right)^{G_{b,a}} \; ,
\label{eq:Ysystem}
\en
where we also set $E= e^{\theta/\mu}$ with $\mu = (M+1)/M(r+1)$. Note that the entirety of $T$ and $Y$ as functions of $E$ means that they are periodic functions of $\theta$, with period $2\pi \mathbbm i\mu$. These equations coincide, in the case of $\mathfrak a_{h-1}$, with the $Y$-system found by Zamolodchikov in \cite{AlZa_91} for certain integrable quantum field theories with $\mathbbm Z_h$ symmetry.

In order to proceed further we need to introduce the \emph{pseudoenergies} $\varepsilon_a$ along with the particular functions $L_a$:

\eq
	\varepsilon_a(\theta) \doteq \ln Y_a(\theta) \; , \quad L_a(\theta) \doteq \ln(1+e^{-\varepsilon_a(\theta)}) \; .
\en
With some elementary manipulations, the pseudoenergies are found to satisfy

\eq
	\varepsilon_a(\theta + \mathbbm i\frac{\pi}{r+1}) + \varepsilon_a(\theta - \mathbbm i\frac{\pi}{r+1}) - \sum_{b=1}^r G_{a,b}\varepsilon_b(\theta) = \sum_{b=1}^r G_{a,b}L_b(\theta) \; .
\label{eq:varepsilonsystem}
\en

This equation possess many solutions and in order to select the desired one we have to specify some properties of the functions involved. We begin by noticing that, since $T_a(E)$ are regular at $E=0$, the functions $Y_a(\theta)$ approach a finite constant value as $\theta\rightarrow -\infty$. This constant can be found by solving the $\theta$-independent version of (\ref{eq:Ysystem}); concentrating on the $\mathfrak a_{h-1}$ case, it is not difficult to prove that the following constants

\eq
	\mathcal Y_a \doteq \lim_{\theta\rightarrow -\infty} e^{\varepsilon_a(\theta)} = \frac{\sin \pi\frac{a}{h+2}\sin\pi\frac{a+2}{h+2}}{\sin^2\frac{\pi}{h+2}} \; ,
\en
are a solution\footnote{This is not the sole solution of the $\theta$-independent Y-system, however it is the one we need. In fact we are searching for the ground state eigenvalue, for which the zeroes of the $T$ functions lie on the negative $E$ axis; as a consequence, the $Y$ functions do not vanish on the real $\theta$ axis. This fact, combined with the behaviour (\ref{eq:largethetavarepsilon}), justifies choosing all the $\mathcal Y_a$ positive.}.

Turning to the large $E$ behaviour, it is not hard to see, using some standard ODE analysis, that

\eq
	\ln Y_a(E) \propto E^\mu \; , \quad \vert E\vert \rightarrow \infty \; , \ \vert \arg(E)\vert < \pi-\delta \; ,
\en
with $\delta$ and arbitrary small real number. Introducing the constant $m_0 L$ we easily see that

\eq
	\varepsilon_a(\theta) \underset{\Re e \theta \rightarrow \infty}{\sim} m_0 L e^\theta \; , \quad \vert \Im m \theta\vert < \pi \frac{h+2}{h} - \delta \; ,
\label{eq:largethetavarepsilon}
\en
and, from the large-$\theta$ limit of (\ref{eq:varepsilonsystem}), that

\eq
	m_0 L = \frac{b_0}{2}\sin \pi\frac{a}{h} \; , \quad b_0\in\mathbbm R\; ,
\en
where $1/2\,\sin\pi a/h$ are the components of Perron-Frobenius eigenvector for the $\mathfrak a_{h-1}$ incidence matrix.

The above properties mean that the ``regularised" $\varepsilon$

\eq
	f_a(\theta) \doteq \varepsilon_a(\theta) - m_0 L e^\theta \; ,
\en
are bounded in the analyticity strip $\vert\Im m \theta\vert < \pi/h$ and satisfy the relations (\ref{eq:varepsilonsystem}). We can thus take their Fourier transform

\eq
	\tilde{f}_a(k) \doteq \lim_{\epsilon\rightarrow 0^+} \int_{-\infty}^{\infty} d\theta f_a(\theta)e^{-\mathbbm ik\theta + \epsilon\theta} \; ,
\en
which turns the (\ref{eq:varepsilonsystem}) into

\eq
	\sum_{b=1}^{h-1} \bigg[\big(2\delta_{a,b} \cosh\pi\frac{k}{h} - G_{a,b}\big)\tilde{f}_b(k)- G_{a,b}\tilde{L}_b(k)\bigg] \; ,
\en
where

\eq
	\tilde{L}_a(k) \doteq \lim_{\epsilon\rightarrow 0^+} \int_{-\infty}^{\infty} d\theta L_a(\theta)e^{-\mathbbm ik\theta + \epsilon\theta} \; .
\en
Finally, solving for $\tilde{f}_a(k)$ and transforming back to the $\theta$ space we arrive at the following nonlinear integral equation

\eq
	\epsilon_a(\theta) = m_a L e^\theta - \frac{1}{2\pi} \sum_{b=1}^{h-1}\int_{-\infty}^\infty \phi_{a,b}(\theta-\theta') L_b(\theta')d\theta' \; ,
\label{eq:NLIE}
\en
where the kernel $\phi_{a,b}(\theta)$ is the Fourier image of the function

\eq
	\tilde{\phi}_{a,b}(k) \doteq -2\pi \sum_{c=1}^{h-1}\left[2\delta_{a,c} \cosh\pi\frac{k}{h}-G_{a,c}\right]^{-1}G_{c,b} \; .
\en
This function can be exactly computed in terms of elementary functions and the result is

\eq
	\phi_{a,b}(\theta) = -\mathbbm i \frac{d}{d\theta}S_{a,b}(\theta) \; ,
\en
with

\eq
	S_{a,b}(\theta) \doteq \prod_{\underset{\textrm{step}\, 2}{x=\vert a-b\vert +1}}^{a+b-1} \{x\} \; , \quad a,b = 1,\ldots ,h-1 \; ,
\en
and

\eq
	\{x\} \doteq (x-1)(x+1) \; , \quad (x)\doteq \frac{\sinh\left(\frac{\theta}{2} + \mathbbm i\pi \frac{x}{2h}\right)}{\sinh\left(\frac{\theta}{2} - \mathbbm i\pi \frac{x}{2h}\right)} \; .
\en

The equations (\ref{eq:NLIE}) can be interpreted in the context of the Thermodynamic Bethe Ansatz \cite{AlZa_90} and are thus often referred to as TBA equations. In this picture the parameter $L$ corresponds to the circumference of an infinite cylinder on which a massive relativistic integrable theory is defined; the functions $S_{a,b}(\theta)$ represent the matrix elements describing the factorised scattering of $r$ particles of mass $m_0$. From this set of equations one can recover the form of the integrals of motion by analysing the large-$\theta$ asymptotic expansion of the fundamental transfer matrix $T_1(\theta)$.\\

So, starting from the six-vertex model and applying the Bethe ansatz, we were 
able to obtain a series of functional relations: the Baxter's T-Q relation 
(\ref{eq:TQrelation}), the quantum Wronskian (\ref{eq:wronsky}) and the fusion 
hierarchy (\ref{eq:fushier}). Then we took the continuum limit, obtaining a set 
of equations describing a $c=1$ CFT defined on a cylinder with twisted boundary 
conditions; these can be transformed into nonlinear integral equations which allow to extract the scaling dimensions and the effective central charge $c_{\textrm{eff}}$ of the CFT. There are, however, other means to derive this set of NLIE; one could, for example, start from the remark that the ultraviolet limit of sine-Gordon model is described by a $c=1$ CFT and apply the Thermodynamic Bethe Ansatz \cite{AlZa_90} to derive the NLIE.

Another possible approach was proposed by V. V. Bazhanov, S. L. Lukyanov and 
A. B. Zamolodchikov in \cite{Bazh_Luky_AZam_96,Bazh_Luky_AZam_97,Bazh_Luky_AZam_99}. They didn't consider the 
unitary $c=1$ conformal theory with twisted boundary conditions, but rather a 
CFT with central charge

\eq
	c=1 - \left(\beta - \frac{1}{\beta}\right)^2 < 1 \quad , \qquad 
0<\beta<1
\label{eq:BLZcentralcharge}
\en
and periodic boundary conditions. Despite this theory being neither unitary nor 
minimal and the fact that, fixed a value of $\beta$, the Hilbert space still 
depends on a free parameter $p$, they were able to recover the NLIE we were 
talking about. In the next subsection we give a brief sketch of their results, 
since they are relevant for the ODE/IM correspondence.

\subsection{The BLZ approach}
\label{subsec:BLZ}

In the series of papers \cite{Bazh_Luky_AZam_96,Bazh_Luky_AZam_97,Bazh_Luky_AZam_99} is shown how, for 
integrable models, it is possible to build the continuum analogues of Baxter's 
$\mathbbm T$ and $\mathbbm Q$ matrices directly employing field-theoretical 
methods (that is, without having to pass through the corresponding lattice 
theory). Their starting point is a non-unitary CFT with central charge 
parametrised in terms of $\beta$ as in (\ref{eq:BLZcentralcharge}).

As is widely known, the spectrum of a CFT is described in terms of a tower of 
states each consisting of a highest-weight state $\vert p \rangle$ and its 
descendants; in the case of this particular theory, the highest-weight states 
have conformal dimension $\Delta_p = (p/\beta)^2 + (c-1)/24$, with $p$ being a 
continuous parameter. For each tower of states, BLZ defined an operator-valued 
entire function $\mathbf T(s,p)$, which is a continuum analogue of the lattice 
transfer matrix $\mathbbm T$; they also introduce the two auxiliary 
operator-valued functions $\mathbf Q_{\pm}(s,p)$, mutually commuting and 
satisfying a T-Q relation

\eq
	\mathbf T(s,p)\mathbf Q_{\pm}(s,p) = \mathbf Q_{\pm}(q^2s,p) + \mathbf 
Q_{\pm}(q^{-2}s,p) \; ,
\en
with $q=e^{\mathbbm i\pi\beta^2}$. Restricting to a given tower of states, it is possible 
to recast this relation in terms of the highest-weight eigenvalues of $\mathbf 
T$ and $\mathbf Q$\footnote{Where unnecessary, we omit the explicit dependence 
on $p$.}

\eq
	T(s)Q_{\pm}(s) = e^{\pm 2 \pi \mathbbm i p}Q_{\pm}(q^2 s)  + e^{\mp 2\pi \mathbbm i 
p}Q_{\pm}(q^{-2}s) \; ,
\label{eq:BLZTQ}
\en
where the eigenvalues are defined as

\begin{align}
	T(s,p) &\doteq \langle p \vert \mathbf T(s,p) \vert p \rangle	
\nonumber
\\
\\	Q_{\pm}(s,p) &\doteq \langle p \vert s^{\mp \frac{\mathbf 
P}{\beta^2}}\mathbf Q_{\pm}(s,p)\vert p \rangle \; ;	\nonumber
\end{align}
here $\mathbf P$ is an operator such that $\mathbf P\vert p \rangle = p \vert p 
\rangle $. This T-Q relation matches perfectly (\ref{eq:continuumTQ}), given we 
set $\phi = \pm 2 \pi p$ and $\eta = \frac{\pi}{2}(2-\beta^2)$ (which imply 
$e^{2\pi \mathbbm i \beta^2} = q^2 = \omega^2 = e^{-4 \mathbbm i \eta})$.

If we let $0 < \beta^2 < 1/2$, the so-called \emph{semiclassical domain}, then 
the following product representation is convergent

\eq
	Q_+(s) = \prod_{k=1}^\infty \left(1-\frac{s}{s_k}\right)
\en
and from the T-Q relation (and the entirety in s of the eigenvalues), we obtain 
a set of BAE

\eq
	\prod_{\ell=1}^\infty \frac{s_{\ell} -q^2s_j}{s_{\ell}-q^{-2}s_j} = 
-e^{4 \pi \mathbbm i p} \; .
\en

One can go further and define the analogues of the functions $t^{(m)}$, starting 
from the identity operator $\mathbf T_0(s) \equiv \mathbbm I$, the $\mathbf 
T$-operator $\mathbf T_{\frac{1}{2}}(s) \equiv \mathbf T(s)$ and the fusion 
relations

\eq
	\mathbf T_j(q s)\mathbf T_j(q^{-1}s) = 1 + \mathbf 
T_{j-\frac{1}{2}}(s)\mathbf T_{j+\frac{1}{2}}(s) \quad , \qquad j= 
\frac{1}{2},1,\frac{3}{2},\ldots \ .
\en
Again, at rational values of $\beta^2$ this hierarchy truncates to a finite set 
of operators and relations, exactly as happens in the lattice case (in fact 
$\beta^2$ rational implies $\eta/\pi$ rational). One can also find an expression 
which directly gives the operators $\mathbf T_j(s)$ in terms of the $\mathbf 
Q_{\pm}(s)$:

\eq
	2 \mathbbm i \sin (2 \pi \mathbf P) \mathbf T_j(s) = \mathbf Q_+(q^{2j+1}s) 
\mathbf Q_-(q^{-2j-1}s)-\mathbf Q_+(q^{-2j-1}s) \mathbf Q_-(q^{2j+1}s) \; ,
\label{eq:partnerrelation}
\en
which, for $j=0$ and evaluated on the state $\vert p \rangle$ gives us the 
quantum Wronskian relation

\eq
	q^{\frac{2 p}{\beta^2}} Q_+(q s)Q_-(q^{-1}s) - q^{-\frac{2 p}{\beta^2}} 
Q_+(q^{-1}s)Q_-(q s) = 2 \mathbbm i \sin (2 \pi p) \; ,
\label{eq:BLZquantwronsk}
\en
which, as a consequence, implies $Q_-(s,p) = Q_+(s,-p)$.

In the following, we will concentrate on the (\ref{eq:BLZTQ}) version of the T-Q 
relation, however, as remarked above, the connection with the continuum limit of 
the six-vertex model is simply given by identifications

\eq
	\beta^2 = 1-2\frac{\eta}{\pi} \; ,\qquad p = \frac{\phi}{2 \pi} \; .
\en
The coincidence is neatly seen if one computes the effective central charge of 
the non-unitary CFT, which, on a cylinder with periodic boundary conditions, is 
simply given by

\eq
	c_{\textrm{eff}} = c-24 \Delta_{\textrm{min}}
\en
so that, for a given tower with highest-weight state $\vert p \rangle$, we have

\eq
	c_{\textrm{eff}}^{(p)} = c- 24 \Delta_p = 1-24 
\left(\frac{p}{\beta}\right)^2 \; ;
\en
using the identifications above we find a perfect match with the effective 
central charge for the twisted six-vertex model (\ref{eq:6Vceff}).

\resection{Ordinary differential equation}
\label{sec:ODE}

Now let us turn our attention to the ODE side of the correspondence. In this 
section we will face spectral problems, some of which of a rather peculiar nature; 
that is, starting from a differential operator (the quantum-mechanical 
Hamiltonian) with some defined boundary conditions, we search for the spectrum 
of its eigenvalues and eigenvectors. One question, given an Hamiltonian, 
immediately arise: is its spectrum real? To some ears, this might sound like a 
trivial question: as everyone knows from a quantum mechanics course, the 
spectrum of any hermitian Hamiltonian is real. Note, however, that we never 
asked our differential operator to be hermitian, nor did we impose any kind of 
complex structure on our spectral problem. Consider for example the following 
Hamiltonian and its related Schr\"odinger equation

\eq
	\mathcal H = p^2 + \mathbbm i x^2 \quad ; \qquad -\frac{d^2}{dx^2} \psi(x) + \mathbbm i 
x^3 \psi(x) = E\psi(x) \; .
\label{eq:BessisZinnHam}
\en
This is a cubic oscillator with purely imaginary coupling and is clearly a 
non-hermitian problem, however, quite astonishingly, the spectrum of this 
Hamiltonian appears to be real and positive. This was conjectured by D. Bessis 
and J. Zinn-Justin in an unpublished paper, on the basis of perturbative and 
numerical studies. C. M. Bender and S. Boettcher later showed how $\mathcal P 
\mathcal T$ symmetry might be the reason behind the strange reality of the 
spectrum of (\ref{eq:BessisZinnHam}) \cite{Bend_Boet_98}. More precisely 
$\mathcal P$, the ``parity", acts by sending $x$ to $-x$ and $p$ to $-p$, while 
$\mathcal T$, the ``time-reversal" maps $x$ to $x$, $p$ to $-p$ and $\mathbbm i$ to $-\mathbbm i$; 
both $\mathcal P$ and $\mathcal T$ preserve the canonical commutation relation 
$[x,p]=\mathbbm i$ of quantum mechanics, even in the case of $x$ and $p$ complex. In 
\cite{Bend_Boet_Meis_99} was shown how $\mathcal P\mathcal T$ invariance forces the 
eigenvalues to be either real or complex-conjugate in pairs, much like the roots 
of a real polynomial and, just like many roots of a real polynomial are complex, 
$\mathcal P\mathcal T$ invariance is not enough to guarantee the reality of the 
Hamiltonian's spectrum.

\subsection{$\mathcal P\mathcal T$ symmetric Hamiltonians}
\label{subsec:PTsymmHam}

In \cite{Bend_Boet_98}, C. M. Bender and S. Boettcher proposed a generalisation of 
the Bessis-Zinn-Justin Hamiltonian (\ref{eq:BessisZinnHam}):

\begin{align}
	\mathcal H_M &= p^2 - (\mathbbm ix)^{2M}	\nonumber
\\	\label{eq:BendBoetHam}
\\ -\frac{d^2}{dx^2}\psi(x) - (\mathbbm ix)^{2M}\psi(x) &= E \psi(x) \qquad  , \qquad 
\psi(x) \underset{x\rightarrow\pm\infty}{\rightarrow} 0 \; ,	\nonumber
\end{align}
which is explicitly $\mathcal P\mathcal T$ symmetric and incorporates in a 
one-parameter family of eigenvalue problems both the Bessis-Zinn-Justin 
Hamiltonian ($M=3/2$) and the harmonic oscillator ($M=1$). One needs to take care 
when considering non-integer values of $2M$, since the factor $-(\mathbbm ix)^{2M}$ is no 
more single valued; this is easily accounted for by placing a branch cut running 
along the positive imaginary $x$-axis. Another problem arises when $M$ reaches the 
value $2$; we will face this point later, for the moment we keep $M<2$.

\begin{figure}[t]
\centering
\includegraphics[scale=0.5]{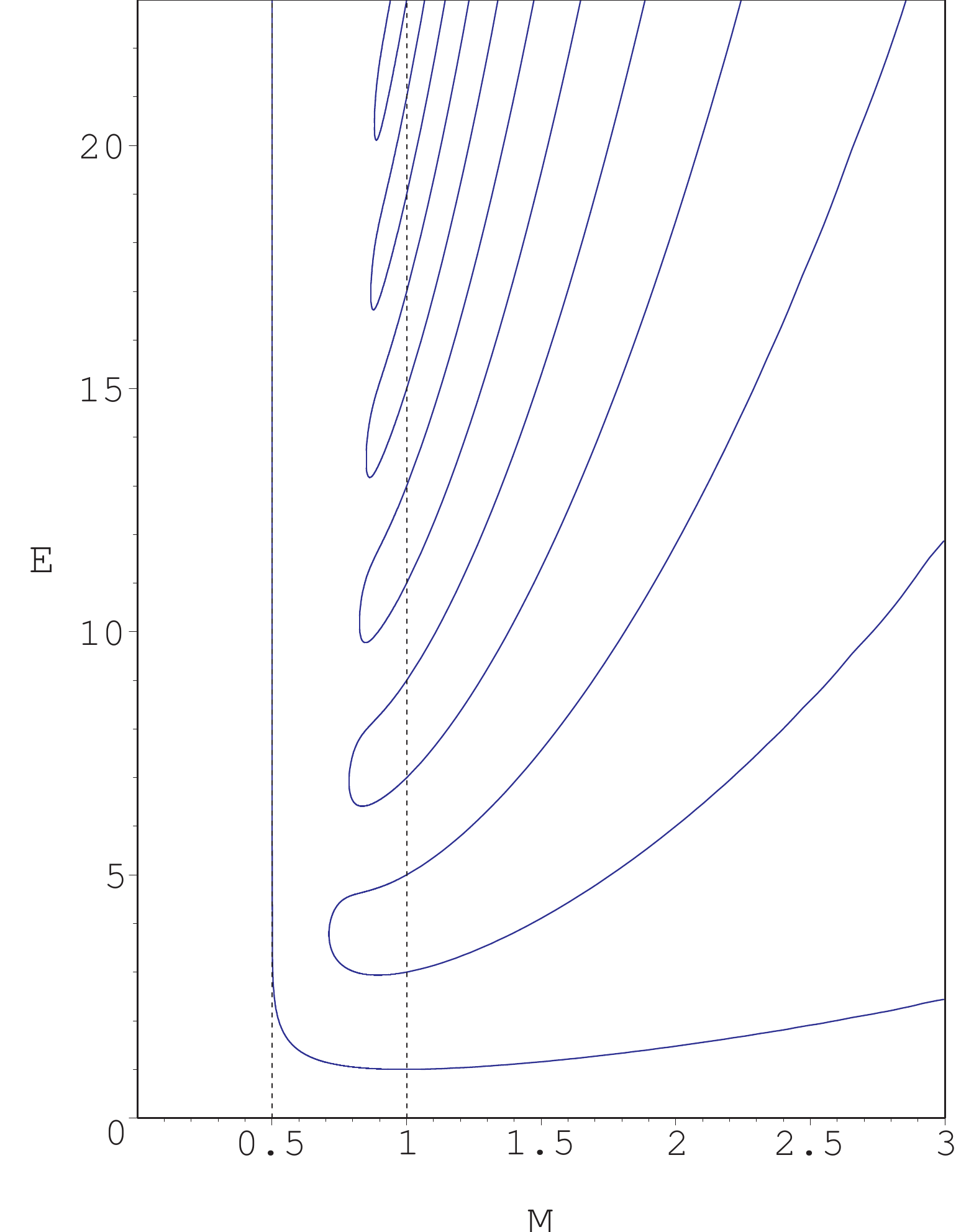}
\caption{Real eigenvalues of $\mathcal H_M$ as a function of $M$.}
\label{fig:BendBoetHamSpectr}
\end{figure}

The spectrum of (\ref{eq:BendBoetHam}) shows a very unusual behaviour when 
plotted against the parameter $M$: in Figure  \ref{fig:BendBoetHamSpectr} we see that, 
as $M$ decreases below $1$, infinitely many eigenvalues pair off and become 
complex and, at $M=0.5$ also the last remaining real eigenvalue diverges, 
leaving the spectrum with only complex eigenvalues. For $M>1$, however, 
numerical results joined with analytical evidences indicate that the spectrum is 
entirely real and positive. The transition to infinitely-many complex 
eigenvalues at $M=1$ can be interpreted as a spontaneous breaking of the 
$\mathcal P\mathcal T$ symmetry.

By adding parameters, it is possible to generalise further the Bessis-Zinn-Justin 
Hamiltonian; we might add an angular-momentum-like term $l(l+1)x^{-2}$, which 
gives \cite{Dore_Tate_99}

\eq
	\mathcal H_{M,l} = p^2 -(\mathbbm{i} x)^{2M} +\frac{l+1}{x^2}l \; .
\label{eq:BLZHam}
\en
When considering the related Schr\"odinger equation, while still imposing 
$\psi(x)\rightarrow 0$ as $x\rightarrow\pm\infty$, we have to specify how the 
wavefunction should be continued around the new singularity at $x=0$; given the 
choice we made of placing a branch cut on the positive imaginary $x$-axis, we 
agree that the continuation has to be done in the lower-half plane. Again, as proved in Appendix B of \cite{Dore_Dunn_Tate_01}, the 
spectrum of (\ref{eq:BLZHam}) is real and positive for $M\geq1$ and $\vert 2l+1\vert < M+1$.

\begin{center}
\begin{figure}[t]
\centering
\includegraphics[scale=0.5]{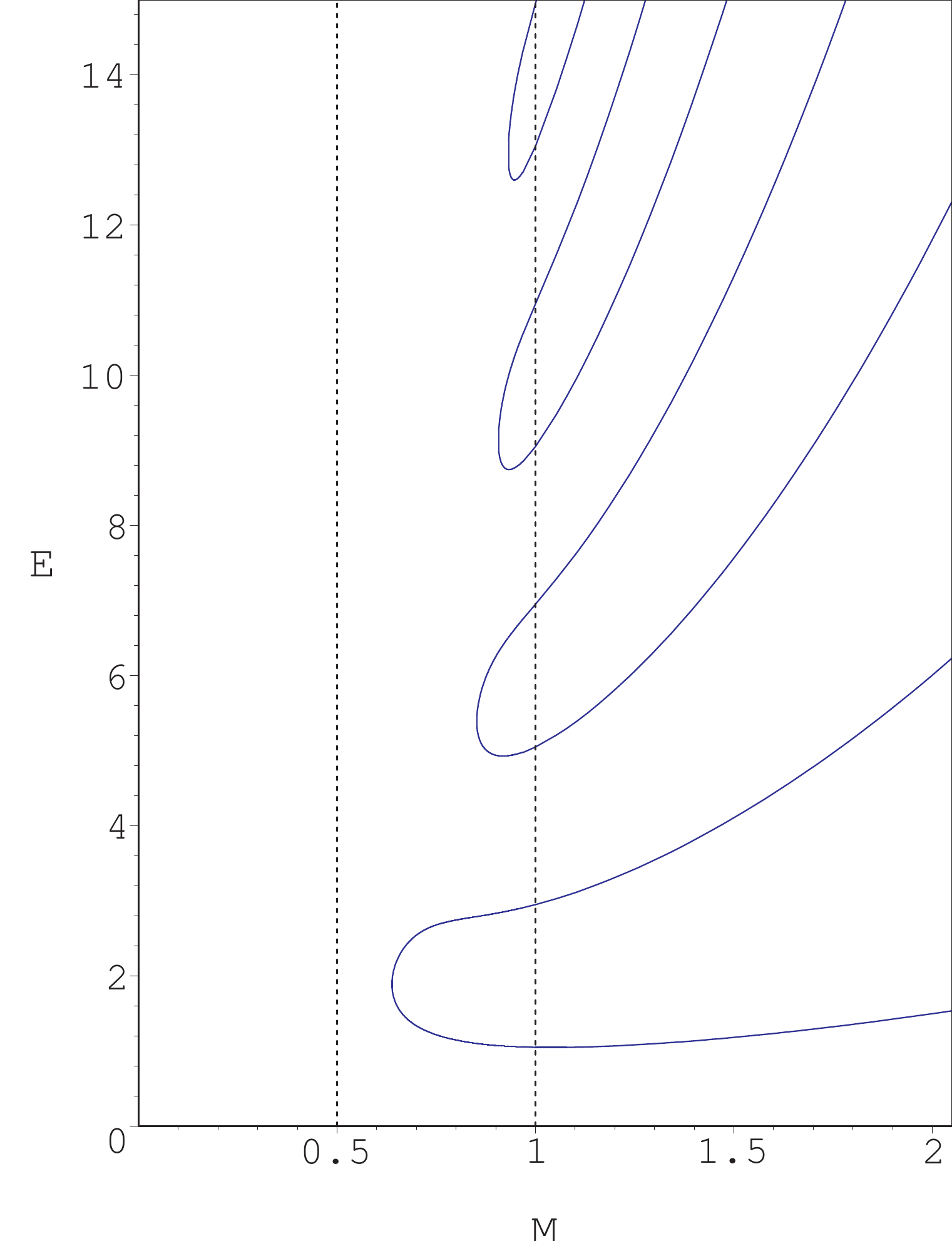}
\caption{Real eigenvalues of $\mathcal H_{M,l}$ as a function of $M$, for $l=-0.025$.}
\label{fig:BLZHamSpectr1}
\end{figure}
\end{center}

\begin{figure}[!t]
\begin{center}
\includegraphics[width=0.29\linewidth]{fm025_review.pdf}
{}~~~~~~~~~~~
\includegraphics[width=0.29\linewidth]{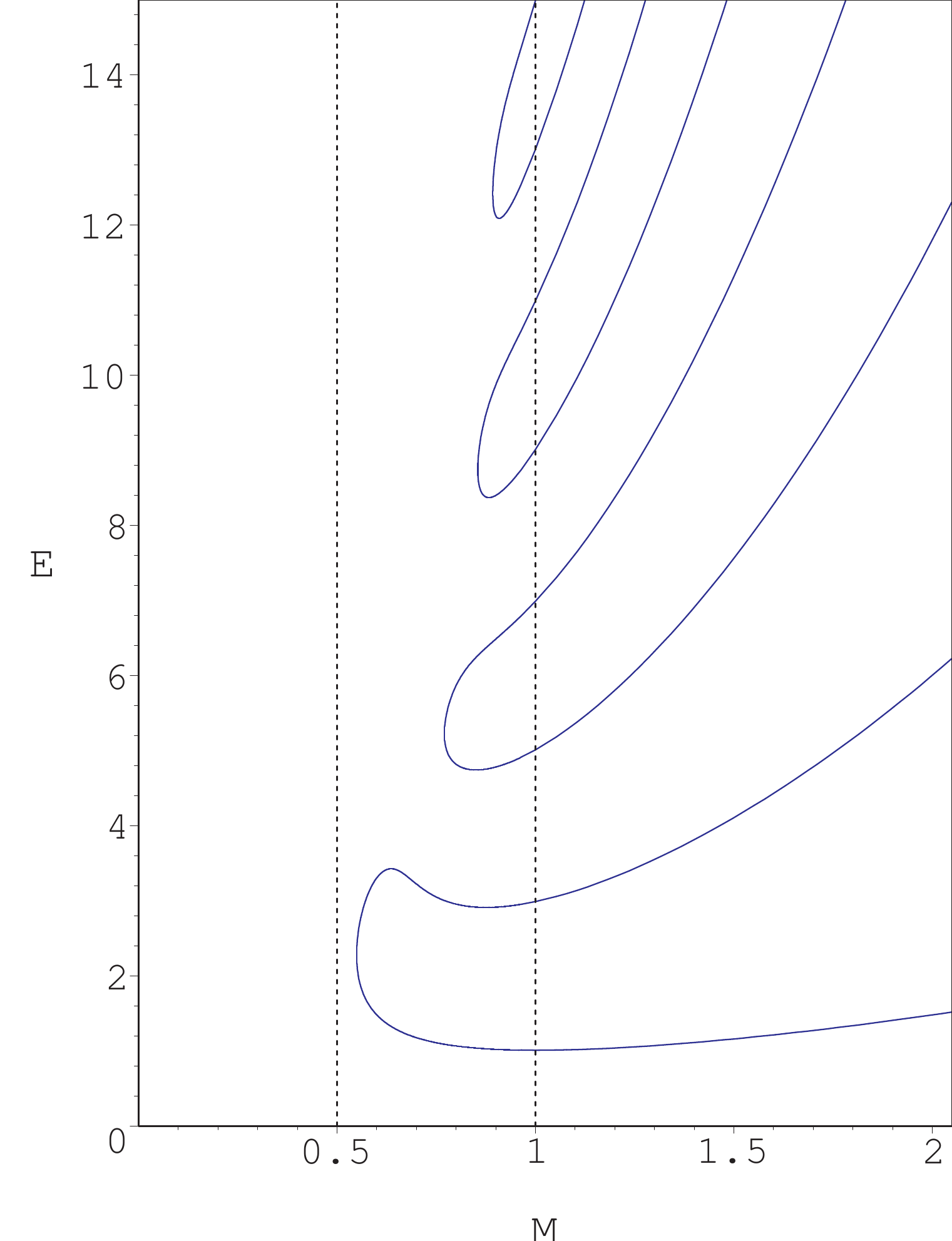}\\[1pt]
\parbox{0.33\linewidth}{~~~~~~~~~~~\small\protect\ref{fig:BLZHamSpectr2}a: $l=-0.025$}~~~~
{}~~~~
\parbox{0.33\linewidth}{~~~~~~~~~~~\small\protect\ref{fig:BLZHamSpectr2}b:
$l=-0.005$}\\[14pt]
\includegraphics[width=0.29\linewidth]{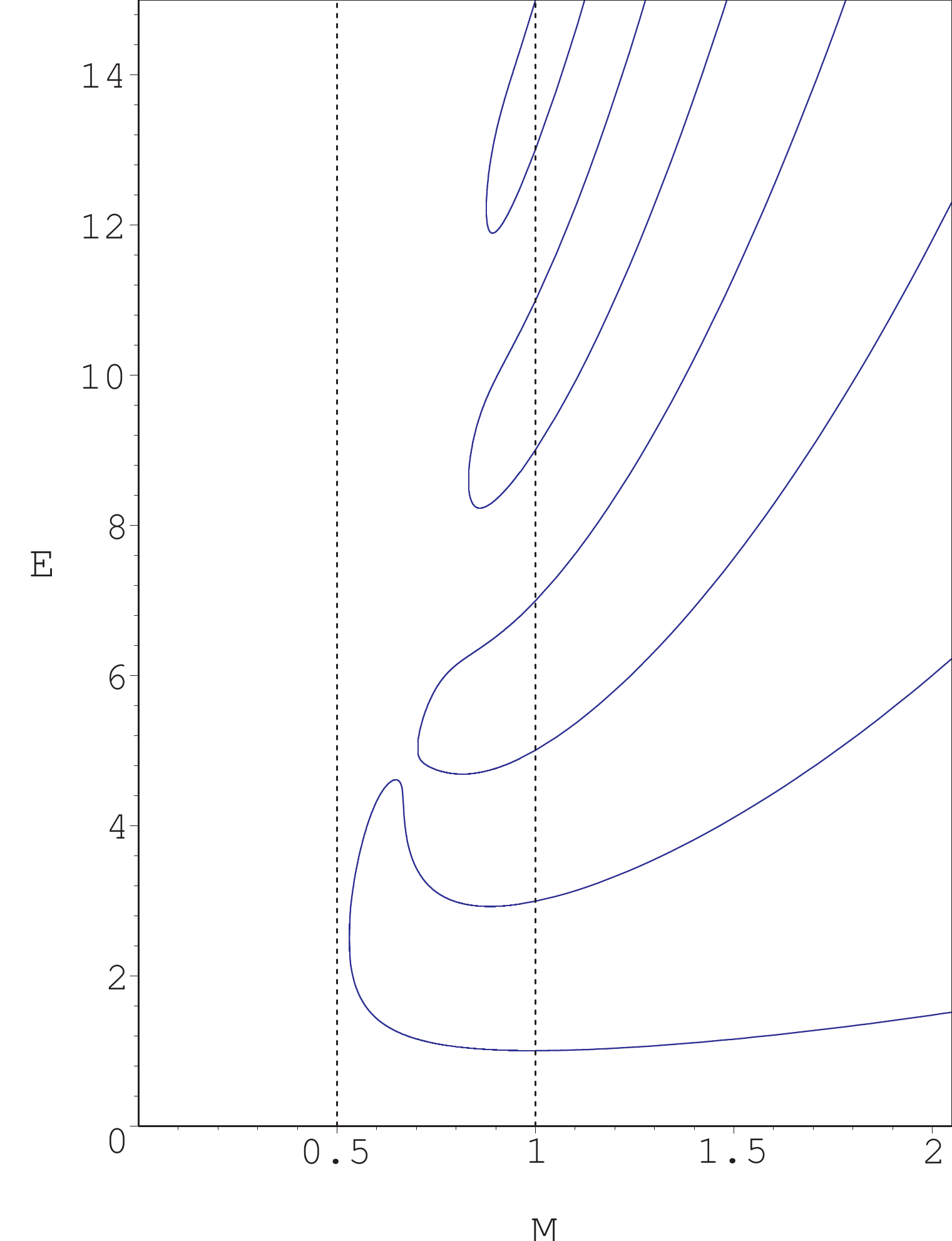}
{}~~~~~~~~~~~
\includegraphics[width=0.29\linewidth]{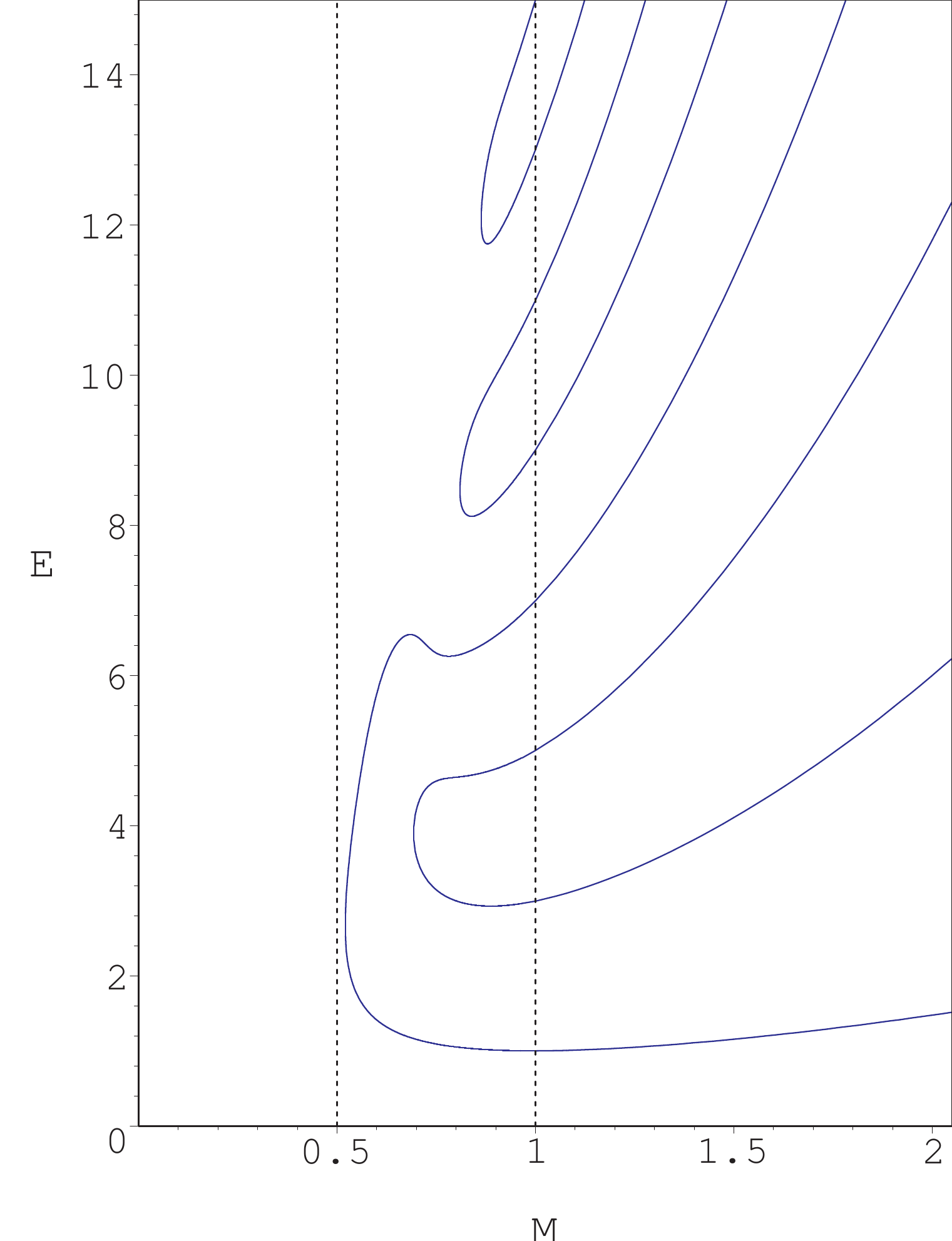}\\[1pt]
\parbox{0.33\linewidth}{~~~~~~~~~~~\small\protect\ref{fig:BLZHamSpectr2}c:
 $l=-0.0025$}~~~~
{}~~~~
\parbox{0.33\linewidth}{~~~~~~~~~~~\small\protect\ref{fig:BLZHamSpectr2}d:
$l=-0.0015$}\\[14pt]
\includegraphics[width=0.29\linewidth]{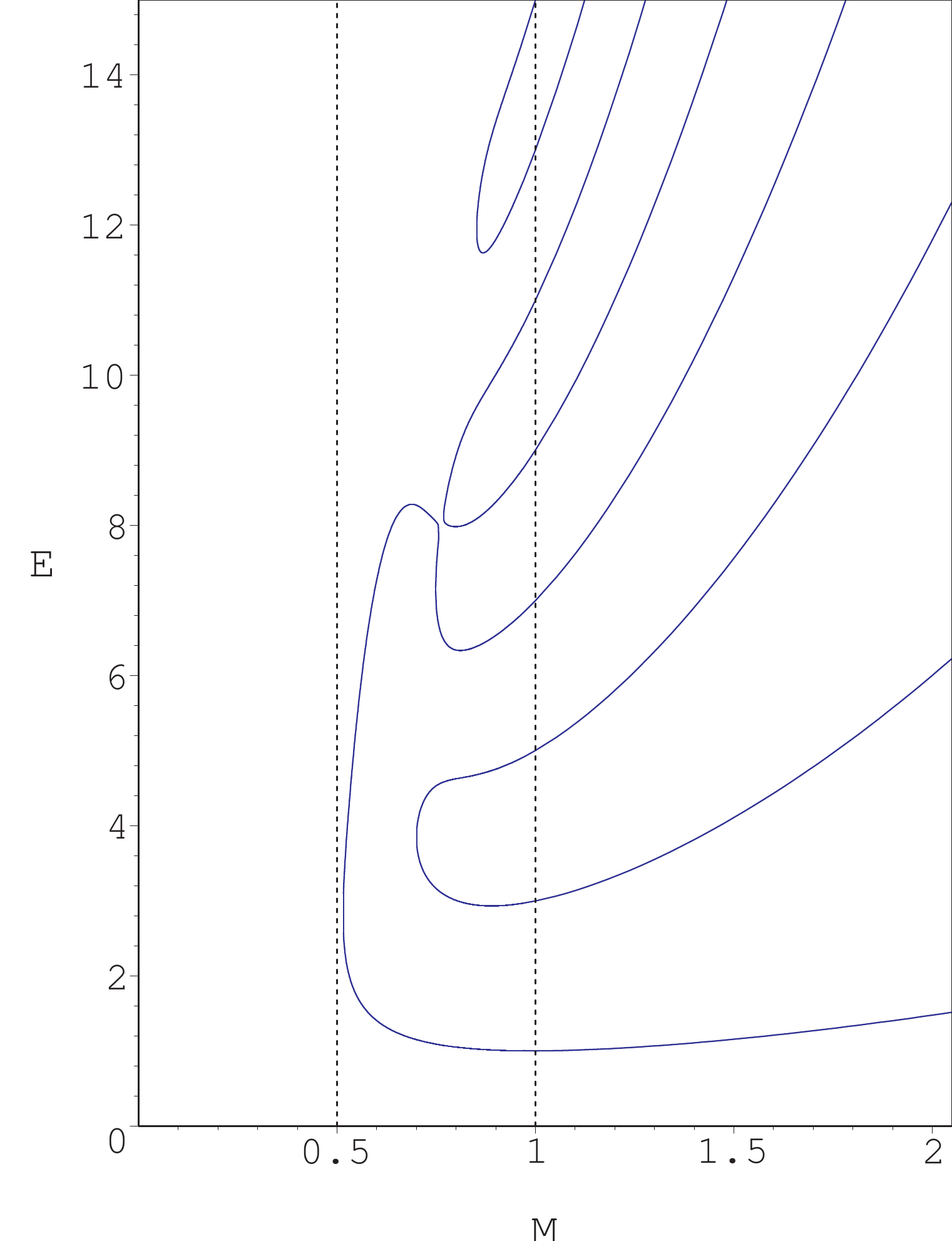} 
{}~~~~~~~~~~~
\includegraphics[width=0.29\linewidth]{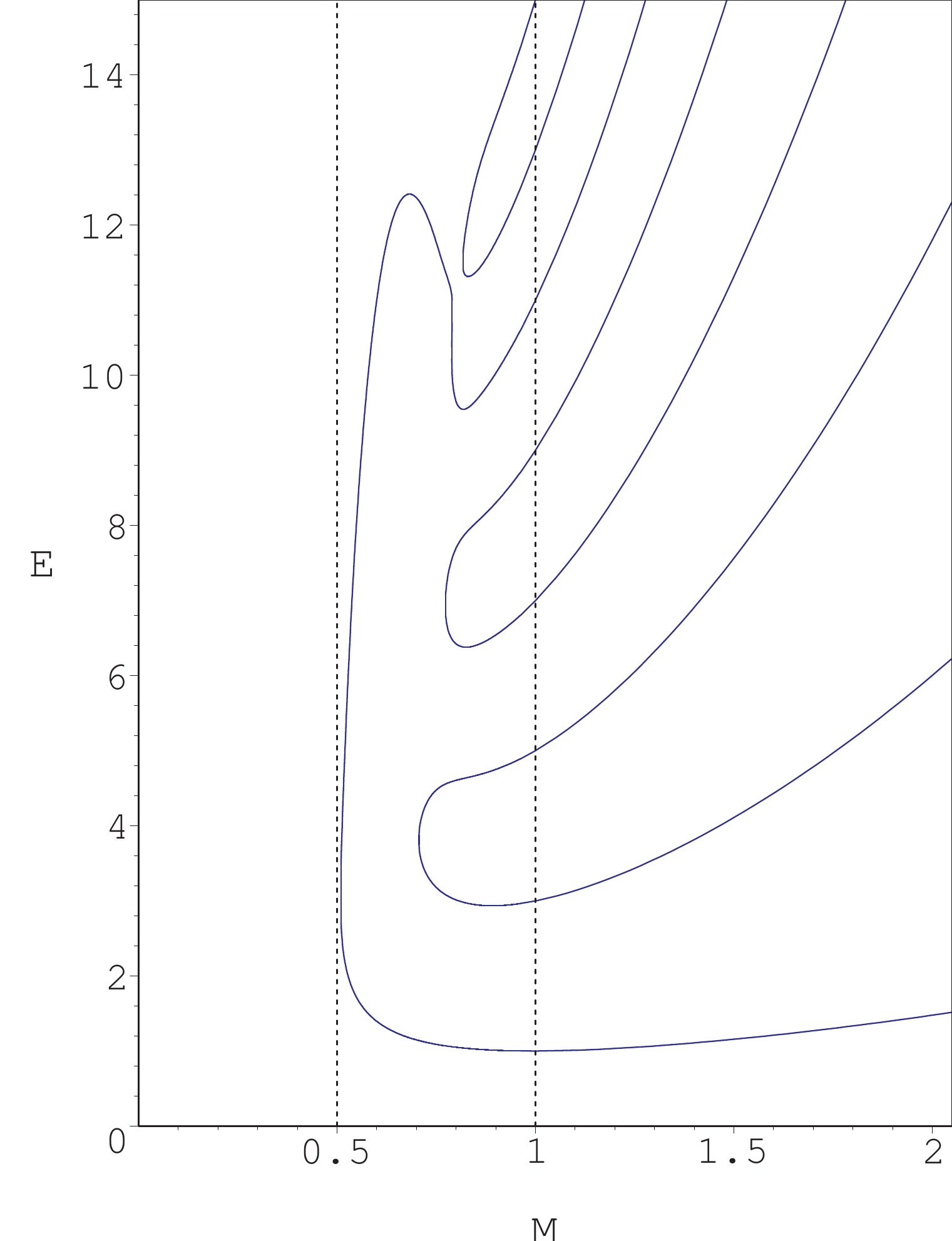}\\[1pt]
\parbox{0.33\linewidth}{~~~~~~~~~~~\small\protect\ref{fig:BLZHamSpectr2}e:
$l=-0.001$} ~~~~
{}~~~~
\parbox{0.33\linewidth}{~~~~~~~~~~~\small\protect\ref{fig:BLZHamSpectr2}f:
$l=-0.0005$}\\[2pt]
\end{center}
\caption{  \protect{ \label{fig:BLZHamSpectr2}}Real eigenvalues of
$\mathcal H_{M,l}$ as functions of $M$,
for various values of $l$.   }
\end{figure}

While a small angular momentum does not significantly alter the spectrum for $M>1$ 
and all the eigenvalues remain real, for $M<1$ there is a remarkable difference, 
in the way they become complex, from the case of (\ref{eq:BendBoetHam}). In Figure  
\ref{fig:BLZHamSpectr1} we see the plot of the spectrum against $M$ for $l= 
-0.025$: the ``connectivity" of the real eigenvalues has been reversed and the 
ground-state one does not diverge anymore. The mechanism which allow for a 
continuous deformation from $l=0$ to $l=-0.025$ might be hard to conceive, but Figure  \ref{fig:BLZHamSpectr2} should clarify the peculiar behaviour.

It is possible to introduce further generalisations, but they will not be relevant 
to our needs. The interested reader can find more details on this generalisations in \cite{Dore_Dunn_Tate_07} and references therein.

\subsection{Eigenvalue problems in the complex plane}
\label{subsec:eigencomplex}

Let us return for a moment to the Bender-Boettcher Hamiltonian 
(\ref{eq:BendBoetHam}): as we hinted, for $M=2$ the na\"ive definition of our 
eigenvalue problem runs into difficulties; let us see what happens. The 
Schr\"odinger equation, in this particular case, becomes

\eq
	-\frac{d^2}{dx^2}\psi(x) -x^4\psi(x) = E\psi(x) \; ,
\en
an ``upside-down" quartic oscillator and a simple WKB analysis\footnote{Which in 
this case simply consists in substituting $\psi(x)=f(x)e^{g(x)}$ in the ODE and 
equating the coefficients.} tells us that, as $x\rightarrow\pm\infty$, 
wavefunctions behave as

\eq
	\psi(x) \sim P(x)^{-\frac{1}{4}} e^{\pm \int\limits^x \sqrt{P(t)} dt} \; ,
\label{eq:WKBexp}
\en
where $P(x)$ is the potential in the Schr\"odinger equation; in our case $P(x) = 
- x^4 - E$ which, as $x\rightarrow\pm\infty$ can be replaced by $-x^4$, giving

\eq
	\psi(x) \underset{x\rightarrow\infty}{\sim} x^{-1} e^{\pm 
\frac{\mathbbm i}{3}x^3} \; .
\en
Thus, instead of the usual exponentially growing or decaying solutions, we see 
that \emph{all} solutions decay, albeit algebraically, moving the problem from 
the so-called \emph{limit-point} to the \emph{limit-circle} case (see 
\cite{Reed_Simo_72,Reed_Simo_75,Rich_78}): this eigenproblem is clearly not a smooth 
continuation from the $M<2$ region.

The key to solve this matter is to consider $x$ as a complex variable and 
examine the behaviour of solutions as $\vert x \vert\rightarrow \infty$ along a 
general ray in the complex x-plane even though our initial problem involved only 
the positive and negative real axes. This method has been discussed by many 
authors, among whom we recall Y. Sibuya \cite{Sibu_75} and C. M. Bender, 
S. Boettcher and A. Turbiner \cite{Bend_Turb_93,Bend_Boet_Meis_99}.

So, in order to remain more general, let us take in consideration the 
Hamiltonian (\ref{eq:BLZHam}) and, since we placed a branch cut on the positive 
imaginary axis, let us set $x= -\mathbbm i \rho e^{\mathbbm i \theta}$. The WKB expansion (\ref{eq:WKBexp}) is valid along 
any ray in the complex plane, as $\vert x \vert\rightarrow\infty$ and, 
substituting the potential 

\eq
P(x) = -(\mathbbm ix)^{2M} +l(l+1)x^{-2}-E \underset{\vert x 
\vert\rightarrow\infty}{\sim} -(\mathbbm ix)^{2M}
\en
one obtains two possible behaviours, as expected from a second-order ODE:

\eq
\psi_{\pm} \underset{\vert x \vert\rightarrow\infty}{\sim} \frac{\exp\left[\pm 
\frac{1}{M+1} e^{\mathbbm i(M+1)\theta}\rho^{M+1}\right]}{P^{\frac{1}{4}}} \; .
\en
Notice how, for almost every value of $\theta$, we have one exponentially 
growing solution and an exponentially decaying one; however, when $\mathfrak 
Re\left[e^{\mathbbm i(M+1)\theta}\right]=0$, we end up with a pair of oscillating 
solutions, neither of which dominates the other. The rays defined by the values

\eq
	\theta_n = \frac{2 n +1}{2M+2}\pi \qquad , \qquad n\in \mathbbm Z
\en
are called \emph{anti-Stokes lines}\footnote{Notice that some authors call these 
rays \emph{Stokes lines}, see, for example, \cite{Berr_89}} and split the 
complex plane in sectors, called \emph{Stokes sectors}; in each of these 
portions of the complex plane ($\theta_{n-1} < \theta < \theta_n$) we have the 
usual decaying/growing pair of solutions with a straightforward and discrete 
spectrum. When moving through an anti-Stoke line, these two solutions swap 
r\^oles, while, whenever $\theta=\theta_n$, our definition of the eigenvalue 
problem becomes much more delicate. This is exactly what happens at $M=2$ for 
our original problem; in fact, increasing $M$ from $1$, the value $M=2$ is the 
first one at which anti-Stokes lines lie on the positive and negative real axes. But now we see how this problem can be averted: 
since all the functions involved in our problem are analytic, nothing stops us 
from deforming in the complex plane the contour on which we examine the 
wavefunction. In particular, before $M$ reaches $2$, one can bend the ends of 
the contour downwards from the real axis, without changing the spectrum as long 
as the asymptotic directions do not cross any anti-Stokes line. Now $M$ can be 
increased above $2$ without any problem.

It is worth noticing that, when $M\neq 2n \; , \; \forall n \in \mathbbm Z^+$, no 
anti-Stokes line rests on the real axis and this last is once again a ``good" 
quantisation contour; however this contour corresponds to a different eigenvalue 
problem, which is not the analytic continuation of the original for $M<2$. More 
generally, one could choose any pair of Stokes sectors in which the 
asymptotes of the contour are to be sent and, a priori, each choice leads to different problems 
(though, as we will see, some of these are related by simple change of variables). 
All of these problems share a common feature: their quantisation contours begin 
and end at $\vert x \vert = \infty$; in WKB method terminology, they relate to 
the so-called \emph{lateral connection problems} \cite{Olve_74}. There is 
another class of natural quantisation contours, namely those joining $x=0$ to 
$\vert x \vert = \infty$, leading to the \emph{radial} (also called 
\emph{central}) \emph{connection problems}, which, when granted suitable 
boundary conditions, also have interesting and discrete spectra. On the 
contrary, the case of contours having both ends at $x=0$ results always in a 
trivial eigenvalue problem.

How come $x=0$ and $\vert x \vert = \infty$ behave so differently as endpoints 
of quantisation contours? The reason is simply that, even considering ODEs with 
angular-momentum-like term $l(l+1)x^{-2}$, the singularity sitting at the origin 
is way milder than that at $\vert x \vert = \infty$; solutions there behave 
algebraically, as $x^{l+1}$ or $x^{-l}$, no matter which direction of approach 
is chosen. For this reason all the complications associated with the Stokes 
sectors and the decaying/growing solutions do not arise and we are left with two 
simple possible boundary conditions: we can either ask solutions to behave as 
$x^{l+1}$ or as $x^{-l}$ (it is understood that the singular solution is defined 
by analytic continuation). On the other hand, near $\vert x \vert = \infty$ one 
can ask the solution to be subdominant in any of the Stokes sectors (which, for 
$M$ irrational, are infinitely many). In more technical words, the ODE possess 
two singular points:

\begin{itemize}
	\item the origin, $x=0$, is a regular singularity; solutions have 
straightforward series expansions in its vicinity which converge in its whole 
neighbourhood and can be analytically continued in a simple way\footnote{The 
cases with $2M$ not being an integer behave essentially the same way as sketched 
above, see for example \cite{Chen_62}. We also omitted some subtleties, more of 
which can be found in \cite{Ince_56}.}.
	\item the infinity, $\vert x \vert = \infty$, is an irregular 
singularity; in its neighbourhood solutions possess asymptotic expansions that 
hold only in selected Stokes sectors, making analytic continuation a very subtle 
issue.
\end{itemize}

Summarising, we have seen how we can associate to a given ODE, many natural 
eigenvalue problems, which fall in two classes:

\begin{itemize}
	\item the lateral connection problems, which are defined by specifying a 
pair of Stokes sectors at infinity and asking for those values of $E$ at which solutions to the ODE exponentially decaying at infinity in both 
sectors exist;
	\item the radial connection problems, which are defined by specifying a 
single Stokes sector at infinity and asking for those values of $E$ at which solutions to the ODE exponentially decaying at infinity in the given 
sector and behaving in one of the two possible ways at the origin exist.
\end{itemize}

Up to now, it seems that all these possible eigenvalue problems are isolated, 
without relation to one another. In the next subsection we will see how it is 
possible, thanks to the works of Y. Sibuya and A. Voros, to put them in relation. 
Remarkably the equations governing these relations turn out to be exactly the 
functional equations we saw emerge in the context of integrable QFT.

\subsection{The Stokes multipliers and relations: a simple example}
\label{subsec:stokesmultrel}

In order to introduce the main ideas with the least complication possible, we 
will keep on working with the Bender-Boettcher problem:

\eq
	-\frac{d^2}{dx^2} \psi(x) -(\mathbbm ix)^{2M}\psi(x) = E\psi(x) \quad , \qquad 
\psi \in L^2(\mathcal C) \; ,
\en
where we agreed to leave the quantisation contour $\mathcal C$ unspecified at 
the moment. It is convenient to eliminate the factor $\mathbbm i$ with the change of 
variables

\eq
	x \rightarrow \frac{x}{\mathbbm i} \quad , \qquad E\rightarrow -E
\en
which moves the branch cut onto the negative real axis and the original 
quantisation contour (the former real axis) on the imaginary axis. The ODE 
becomes

\eq
	\left[-\frac{d^2}{dx^2}+x^{2M}-E\right]\psi(x) = 0 \; .
\label{eq:realODE}
\en

Now we will rely largely on the works by Y. Sibuya and P. F. Hsieh 
\cite{Sibu_75,Hsie_Sibu_66} to study our differential equation in the complex 
plane. The starting point is the following result

\newtheorem*{theo}{Theorem}

\begin{theo}
	The ODE (\ref{eq:realODE}) admit a basic solution $y(x,E)$ such that
	\begin{enumerate}
		\item it is an entire function of $x$ and $E$ (if $2M\notin 
\mathbbm Z$, because of the multivalued potential, $x$ has to be considered as a 
coordinate on a cover of $\mathbbm C \backslash \lbrace0\rbrace$, see 
\cite{Taba_99});
		\item it has the following asymptotic behaviour for $\vert 
\textrm{arg}(x)\vert < 3 \pi/(2M+2)$
			\begin{align}
				y &\underset{\vert x \vert 
\rightarrow\infty}{\sim} \frac{1}{\sqrt{2 \mathbbm i}}x^{-\frac{M}{2}}e^{-\frac{1}{M+1} 
x^{M+1}} \; ,	\nonumber
\\	\label{eq:asymptbasic}
\\
				y' &\underset{\vert x \vert 
\rightarrow\infty}{\sim} -\frac{1}{\sqrt{2 \mathbbm i}}x^{\frac{M}{2}}e^{-\frac{1}{M+1} 
x^{M+1}} \; ;	\nonumber
			\end{align}
			(Note that for $M<1$ one must make some small 
modifications, see \cite{Dore_Mill_Tate_05}).
	\end{enumerate}
	The two properties above fix $y(x,E)$ uniquely.
\end{theo}

\begin{figure}[h]
\centering
\includegraphics[scale=1]{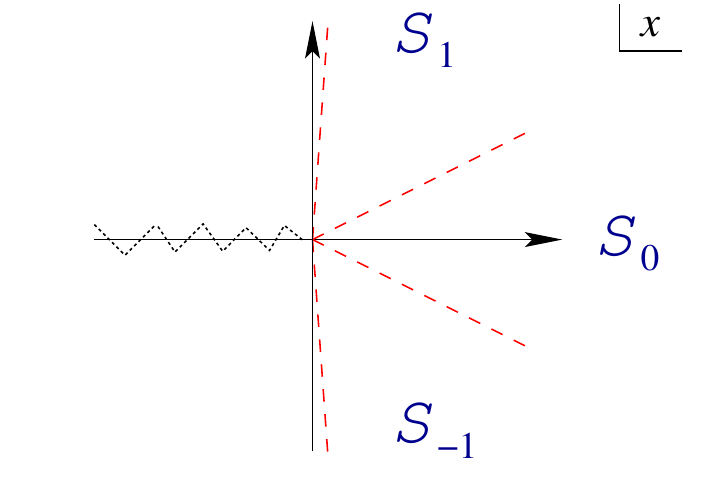}
\caption{Stokes sectors for the ODE (\ref{eq:realODE}) with $M=2.1$.}
\label{fig:realStokessectors}
\end{figure}

We will not give the proof of this theorem, as it can be found in the book by 
Y. Sibuya \cite{Sibu_75}; however we do wish to mention that the second property can 
be obtained via the WKB approximation presented in the preceding subsection, by 
taking care of the shift $x\rightarrow-\mathbbm ix$. The anti-Stokes lines are now

\eq
	\textrm{arg}(x) = \frac{2n+1}{2M+2}\pi \quad , \qquad n\in \mathbbm Z
\en
and between them lie the Stokes sectors, which we can define as

\eq
	\mathcal S_k \doteq \left\vert \textrm{arg}(x) -\frac{2 \pi k}{2M+2} 
\right\vert < \frac{\pi}{2M+2} \quad , \qquad k\in \mathbbm Z \; ;
\en
in Figure  \ref{fig:realStokessectors} three of these sectors are shown for $M$ slightly 
bigger than $2$.

The asymptotics (\ref{eq:asymptbasic}) matches the result of a WKB approximation 
in the portion of the plane $\mathcal S_{-1} \cup \mathcal S_0 \cup \mathcal 
S_1$; as we will see, in order to correctly derive the asymptotic behaviour of the 
solution $y(x,E)$ in the rest of the complex plane, one must take care of the 
so-called \emph{Stokes phenomenon}. From here on we will call \emph{dominant} in 
any given sector a solution growing exponentially faster than any other in that 
sector; conversely the solution decaying faster is called \emph{subdominant}. 
One immediately notices (since there are only two solutions and we already know 
that, in each sector, one of them grows and one decays) that the basic solution 
$y(x,E)$ is subdominant $\mathcal S_0$ and dominant in $\mathcal S_{\pm 1}$.

Now, having a solution of the ODE in our hands, we can generate a whole family 
of solutions exploiting a trick introduced by Y. Sibuya (sometimes it is given the name 
of \emph{Symanzik rescaling}); consider the function $\hat y(x,E) \doteq y(a 
x,E)$ for some fixed $a\in\mathbbm C$: it is easy to see that it satisfies the 
following ODE

\eq
	\left[-\frac{d^2}{dx^2} + a^{2M+2}x^{2M} -a^2 E\right]\hat y(x,E) = 0 \; 
.
\en
If we choose $a$ such that $a^{2M+2} = 1$, by shifting $E\rightarrow a^{-2}E$, 
we see that $\hat y(x,a^{-2}E)$ solves (\ref{eq:realODE}). Thus, the following 
theorem is easily demonstrated

\begin{theo}
	Given a basic solution $y(x,E)$ to the ODE (\ref{eq:realODE}), the 
following functions
	\eq
		y_k(x,E) \doteq \omega^{\frac{k}{2}} 
y(\omega^{-k}x,\omega^{2k}E) \quad , \qquad \omega \doteq e^{\frac{2 \pi 
\mathbbm i}{2M+2}}
	\en
	satisfy the properties
	\begin{enumerate}
		\item $y_k \; , \ \forall k\in\mathbbm Z$ solves 
(\ref{eq:realODE});
		\item in the portion of the plane $\mathcal S_{k-1} \cup 
\mathcal S_k \cup \mathcal S_{k+1}$, the following asymptotics holds
		\begin{align}
				y_k &\underset{\vert x \vert 
\rightarrow\infty}{\sim} \frac{\omega^{\frac{M+1}{2}k}}{\sqrt{2 
\mathbbm i}}x^{-\frac{M}{2}}e^{-\frac{\omega^{-(M+1)k}}{M+1} x^{M+1}} \; ,	
\nonumber
\\	\label{eq:asymptrotated}
\\
				y_k' &\underset{\vert x \vert 
\rightarrow\infty}{\sim} -\frac{\omega^{-\frac{M+1}{2}k}}{\sqrt{2 
\mathbbm i}}x^{\frac{M}{2}}e^{-\frac{\omega^{-(M+1)k}}{M+1} x^{M+1}} \; ;	
\nonumber
			\end{align}	
		\item $y_k$ is, up to a constant, the unique solution to 
(\ref{eq:realODE}) subdominant in $\mathcal S_k$;
		\item the functions $y_k$ and $y_{k+1}$ are linearly independent 
for all $k\in \mathbbm Z$, thus each pair $\lbrace y_k,y_{k+1}\rbrace$ is a basis 
of solutions for (\ref{eq:realODE}).
	\end{enumerate}
\end{theo}
The last three properties are easily demonstrated by making use of the asymptotic 
of $y(x,E)$ (\ref{eq:asymptbasic}) and the fact that $(\omega^{-k})^{2M+2} = 1 
\; , \ \forall k \in \mathbbm Z$

Now, the fact that the pairs $\lbrace y_k,y_{k+1} \rbrace$ are a basis of our ODE 
means that the following relation

\eq
	y_{-1}(x,E) =C(E)y_0(x,E) + \widetilde C(E)y_1(x,E)
\label{eq:stokesrelation}
\en
must hold for some appropriate functions $C(E)$ and $\widetilde C(E)$. This is 
an example of \emph{Stokes relation} and the coefficients $C(E)$ and $\widetilde 
C(E)$ are called \emph{Stokes multipliers}. These can be expressed in terms of 
Wronskians, where the wronskian of two functions is defined in the usual way

\eq
	W[f,g](x) \doteq f(x)g'(x) - f'(x)g(x) \; .
\en
Recall that, given two solutions $f(x)$ and $g(x)$ of a second-order ODE, their 
Wronskian vanishes iff $f$ and $g$ are proportional; moreover if the first 
derivative term in the ODE vanishes, then the Wronskian of two solutions is 
independent of $x$, thanks to Abel's identity. For notational convenience we set

\eq
	W_{k_1,k_2}(E) \doteq W[y_{k_1},y_{k_2}](E) \equiv 
y_{k_1}(x,E)y_{k_2}'(x,E) - y_{k_1}'(x,E)y_{k_2}(x,E)
\en
and remark the following properties

\eq
	W_{k_1+1,k_2+1}(E) = W_{k_1,k_2}(\omega^2E) \qquad ; \qquad W_{0,1}(E) = 
1 \; .
\en

If we take the Wronskians $W_{-1,1}$ and $W_{-1,0}$ and make use of the Stokes 
relation (\ref{eq:stokesrelation}), we find

\eq
	C(E) = \frac{W_{-1,1}(E)}{W_{0,1}(E)} = W_{-1,1}(E) \qquad , \qquad 
\widetilde C(E) = -\frac{W_{-1,0}(E)}{W_{0,1}(E)} = -1
\en
so that we can rewrite the Stokes relation as

\eq
	C(E) y_0(x,E) = y_{-1}(x,E) + y_1(x,E) \; ,
\en
which, in terms of the basic solution $y(x,E)$, reads

\eq
	C(E) y(x,E) = \omega^{-\frac{1}{2}}y(\omega x,\omega^{-2}E) + 
\omega^{\frac{1}{2}}y(\omega^{-1}x,\omega^2E) \; .
\label{eq:stokesrelation2}
\en
This looks very similar to a T-Q relation, were it not for the $x$-dependence of 
the solution $y(x,E)$; but we can easily fix this by sending $x$ to zero, or, 
equivalently, by taking an $x$ derivative, which swaps the phase factors 
$\omega^{\pm1/2}$, and then set $x$ to zero. In formulae, defining

\eq
	D_-(E) \doteq y(0,E) \qquad , \qquad D_+(E) \doteq y'(0,E) \; ,
\en
the Stokes relation (\ref{eq:stokesrelation2}) becomes

\eq
	C(E)D_{\pm}(E) = \omega^{\pm\frac{1}{2}}D_{\pm}(\omega^{-2}E) + 
\omega^{\mp\frac{1}{2}}D_{\pm}(\omega^2E) \; ,
\en
which matches exactly the form of the T-Q relations (\ref{eq:continuumTQ}) and 
(\ref{eq:BLZTQ}) given that we set $\phi = 2 \pi p = \pi/(2M+2)$.

Even though we have been quite sketchy, we already see how concepts of ODE are 
related to those of IM:

\begin{center}
\begin{tabular}{|l | l|}
	\hline
	Six-vertex model with twist & Schr\"odinger equation with \\
	$\phi = 2 \pi p = \pi/(2M+2)$ & homogeneous potential $x^{2M}$ \\ \hline 
& \\
	Spectral parameter & Energy \\ & \\
	Anisotropy & Degree of potential \\ & \\
	Transfer Matrix & Stokes multiplier $C$ \\ & \\
	$Q$ operator & Value of $y(x,E)$ at $x=0$: $D_-(E)$ \\ & \\ \hline
\end{tabular}
\end{center}
If we were to replace $y$ by $y'$ on the last line, then we should have to change the 
twist to $\phi = - \pi/(2M+2)$.

A question arises naturally: what kind of objects exactly are $C$ and $D$, from the ODE point of view? It is not difficult to realise that they are \emph{spectral determinants} associated to particular eigenvalue 
problems, that is functions vanishing exactly at the eigenvalues of the 
associated problem: they can be regarded as infinite-dimensional analogues of 
the characteristic polynomial $\det(M-\lambda \mathbbm I)$ of a 
(finite-dimensional) matrix $M$. In order to see this, recall that $C(E)$ is 
equal to the Wronskian $W_{-1,1}(E)$ and thus vanishes if and only if $E$ is 
such that $y_{-1}$ and $y_1$ are linearly dependent; but this is true only if 
the ODE has a solution decaying simultaneously in the two sectors $\mathcal 
S_{-1}$ and $\mathcal S_1$ and this is exactly the lateral eigenvalue problem we 
discussed in \ref{subsec:eigencomplex} (taking in account the redefinition of 
$x$ and $E$). This is enough to deduce that $C(E)$ (to be precise $C(-E)$, given 
the redefinition) is precisely the spectral determinant for the Bender-Boettcher 
problem, up to a factor of an entire function with no zeroes; this 
ambiguity can be eliminated by using the Hadamard's factorisation theorem, see 
\cite{Dore_Tate_99} for details. By definition, the zeroes of $D_-(E)$ are those 
values of $E$ at which the function $y$, which is vanishing at $x=\infty$, also 
vanishes at $x=0$; likewise the zeroes of $D_+(E)$ correspond to points at 
which $y$ has vanishing first derivative at the origin. Thus also $D_{\pm}(E)$ are 
spectral determinants, but for the radial version of the Bender-Boettcher 
problem; note that their vanishing corresponds to the existence of normalisable 
wavefunctions, for the ODE in the full real axis, which are odd (for $D_-$) or 
even (for $D_+$).

Looking back at the table above, it is natural to ask: why should one particular 
value of the twist in the six-vertex model be singled out when making this 
connection with ODE? Or, better, is there a generalisation of our ODE allowing 
us to make a connection with a six-vertex model possessing a generic twist? The 
answer to this question was given, shortly after the original observation in 
\cite{Dore_Tate_99}, by V. V. Bazhanov, S. L. Lukyanov and A. B. Zamolodchikov in 
\cite{Bazh_Luky_AZam_01}, where they included in the ODE the 
angular-momentum-like term $l(l+1)x^{-2}$ showing how this addition allowed $Q$ 
operators for other values of the twist to be matched. In the next section we will 
briefly review these results, fill some gaps we left in this simplified 
discussion and finally give a complete mapping between the IM and the ODE.

\resection{Applying some glue}
\label{sec:corresp}

\subsection{The BLZ problem}
\label{subsec:BLZprob}

Let us now restore the angular-momentum-like term and consider the following 
eigenvalue problem

\eq
	\left[-\frac{d^2}{dx^2} + x^{2M} + \frac{l(l+1)}{x^2}\right]\Phi(x) = 
E\,\Phi(x) \; ,
\label{eq:BLZproblem2}
\en
where we already performed the $\pi/2$ rotation $x\rightarrow-\mathbbm i x \ , \ E 
\rightarrow -E$. This generalisation of the Bender-Boettcher problem was first 
studied by V. V. Bazhanov, S. L. Lukyanov and A. B. Zamolodchikov in 
\cite{Bazh_Luky_AZam_01}, hence we refer to it as \emph{BLZ equation} or, when 
providing boundary conditions, \emph{BLZ problem}.

Solutions to (\ref{eq:BLZproblem2}) behave, at the origin, as linear 
combinations of $x^{l+1}$ and $x^{-l}$ and a natural eigenproblem for this 
equation asks for values of $E$ such that a solution decaying at 
$x\rightarrow\infty$ and behaving as $x^{l+1}$ at $x\rightarrow 0$ exists; in WKB 
language, as we saw above, this is a radial problem. For $\mathfrak Re l> -1/2$ 
the condition at the origin is equivalent to the request that the dominant 
behaviour $x^{-l}$ be absent; outside this region the problem can be defined by 
analytic continuation.

Following the tracks of the simple example above, we apply Sibuya's trick 
starting from the basic solution $y(x,E,l)$, uniquely determined by its 
asymptotic behaviour

\eq
	y(x,E,l) \underset{\vert x \vert \rightarrow \infty}{\sim} 
\frac{x^{-\frac{M}{2}}}{\sqrt{2\mathbbm i}} e^{-\frac{1}{M+1}x^{M+1}} \; ,\qquad 
\textrm{arg}(x) < \frac{3\pi}{2M+2} \; ,
\en
and generating the family of functions

\eq
	y_k(x,E,l) \doteq \omega^{\frac{k}{2}} 
y(\omega^{-k}x,\omega^{2k}E,l) \; ,\qquad \omega \doteq e^{\frac{2 \pi 
\mathbbm i}{2M+2}} \; ,
\en
all of which, for $k\in \mathbbm Z$, solve (\ref{eq:BLZproblem2}). Just as 
before, any pair $\lbrace y_k,y_{k+1}\rbrace$ is a basis of the two-dimensional 
space of solutions and thus we can write $y_{-1}$ in terms of $y_0$ and $y_1$; 
rearranging we obtain:

\eq
	C(E,l) y_0(x,E,l)=y_{-1}(x,E,l) + y_1(x,E,l) \; ,
\label{eq:BLZStokesrel}
\en
where the Stokes multiplier $C(E,l)$ takes again the simple form

\eq
	C(E,l) = \frac{W[y_{-1},y_1]}{W[y_0,y_1]} = W[y_{-1},y_1]
\en

Up to here everything worked out the same way as in the $l=0$ case, but now we 
face a complication: the addition of the angular momentum term means that we 
cannot simply set $x=0$ in (\ref{eq:BLZStokesrel}) in order to eliminate the $x$ 
dependence. Instead we should ``project" the solution $y$, determined by its 
$\vert x \vert \rightarrow \infty$ asymptotics, onto another solution, defined 
by its $x\rightarrow 0$ behaviour, that is

\eq
	\psi(x,E,l) \underset{x\rightarrow 0}{\sim} x^{l+1} + O(x^{l+3}) \; .
\en
We can define a second solution by remarking that the equation 
(\ref{eq:BLZproblem2}) - but not the boundary conditions! - is invariant under 
the analytic continuation $l\rightarrow -l-1$, meaning that also 
$\psi(x,E,-l-1)$ solves our ODE. Moreover, the pair of solutions

\begin{align}
	\psi_+(x,E,l) &\doteq \psi(x,E,l)	\nonumber
\\	\label{eq:psibasis}
\\
	\psi_-(x,E,l) &\doteq \psi(x,E,-l-1)	\nonumber
\end{align}
are linearly independent (there are isolated values of $l$ which pose some 
problems, we will return to this later), since $\psi_- \sim x^{-l}$ near the 
origin.

If we take the Wronskian of $y_0$ with $\psi_{\pm}$ and use the 
(\ref{eq:BLZStokesrel}), we find an $x$-independent equation

\eq
	C(E,l)W[y_0,\psi_{\pm}](E,l) = 
W[y_{-1},\psi_{\pm}](E,l)+W[y_1,\psi_{\pm}](E,l) \; .
\en
In order to relate the Wronskians in the right-hand side to that in the 
left-hand side, we define the functions

\eq
	\psi_k(x,E,l) \doteq \omega^{\frac{k}{2}} 
\psi(\omega^{-k}x,\omega^{2k}E,l) \; ,
\en
which also solve the ODE. Considering the $x\rightarrow 0$ behaviour, we find 
that

\eq
	\psi_k(x,E,l) = \omega^{-\frac{2l+1}{2}k}\psi(x,E,l) \; .
\en
and, by using the easily demonstrated relation $W[y_k,\psi_k](E,l) = 
W[y,\psi](\omega^{2k}E,l)$ we arrive at

\eq
	W[y_k,\psi](E,l) = \omega^{\frac{2l+1}{2}k}W[y,\psi](\omega^{2k}E,l) \; 
.
\en
Finally, setting

\eq
	D_{\pm}(E,l) \doteq W[y,\psi_{\mp}](E,l) \quad \Rightarrow \  
D_+(E,l) \equiv D_-(E,-l-1)
\en
we obtain the T-Q relation

\eq
	C(E,l)D_{\pm}(E,l) = \omega^{\pm\frac{2l+1}{2}}D_{\pm}(\omega^{-2}E,l) + 
\omega^{\mp\frac{2l+1}{2}}D(\omega^2,l) \; .
\label{eq:BLZCDrelation}
\en

If we set

\eq
	\beta^2 = \frac{1}{M+1} \quad , \qquad p = \frac{2l+1}{4M+4}
\label{eq:BLZsubstitutions}
\en
then the match between (\ref{eq:BLZCDrelation}) and (\ref{eq:BLZTQ}) is perfect, 
with the following correspondence

\begin{align}
	T \quad &\leftrightarrow \quad C	\nonumber
\\
	Q_{\pm} \quad &\leftrightarrow \quad D_{\mp} \; .	\nonumber
\end{align}
We can also map the relation (\ref{eq:BLZCDrelation}) to the continuum limit 
form of Bethe ansatz equations (\ref{eq:continuumTQ}) by identifying $t_0$ with 
$C$ and $q_0(\pm \phi)$ with $D_{\pm}$ and setting the parameters $\eta = \pi 
M/(2M+2)$ and $\phi = \pi(2l+1)/(2M+2)$.

The identification we made between objects from the ODE world and objects 
arising in the IM context is still lingering at a formal level: if we want an 
exact mapping we have to take into consideration the analytical properties of 
these functions. We will focus on $D_-(E,l) \equiv D(E,l)$ since the properties of 
$D_+$ can be easily deduced from those. The following properties hold:

\begin{enumerate}
	\item $C$ and $D$ are entire functions of $E$;
	\item The zeroes of $D$ are all real and, if $l>-1/2$, they are all 
positive;
	\item The zeroes of $C$ are all real and, if $-1-M/2<l<M/2$, they are 
all negative;
	\item If $M>1$ the large-$E$ asymptotic of $D$ is
		\eq
			\ln D(E,l) \underset{\vert E 
\vert\rightarrow\infty}{\sim} \frac{a_0}{2}(-E)^{\mu} \; ,\qquad 
\vert\textrm{arg}(-E)\vert < \pi \; ,
		\en
		with
		\eq
			a_0 = -\Gamma(-\mu) 
\frac{\Gamma(\mu+\frac{1}{2})}{\sqrt{\pi}} \; ,\qquad \mu = \frac{M+1}{2M} \; 
;
		\en
	\item The zero-energy value of $D$ is
		\eq
			D(0,l) = \frac{\Gamma(1+\frac{2l+1}{2M+2})}{\sqrt{2 \pi 
i}}(2M+2)^{\frac{2l+1}{2M+2}+\frac{1}{2}} \; ;
		\en
	\item The function $D(E,l)$ can be simply represented as a product over 
its zeroes $E_k$:
		\eq
			D(E,l) = D(0,l) \prod_{k=1}^\infty 
\left(1-\frac{E}{E_k}\right) \; .
		\en
\end{enumerate}
The proofs of these properties can be found in \cite{Dore_Dunn_Tate_07}.

The corresponding properties of $T(s)$ and $Q_+(s)$, found, along with their 
proofs, in \cite{Bazh_Luky_AZam_97}, are, for $\beta^2$ in the semiclassical 
domain ($0<\beta^2<1/2$):

\begin{enumerate}
	\item $T$ and $Q_+$ are entire functions of $s$ with an essential 
singularity at infinity on the real axis;
	\item The zeroes of $Q_+(s,p)$ are all real and, if $2p>-\beta^2$, they 
are all strictly positive;
	\item The zeroes of $T(s,p)$ are all real and, if $\vert p \vert<1/4$, 
they are all negative;
	\item The large-$s$ asymptotics of the functions read
		\begin{align}
			\ln T(s,p) &\underset{\vert s \vert \rightarrow 
\infty}{\sim} 2\sqrt{\pi} 
\frac{\Gamma(1-\mu)}{\Gamma(\frac{3}{2}-\mu)}\Gamma\left(\frac{1}{2\mu}\right)^{
2\mu} s^\mu \; ,	\nonumber
		\\
		\\	\ln Q_{\pm}(s,p) &\underset{\vert s \vert \rightarrow 
\infty}{\sim} a_0 (M+1)\Gamma\left(\frac{1}{2\mu}\right)^{2 \mu} (-2)^\mu \; ,	
\nonumber
		\end{align}
		for $\textrm{arg}(-s)<\pi$, with $\mu=1/(2-2\beta^2)$ and $a_0$ 
defined above;
	\item The zero-$s$ value of $Q_+$ is
		\eq
			Q_{\pm}(0) = 1 \; ;
		\en
	\item The function $Q_{\pm}(s)$ can be represented as a product over its 
zeroes $s_k$:
		\eq
			Q_{\pm}(s) = \prod_{k=1}^\infty 
\left(1-\frac{s}{s_k}\right)
		\en
\end{enumerate}

With the substitutions (\ref{eq:BLZsubstitutions}) and comparing the properties 
we find that there is a perfect match between the $T-Q$ functions and the $C-D$ 
ones if we set

\begin{align}
	Q_{\pm}(s,p) &= \gamma_{\mp}D_{\mp}(\frac{s}{v}, \frac{2p}{\beta^2} 
-\frac{1}{2}) \; ;	\nonumber
\\
\\
	T(s,p) &= C(\frac{s}{v},\frac{2p}{\beta^2}-\frac{1}{2}) \; ,	
\nonumber
\end{align}
where we agree that $M+1 = \beta^{-2}$ and we introduced

\eq
	v = \frac{\Gamma\left(\frac{1}{2\mu}\right)^{-2}}{(2M+2)^\frac{1}{\mu}} 
\qquad , \qquad \gamma_{\mp} = D_{\mp}(0,\frac{2p}{\beta^2}-\frac{1}{2})^{-1} \; 
.
\en

\subsection{The fusion hierarchy}
\label{subsec:fusionhierarchy}

Aside from the T-Q relation, as we have seen in \ref{subsec:fusion}, there is 
a whole hierarchy of functional relations in integrable models; now that we have 
found a precise mapping between the T-Q relation and a Stokes relation, it is 
natural to ask whether this hierarchy admits an analogue on the ODE side: indeed this 
turns out to be the case \cite{Dore_Tate_99}, let us see how.

Since the pair of solutions $\lbrace y_k,y_{k+1}\rbrace$ are a basis in the 
space of solutions, we can safely write

\eq
	y_{k-1} = C_k^{(r)}y_{k+r-1} + \widetilde C_k^{(r)} y_{k+r} \quad , 
\qquad \forall r \in \mathbbm Z \; .
\label{eq:generalbasisexpansion}
\en
From these relations we see that a ``change of basis", from $\lbrace 
y_{k+r-1},y_{k+r}\rbrace$ to $\lbrace y_{k-1},y_k\rbrace$ can be encoded in a 
$2\times2$ matrix $\mathbf C_k^{(r)}$:

\eq
	\binom{y_{k-1}}{y_k} = \mathbf C_k^{(r)} \binom{y_{k+r-1}}{y_{k+r}} 
\; ,\qquad \mathbf C_k^{(r)} = \left(\begin{array}{c c}
		C_k^{(r)} & \widetilde C_k^{(r)} \\
		C_{k+1}^{(r-1)} & \widetilde C_{k+1}^{(r-1)}
	\end{array}\right) \; ,
\en
which depends on $E$ and $l$ but not on $x$. One easily shows that the following 
properties hold

\begin{align}
	\mathbf C_k^{(r)}(E,l) &= \mathbf C_{k-1}^{(r)} (\omega^2 E,l) \; ,	
\nonumber
	\\	\nonumber
	\\
	\mathbf C_k^{(0)} \equiv \mathbbm I_2 \quad &, \quad \mathbf C_k^{(1)} = 
\left(\begin{array}{c c}
		C_k^{(1)} & -1 \\
		1 & 0
	\end{array}\right) \; ,	\label{eq:Crnproperties}
	\\	\nonumber
	\\
	\mathbf C_k^{(r)}\mathbf C_{k+r}^{(n)} &= \mathbf C_k^{(r+n)} \; .	
\nonumber
\end{align}
By using these properties with $r=1$ we find

\begin{align}
	&C_k^{(1)} C_{k+1}^{(n)} -C_{k+2}^{(n-1)} = C_k^{(n+1)} \; ,	
\nonumber
	\\	\label{eq:twoCrelation}
	\\
	&C_k^{(1)}\widetilde C_{k+1}^{(n)} - \widetilde C_{n+2}^{(n-1)} = 
\widetilde C_k^{(n+1)} \; ,
\end{align}
which, combined with the ``initial conditions" above, give

\eq
	\widetilde C_k^{(n)} = -C_k^{(n-1)} \; .
\en
Now, setting

\eq
	C^{(n)}(E) \doteq C_0^{(n)}(\omega^{1-n}E) \; ,
\en
we easily see that (\ref{eq:twoCrelation}) can be rewritten as

\eq
	C(E)C^{(n)}(\omega^{n+1}E) = C^{(n-1)}(\omega^{n+2}E) + 
C^{(n+1)}(\omega^nE) \; ,
\label{eq:ODEcontfushier1}
\en
which matches the second fusion relation (\ref{eq:contfushier}), given we make 
the identification

\eq
	C^{(n)}(E) = t_{\frac{n}{2}}(vE) \; .
\en

Now, if we take the Wronskian of (\ref{eq:generalbasisexpansion}) with $y_{k+r}$ 
and $y_{k+r-1}$ we find

\eq
	C_k^{(r)} = W_{k-1,k+r} \qquad , \qquad \widetilde C_k^{(r)} = 
-W_{k-1,k+r-1} \; ,
\en
from which we deduce

\eq
	C_k^{(r)} = C_{k+r+1}^{(-r-2)} \; .
\en
This relation, combined with the properties (\ref{eq:Crnproperties}) for $n=-r$, 
implies

\eq
	C^{(r-1)}(\omega^{-1}E)C^{(r-1)}(\omega E) - C^{(r)}(E)C^{(r-2)}(E) =1 
\; ,
\label{eq:ODEcontfushier2}
\en
that reproduces the first fusion relation (\ref{eq:contfushier}), given we make 
the identification above.

\begin{figure}[h]
\centering
\includegraphics[scale=1]{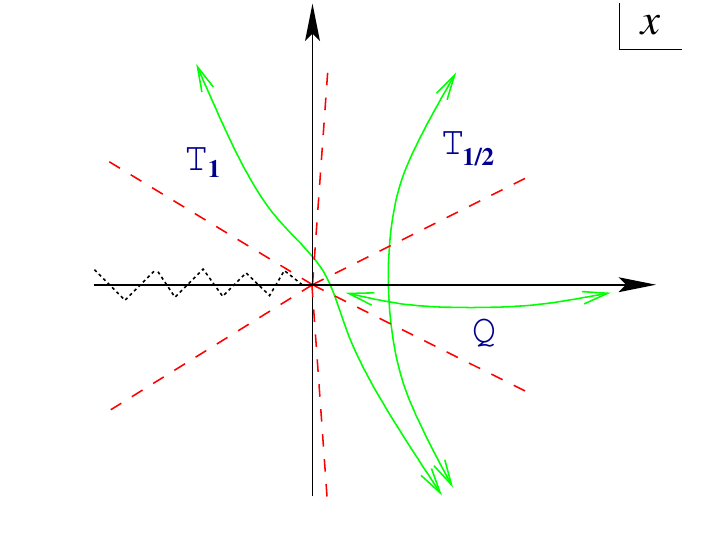}
\caption{Some of the possible quantisation contours and their associated spectral determinants.}
\label{fig:Tasspectral}
\end{figure}

An interesting fact is that we can express $T_{n/2}(vE)$ in terms of Wronskians:

\eq
	T_{\frac{n}{2}}(vE) = C^{(n)}(E) = W_{-1,n}(\omega^{1-n}E) \; ;
\en
this shows how also the fused transfer matrices can be interpreted as spectral 
determinants. In particular, $T_{n/2}(vE)$ vanishes iff $E$ is such that the ODE 
has a nontrivial solution simultaneously decaying at $\vert x\vert \rightarrow 
\infty$ in sectors $\mathcal S_{-1}$ and $\mathcal S_n$; in other words the 
zeroes of $T_{n/2}$ are the eigenvalues for the lateral problem defined in the 
sectors $\mathcal S_{-1}$ and $\mathcal S_n$. The Figure  \ref{fig:Tasspectral} 
illustrates this fact.

The truncation of the fusion hierarchy now has a clear interpretation: whenever 
$M\in \mathbbm Q$, the functions $y_k$ are (quasi-)periodic in $k$; this 
periodicity is a consequence of the fact that solutions to the ODE live on a 
finite cover of $\mathbbm C\backslash\lbrace0\rbrace$\footnote{Actually this is 
exact only if $l(l+1)=0$; in other cases one must take care of the monodromy 
around $0$, but, essentially, the story remains the same.}.

As an example, take $2M \in \mathbbm Z^+$ and $l(l+1)=0$; in these cases all the 
solutions of the ODE are single-valued functions of $x$ and the sectors 
$\mathcal S_{n+2(M+1)}$ coincide with $\mathcal S_n$. This implies that both 
$y_{n+2(M+1)}$ and $y_n$ are subdominant in $\mathcal S_n$ and, thus, are 
proportional. We conclude that

\eq
	C^{(2M)}(E)=1 \qquad , \qquad C^{(2M+1)}(E)= 0
\en
and the relation (\ref{eq:ODEcontfushier2}) truncates to

\eq
	C^{(r)}(\omega^{-1}E)C^{(r)}(\omega E) = 1 + \prod_{n=1}^{2M-1} 
\left[C^{(n)}(E)\right]^{G_{nr}}
\en
matching perfectly the T-system (\ref{eq:TSystem}).

\subsection{One more functional relation}
\label{subsec:morefuncrel}

Lastly we wish to discuss one final set of functional relations, that is the 
quantum Wronskian (\ref{eq:BLZquantwronsk}) and its partner relations 
(\ref{eq:partnerrelation}), expressing the fused transfer matrices $T$ in terms 
of the operators $Q$. Let us return for a moment to the solutions $\psi_{\pm}$, 
defined in (\ref{eq:psibasis}) by means of their behaviour around $x=0$; their 
Wronskian is easily calculated in the neighbourhood of $x=0$

\eq
	W[\psi_-,\psi_+] = 2l + 1 \; .
\label{eq:psi-+wronsk}
\en
Since $\lbrace \psi_+,\psi_-\rbrace$ is a basis, we can write, remembering that 
$D_{\mp}=W[y,\psi_{\pm}]$, the following relation

\eq
	(2l+1)y(x,E,l) =D_-(E,l)\psi_-(x,E,l) - D_+(E,l)\psi_+(x,E,l) \; .
\en
Note that there's a problem with this expansion at $l=-1/2$, easily understood 
since, at this point, the solutions $\psi_+$ and $\psi_-$ coincide and thus form 
no longer a basis. Truth be told, this is not the sole point where difficulties 
arise; in fact, while for $\mathfrak Re\, l >-1/2$ the solution $\psi_+(x,E,l)$ 
can be proven to exist (see, for example, \cite{Chen_62,Newt_64,Squi_63}), it is 
not so in the left half-plane. This is the reason why the second solution 
$\psi_-(x,E,l)$ had to be defined through analytic continuation; still, at 
isolated values of $l$ in the half-plane $\mathfrak Re\, l<-1/2$, poles may 
arise and make $\psi_-$ ill-defined. Even though it is possible to regularise 
$\psi_-$ multiplying it by an appropriate factor, this inevitably inserts spurious zeroes in the Wronskian (\ref{eq:psi-+wronsk}) and the points where 
$\psi_-$ failed to exist turn into points where the regularised $\widetilde 
\psi_-$ becomes dependent of $\psi_+$; for the simple power-law potential 
$x^{2M}$ these points are \cite{Dore_Tate_99}

\eq
	l+\frac{1}{2} = \pm (m_1 + (M+1) m_2) \quad , \qquad m_1,m_2 \in \mathbbm 
Z^+ \; ,
\en
corresponding to values of the twist

\eq
	2 p = \pm (m_1\beta^2 + m_2) \; .
\en
Notice that at the points

\eq
	2p = \pm m_2 \quad , \qquad m_2\in \mathbbm Z^+
\en
the quantum Wronskian (\ref{eq:BLZquantwronsk}) vanishes, while at

\eq
	2p = -\beta^2 - m_2 \quad , \qquad m_2 \in \mathbbm Z^+
\en
there is a normalisation problem for $Q_+(s,p)$, since a zero level ($s_k=0$ for 
some $k$) arises.

The problem described here can be cured by means of a limiting procedure 
\cite{Dore_Tate_99}, however, we will agree to pick $l$ such that these subtleties 
do not present themselves and $\lbrace \psi_+,\psi_-\rbrace$ does indeed provide 
a basis for the space of solutions. Defining the pairs

\eq
	\psi^{\pm}_k(x,E,l) = \omega^{\frac{k}{2}}\psi_{\pm}(\omega^{-k}x, 
\omega^{2k}E,l) \quad , \qquad k \in \mathbbm Z \; ,
\en
whose Wronskian are calculated straightforwardly

\eq
	W[\psi_k^+,\psi_j^+] = W[\psi_k^-,\psi_j^-] = 0 \; ,\qquad 
W[\psi_k^+,\psi_j^-] = (2l+1)\omega^{(k-j)\frac{2l+1}{2}} \; ,
\en
we can expand the ``rotated" solutions $y_k$ as

\eq
	(2l+1)y_k(x,E,l) = D_-(\omega^{2k}E,l)\psi_k^-(x,E,l) - 
D_+(\omega^{2k}E,l)\psi_k^+(x,E,l) \; ;
\en
substituting this expansion in $W[y_{-1},y_0] =1$ we obtain

\eq
	\omega^{-\frac{2l+1}{2}}D_-(\omega^{-1}E)D_+(\omega E) - 
\omega^{\frac{2l+1}{2}}D_-(\omega E)D_+(\omega^{-1}E) = 2l+1 \; ,
\en
which corresponds to the quantum Wronskian relation (\ref{eq:BLZquantwronsk}). 
If we now take the Wronskian of $y_{-1}$ with $y_n$ we easily obtain

\begin{align}
	(2l+1) C^{(n)}(E) = 
&\omega^{-(n+1)\frac{2l+1}{2}}D_-(\omega^{-n-1}E)D_+(\omega^{n+1}E) \\	
\nonumber
	&- \omega^{(n+1)\frac{2l+1}{2}}D_-(\omega^{n+1}E)D_+(\omega^{-n-1}E) \; 
,
\end{align}
corresponding to the relation (\ref{eq:partnerrelation}).

\subsection{The correspondence dictionary}
\label{subsec:dictionary}

In the table below we summarise the mapping between objects living in the world 
of integrable models and objects belonging to the realm of differential 
equations.

\begin{center}
\begin{tabular}{|l | l|}
	\hline
	Integrable model & Schr\"odinger equation \\
	\hline & \\
	Spectral parameter & Energy \\ & \\
	Anisotropy & Degree of potential \\ & \\
	Twist parameter & Angular momentum \\ & \\
	Fused transfer matrices & Spectral determinants for lateral problems \\
	 & at $\vert x \vert = \infty$ \\ & \\
	$Q$ operators & Spectral determinants for radial problems linking \\
	 & $\vert x \vert = \infty$ with $\vert x \vert =0$ \\ & \\
	Truncation of the & Solutions on finite covers of $\mathbbm 
C\backslash\lbrace0\rbrace$ \\
	fusion hierarchy & \\ & \\ \hline
\end{tabular}
\end{center}

We might go further in the generalisation and consider the Schr\"odinger equation 
with anisotropic potential or even more complicated generalisations; however 
what has been exposed in this already long chapter, is more than sufficient to 
the needs of the next sections. Thus we suggest the interested reader to refer 
to the review \cite{Dore_Dunn_Tate_07} and references therein.

We finish this section with a table recollecting the notations used for the 
various objects appearing in the lattice model, in the continuum model and in 
the Schr\"odinger equation; the entries of this table are in correspondence with 
those of the preceding one:

\begin{center}
\begin{tabular}{|c | c| c |}
	\hline
	Lattice integrable & Continuum integrable & Schr\"odinger equation \\
	model & model &  \\ \hline  & & \\
	$\nu$ & $s$ & $E$ \\ & & \\
	$\eta$ & $\beta$ & $M$ \\ & & \\
	$\phi$ & $p$ & $l$ \\ & & \\
	$t^{(m)}$ & $T_m$ & $C^{(2m)}$ \\ & & \\
	$q_0(\nu,\pm\phi)$ & $Q_{\pm}$ & $D_{\mp}$ \\ & & \\ \hline
\end{tabular}
\end{center}

\chapter[The Toda Field Theories]{The Toda Field Theories}
\label{chap:Toda}
\markboth{Chapter 2 - The Toda Field Theories}{}

In the previous chapter we have seen how it is possible to give to objects appearing in 
the analysis of the six-vertex model an alternative interpretation in terms of 
eigenvalue problems for a certain ODE. It is natural to ask wether this duality 
is restricted to the sole case of the six-vertex model or it is the hint of a more 
general feature of integrable systems. In fact, in the years that followed its 
first discovery, the ODE/IM correspondence was successfully extended to a number 
of integrable models, a partial list of which can be found in 
\cite{Dore_Dunn_Tate_07}. Actually, it seems that this correspondence represents 
a somewhat founding property of integrable models.

A very interesting class of integrable models are the so-called \emph{Toda field 
theories}. In this chapter we will first introduce these models and review their properties. Their classical integrability relies on the existence of a pair of operators, called \emph{Lax pair}, whose commutation implies the equations of motion of the model. These operators can be regarded as holomorphic and anti-holomorphic components of a complex Lie algebra-valued connection (covariant derivative). The vanishing of their commutator is then interpreted as a flatness (or zero curvature) condition for this connection. Associated to the Lax connection is a consistency condition, which takes the form of a pair of first order linear vector differential equations, one holomorphic and the other anti-holomorphic. The monodromy properties of the solutions to this pair of equations fully encodes the information on the classical integrability of the model. We will then deform our model by introducing a pair of potentials in the equations of motion. The effect of these potentials is to introduce a highly non-trivial monodromy around the point at infinity in the complex plane which, in turn, will be a necessary for the construction of the functional relations amongst the spectral determinants.

\section{Definition and properties}
\label{sec:defandprop}

The Toda field theories are named after Morikazu Toda who introduced in 
\cite{Toda_67} a simple model for a one-dimensional crystal. This model, that would
later be given the name of \emph{Toda lattice}, describes a chain of 
particles with nearest-neighbour interaction dictated by the equations of motion

\eq
	\dot p_n(t) = e^{-[q_n(t)-q_{n-1}(t)]} - e^{-[q_{n+1}(t)-q_n(t)]} \quad 
, \qquad \dot q_n(t) = p_n(t) \; ,
\en
where $q_n(t)$ is the displacement of the $n$-th particle from its equilibrium 
position and $p_n(t)$ is the momentum of the said particle (we have set the mass 
$m=1$); the dot represents the time derivative. This model is a prototypical 
example of a completely integrable model with soliton solutions and can be solved 
by means of the inverse scattering transform. A suggestive way to write the EoMs 
of this model is the following:

\eq
	\ddot q_n(t) = - \sum_{i=1}^r \alpha_i^n \exp\left[\sum_{k=1}^r 
\alpha_i^k q_k(t)\right] \; ,
\en
with the $r$-dimensional vectors $\boldsymbol{\alpha}_i$ being the simple roots 
of the Lie algebra $\mathfrak a_r$. This way of writing the EoMs allows to perform an 
immediate generalisation of the Toda lattice, namely by letting 
$\boldsymbol{\alpha}_i$ be the simple roots of a generic semi-simple Lie algebra 
$\mathfrak g$ \cite{Kons_79,Bogo_76}, not necessarily finite-dimensional.

A further generalisation is possible: let us simply introduce a further dimension 
$x$ and modify the EoMs to be

\eq
	\left( \partial_t^2 - \partial_x^2 \right)\eta^k(t,x) = - \sum_{i=1}^r 
\alpha^k_i e^{\boldsymbol{\alpha}_i \cdot \boldsymbol{\eta}(t,x)} \; ,
\label{eq:generTodalattice}
\en
where the vector field $\boldsymbol{\eta}(t,x)$ generalises to two dimensions 
the variables $q_n(t)$. By letting $\boldsymbol{\alpha}_i$ be the simple roots 
of whatever Lie algebra, we have obtained a whole family of 2D models which have 
been shown to be integrable \cite{Mikh_79,Mikh_Olsh_Pere_81}. The model 
described by the equation (\ref{eq:generTodalattice}) are called 
\emph{generalised 2D Toda lattices} or \emph{2D Toda field theories}.

Summarising and restoring the bare mass $m$ and the coupling constant $\beta$ 
that we had set to $1$, a Toda field theory over a semi-simple Lie (or 
Lie-Ka\v{c}-Moody) algebra $\mathfrak g$ is a theory of $r=\textrm{rk}(\mathfrak 
g)$ scalar fields in two-dimensional Minkowski space-time, which we collect in a 
vector $\boldsymbol{\eta}$. The classical field theory is determined by the 
Lagrangian density

\eq
	\mathcal L = 
\frac{1}{2}(\partial_{\mu}\boldsymbol{\eta},\partial^\mu\boldsymbol{\eta}) 
-\frac{m^2}{\beta^2} \sum_i^r n_i e^{\beta 
(\boldsymbol{\alpha}_i,\boldsymbol{\eta})} \; ,
\label{eq:Todalagrangian}
\en
with $m$ and $\beta$ being real (classically irrelevant) constants, $\balpha_i$ 
the simple roots of the algebra $\mathfrak g$ and $n_i$ a set of integer called 
\emph{Coxeter labels}\footnote{These numbers are characteristic of each type of 
algebra; they are tabulated in many places, see for example 
\cite{Kac_83,Fuch_92}.}; the field $\boeta \equiv \boeta(t,x)$ lives in the 
Cartan subalgebra $\mathfrak h \subset \mathfrak g $ and the scalar product 
$(\cdot , \cdot)$ is built on the Killing form of $\mathfrak g$. In 
(\ref{eq:Todalagrangian}) we have left the summation inferior limit unspecified, 
in fact the index $i$ can run either from $1$ to $r$ or from $0$ to $r$; these 
two choices give rise to rather different theories:

\begin{itemize}
	\item letting the lower limit of the summation be $1$ means that our 
model encodes solely data from the finite-dimensional semi-simple Lie algebra 
$\mathfrak g$. In this case the theory is conformal (meaning it has no mass 
scale) both classically and after quantisation and will be referred to as 
\emph{conformal Toda field theory} (CTFT);
	\item if we allow the term $i=0$ to be in the sum, by letting
		\eq
			\balpha_0 = -\sum_{i=1}^r n_i \balpha_i \; ,
		\en
		then the theory will inherit information from the affine algebra 
$\hat{\mathfrak g}$ associated with $\mathfrak g$. The presence of the root 
$\balpha_0$ breaks the conformal symmetry introducing a mass scale, even though 
the theory remains classically integrable\footnote{Actually, by letting the 
scalar field live no longer in an Euclidean space (the Cartan subalgebra of a 
finite-dimensional Lie algebra), but rather in a space with signature $(r-1,1)$ 
(the full Cartan subalgebra of the affine Lie algebra), it is possible to restore 
the conformal symmetry; however, this procedure introduces spurious unphysical 
degrees of freedom and the energy is no longer a positive definite functional of 
the field components \cite{Corr_94,Babe_Bono_91,Bono_92}.}.
\end{itemize}
In the following we will be concerned exclusively with 
affine Toda field theories (ATFT).

The interest in this class of theories, put aside the fact that they are 
immediate generalisations of the Liouville and sin(h)-Gordon model, sparked as a 
consequence of an observation made in \cite{AZam_89_2} by A. B. Zamolodchikov; he 
suggested a particular approach to the study of conformal field theory, in the 
specific case the $c=1/2$ CFT, corresponding to the Ising model at the critical 
point, and revealed, through indirect arguments, an integrable structure, 
apparently connected with the Lie algebra $\mathfrak e_8$. In short, he sought a 
set of conserved quantities and used them as a guide to set up a minimal 
solution to the exact $S$-matrix bootstrap; for the $c=1/2$ CFT he was successful 
in finding a set of conserved quantities with spins coprime to $30$ (\emph{i.e.} 
the exponents of $\mathfrak e_8$ modulo its Coxeter number) and in building the 
minimal solution, which consists in a theory of $8$ scalar particles. It is well 
known that the $c=1/2$ CFT can be obtained as the coset model $\mathfrak 
e_8^{(1)}\times \mathfrak e_8^{(1)} /\mathfrak e_8^{(2)}$ and it seems that the 
perturbation by the operator of dimension $(1/16, 1/16)$ reveals the $\mathfrak 
e_8$ structure. Subsequently concrete connections between the perturbed 
conformal field theories and affine Toda field theories have been suggested and 
proved by many authors, among which we wish to remember T. J. Hollowood and P. 
Mansfield \cite{Holl_Mans_89}, H. W. Braden, E. Corrigan, P. E. Dorey and R. 
Sasaki \cite{Brad_Corr_Dore_Sasa_89_1,Brad_Corr_Dore_Sasa_89_2,Brad_Corr_Dore_Sasa_90}, C. Destri and H. J. 
de Vega \cite{Dest_Vega_89,Dest_Vega_91_1,Dest_Vega_91_2}, P. Christe and G. Mussardo 
\cite{Chri_Muss_90}.

Let us now review some facts about ATFT. The classical equations of motion 
corresponding to the Lagrangian (\ref{eq:Todalagrangian}) are

\eq
	\left(\partial_t^2-\partial_x^2\right)\boeta + \frac{m^2}{\beta} 
\sum_{i=0}^r n_i \balpha_i e^{\beta (\balpha_i,\boeta)} = 0 \; ,
\label{eq:TodaEom}
\en
and, expanding the interaction term in the Lagrangian around the minimum $\boeta 
= \bzero$

\begin{align}
	V(\boeta) &\equiv \frac{m^2}{\beta^2}\sum_{i=0}^r n_i 
e^{\beta(\balpha_i,\boeta)} = \frac{m^2}{\beta^2} \sum_{i=0}^r n_i +\nonumber
	\\
	\\ & + \frac{m^2}{2} \sum_{i=0}^r n_i \alpha_i^a \alpha_i^b \eta_a 
\eta_b + \frac{m^2 \beta}{6} \sum_{i=0}^r n_i \alpha_i^a \alpha_i^b \alpha_i^c 
\eta_a \eta_b \eta_c + \cdots \  ,	\nonumber
\end{align}
we can extract the mass matrix and the three-point coupling

\eq
	\left(M^2\right)^{a b} = m^2 \sum_{i=0}^r n_i \alpha_i^a \alpha_i^b 
\; ,\qquad c^{a b c} = m^2\beta \sum_{i=0}^r n_i \alpha_i^a \alpha_i^b 
\alpha_i^c \; .
\en
The magnitude of the non-vanishing three point couplings is related to the 
eigenvalues $\lbrace m_a \rbrace_{a=1}^r$ of the mass matrix by the 
quasi-universal formula\footnote{There are slight modifications for the cases 
where $\hat{\mathfrak g}$ is an untwisted non simply-laced affine algebra.}

\eq
	\vert c^{a b c} \vert = \frac{2 \beta}{\sqrt{h}} m_a m_b \sin \theta_{a 
b}^c \; ,
\en
where $h = \sum_{i=0}^r n_i$ is the Coxeter number of $\hat{\mathfrak g}$ and 
$\theta_{a b}^c$, called \emph{fusion angle}, is supplementary to the angle $ 
\overline{\theta}_{a b}^c = \pi - \theta_{a b}^c$ between $m_a$ and $m_b$ in the 
triangle with sides having length equal to the three masses $m_a$, $m_b$ and 
$m_c$.

\begin{figure}[h]
\centering
\includegraphics[scale=0.5]{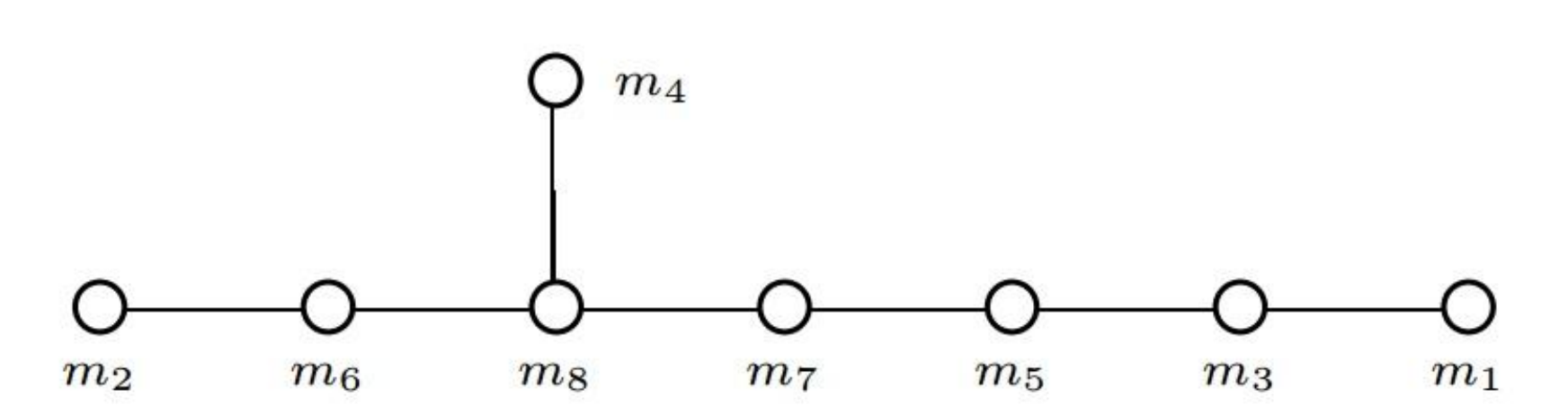}
\caption{Assignement of the masses to the nodes of the Dynkin diagram.}
\label{fig:e8masses}
\end{figure}

The mass matrix was diagonalised, for most cases, quite some time ago 
\cite{Mikh_Olsh_Pere_81,Wils_81,Oliv_Turo_83_1,Oliv_Turo_83_2} and can be found, along with the 
three point couplings, in many papers, amongst which we suggest 
\cite{Brad_Corr_Dore_Sasa_90}; however we wish to remark a nice property of the 
eigenvalues of the mass matrix. Exception be made for the twisted algebras, it is 
possible to choose an ordering of the masses such that $\mathbf{m} = 
(m_1,m_2,\ldots,m_r)$ and

\eq
	C \mathbf{m} = \left(2 \sin \frac{\pi}{2 h}\right)^2 \mathbf{m} \; ,
\label{eq:cartaneigenvaluemass}
\en
with $C$ being the Cartan matrix of the algebra $\mathfrak g$; in words the mass 
matrix eigenvalues $m_a^2$ are the square of the components of the 
lowest-eigenvalue eigenvector of the Cartan matrix. This is quite remarkable 
since it allows the particles to be assigned unambiguously, up to mass 
degeneracies, to the nodes of the Dynkin diagram for $\mathfrak g$. Curiously 
enough, the masses are assigned following the ``weight" ordering in terms of the 
dimension of the fundamental representations assigned to the nodes. For example, 
the $a_n^{(1)}$ masses (they are $m_a = 2 m \sin\frac{\pi a}{h}$) 
increase from the ends of the Dynkin diagram towards the centre and are doubly 
degenerate, corresponding to the $\mathbbm Z_2$ symmetry of the diagram; for 
$\mathfrak e_8^{(1)}$ the masses are assigned as shown in Figure  \ref{fig:e8masses}.

Even more remarkably, for the \textit{ADE} series of simply-laced algebras (and, 
strangely enough, for one of the twisted cases $\hat{\mathfrak a}_{2n}^{(2)}$), the 
classical mass ratios are preserved in the perturbative theory, at least to 
one-loop order \cite{Brad_Corr_Dore_Sasa_90,Chri_Muss_90}; this is suggestive of 
a generalisation of the equation (\ref{eq:cartaneigenvaluemass}):

\eq
	C \mathbf{q}^{(s)} = \left(2 \sin \frac{s \pi}{2 h}\right)^2 
\mathbf{q}^{(s)}
\en
where the components of $\mathbf{q}^{(s)} = (q_1^{(s)},q_2^{(s)}, \ldots 
,q_r^{(s)})$ are the eigenvalues of the conserved charges $Q_s$ corresponding to 
single-state particles $\vert a \rangle \; , \; a = 1,\ldots,r$ with rapidities 
$\theta_a$:

\eq
	Q_s \vert a\rangle = q_a^{(s)} e^{s\, \theta_a} \vert a \rangle \quad ; 
\quad q_a^{(\pm 1)} = m_a \; .
\label{eq:conservedchargeseigenvalues}
\en
This fact is actually true in the quantum theory, in the sense that it is 
consistent with other known facts.

\section{Integrability, Lax pair and associated linear system}
\label{sec:integrabilityandLax}
\setcounter{equation}{0}

In the definition of the conserved charges eigenvalues 
(\ref{eq:conservedchargeseigenvalues}), we silently assumed that the operators 
$Q_s$ are conserved and in involution after quantisation of the theory. While, a 
priori, we have got nothing that assure us of the conservation at a quantum level 
of the charges $Q_s$, these are granted to be classically conserved by the 
classical integrability of the ATFT, which means, among other facts, that the 
equations of motion (\ref{eq:TodaEom}) can be solved explicitly by means of the 
\emph{inverse scattering method}. This technique relies essentially on the 
existence of two matrix operators $\, \mathcal U(\boeta\vert t,x;\lambda) = 
\partial_t + U(\boeta\vert t,x;\lambda)$ and $\mathcal V(\boeta\vert 
t,x;\lambda) = \partial_x + V(\boeta\vert t,x;\lambda)$, called \emph{Lax pair}, 
which depend on the field $\boeta$, its derivatives and a spectral parameter 
$\lambda$ and are built in such a way that they reproduce the EoMs through a 
\emph{zero curvature condition}

\eq
	\left[\mathcal U(\boeta\vert t,x;\lambda),\mathcal V(\boeta\vert 
t,x;\lambda)\right] = 0 \quad \Leftrightarrow \quad 
\left(\partial_t^2-\partial_x^2\right)\boeta + \frac{m^2}{\beta} \sum_{i=0}^r 
n_i \balpha_i e^{\beta (\balpha_i,\boeta)} = 0 \; ,
\label{eq:ZCC}
\en
for any values of $\lambda$, where $\left[\cdot \,,\cdot\right]$ denotes the 
usual commutator. It is rather straightforward to verify that the operators can 
be chosen to be

\begin{align}
	U(\boeta\vert t,x;\lambda) &\doteq \frac{\beta}{2} 
\left(\partial_x\boeta,\mathbf{h}\right) + \sum_{i=0}^r \mu_i \left(\lambda e_i 
- \frac{1}{\lambda}f_i\right) e^{\frac{\beta}{2}(\balpha_i,\boeta)}	
\nonumber
	\\	\label{eq:UVoperators}
\\	V(\boeta\vert t,x;\lambda) &\doteq \frac{\beta}{2} 
\left(\partial_t\boeta,\mathbf{h}\right) + \sum_{i=0}^r \mu_i \left(\lambda e_i 
+ \frac{1}{\lambda}f_i\right) e^{\frac{\beta}{2}(\balpha_i,\boeta)} \; ,	
\nonumber
\end{align}
where the coefficients are chosen as $\mu_i^2 = m^2 \vert\balpha_i\vert^2 
n_i/8$. The components of the vector $\mathbf{h}$ are the generators of the 
Cartan subalgebra of $\mathfrak g$, while the objects $e_i$ and $f_i$ are the 
positive and negative step operators corresponding to the simple root 
$\balpha_i$; thus, they satisfy the Cartan-Weyl relations

\begin{align}
	\left[\mathbf{h},e_i\right] = \balpha_i e_i \; ,\qquad 
\left[\mathbf{h},f_i\right] = -\balpha_i f_i \; ,\qquad \left[e_i,f_j\right] 
&= \delta_{ij} \frac{2(\balpha_i,\mathbf{h})}{\vert\balpha_i\vert^2} \; .
\end{align}

The zero curvature condition (\ref{eq:ZCC}) can be interpreted as a consistency 
condition for the following linear system (we omit the explicit dependence on 
the field $\boeta$)

\begin{align}
	\left[\partial_t + \mathscr R^d U(t,x;\lambda)\right] \bPsi(t,x;\lambda) 
&= 0 \; ,	\nonumber
\\
\\	\left[\partial_x + \mathscr R^d V(t,x;\lambda)\right] \bPsi(t,x;\lambda) 
&= 0 \; .	\nonumber
\end{align}
Here $\mathscr R^d U$ and $\mathscr R^d V$ denote the $d$-dimensional 
representations of the abstract operators (\ref{eq:UVoperators}), while 
$\bPsi(t,x)$ is an auxiliary $d$-dimensional vector field\footnote{Even though 
we will only consider finite dimensional representations, nothing forces us to set 
$d<\infty$.}. From now, with a slight abuse of notation, we will write $U$ and $V$ 
to denote both their abstract definition and their $d$-dimensional 
representation. It is useful to consider translations of this vector field along 
the $x$ direction, at fixed time $t$ (we suppress the explicit dependence on 
time)

\eq
	\bPsi(x;\lambda) = \mathcal T(x,x';\lambda) \bPsi(x';\lambda) \; ,
\label{eq:psitranslation}
\en
where the \emph{transition matrix} $\mathcal T(x,x';\lambda)$, performing the 
translation, is defined on an interval $[x',x] \; , \; x \geq x'$ and satisfies 
the following requirements

\eq
	\left[\partial_x + V(x;\lambda)\right] \mathcal T(x,x';\lambda) =0 \quad 
, \quad \mathcal T(x,x;\lambda)=0 \; ,
\en
which have the following formal solution:

\eq
	\mathcal T(x,x';\lambda) = \mathcal P 
\exp\left\lbrace-\int\limits_{x'}^x V(\xi;\lambda) d\xi\right\rbrace
\en
with $\mathcal P$ denoting the path ordering of non-commutative factors. The 
transition matrix possesses also the following group-like property, which 
follows quite obviously from the definition (\ref{eq:psitranslation}):

\eq
	\mathcal T(x,x'';\lambda) \mathcal T(x'',x';\lambda) = \mathcal 
T(x,x';\lambda) \; ,\qquad x\geq x'' \geq x' \; .
\en
The transition matrix for the $x$-axis $\mathbbm T(\lambda) \doteq \mathcal 
T(\infty,-\infty;\lambda)$ is called \emph{monodromy matrix}; since, as it is 
straightforward to see

\eq
	\partial_t \mathcal T(x,x';\lambda) = \mathcal 
T(x,x';\lambda)U(x';\lambda)-U(x;\lambda)\mathcal T(x,x';\lambda) \; ,
\en
if we require $\partial_x \boeta \rightarrow 0$ as $x\rightarrow \pm \infty$ and 
 $\boeta(\infty) = \boeta(-\infty) + 2 \bkappa \ , \ (\bkappa,\balpha_i)\in 
\mathbbm Z$ it follows that the trace of the monodromy matrix is time 
independent:

\eq
	M(\lambda) \doteq \tr \mathbbm T(\lambda) \quad , \qquad \partial_t\, 
M(\lambda) = 0 \; .
\en

A very important fact about the Lax pair is the possibility of projecting them 
into the Cartan subalgebra $\mathfrak h$ by means of a gauge transformation, 
effectively making them diagonal \cite{Oliv_Turo_85,Oliv_Turo_86}; after the 
gauge transformation the potential $V(x;\lambda)$ takes the form\footnote{An 
alternative diagonalisation exists for which the gauge transformed Lax operator 
$v(x;\lambda)$ is expressed in terms of $\mathcal F = \sum_{i=0}^r \mu_i f_i$ 
and positive powers of $\lambda$.}

\eq
	v(x;\lambda) = \lambda \mathcal E + \sum_{s\in \textit{ex}\,(\mathfrak 
g)} \lambda^{-s}\widetilde h_sI_0^{(s)} \; ,\qquad \mathcal E = \sum_{i=0}^r 
\mu_i e_i \; ,
\en
where we denoted with $\textit{ex}\,(\mathfrak g)$ the set of $r$ exponents of 
the algebra $\mathfrak g$ modulo the Coxeter number $h$ and $\widetilde h_s 
\equiv \widetilde h_{s+ nh} \; , \; \forall n\in \mathbbm Z$ are appropriate 
elements of the Cartan subalgebra. Since now the zero-curvature condition reads

\eq
	\partial_t v(x,t;\lambda) = \partial_x u(x,t;\lambda) \; ,
\en
the following objects constitute indeed a set of infinitely many conserved 
charges

\eq
	Q_s = \int\limits_{-\infty}^\infty dx I_0^{(s)} \quad , \qquad 
\partial_t \, Q_s = 0 \; .
\en
The label $s$ is called spin of the charge, in fact, under a Lorentz 
transformation, the spectral parameter scales as $\lambda \rightarrow l \lambda$ 
which reflect into a scaling of the conserved charges $Q_s$ by a factor $l^s$. 
The quantities $Q_{\pm 1}$ correspond to the light-cone coordinates of the 
energy-momentum tensor.
These charges are in relation with the logarithm of the monodromy matrix trace 
$\ln M(\lambda)$ through the so-called \emph{trace identities} 
\cite{Kore_Bogo_Izer_93}.

\section{The auxiliary differential equation}
\label{sec:auxdiffeq}
\setcounter{equation}{0}

In order to simplify the subsequent discussion, it is better to adopt light-cone 
coordinates

\begin{align}
	z \equiv \rho e^{\mathbbm i \phi} \doteq (x + t) \quad &\Rightarrow \quad 
\partial \doteq \frac{\partial}{\partial z} = \frac{1}{2} (\partial_x + 
\partial_t) \; ,	\nonumber
\\
\\	\bz \equiv \rho e^{-\mathbbm i \phi} \doteq (x - t) \quad &\Rightarrow \quad 
\bp \doteq \frac{\partial}{\partial \bz} = \frac{1}{2} (\partial_x - 
\partial_t) \; ,	\nonumber
\end{align}
which we will consider as independent coordinates in $\mathbbm C^2$, reserving 
ourselves the right to fix the ``real slice" by setting $\bz = z^*$ as needed; 
we will call these $z$ and $\bz$ holomorphic and anti-holomorphic coordinates, 
respectively. The equations of motion take the following form

\eq
	\partial\bp\boeta(z,\bz) = \frac{m^2}{4\beta}\sum_{i=0}^r n_i \balpha_i 
e^{\beta (\balpha_i,\boeta)} \; ;
\label{eq:complexTodaEOM}
\en
revealing a symmetry under rescaling:

\eq
	(z,\bz) \rightarrow (\gamma z, \gamma^{-1} \bz) 
\; ,\qquad \forall \gamma\in\mathbbm C \; ,
\label{eq:rescaling1}
\en
which on the real slice means $(\rho,\phi)\rightarrow(\vert\gamma\vert\, \rho,\phi + \textrm{arg}(\gamma))$. It follows that, starting from a given solution $\boeta(z,\bz)$ to (\ref{eq:complexTodaEOM}) one can generate a one-parameter family of solutions $\boeta_\gamma(z,\bz)$ by analytic continuation:

\eq
	\boeta_\gamma(z,\bz) \doteq \boeta(\gamma z, \gamma^{-1} \bz) \; .
\en

The Lax operators in light-cone coordinates are $\mathcal L =\partial + L 
\doteq (\mathcal V + \mathcal U)/2$ and $\overline{\mathcal L} = \partial + \overline 
L \doteq (\mathcal V - \mathcal U)/2$, with

\begin{align}
	L(z,\bz;\lambda) &= \frac{\beta}{2} (\partial\boeta\, ,\mathbf{h}) + 
\lambda \sum_{i=0}^r\mu_i e^{\frac{1}{2}\beta(\balpha_i,\boeta)} e_i \; ,	
\nonumber
\\	\label{eq:complexLaxpair}
\\	\overline L(z,\bz;\lambda) &= -\frac{\beta}{2} (\bp\boeta\, ,\mathbf{h}) + 
\lambda^{-1} \sum_{i=0}^r\mu_i e^{\frac{1}{2}\beta(\balpha_i,\boeta)} f_i \; .	
\nonumber
\end{align}
The zero curvature condition remains formally the same $[\mathcal 
L,\overline{\mathcal L}] = 0$ and also the the linear system

\eq
	\left\lbrace\begin{array}{l}
		\mathcal L(z,\bz;\lambda) \, \bPsi(z,\bz;\lambda) =0 \\
		\\
		\overline{\mathcal L}(z,\bz;\lambda) \, \bPsi(z,\bz;\lambda) =0 
	\end{array}\right. \; .
\label{eq:complexlinearsystem}
\en

Let us concentrate on the holomorphic part of this system; in order to simplify 
the action of $\mathcal L$ on the vector $\bPsi$, we consider the following 
function of the field $\boeta$

\eq
	\Omega = e^{-\frac{1}{2} \beta (\mathbf{h},\boeta)}
\en
which lives in the \emph{universal enveloping algebra} $U(\mathfrak g)$ of the 
algebra $\mathfrak g$. In particular this function is an element of the group 
$G$ associated with the algebra $\mathfrak g$ and we know by Hadamard lemma that, 
if $X$ and $Y$ are elements of $\mathfrak g$ such that $[X,Y] = 
\textrm{ad}(X)Y$, then

\eq
	e^{X} Y e^{-X} = e^{\textrm{ad}(X)}Y\; ,
\en
meaning

\eq
	\Omega e_i \Omega^{-1} = e^{-\frac{1}{2}\beta (\balpha_i,\boeta)} e_i 
\; ,\qquad \Omega \mathbf{h} \Omega^{-1} = \mathbf{h} \; .
\en
Setting with $\hat{\bPsi} = \Omega \bPsi$, the holomorphic part of the linear 
system becomes

\eq
	\hat{\mathcal L}\hat{\bPsi} = 0 \; ,\qquad \hat{\mathcal L} = 
\partial + \beta(\partial\boeta,\mathbf{h}) + \lambda\, \mathcal E \; .
\label{eq:simplifiedholosyst}
\en
Following the same tracks, it is possible to simplify the antiholomorphic part of 
the problem by using $\Omega^{-1}$ instead of $\Omega$.

Up to now, we have always worked with abstract operators. The time has come to 
choose a representation of our algebra and we decide to consider a d-dimensional 
one $\mathscr R^d(\mathfrak g)$; the Cartan subalgebra $\mathbf{h}$ is then 
represented by $d\times d$ diagonal matrices $\left[\mathscr R^d (h_i)\right]_{j 
k} = h^i_j \delta_{jk}$, where $h^i_j \doteq \left[\mathscr 
R^d(h_i)\right]_{jj}$. Thus we can rewrite the holomorphic linear system in the 
following more compact form

\eq
	\Delta[\kappa_j]\psi_j + \lambda\left[\mathcal E\hat{\bPsi}\right]_j = 0 
\; ,\qquad \forall j=1,\ldots,d \; ,
\label{eq:holosyst}
\en
where $\mathcal E$ is now a $d\times d$ matrix and $\hat{\bPsi} = 
(\psi_1,\ldots,\psi_d)^{T}$ a $d$-dimensional column vector; we also introduced 
the differential operator $\Delta[f]$ and the fields $\kappa_j$ defined as 
follows

\eq
	\Delta[f] \doteq \partial + (\partial f) \equiv e^{-f}\partial e^{f} 
\; ,\qquad \kappa_j \doteq \beta \sum_{k=1}^r \eta^k h_j^k \; .
\label{eq:diffop}
\en

The system (\ref{eq:holosyst}) admits an interesting graphical representation; 
first of all, let us rewrite it by inverting the differential operator 
$\Delta[\kappa_j]$:

\eq
	\psi_j = -\lambda \,\mathcal I_j \left[\mathcal E\hat{\bPsi}\right]_j \; 
.
\label{eq:holoinvsyst}
\en
The integral operator $\mathcal I_j$ is defined as the inverse of 
$\Delta[\kappa_j]$, that is $\mathcal I_j \Delta[\kappa_j] = 
\Delta[\kappa_j]\mathcal I_j = \mathbbm I$ and it can be formally represented as

\eq
	\mathcal I_j = e^{-\kappa_j} \partial^{-1} e^{\kappa_j} \; ,\qquad 
\partial^{-1}\partial = \partial\partial^{-1} = \mathbbm I \; .
\en
Now we associate to the system (\ref{eq:holoinvsyst}) a graph $\mathcal G$ with 
$d$ nodes and place a directed edge from the node $j$ towards the node $k$ iff 
$\mathcal E_j^k \neq 0$. To an edge going from $j$ to $k$ we associate an 
``hopping operator" $H_j^k \doteq -\lambda \mathcal I_j \mathcal 
E_j^k$ and we can express any vector element $\psi_j$ in term of any other 
$\psi_k$, given the nodes $j$ and $k$ are connected in the graph (we use 
Einstein summation convention):

\eq
	\psi_j = \Lambda_j^k\psi_k \; ,\qquad \Lambda_j^k \doteq 
\sum_{\underset{j\rightarrow k}{\textrm{paths}}}\left(\prod_{\langle l,m\rangle} 
H_l^m\right) \; .
\label{eq:pathrepresentation}
\en
The sum is over all the possible paths connecting the node $j$ to the node $k$ 
while the product is over all the edges in the path. This ``path representation" 
allows to reduce the number of equations in the system, as the 
(\ref{eq:pathrepresentation}) already takes into account all the equations that 
``lie" between $j$ and $k$.

An important fact about Ka\v{c}-Moody algebras is that, for the particular representations we are going to be considering, their matrices $\mathcal 
E$ are such that these diagrams happen to be cyclic\footnote{This is a 
consequence of the structure of the operators $e_0$.}; this means that the 
system (\ref{eq:holoinvsyst}) can be reduced to a single equation involving only 
one of the vector components.

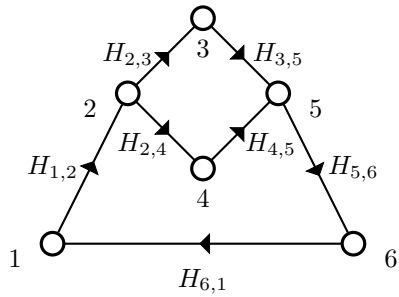
\begin{figure}[h!]
\begin{center}
\begin{tikzpicture}

\node[dynkin node] at (0,0){};
\node[] at (-0.5,-0.2){$1$};
\node[dynkin node] at (1,2){};
\node[] at (0.5,1.8){$2$};
\node[dynkin node] at (2,3){};
\node[] at (2,2.6){$3$};
\node[dynkin node] at (2,1){};
\node[] at (2,0.6){$4$};
\node[dynkin node] at (3,2){};
\node[] at (3.5,1.8){$5$};
\node[dynkin node] at (4,0){};
\node[] at (4.5,-0.2){$6$};

\node[] at (2,0){\tikz \draw[-triangle 90] (0,0) -- +(-.1,0);};
\node[] at (2,-0.5){$H_{6,1}$};
\node[] at (0.5,1){\tikz \draw[-triangle 90] (0,0) -- +(.1,.2);};
\node[] at (0,1){$H_{1,2}$};
\node[] at (1.5,2.5){\tikz \draw[-triangle 90] (0,0) -- +(.1,.1);};
\node[] at (1,2.5){$H_{2,3}$};
\node[] at (1.5,1.5){\tikz \draw[-triangle 90] (0,0) -- +(.1,-.1);};
\node[] at (1.2,1.3){$H_{2,4}$};
\node[] at (2.5,2.5){\tikz \draw[-triangle 90] (0,0) -- +(.1,-.1);};
\node[] at (3,2.5){$H_{3,5}$};
\node[] at (2.5,1.5){\tikz \draw[-triangle 90] (0,0) -- +(.1,.1);};
\node[] at (2.9,1.3){$H_{4,5}$};
\node[] at (3.5,1){\tikz \draw[-triangle 90] (0,0) -- +(.1,-.2);};
\node[] at (4,1){$H_{5,6}$};

\draw[scale=1,domain=0.15:3.85,smooth,variable=\x, color=black,thick,-] plot ({\x},{0});
\draw[scale=1,domain=0.075:0.925,smooth,variable=\x, color=black,thick,-] plot ({\x},{2*\x});
\draw[scale=1,domain=1.1:1.9,smooth,variable=\x, color=black,thick,-] plot ({\x},{1+\x});
\draw[scale=1,domain=1.1:1.9,smooth,variable=\x, color=black,thick,-] plot ({\x},{3-\x});
\draw[scale=1,domain=1.1:1.9,smooth,variable=\x, color=black,thick,-] plot ({1+\x},{\x});
\draw[scale=1,domain=1.1:1.9,smooth,variable=\x, color=black,thick,-] plot ({1+\x},{4-\x});
\draw[scale=1,domain=0.075:0.925,smooth,variable=\x, color=black,thick,-] plot ({4-\x},{2*\x});

\end{tikzpicture}
\end{center}
\caption{Diagram corresponding to the system (\ref{eq:exsystem})}
\label{fig:example1}
\end{figure}

To make things more clear, let us present a toy example. Consider the following system

\eq
	\left\lbrace\begin{array}{l}
		\psi_j = H_j^{j+1} \psi_{j+1} \quad , \ j= 1,4,5,6 \\ 
\\
		\psi_3 = H_3^5 \psi_5 \\ \\
		\psi_2 = H_2^3 \psi_3 + H_2^4 \psi_4
	\end{array}\right. \; ;
\label{eq:exsystem}
\en
we can represent it with the diagram in \ref{fig:example1}. As we see from 
the picture, starting from node $1$, there are two possible closed paths:

\begin{align}
	\mathcal P_1 \, : \; 1\rightarrow 2 \rightarrow 3 \rightarrow 5 
\rightarrow 6 \rightarrow 1	\nonumber
\\	\nonumber
\\	\mathcal P_2 \, : \; 1\rightarrow 2 \rightarrow 4 \rightarrow 5 
\rightarrow 6 \rightarrow 1	\nonumber
\end{align}
which translate into the following

\eq
	\Lambda_1^1 = \sum_{i=1}^2 \left(\prod_{\langle i,j\rangle \in \mathcal 
P_i} H_i^j\right) = H_1^2 (H_2^3H_3^5 + 
H_2^4H_4^5)H_5^6H_6^1 \; .
\en
These two paths are the only possible closed ones and this means that they take 
in account all the equations; in other words the following equation

\eq
	\psi_1 = -\frac{\lambda^5}{32}\left[\mathcal I_1\mathcal I_2(\mathcal 
I_3 + \mathcal I_4)\mathcal I_5\mathcal I_6\right] \psi_1 \; ,
\label{eq:exinvertsyst}
\en
where we supposed that $\mathcal E_i^j=1/2$, when they are non-vanishing, is 
equivalent to the system (\ref{eq:exsystem}). Noticing that the sum of two 
integral operators can be rewritten as

\eq
	\mathcal I_j + \mathcal I_j = \mathcal I_i(\Delta[\kappa_i] + 
\Delta[\kappa_j])\mathcal I_j = 2\,\mathcal I_i \Delta\left[\frac{\kappa_i + 
\kappa_j}{2}\right]\,\mathcal I_j \; ,
\en
we can easily invert the equation (\ref{eq:exinvertsyst}), obtaining

\eq
	\psi_1 = -\frac{16}{\lambda^5} 
\Delta[\kappa_6]\Delta[\kappa_5]\Delta[\kappa_4]\left(\Delta\left[\frac{
\kappa_3+\kappa_4}{2}\right]\right)^{-1}\Delta[\kappa_3]\Delta[\kappa_2]\Delta[
\kappa_1] \psi_1 \; ,
\en
which is a $5$-th order equation, as expected, since the shortest closed path is 
$5$ steps long. Let us notice also that the equation contains an integral operator 
which corresponds to the bifurcation at the node $2$ of the diagram. We might 
further simplify the pseudo-differential equation by exploiting the 
representation $\Delta[\kappa_j]^{\pm1} = 
e^{-\kappa_j}\partial^{\pm1}e^{\kappa_j}$, but it is pointless to proceed with 
this uninteresting example.

Before turning to the analysis of the equations for the various algebras, we 
still have to apply a modification to the Toda equation of motion which, as 
we will see, will pave the way to the correspondence.

\section{The modified equations of motion}
\label{sec:modEOM}
\setcounter{equation}{0}

Following the tracks of \cite{Luky_AZam_10}, we introduce a modified version of 
the equations of motion for the Toda models; we will refer to them as 
\emph{modified Toda equations}. They are

\eq
	\partial\bp\boeta = \frac{m^2}{4 \beta} \sum_{i=0}^r n_i 
\widetilde{\balpha}_i e^{\beta (\balpha_i,\boeta)} \; ,\qquad 
\widetilde{\balpha}_i =	\left\lbrace\begin{array}{l r}
				\balpha_i & \forall i \neq 0 \\ \\
				p_M(z;s)p_M(\bz;s)\balpha_0 & i = 0
			\end{array}\right. \; ,
\label{eq:modTodaEOM}
\en
where $p_M(x;s) = x^{h^\vee M} - s^{h^\vee M}$ and $h^\vee$ is the dual Coxeter number of the 
algebra $\mathfrak g$ (which, for simply-laced algebras corresponds to the Coxeter number $h$ introduced above; in the following we will simplify the 
notation: $p\equiv p_M(z;s)$ and $\overline p \equiv p_M(\bz;s)$.

The modified equation (\ref{eq:modTodaEOM}) can be obtained from the original 
one (\ref{eq:complexTodaEOM}) by means of the following change of variable

\eq
	z\rightarrow w(z) \doteq \int\limits^z p(z')^{1/h^\vee} dz' \quad , 
\quad \bz \rightarrow \overline w(\bz) \doteq \int\limits^{\bz} p(\bz')^{1/h^\vee} 
d\bz'
\en
combined with a shift of the field

\eq
	\boeta(z,\bz) \rightarrow \boeta(z,\bz) -\widetilde\boeta(z,\bz) \; ,
\en
where the functions $\widetilde\boeta(z,\bz)$ have to satisfy the following 
relations

\eq
	\partial\bp \widetilde\boeta(z,\bz) = 0 \; ,\qquad 
(\balpha_i,\widetilde\boeta) = \frac{\ln(p\,\overline p)}{h^\vee\beta} \ , \ \forall i\neq 
0 \; .
\en
The form of these functions depends on the choice for the simple root of the 
algebra.

The presence of the potentials $p$ and $\overline p$ in modified Toda equation 
(\ref{eq:modTodaEOM}) break the original rescaling symmetry (\ref{eq:rescaling1}), leaving behind a 
residual discrete symmetry:

\eq
	(z,\bz) \rightarrow (\omega z, \omega^{-1} \bz) \quad \Rightarrow \quad (\rho,\phi) \rightarrow (\rho, \phi + \frac{2 
\pi}{h M}) \ ,
\label{eq:brokenAnalCont}
\en
where $\omega \doteq e^{\frac{2\pi \mathbbm i}{h^\vee M}}$. The solutions to the modified Toda equation which will be relevant for us have 
to respect this discrete symmetry; moreover, since we will be interested in the 
radial spectral problem (see \ref{subsec:eigencomplex}), we want our solutions 
to be ``well-behaved" as $\rho \rightarrow \infty$ and satisfy some asymptotics 
for $\rho \rightarrow 0$. The exact conditions are

\begin{enumerate}
	\item	\textbf{periodicity}:
		\eq
			\boeta(\rho,\phi) = \boeta(\rho,\phi + \frac{2 \pi}{h^\vee M}) \ ,
		\label{eq:cond1oneta}
		\en
		or, more precisely, the solutions $\boeta(\rho,\phi)$ are 
single-valued functions on a cone with apex angle $\frac{2 \pi}{h^\vee M}$:
		\eq
			\mathcal C_{\frac{2 \pi}{h^\vee M}} \, : \quad \phi \sim \phi + \frac{2 \pi}{h^\vee M} \ , \  0 \leq\rho<\infty \; ;
		\en
	\item	\textbf{real-valuedness and finiteness}\\
		the solutions $\boeta(\rho,\phi)$ have to be real-valued for 
real $\rho$ and $\phi$, and be finite everywhere on the cone $\mathcal 
C_{\frac{2 \pi}{h^\vee M}}$, except for the apex $\rho = 0$;
	\item	\textbf{large-$\rho$ asymptotic}
		\eq
			\boeta(\rho,\phi) \underset{\rho\rightarrow\infty}{\sim} 
-2 \bnu^{\infty} \ln \rho \; ;
		\label{eq:etainfasymptotic}
		\en
	\item	\textbf{small-$\rho$ asymptotic}
		\eq
			\boeta(\rho,\phi) \underset{\rho\rightarrow0}{\sim} -2 
\bnu^{0} \ln \rho
		\label{eq:eta0asymptotic}
		\en
\end{enumerate}
The parameters $\bnu^\infty$ and $\bnu^0$ have to satisfy the following constraints:

\begin{align}
	(\balpha_j -\balpha_0)\cdot \bnu^{\infty} = -h^\vee M \; , \quad \forall j = 1, \ldots, r \; ,
	\\	\nonumber
	\\
	\vert \balpha_j\cdot \bnu^0\vert < 1 \; , \quad \forall j=0,\ldots ,r \; . \label{eq:nu0constraint}
\end{align}

Starting from the asymptotic (\ref{eq:eta0asymptotic}), it is possible to develop 
an expansion for $(z,\bz)\sim(0,0)$ of the form

\begin{align}
	\boeta &= -\bnu^0 \ln(z\,\bz) + \boeta_0 + \sum_{k=1}^\infty 
\bgamma_k\left(z^{h^\vee M k} + \bz^{h^\vee M k}\right)+ \frac{m^2}{4 \beta}\sum_{i=1}^r 
n_i \balpha_i \times 	\nonumber
\\	\label{eq:eta00expansion}
\\	& \times\left(\frac{e^{-\beta(\balpha_i,\boeta_0)} (z\,\bz)^{1- 
\beta(\balpha_i,\bnu^0)}}{\left[1- \beta(\balpha_i,\bnu^0)\right]^2} + s^{2 h 
M}\frac{e^{-\beta(\balpha_0,\boeta_0)}(z\,\bz)^{1- 
\beta(\balpha_0,\bnu^0)}}{\left[1-\beta(\balpha_0,\bnu^0)\right]^2}\right) + 
\cdots \  ,	\nonumber
\end{align}
where $\boeta_0$ and $\{\bgamma_k\}$ are integration constants which determine 
completely all the terms of this expansion. It has to be remarked that these 
constants are not new parameters, but they have to be fixed by demanding 
consistency with the properties (1-3) listed above.

The expansion (\ref{eq:eta00expansion}) remains valid also if we regard $z$ and 
$\bz$ as independent variables, thus it is admissible to perform the so-called 
\emph{light-cone} limit: $\bz \rightarrow 0$ with fixed $z$; in this limit we 
have

\eq
	\partial\boeta \underset{\bz\rightarrow0}{\sim} -\frac{\bnu^0}{z} + \partial\bgamma(z) \; ,\qquad \bgamma(z) = \sum_{k=1}^\infty \bgamma_k z^{h^\vee M k} \; ,
\en

Everything we said in the preceding subsection apply to the modified Toda 
equation with the sole change $\balpha_i \rightarrow \widetilde\balpha_i$; this 
means that the holomorphic system now reads

\eq
	\hat{\mathcal L} \,\hat\bPsi(z,\bz) = 0 \quad , \qquad \hat{\mathcal 
L}_{jk} = \Delta[\kappa_j]\delta_{jk} + \lambda \widetilde{\mathcal E}_{jk} 
\quad , \qquad \widetilde{\mathcal E} = \sum_{i=0}^r \mu_i \mathscr 
R^d(\widetilde e_i) \; ,
\en
where $\widetilde e_i = e_i \  , \ \forall i \neq 0$ and $\widetilde e_0 = p\, 
e_0$. It is possible to apply the graphical method as sketched above, with the 
only care that, unlike the unmodified case, the matrix $\widetilde{\mathcal 
E}$ now contains a dependence on $(z,\bz)$, given by the potentials in $\widetilde 
e_0$ and thus does not commute with the operators $\mathcal I$.

Up until now we have considered only the holomorphic part of the system since the 
discussion for the anti-holomorphic part is essentially identical; the only 
difference consists in the definition of the vector $\hat \bPsi$ which, for the 
holomorphic part was $\Omega \bPsi$ while for the anti-holomorphic part has to 
be $\Omega^{-1}\bPsi$ in order for the form of $\overline{\mathcal L}$ to simplify. 
This means that we end up with two equations

\begin{align}
	\psi &= \Lambda_1^1 \psi \; ,\qquad \Lambda_1^1 = \sum_{\mathcal L_1} 
\prod_{\langle j,k\rangle\in \mathcal L_1} (-\lambda\mathcal I_j 
\widetilde{\mathcal E}_j^k) \; ,\qquad \psi = e^{-\frac{\kappa_1}{2}} \bPsi_1 
\; ,	\nonumber
\\	\label{eq:pairofpseudodiffeq}
\\	\overline\psi &= \overline\Lambda_1^1 \overline\psi \; ,\qquad \overline\Lambda_1^1 = 
\sum_{\overline{\mathcal L}_1} \prod_{\langle j,k\rangle\in \overline{\mathcal L}_1} 
(-\lambda^{-1}\overline{\mathcal I}_j \widetilde{\mathcal F}_j^k) \; ,\qquad 
\overline\psi = e^{\frac{\kappa_1}{2}} \bPsi_1 \; ,	\nonumber
\end{align}
where $\mathcal L_1$ is the set whose elements are all the possible closed loops 
of $\mathcal G$ starting at node $1$; similarly one defines $\overline{\mathcal 
L}_1$\footnote{Notice that $\overline{\mathcal G}$, whose closed loops are elements 
of $\overline{\mathcal L}$, is not necessarily the same graph as $\mathcal G$!}. The 
antiholomorphic integral operators are $\overline{\mathcal I_j} = e^{-\kappa_j}\bp 
e^{\kappa_j}$ and $\widetilde{\mathcal F} = \sum_{i=0}^r\mu_i\mathscr 
R^d(\widetilde f_i)$ where $\widetilde f_i = f_i \  , \ \forall i\neq 0$ and 
$\widetilde f_0 = \overline p\, f_0$. From the same definition of the functions 
$\psi$ and $\overline\psi$ we see that they must satisfy the condition

\eq
	\overline\psi(z,\bz) = e^{\kappa_1}\psi(z,\bz) \quad \Rightarrow \quad 
\Lambda_1^1 \psi(z,\bz) = e^{\kappa_1} \overline{\Lambda}_1^1 e^{-\kappa_1} 
\psi(z,\bz) \; ;
\label{eq:compcondonfunctions}
\en
this relation will be useful later, when analysing the large-$\rho$ asymptotics.
\\

So, recollecting, we started from the equations of motion for the Toda model in 
light-cone coordinates (\ref{eq:complexTodaEOM}) and we deformed it by inserting 
(in front of the exponential containing $\balpha_0$, the root associated with 
the affine structure of $\mathfrak g$) the potentials $p_M(z;s)$ and 
$p_M(\bz;s)$, obtaining (\ref{eq:modTodaEOM}); we asked for the solutions to 
this modified EoM to satisfy some particular conditions (\ref{eq:cond1oneta} - 
\ref{eq:eta0asymptotic}). Then we translated the EoM into a zero curvature 
condition for two operators (\ref{eq:complexLaxpair}) which, in turn, can be 
read as the compatibility condition for a linear system 
(\ref{eq:complexlinearsystem}) involving an auxiliary vector function. By 
slightly manipulating it we are able, through a graphical representation, to 
recast the linear system into a pair of pseudo-differential equations for two 
functions (\ref{eq:pairofpseudodiffeq}), constrained by the relation 
(\ref{eq:compcondonfunctions}) which play the role of the compatibility 
condition for the linear system.

\chapter[Representations and $\psi$-systems]{Fundamental Representations, Generalised Quantum Wronskians and the $\psi$-Systems}
\label{chap:repthe}
\markboth{Chapter 3 - Fundamental Representations and the $\psi$-systems}{}

The linear problem (\ref{eq:complexLaxpair}) is fulfilled for an arbitrary representation of  $\hat \mg$, be it finite or infinite-dimensional, irreducible or not. Since every reducible representation can be obtained as suitable tensor products of irreducible ones, we will concentrate on the latter. However, in order to proceed to the construction of the spectral determinants and the 
functional relations between them we need the operators 
$\mathbf h$, $e_i$ and $f_i$ to be $d\times d$ matrices. In other words we are searching for irreducible finite-dimensional representations of the affine algebra $\hat{\mathfrak g}$. As proved in \cite{Char_Pres_87}, the only finite-dimensional irreducible modules of an affine algebra $\hat \mg$ are the so-called 
evaluation modules. They are defined as follows.
\begin{Def}[Evaluation $\hat \mg$-modules]
Let $\mg$ be a simple Lie algebra and $\hat \mg = \mg \otimes \mathbb C[t,t^{-1}] \,\oplus\,\mathbb C_K$ be the corresponding untwisted affine algebra,
where $\mathbb  C[t,t^{-1}]$ denotes the ring
of formal power series in the variables $t$ and $t^{-1}$.
We use the notation $a(n) = a\otimes t^n$ for elements of $\mg\otimes \mathbb C[t,t^{-1}]$. Let moreover $V$ be an irreducible $\mg$-module and $\zeta\in\mathbb C$ an arbitrary complex number, then $V$ equipped with the following action of $\hat \mg$
\eq
\left\{\begin{array}{l}
a(n)\cdot v = \zeta^n\, a\cdot v\nonumber \\
\\
K\cdot v = 0 \nonumber
\end{array}\right., \quad \forall a\in \mg\, , \ \forall n\in\mathbb Z, \ \forall v\in V,
\en
is a finite-dimensional, irreducible $\hat \mg$-module of level zero, called ``evaluation module'' and denoted as $V(\zeta)$.
\end{Def}
Thus, for any irreducible module $V$ of $\mg$ we have a whole family of evaluation $\hat \mg$-modules, parametrised by a complex number $\zeta$. These evaluation modules $V(\zeta)$ have the same structure as the $\mg$-module  $V$, the only difference is that they  include the action of the $\mathfrak{sl}_2$ triple $\{h_0(1),e_0(1),f_0(1)\}$ associated to the additional root $\balpha_0 = -\btheta$  where $\btheta$ is the longest root of $\mg$. 

In this chapter we first analyse the structure of fundamental representations and their tensor product and define the concept of \emph{projected isomorphism}. We will see how the fundamental representations of any Lie algebra $\mathfrak g$ are connected one another by a set of projected isomorphisms. Finally we will show how this set of isomorphisms implies a system of functional relations, called $\psi$-system, for  solutions of the linear system (\ref{eq:complexLaxpair}).


\section{Product representations and weak isomorphisms}
\label{sec:productrepandisomorphism}

Consider a Lie algebra $\mg$ with  $\textrm{rank}(\mg) = r$, whose sets of simple roots and co-roots are
\eq
\Phi_s = \{\boldsymbol{\alpha}_i \ \vert\  i = 1,\ldots,r\}~,~~
\Phi_s^\vee = \left \{\boldsymbol{\alpha}_i^\vee = 2\frac{\boldsymbol{\alpha}_i}{\vert\boldsymbol{\alpha}_i\vert^2}\ \vert\  i=1,\ldots,r \right \}~.
\en
It is well known that both these sets provide a basis for the root space and with them we can build the Cartan matrix
\eq
C_{i j} = \balpha_i \cdot\balpha^{\vee}_j,
\en
and the Chevalley basis\footnote{Note that the Cartan-Weyl basis $\{h_i,e_i,f_i\}$
defined in the preceding chapter is related to this one by the equalities $e_i
= E^i_+$, $f_i = E^i_-$ and $W^i =
\balpha^{\vee}_i\cdot\mathbf{h}$.} of generators of $\mg$:
$\{W^i,E^i_{\pm}\}_{i=1}^r$, satisfying the following commutation relations
\eq
\begin{array}{l}
[W^i,W^j] = 0,~~~~~~
[W^i,E^j_{\pm}] = \pm C_{j i}\,E^j_{\pm}, ~~~~~~
[E^i_+,E^j_-] = \delta^{ij}W^i,\\[0.3cm]
\big(Ad_{E^i_{\pm}}\big)^{1-C_{j i}}\,E^j_{\pm} = 0,~~~\forall i, j = 1,\ldots,r,\quad i\neq j.
\end{array}
\en
Defining the $0$-th root as minus the longest root, that is $\balpha_0 =
-\btheta = \sum_{i=1}^r n_i\balpha_{i}$, we introduce the associated
operators $W^0$ and $E^0_\pm$, satisfying the relations:
\eq
\begin{array}{l}
[W^0,W^j] = 0, ~~~~~~
[W^a,E^b_{\pm}] = \pm \hat C_{b a}\,E^b_{\pm}, ~~~~~~
[E^0_+,E^0_-] = W^0,\ \\[0.3cm]
\big(Ad_{E^j_{\pm}}\big)^{1-\hat C_{0 j}}\,E^0_{\pm} = 0,\;\;\forall j =
1,\ldots,r, \quad \forall a,b = 0,1,\ldots ,r, \end{array}
\en
where $\hat C$ is the affine Cartan matrix. 
The structure of an untwisted affine algebra $\hat\mg$ associated to a simple Lie algebra $\mg$ is completely 
determined by its affine Chevalley basis $\{\hat W^i,\hat E^i_{\pm}\}_{i=0}^r$, satisfying exactly the same relations as above and defined as follows:
\begin{align}
&\hat W^i = W^i\otimes 1, \, \forall i\neq 0, \quad \hat W^0 = K -(\btheta \cdot \mathbf{h})\otimes 1, \nonumber
\\[0.3cm]
&\hat E^i_{\pm} = E^i_{\pm} \otimes 1, \, \forall i\neq 0, \quad \hat E^0_{\pm} = E^{\mp\btheta}\otimes t. \nonumber
\end{align}
When considering an evaluation module $V(\zeta)$, then these operators can be expressed as
\begin{align}
 \hat W^0 = -(\btheta \cdot \mathbf{h}), ~~~ \hat E^0_{\pm} = \zeta\, E^{\mp\btheta},~~~ \hat W^i = W^i, \quad \hat E^i_{\pm} = E^i_{\pm}, ~~~ \forall i\neq 0. \nonumber
\end{align}
Thus we can reduce the study of the representations associated to evaluation $\hat\mg$-modules $V(\zeta)$ to the analysis of irreducible (that is, highest-weight) $\mg$-representations $V$. Let us recall a few facts about them; we will be rather sketchy since the concepts and methods introduced are part of the standard theory of representations. The interested reader is addressed to classical textbooks on the subject, such as \cite{Fuch_92}.

Let $\Phi_s^\ast$ be the space of 1-forms on the root space of $\mg$, which is called the \emph{weight space}; it can be described by introducing 
the \emph{fundamental weights} $\bLambda_{(i)}$:
\eq
\Phi_f^\ast = \{\bLambda_i \; : \;  \bLambda_{(i)} \cdot \balpha^{\vee}_j  = \delta_{i j}, \quad \forall \balpha^{\vee}_j\in\Phi_s^\vee \ \vert \ i= 1,\ldots,r\}.
\en
This particular basis for the weight space is called \emph{Dynkin basis} and the components of any weight in this basis are 
named \emph{Dynkin labels}. It turns out to be extremely convenient to express the weights in Dynkin labels when building representations, 
as we will shortly see.
Since the root and the weight spaces can be identified via the scalar product we can expand the simple roots in the Dynkin basis,
discovering that  their Dynkin labels are nothing else but the rows of the Cartan matrix:
\eq
\balpha_i = \sum_{j=1}^r C_{i j} \bLambda_{(j)}.
\en
The representations we shall deal with, are the irreducible modules over a \emph{finite dimensional} vector space $V$; we denote them as $R_{\bLambda}(\mg)$. 
As is well known, these modules possess a unique highest weight of the form
\eq
\bLambda = \sum_{i=1}^r \Lambda^i \bLambda_{(i)} \quad \backslash \quad \Lambda^i \in \mathbb N, \ \forall i = 1, \ldots,r,
\label{eq:highestweightform}
\en
whose associated vector $v_{\bLambda}$ is annihilated by any positive step operator $E^i_+$:
\eq
R_{\bLambda}(E^i_+)\, v_{\bLambda}= 0 \  , \quad \forall i = 1,\ldots,r.
\en
Then, an arbitrary  vector in $V$ can be obtained by acting on $ v_{\bLambda}$ with an  appropriate
element of the enveloping algebra of $\mg_- = \{E^i_- \; \vert \; i=1,\ldots,r\}$:
\eq
\forall v \in V \; , \; \exists \{i_1,\ldots,i_\ell\} \quad \backslash \quad  v = 
R_{\bLambda}(E^{i_1}_-) R_{\bLambda}(E^{i_2}_-)\cdots R_{\bLambda}(E^{i_\ell}_-) v_{\bLambda}.
\en
It is important to remark that to each weight (\ref{eq:highestweightform}) corresponds uniquely an irreducible finite-dimensional module and vice-versa.
The highest-weight modules associated to weights $\bLambda_{(i)}$ are of particular relevance. 
These are called \emph{fundamental representations} and 
can be put in correspondence with the nodes of the Dynkin diagram in the same ways as the simple roots. 
Their interest lies in the fact that amongst them there are the simplest irreducible representations from which all others can be obtained through, appropriately projected, \emph{Kronecker products}. These lowest-dimensional irreducible representations are listed, along with their dimension, in Table \ref{tab:lowdimirrep}

\begin{table}[t!]
\begin{center}
\begin{tabular}{| c || c | c | }
 \hline
 $\mg$ & $\bLambda$ & $\textrm{dim}(\bLambda)$ \\
 \hline\hline
 $a_r$ & $\bLambda_1 \; \textrm{or} \; \bLambda_r\cong \bLambda_1^\dagger$ & $r+1$\\
  \hline
  $b_r$ & $\bLambda_r$ & $2^r$\\
  \hline
  $c_r$ & $\bLambda_1$ & $2r$\\
  \hline
  $d_r$ & $\bLambda_1 \; \textrm{and} \; \bLambda_r,\bLambda_{r-1}(\cong \bLambda_r^\dagger)$ & $2r \; \textrm{and} \; 2^{r-1}$\\
  \hline
  $e_6$ & $\bLambda_1\; \textrm{or} \; \bLambda_5\cong \bLambda_1^\dagger$ & $27$\\
  \hline
  $e_7$ & $\bLambda_1$ & $56$\\
  \hline
  $e_8$ & $\bLambda_1$ & $248$\\
  \hline
  $f_4$ & $\bLambda_4$ & $26$\\
  \hline
  $g_2$ & $\bLambda_2$ & $7$\\
  \hline
\end{tabular}
\end{center}
\caption{Lowest-dimensional irreducible representations.}
\label{tab:lowdimirrep}
\end{table}

The use of Dynkin basis provides also a very straightforward algorithm to build the 
\emph{weight space} corresponding to any given highest-weight, that is the tower of states obtained from 
$\vert v_{\bLambda} \rangle$. The first step is to write the highest weight as a row vector $\bLambda = [\Lambda^1,\Lambda^2,\ldots ,\Lambda^r]$, then one can 
obtain all the weights in the representations by repeatedly subtracting the simple roots, which are nothing else but the 
rows of the Cartan matrix. 
The only care one must take is to check that, at each step, the vector $\blambda$ obtained after the subtraction is indeed a weight.  
This is most easily done by expanding $\blambda$ in the simple root basis and computing the integer
\eq
m_j =\blambda \cdot  \balpha^{\vee}_j\;,
\en
if $m_j>0$, then $\{\blambda - n \balpha_j \}_{n=1}^{m_j}$ are admissible weights.

As mentioned above, it is possible to obtain irreducible representations from fundamental ones by performing Kronecker products and 
appropriate projections. Let us clarify this statement. 
Suppose we have two finite-dimensional vector spaces 
$V^{\bLambda}$ and $V^{\bPi}$, with basis $\{\phi_i^{\bLambda}\}_{i=1}^{d_{\bLambda}}$ 
and $\{\phi_i^{\bPi}\}_{i=1}^{d_{\bPi}}$, 
on which the algebra $\mg$ acts as irreducible modules 
$ R_{\bLambda}(\mg)$ and $R_{\bPi}(\mg)$ 
with highest weights, respectively, $\bLambda$ and $\bPi$. 
It is straightforward to verify that the following representation
\eq
R_{\bLambda \otimes \bPi}(A,B) \ = \ R_{\bLambda}(A) \oplus_K R_{\bPi}(B) \ , \quad \forall A,B \in 
\mg,
\en
where the \emph{Kronecker sum} is defined as
$ g \oplus_K f = g \otimes \mathbb I + \mathbb I \otimes f$,
correctly realises\footnote{In fact, since the operation 
$\oplus_K$ preserves the structure of commutator, the Kronecker sum of two representations 
of the Chevalley basis is still a Chevalley basis.} 
the algebra $\mg$ as a module on the vector space 
$V^{\bLambda}\otimes V^{\bPi}$. The dimension of the product module is 
the product $d_{\bLambda}d_{\bPi}$ while the weights are given by the sums of the 
weights: $\{\blambda_i+\bpi_j\}_{i=1\ldots d_{\blambda}}^{j=1\ldots d_{\bpi}}$. 
Notice that the particular weight $\bLambda + \bPi$ is always a highest weight. 
In general, the module $R_{\bLambda \otimes \bPi}(\mg)$ is not irreducible, 
however, from representation theory, we know that it can be written as the direct sum of highest weight submodules
\eq
R_{\bLambda\otimes\bPi}(\mathfrak g) = \bigoplus_{i}\,R_{\boldsymbol{\Sigma}_i}(\mathfrak g)\;,
\en
where the sum runs on some weights $\boldsymbol{\Sigma}_i$ which can, in practice, be computed. Let $\boldsymbol{\Sigma}_j$ be one of these weights, then we can define a projector $\mathcal P_{\boldsymbol{\Sigma}_j}$ on the representation $R_{\boldsymbol{\Sigma}_j}(\mathfrak g)$:
\eq
\mathcal P_{\boldsymbol{\Sigma}_j}\Big(R_{\bLambda\otimes\bPi}(\mathfrak g)\Big) = R_{\boldsymbol{\Sigma}_j}(\mathfrak g)\;.
\en
When the factors of a Kronecker product are two copies of the same module, 
the following decomposition of the product representation is obtained:
\eq
R_{\bLambda \otimes \bLambda} = R_{\bLambda \times \bLambda}  
\oplus R_{\bLambda \wedge \bLambda}.
\label{two}
\en

\begin{table}[t!]
\begin{center}
\begin{tabular}{| c || l   c | }
\hline
$\mg$ & & \\
\hline\hline
& $\bLambda_{(1)} \wedge \bLambda_{(1)} \simeq \bLambda_{(2)}$, &\\
$A_r$ & $\bLambda_{(i)} \wedge \bLambda_{(i)} \simeq \bLambda_{(i-1)}\times\bLambda_{(i+1)}$, & $1<i<r$\\
& $ \bLambda_{(r)} \wedge \bLambda_{(r)} \simeq \bLambda_{(r-1)}$. &\\
\hline
  & $\bLambda_{(1)} \wedge \bLambda_{(1)} \simeq \bLambda_{(2)}$, &\\
 $D_r$ & $\bLambda_{(i)} \wedge \bLambda_{(i)} \simeq \bLambda_{(i-1)}\times\bLambda_{(i+1)}$, & $1<i<r-2$\\
 & $ \bLambda_{(r-2)} \wedge \bLambda_{(r-2)} \simeq \bLambda_{(r-3)}\times\bLambda_{(r-1)}\times\bLambda_{(r)}$, &\\
 & $\bLambda_{(r-1)}\wedge\bLambda_{(r-1)} \simeq \bLambda_{(r)}\wedge\bLambda_{(r)} \simeq \bLambda_{(r-2)}$. &\\
  \hline
   & $\bLambda_{(1)} \wedge \bLambda_{(1)} \simeq \bLambda_{(2)}$, &\\
 & $\bLambda_{(i)} \wedge \bLambda_{(i)} \simeq \bLambda_{(i-1)}\times\bLambda_{(i+1)}$, & $i=2,4$\\
 $E_6$ & $ \bLambda_{(3)} \wedge \bLambda_{(3)} \simeq \bLambda_{(2)} \times \bLambda_{(4)} \times \bLambda_{(6)}$, &\\
 & $\bLambda_{(5)} \wedge \bLambda_{(5)} \simeq \bLambda_{(4)}$, &\\
 & $\bLambda_{(6)} \wedge \bLambda_{(6)} \simeq \bLambda_{(3)}$. &\\
  \hline
     & $\bLambda_{(1)} \wedge \bLambda_{(1)} \simeq \bLambda_{(2)}$, &\\
 & $\bLambda_{(i)} \wedge \bLambda_{(i)} \simeq \bLambda_{(i-1)}\times\bLambda_{(i+1)}$, & $i=2,3,5$\\
 $E_7$ & $ \bLambda_{(4)} \wedge \bLambda_{(4)} \simeq \bLambda_{(3)} \times \bLambda_{(5)} \times \bLambda_{(7)}$, &\\
 & $\bLambda_{(6)} \wedge \bLambda_{(6)} \simeq \bLambda_{(5)}$, &\\
 & $\bLambda_{(7)} \wedge \bLambda_{(7)} \simeq \bLambda_{(4)}$. &\\
  \hline
     & $\bLambda_{(1)} \wedge \bLambda_{(1)} \simeq \bLambda_{(2)}$, &\\
 & $\bLambda_{(i)} \wedge \bLambda_{(i)} \simeq \bLambda_{(i-1)}\times\bLambda_{(i+1)}$, & $i=2,3,4,6$\\
 $E_8$ & $ \bLambda_{(5)} \wedge \bLambda_{(5)} \simeq \bLambda_{(4)} \times \bLambda_{(6)} \times \bLambda_{(8)}$, &\\
 & $\bLambda_{(7)} \wedge \bLambda_{(7)} \simeq \bLambda_{(6)}$, &\\
 & $\bLambda_{(8)} \wedge \bLambda_{(8)} \simeq \bLambda_{(5)}$. &\\
  \hline
\end{tabular}
\end{center}
\caption{Set of projected isomorphisms between products of fundamental representations for the ADE Lie algebras.}
\label{tab:weakisomorphisms}
\end{table}
In (\ref{two})  the two -non necessarily irreducible- modules in the direct sum are 
the symmetric and the antisymmetric parts of the  product representation, 
acting on the vector spaces
\begin{align}
V^{\bLambda}_S = V^{\bLambda} \times V^{\bLambda},~~ &\textrm{with basis} \quad \Big\{\phi_i \times \phi_j = 
\frac{1}{2}(\phi_i\otimes\phi_j + \phi_j\otimes\phi_i)\Big\}_{i\leq j},
\\
V^{\bLambda}_A = V^{\bLambda}\wedge V^{\bLambda},~~ &\textrm{with basis} \quad \Big\{\phi_i \wedge \phi_j = 
\frac{1}{2}(\phi_i\otimes\phi_j - \phi_j\otimes\phi_i)\Big\}_{i< j}.
\end{align}
The weights of these two representations are 
$\{\blambda_i + \blambda_j\}_{i\leq j}$ and $\{\blambda_i + \blambda_j\}_{i<j}$, 
respectively. Here too of particular relevance are the highest 
weights $2\bLambda$ and $\bLambda + \bLambda'$, where $\bLambda'$ is the
next-to-highest weight in $R_{\bLambda}$. 
For our purposes, the only relevant submodules are the ones corresponding to these particular weights and we will be particularly interested in antisymmetric products; thus we define the following distinguished projector:
\eq
\mathcal P_{\bLambda}^{\wedge}\Big(R_{\bLambda\wedge\bLambda}\Big) = R_{\bLambda+\bLambda'}\;.
\en
When the representations involved in these 
products are fundamental ones it is not hard to check that, for the algebras of the ADE series, the structure of the weight space implies 
\eq
\mathcal P_{\bLambda_{(i)}}^{\wedge}\Big(R_{\bLambda_{(i)} \wedge \bLambda_{(i)} }\Big) =  \mathcal P_{ \sum_{(m \sim i)} \bLambda_{(m)}}\Big(R_{ \bigotimes_{(m \sim i)} \bLambda_{(m)}}\Big)\;,
\label{eq:corepsisystem0}
\en
where the index $m$ of the Kronecker product runs over the neighbours of the node $i$ in the Dynkin diagram.
From now on we will lighten the notation by writing simply $\bLambda_{(i)}$ to 
denote both the $i$-th fundamental representation $R_{\bLambda_{(i)}}$ and the associated highest weight. We will also denote the equality (which, more precisely, is an isomorphism) between projected representations with the symbol $\simeq$, so that relation  (\ref{eq:corepsisystem0}) becomes
\eq
\bLambda_{(i)} \wedge \bLambda_{(i)}\simeq \bigotimes_{(m \sim i)} \bLambda_{(m)}\;.
\label{eq:corepsisystem}
\en 
Although it seems rather innocuous, the set of projected isomorphisms (\ref{eq:corepsisystem}) in any ADE algebra $\mg$ actually 
represents the core of the $\psi$-system 
associated to the Affine Toda Field Theory based on $\hat \mg$.  A first clue of this fact comes by specialising (\ref{eq:corepsisystem}) to the various algebras as in 
Table \ref{tab:weakisomorphisms} and by comparing the forms obtained with those of the $\psi$-systems in \cite{Dore_Dunn_Maso_Suzu_Tate_07}.

\begin{figure}[h!]
\begin{center}
\includegraphics[width=1.8cm]{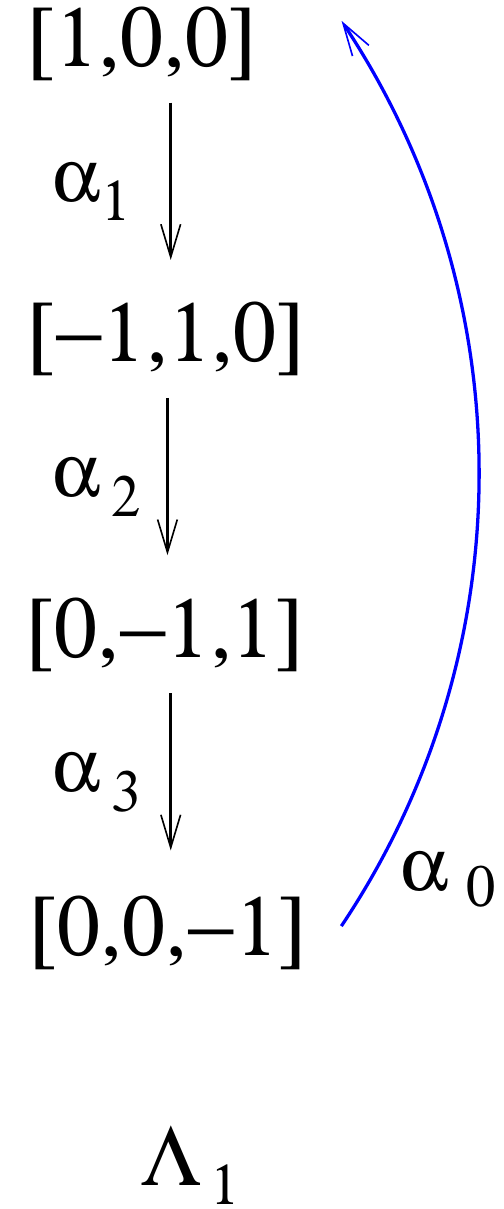} 
\hspace{2.cm}
\includegraphics[width=2.5cm]{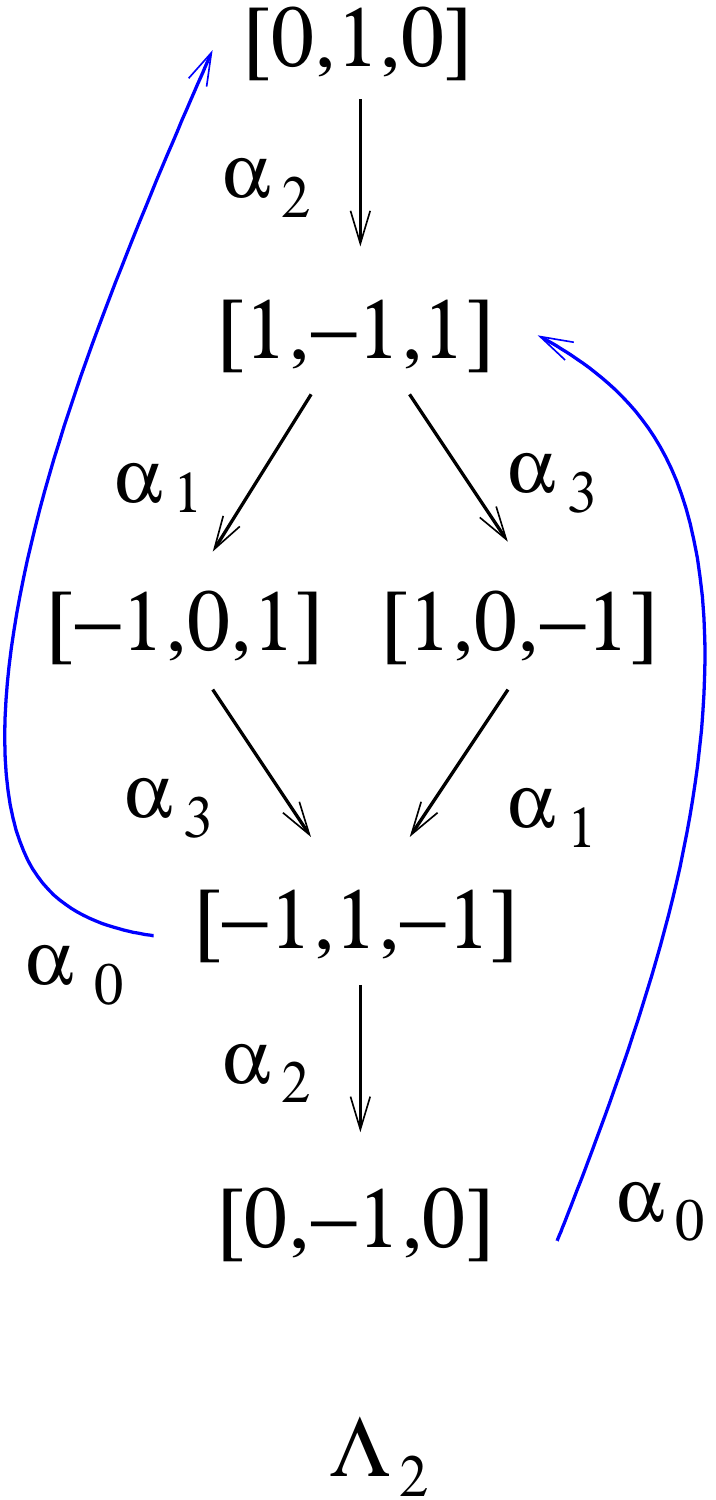} 
\hspace{2.cm}
\includegraphics[width=1.8cm]{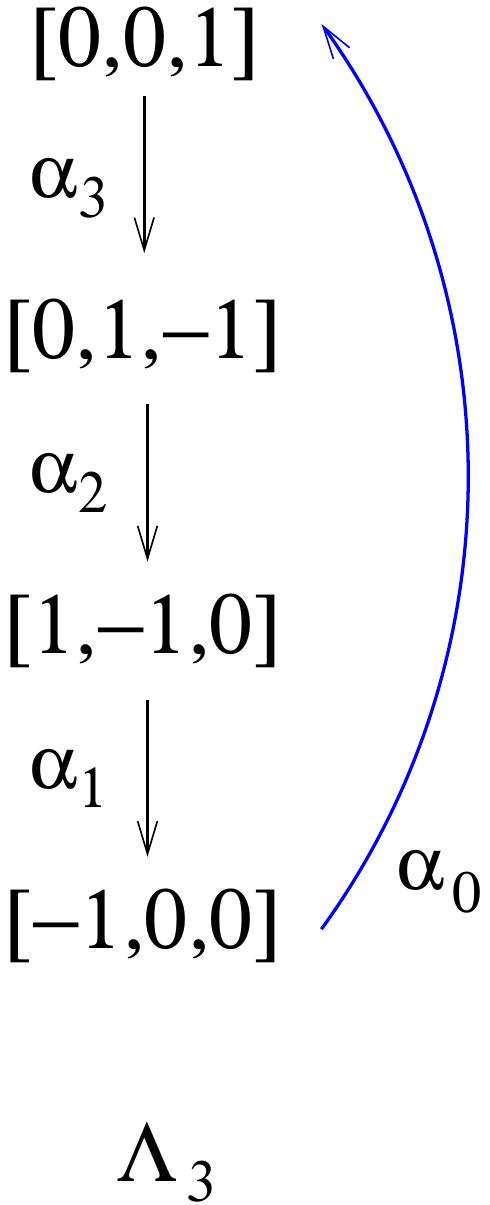}
\end{center}
\caption{The $\mathfrak a_3$ weight systems.}
\label{fig:a3E1}
\end{figure}

After this excursus of representation theory for simple algebras, let us return to evaluation modules. We notice that, although the underlying space is the same, in general $V^{\bLambda_{(i)}}(\zeta_1) \neq V^{\bLambda_{(i)}}(\zeta_2)$ if $\zeta_1 \neq \zeta_2$. Moreover, having a close look at the tensor products, we see that (\ref{eq:corepsisystem}) is not fulfilled anymore for general evaluation modules. 
As a concrete example, consider the weight systems  $\bLambda_{(1)}(A_3)$ and $\bLambda_{(2)}(A_3)$  represented in  Figure~ \ref{fig:a3E1}.

Denoting  $\{v_i\}_{i=1}^4$ and $\{w_i\}_{i=1}^6$ the vectors in the two representations, ordered from top to bottom and from left to right, we have the following identifications:
\begin{align}
v_1\wedge v_2 = w_1, \quad v_1\wedge v_3 = w_2, \quad v_1\wedge v_4 = w_4,\nonumber \\
v_2\wedge v_3 = w_3, \quad v_2\wedge v_4 = w_5, \quad v_3\wedge v_4 = w_6.
\end{align}
Let us now act with $\zeta_1\,E^{\btheta}\oplus_K\zeta_2\,E^{\btheta}$ (respectively $\zeta\, E^{\btheta}$) on the left (respectively right) sides of these equalities, the result is
\begin{align}
&0=0, \quad 0=0\, , \hspace{4.3cm} (\zeta_2-\zeta_1)v_1\otimes v_1=0,\nonumber
\\
&0=0, \quad \zeta_2 v_2\otimes v_1-\zeta_1 v_1\otimes v_2 =\zeta w_1, \quad \quad \zeta_2 v_3\otimes v_1-\zeta_1 v_1\otimes v_3 =\zeta w_2.
\end{align}
Therefore, in order for these equalities to be satisfied, we must set $\zeta_1 = \zeta_2= -\zeta$. Let us write
\eq
\bLambda_{(i)}^{[k]} = V^{\bLambda_{(i)}}(e^{2\pi i k}),
\en
with this notation we  see that 
\eq
\bLambda_{(1)}^{[k+\frac{1}{2}]}(A_3^{(1)})\, \wedge \, \bLambda_{(1)}^{[k-\frac{1}{2}]}(A_3^{(1)}) \simeq \bLambda_{(2)}^{[k]}(A_3^{(1)}).
\en
It is possible to check that, in general, for the untwisted affine algebras, the following projected isomorphisms are satisfied
\eq
\bLambda_{(i)}^{[k-\frac{1}{2}]} \wedge \bLambda_{(i)}^{[k+\frac{1}{2}]}\simeq \bigotimes_{(m \sim i)} \bLambda_{(m)}^{[k]}.
\label{eq:corepsisystem_rotated}
\en

What is now left to do in order to show that the set of weak isomorphism (\ref{eq:corepsisystem}) represents the very 
structure of the $\psi$-system, is to find a way to relate the functions $\psi^j$ and the differential equations 
(\ref{eq:holosyst}) they satisfy to the weight system in a given representation. This turns out to be a rather easy task. 

\section{From weak isomorphisms to the $\psi$-systems}
\label{sec:isomorphismtopsisystem}
\setcounter{equation}{0}

Let us consider an untwisted Lie algebra $\hat \mg$ in the 
ADE series, the Lax operators $\mathcal L^{(n)}_{[k]}$ and the solutions to the corresponding linear system $\Psi^{(n)}_{[k]}$, obtained through the analytic continuation (\ref{eq:brokenAnalCont}):
\eq
\lambda  \rightarrow \omega^{-k} \lambda,~
(z,\bz;s,\bar s) \rightarrow ( \omega^k z,  \omega^{-k} \bz;\omega^k s,  \omega^{-k} \bar s)~
\en
with $k \in \mathbbm R$  and $\omega=e^{\frac{2\pi \mathbbm i}{h^\vee\,M}}$,
\eq
\mL^{(n)}(z,\bz; s,\bar s;\lambda) \rightarrow \;  \mL^{(n)}_{[k]}(z,\bz; s,\bar s;\lambda) =  
\omega^k \; \mL^{(n)}( \omega^k z,  \omega^{-k} \bz;\omega^k s,  \omega^{-k} \bar s,  \omega^{-k}  \lambda ).
\label{rot1}
\en
Here $\mathcal L^{(n)}$ is the realisation in the module $\bLambda_{(n)}$ of the abstract holomorphic
Lax operator (\ref{eq:simplifiedholosyst}), with an analog 
definition for the  antiholomorphic operators $\bar{\mathcal L}^{(n)}$. 

Restricting the attention to a specific periodic field configuration, such that
\eq
\boeta_{\omega}(z,\bz) = \boeta(\omega z, \omega^{-1} \bz) = 
\boeta(z,\bz),
\label{per1}
\en
then
\eq
\mL^{(n)}_{[k]}(z,\bz; s,\bar s;\lambda) = \partial -\partial\boeta\cdot\mathbf{h} + \lambda\sum_{i=1}^r\mu_i E^i_+ +e^{2\pi \mathbbm i k}\mu_0 E^0_+,
\label{eq:k-rot_e0_lax}
\en
with $\mL^{(n)}_{[k]} = \mL^{(n)}_{[k \; \text{mod} \;1]}$, while, in general  
\eq
\Psi^{(n)}_{[k]}(z, \bz, \lambda) = \omega^{r k/2} \, \Psi^{(n)}_{[0]}(\omega^k z,   \omega^{-k} \bz,  \omega^{-k} \lambda), 
\en
and $ \Psi^{(n)}_{[k \,\text{mod} \;1]}(z, \bz, \lambda)$ are linearly independent. 
Thus,  we have a whole family of  \emph{rotated} linear systems
\eq
\mathcal L^{(n)}_{[k \,\text{mod}\; 1]}\,\Psi^{(n)}_{[k]} = 0,
\label{eq:rotatedlinearproblem}
\en
and, for any $k \in [0,1)$, we have a countably infinite set of solutions $\{\Psi^{(n)}_{[k+m]}\}$ with $m \in \mathbbm Z$.
As will be shown explicitly in Section \ref{sec:arcase}, the solutions $\Psi^{(n)}_{[0]}=\Psi^{(n)}$ 
display, in the  $\rho\rightarrow \infty$ limit, 
a behaviour known as \emph{Stokes phenomenon}, with an asymptotic behaviour depending in a discontinuous
way from $\text{arg}(z)$. 
From the definition of the evaluation $\hat\mg$-modules and (\ref{eq:k-rot_e0_lax}) we see that $\mathcal L^{(n)}_{[t]}$ represents the action of $\mathcal L^{(n)}$ on $\bLambda_{(n)}^{[t]}$.

Since  $\Psi^{(n)}_{[k]}$ lives in the vector space $V^{\bLambda_{(n)}}\cong\mathbb C^{d_{\bLambda_{(n)}}}$, 
it is possible to expand it on the weight basis $\{\phi^{\bLambda_n}_i\}_{i=1}^{d_{\bLambda_{(n)}}}$, 
with $\phi_1^{\bLambda_{(n)}}$ being the highest weight vector of the module $\bLambda_n$
\eq
\Psi^{(n)}_{[k]}(z,\bz) = \sum_{i=1}^{d_{\bLambda_{(n)}}} \,\psi^{(n),i}_{[k]}(z,\bz)\phi_i^{\bLambda_{(n)}},~
k \in \Rth.
\label{eq:weightbasisexpansion}
\en
From (\ref{eq:weightbasisexpansion}), we see that the operator $\mathcal L^{(n)}_{[t]}$ acts on $\Psi^{(n)}_{[t+k]}$ in a  
simple way. In fact its diagonal part acts on the components $\psi^{(n),i}_{[t+k]}(z,\bz)$ as the operator 
$\Delta[\kappa^{(n)}_i]$, with $\kappa^{(n)}_i = \sum_{\ell=1}^r \eta^\ell\big[ \bLambda^{[t]}_{(n)}(h_\ell)
\big]_{ii}$, 
while the off-diagonal part acts directly on the basis $\{\phi^{\bLambda_{(n)}}_i\}$ as the operator 
$\tilde{\mathcal E}_{[t]}^{(n)}  = \bLambda^{[t]}_{(n)}(\tilde{\mathcal E})$.

The latter  action can be neatly represented in the state tower: each arrow 
corresponds to the action of an operator $\bLambda^{[t]}_{(n)}(E^i_-)$, for a certain $i$, 
which corresponds to the inverse of $\bLambda^{[t]}_{n}(E^i_+)$. This means that 
$\tilde{\mathcal E}_{[t]}^{(n)}$ acts on the weight basis by sending (multiplicative factors apart) 
$\phi_i^{\bLambda_{(n)}}$ to its neighbour(s) $\phi_j^{\bLambda_{(n)}}$ in the state tower 
against the direction of the arrows. Finally we  arrive at the same graphical description found in Section \ref{sec:auxdiffeq}:
\eq
\psi^{(n),i}_{[t+k]} = \,\Lambda_{[t],j}^{(n),i} \,\psi_{[t+k]}^{(n),j},~~
\Lambda_{[t],j}^{(n),i} = \sum_{\underset{i\rightarrow j}{\textrm{paths}}} 
\left(\prod_{\langle l,m\rangle} \,H_{[t],l}^{(n),m}\right),
\en
where
\eq
H_{[t],l}^{(n),m} = \lambda \Delta^{-1}[\kappa^{(n)}_l] ~ \tilde{\mathcal E}_{[t],l}^{(n),m}\;,
\en
and the graph corresponds to the state tower.

The next step is to  study the  tensor products of vectors $\Psi$.
Consider the following vector
\eq
\Psi^{(n)}_{[k]} \otimes \,\Psi^{(m)}_{[k']} = \sum_{i,j} \,\psi_{[k],[k']}^{(n,m),i,j} \phi^{\bLambda_{(n)}}_i \otimes \phi^{\bLambda_{(m)}}_j,
\label{eq:tensorpsi}
\en
with $k$ and $k'$  generic real numbers, chosen so that the corresponding
Lax operators are related to identical periodic field solution $\boeta$, that is $k= k' \;(\text{mod} \;1)$. Its components are
\eq
\psi_{[k],[k']}^{(n,m),i,j} = \,\psi_{[k]}^{(n),i} \,\psi_{[k']}^{(m),j}.
\en
It is clear that (\ref{eq:tensorpsi}) lives in $\mathbb C^{d_{\bLambda_{(n)}} \,d_{\bLambda_{(m)}}}$ and 
$\hat\mg$ acts on it as the module  $\bLambda^{[t]}_{(n)} \otimes \bLambda^{[t']}_{(m)}$, where $t= k-\lfloor k \rfloor$ is the non-integral part of $k$ and same for $t'$. 
More interesting is the fact that the first component 
$\psi_{[k],[k']}^{(n,m),1,1}$ corresponds to the highest-weight 
vector $\phi_1^{\bLambda_n}\otimes\phi_1^{\bLambda_m}$, meaning that,
when acting with the projector $\mathcal P_{\bLambda_{(n)}+\bLambda_{(m)}}$  on (\ref{eq:tensorpsi}), the vector
\eq
\Psi^{(n\oplus m)}_{[k,k']}= \sum_{i=1}^{d_{\bLambda_{n\oplus m}}} \,\psi^{(n\oplus m),i}_{[k,k']} \phi_i^{\bLambda_{n\oplus m}},
\en
is obtained,  whose first component is the same as that of (\ref{eq:tensorpsi}):
\eq
\psi^{(n\oplus m),1}_{[k,k']} = \,\psi_{[k]}^{(n),1} \,\psi_{[k']}^{(m),1}.
\en
Here too, for $n=m$, we can split the tensor product into symmetric and antisymmetric parts
\eq
\Psi^{(n)}_{[k]} \otimes \,\Psi^{(n)}_{[k']} = \;\Psi^{(n)}_{[k]} \times \,\Psi^{(n)}_{[k']} + \;\Psi^{(n)}_{[k]} \wedge \,\Psi^{(n)}_{[k']},
\en
and  deduce  that the first components are the same as those of the vectors projected with $\mathcal P_{2\bLambda}$ and $\mathcal P^{\wedge}_{\bLambda}$:
\begin{align}
\psi^{(n \oplus \, n)_+,1}_{[k,k']} &= \,\psi_{[k]}^{(n),1} \,\psi_{[k']}^{(n),1}, \\
\psi^{(n \oplus n)_-,1}_{[k,k']} &= \,\psi_{[k]}^{(n),1} \,\psi_{[k']}^{(n),2} - \,\psi_{[k]}^{(n),2} \,\psi_{[k']}^{(n),1} = \det \left\vert \begin{array}{c c}
\,\psi_{[k]}^{(n),1} & \,\psi_{[k']}^{(n),1} \\
\,\psi_{[k]}^{(n),2} & \,\psi_{[k']}^{(n),2}
\end{array}
\right\vert.
\label{latt}
\end{align}
Considering  that the first two weight vectors in {\bf{any}} 
fundamental representation are always uniquely 
defined\footnote{In other words the state towers for the fundamental representations do not present bifurcations at the first step.},
and denoting with $\balpha_{i,j}$  the simple root which sends the vector $i$ in the vector $j$ in the state tower,
we can write 
\eq
\Delta[\kappa_1^{(n)}] \,\psi_{[k]}^{(n),1} =  \lambda \mu_{\balpha_{1,2}} \,\psi_{[k]}^{(n),2}.
\en
This leads  to
\eq
\psi^{(n\oplus n)_-,1}_{[k,k']} = -\frac{1}{\lambda\,\mu_{\balpha_{1,2}}} W^{(2)}\Big[\, \psi_{[k]}^{(n),1},\, \psi_{[k']}^{(n),1}\Big],
\label{eq:firstwronsky}
\en
where we have introduced the wronskian $W^{(2)}[f,g] = f\partial g - g\partial f$.
If we now consider  (\ref{eq:corepsisystem_rotated}) with $k=0$, we get the following weak isomorphism amongst vectors $\Psi$
\eq
\Psi^{(a)}_{[-\frac{1}{2}]} \wedge \,\Psi^{(a)}_{[\frac{1}{2}]}\simeq \bigotimes_{b=1}^r \left(\,\Psi^{(b)}_{[0]}\right)^{A_{ab}},~k= k'= k'' \, (\text{mod} \; 1).
\label{psi00}
\en
By correctly redefining the highest weight components so that 
the prefactor of the wronskian in (\ref{eq:firstwronsky}) is reabsorbed
\eq
\psi^{(a)}_{[k]}  \rightarrow c_a  \psi^{(a)}_{[k]}, \, \, \,  (c_a)^2= -\frac{1}{\lambda\,\mu_{\balpha_{1,2}}}  \prod_b (c_b)^{A_{ab}},
\label{eq:normalisation}
\en
we can rewrite \ref{psi00} as follows:
\eq
W^{(2)}[\psi^{(a)}_{[-\frac{1}{2}]}, \psi^{(a)}_{[\frac{1}{2}]}] = \prod_b \left(\,\psi^{(b)}_{[0]}\right)^{A_{ab}}\;,
\label{eq:genpsisystem}
\en
finding that its form matches exactly that of the $\psi$-system.
In (\ref{psi00}, \ref{eq:normalisation}, \ref{eq:genpsisystem}) $\psi^{(n)}_{[k]} = \,\psi_{[k]}^{(n),1}$ and $A_{ab}$ is the incidence matrix of the Dynkin diagram. 

Considering the state tower for the representation $\bLambda_1$ one finds 
that the multivectors $\Psi^{(n)}_{[k]}$ can be built out of the lowest ones as 
\eq
\Psi^{(n)}_{[k+\frac{n+1}{2}]} = \bigwedge_{\ell =1}^n \left( \,\Psi^{(1)}_{[k+\ell]}\right),~~ \quad n\leq N.
\label{eq:multivectordefinition}
\en
This relation is valid until one arrives either at the end of the tower or at a bifurcation, which means
\eq
N=\left\lbrace\begin{array}{l r}
r+1, & \text{for} \;\quad \mg=A_r,\\
r-2, & \text{for} \;\quad \mg=D_r,\\
r-3, & \text{for} \;\quad \mg=E_r.\\
\end{array}\right.
\label{eq:limitofproducts}
\en
For the vectors associated to the nodes after the bifurcations, the relation (\ref{eq:multivectordefinition}) has to be modified, 
so that for $D_r$ one has
\begin{align}
\bigwedge_{\ell =1}^{r-1} \left( \,\Psi^{(1)}_{[k+\ell]} \right)&= \,\Psi^{(r-1)}_{[k+\frac{r}{2}]} \otimes \;\Psi^{(r)}_{[k+\frac{r}{2}]}, \label{eq:DrPsispin1} \\
\bigwedge_{\ell =1}^{r} \left( \,\Psi^{(1)}_{[k+\ell]} \right) &= \,\Psi^{(r-1)}_{[k+\frac{r}{2}]} \otimes \;\Psi^{(r-1)}_{[k+\frac{r}{2}+1]} +
\,\Psi^{(r)}_{[k+\frac{r}{2}]} \otimes \;\Psi^{(r)}_{[k+\frac{r}{2}+1]}. \label{eq:DrPsispin2}
\end{align}
For the $E_r$, instead, the following relations are valid
\begin{align}
\bigwedge_{\ell =1}^{r-2} \left( \,\Psi^{(1)}_{[k+\ell]} \right)&= \,\Psi^{(r-2)}_{[k+\frac{r-1}{2}]} \otimes \;\Psi^{(r)}_{[k+\frac{r-1}{2}]},
\\
\bigwedge_{\ell =1}^{r-1} \left( \,\Psi^{(1)}_{[k+\ell]} \right) &= \,\Psi^{(r-1)}_{[k+\frac{r}{2}]} \otimes 
\;\Psi^{(r)}_{[k+\frac{r-1}{2}]}\otimes \;\Psi^{(r)}_{[k+\frac{r+1}{2}]}  \\	\nonumber
&+ \,\Psi^{(r-2)}_{[k+\frac{r-1}{2}]} \otimes \;\Psi^{(r-2)}_{[k+\frac{r+1}{2}]}.
\end{align}

In conclusion, the main results of this chapter are:
\begin{itemize}
\item
Given an algebra $\mg$ with rank $r$ of the ADE series, we associate to each node $n$ of the Dynkin diagram a set of 
vector-valued functions $\Psi^{(n)}_{[k]}$ with $k\in\frac{1}{2}\mathbbm Z$. Imposing the periodicity requirement (\ref{per1}) on $\boeta$,
these functions live in the vector space 
associated to the $\bLambda_n$ representation of $\mg$ and are solutions of the linear problem 
\eq
\mathcal L^{(n)}_{[k \,\text{mod}\; 1]}\,\Psi^{(n)}_{[k]} = 0~,
\en
with $\mathcal L^{(n)}_{[k]}$ being the realisation in $\bLambda_n$ of the rotated Lax operator  (\ref{rot1}).

This linear system can be rewritten as a pseudo-differential equation, for the first component 
$\psi^{(n)}_{[k]}$, sided by a set of $d_{\bLambda_n}-1$ coupled differential equations of degree $1$ for the other components. 
The information about the structure of this set of equations is entirely encoded inside the weight system of the considered representation.
\item
The vectors $\Psi^{(n)}_{[k]}$, living in different vector spaces, are related by the set of relations (\ref{psi00})
that, specialised on the first components, produce the $\psi$-system \cite{Dore_Dunn_Maso_Suzu_Tate_07}
\eq
W^{(2)}[\psi^{(a)}_{[-\frac{1}{2}]},\psi^{(a)}_{[\frac{1}{2}]}] = \prod_{b=1}^r \left(\psi^{(b)}_{[0]}\right)^{A_{ab}}~.
\en
\item
More generally the vectors $\Psi^{(n)}_{[k]}$ satisfy certain relations amongst them which are consequences of the projected 
isomorphisms between product of representations. 
An example of this fact is the following 
\eq
\Psi^{(n)}_{[k+\frac{n+1}{2}]} = \bigwedge_{\ell =1}^n \Psi^{(1)}_{[k+\ell]} \; , \quad n\leq N~,
\en
with $N$ given in (\ref{eq:limitofproducts}).
\end{itemize}

\chapter[The ODE/IM Correspondence for ToFT]{The ODE/IM Correspondence for Toda Field Theories}
\label{chap:ODE/IM_Toda}
\markboth{Chapter 4 - The ODE/IM Correspondence for Toda Field Theories}{}

In this chapter we will investigate the linear problems associated to Toda field theories based on the algebras $\hat{\mathfrak a}_{r}$ and $\hat{\mathfrak d}_r$. First of all, we will proceed with the derivation of the differential equation associated to the fundamental representation and the construction of particular solutions specified by their behaviour in the vicinity of singular points of the differential equation: the origin and the point at infinity. While the former will turn out to be a Fuchsian singularity (also called a regular singularity), meaning the solutions display a local power-law behaviour, the nature of the singular point at infinity is radically different: it is an irregular singularity. The consequence of this fact is that the limiting behaviour of solutions to the linear problem will depend non-trivially on the direction of approach to the singularity. As a consequence, a given solution will depend discontinuously on the phase of the variable $\textrm{arg}(z)$ and the asymptotic series representation of the solution will be valid only in some given sector of the complex plane. This property is known as \emph{Stokes phenomenon} and, as we will see, is a fundamental ingredient for the construction of the Bethe Ansatz Equations\footnote{Note that this is the principal reason for deforming the EoMs with the potentials $p(z)$ and $\bar p(\bar z)$: their presence introduce an irregular singularity at $\rho\rightarrow\infty$ in the associated linear problem.}. By making use of the results of the preceding chapter, we will show how the eigenvalues associated to particular central problems for the linear system (\ref{eq:complexlinearsystem}), satisfy a set of algebraic equations, which are nothing else but the Bethe Ansatz Equations for the algebras $\mathfrak a_r$ and $\mathfrak d_r$. Finally we will explicitly work out the simple examples $\mathfrak a_3\cong \mathfrak d_3$ and $\mathfrak d_4$.

Note that, in order to lighten the notation, we will set $\beta =1$ and $m=2$ from now on.

\section{The $\hat{a}_{r}$ case}
\label{sec:arcase}
The algebra $\hat{\mathfrak a}_{r}$ corresponds to  
following Dynkin diagram and  Cartan matrix:

\begin{figure}[h!]
\begin{center}
\begin{tikzpicture}

\node[dynkin node] at (0,0){};
\node[] at (0,-0.5){$1$};
\draw[scale=1,domain=0.15:0.85,smooth,variable=\x, color=black,thick,-] plot ({\x},{0});
\node[dynkin node] at (1,0){};
\node[] at (1,-0.5){$2$};
\draw[scale=1,domain=1.15:1.85,smooth,variable=\x, color=black,thick,-] plot ({\x},{0});
\node[dynkin node] at (2,0){};
\node[] at (2,-0.5){$3$};
\draw[scale=1,domain=2.15:2.55,smooth,variable=\x, color=black,thick,-] plot ({\x},{0});
\node[] at (3,0) {$\cdots$};
\draw[scale=1,domain=3.35:3.85,smooth,variable=\x, color=black,thick,-] plot ({\x},{0});
\node[dynkin node] at (4,0){};
\node[] at (4,-0.5){$r-1$};
\draw[scale=1,domain=4.15:4.85,smooth,variable=\x, color=black,thick,-] plot ({\x},{0});
\node[dynkin node] at (5,0){};
\node[] at (5,-0.5){$r$};
\node[dynkin node] at (2.5,1.25){};
\node[] at (2.5,1.75){$0\equiv r+1$};
\draw[scale=1,domain=0.15:2.35,smooth,variable=\x, color=black,thick,-] plot ({\x},{0.5*\x});
\draw[scale=1,domain=2.35:4.55,smooth,variable=\x, color=black,thick,-] plot ({0.3+\x},{2.35-0.5*\x});

\node[] at (5.75,0) {$,$};
\node[] at (10,0.75) {$\hat C = \left( \begin{array}{cccccc}2&-1&0&\cdots&0&-1  \\-1&2&-1&\cdots&0&0  \\
	                                            0&-1&2&\cdots&0&0  \\
				\vdots&\vdots&\vdots&\ddots&\vdots&\vdots\\ 
				                      0&0&\cdots&2&-1&0 \\0&0&\cdots&-1&2&-1 \\
				                      -1&0&\cdots&0&-1&2 
				                      \end{array} \right)$};
\end{tikzpicture}
\end{center}
\caption{Dynkin diagram and Cartan matrix for the affine algebra $\hat{\mathfrak a}_r$}
\end{figure}

The simple roots for this class of algebras can be represented as vectors in an 
$(r+1)$-dimensional space with orthonormal canonical basis $(\bepsilon_i)^j =
\delta_i^j$ as
\eq
\balpha_i = \bepsilon_i - \bepsilon_{i+1} \; ,
\en
where we impose periodicity on the indices, i.e. $\bepsilon_{r+2} \equiv 
\bepsilon_1$ and set  $\bepsilon_0 \equiv \bepsilon_{r+1}$. The added dimension 
reflects itself in the field theory into the presence of an additional field 
$\eta^{r+1}$, this is taken into account by asking for the sum of all the fields to 
vanish:
\eq
\sum_{a=1}^{r+1} \eta^a(z,\bz) = 0 \; ,\qquad \forall (z,\bz)\in \mathbbm 
C^2 \; .
\en
The simple Lie algebra $\mathfrak a_{r}$ is simply-laced as one can see from the 
fact that all the roots have the same length $\vert\balpha_i\vert^2 =2 \; , \; 
\forall i$. All the Dynkin labels are equal to one: $n_i =1 \; , \; \forall i$, meaning $h^\vee=r+1$.

Let us consider first the lowest-dimensional fundamental representation, $\bLambda_1$: its dimension is 
$d_{\bLambda_1}=  r+1$ and the Cartan subalgebra and step operators are representable as
\eq
(h_i)_j^k = \delta_{i,j}\delta_j^k \; ,\quad (e_i)_j^k = 
\delta_{i,j}\delta_{j+1}^k \;, \quad  (f_i)_j^k = \delta_i^k\delta_j^{k+1} \; , \qquad   \forall i,j=1,\ldots,r+1\; .
\en
\begin{minipage}{\textwidth}
  \begin{minipage}[b]{0.5\textwidth}
  \vspace{0.25\textwidth}
  \centering
    \includegraphics[width=2.8cm]{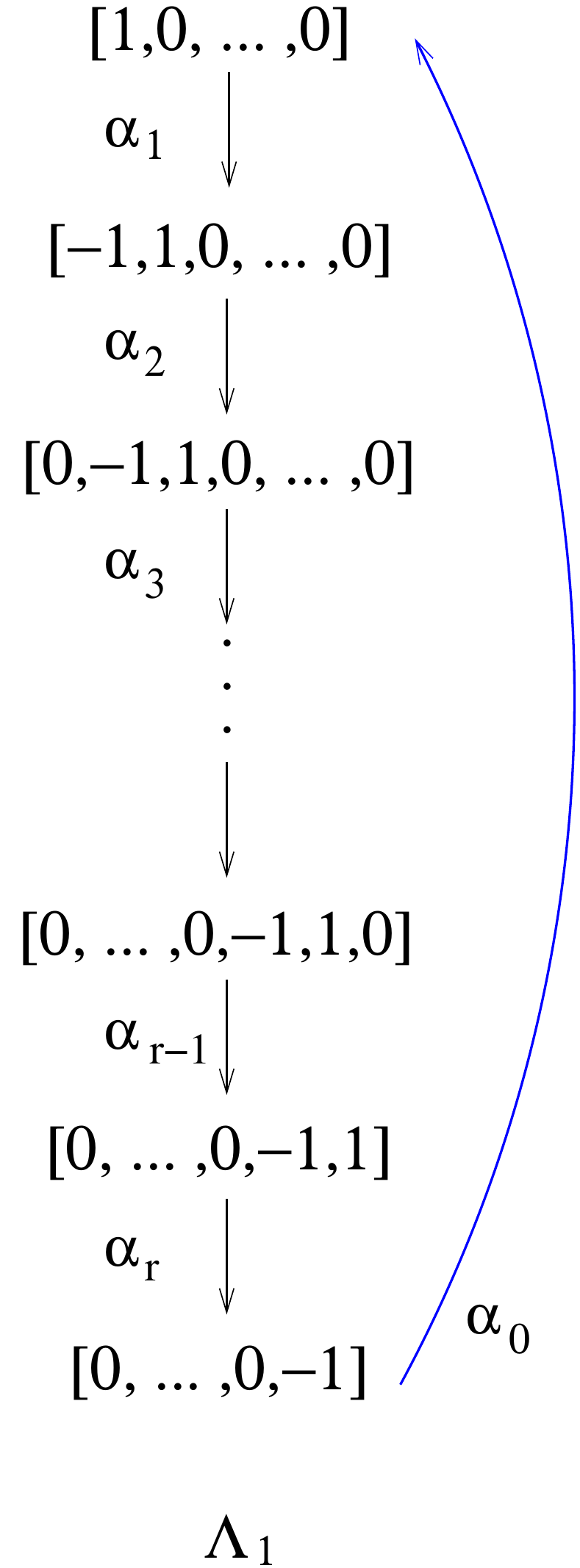}
    \captionof{figure}{The $\hat{\mathfrak a}_r$ weight diagram.}
    \label{Arweight}
  \end{minipage}
\hspace{0.05\textwidth}
  \begin{minipage}[b]{0.4\textwidth}
    \centering
 \begin{tabular}{|c | c|}
\hline 
& \\
Definition & Value in $\bLambda_1$ \\
& \\
\hline & \\
$\mu_j^2 = \frac{n_j}{2}\vert\balpha_j\vert^2$ & $1$ \\ & \\
$\widetilde{\mathcal E}_{i}^{j}$  & $ 
\delta_{i+1}^j \left(1+(p-1)\delta_{i,0}\right)$ \\ & \\
$\widetilde{\mathcal F}_{i}^{j}$  & $ 
\delta_i^{j+1} \left(1+(\bar p-1)\delta_0^j\right)$ \\ & \\
$\kappa_j = \sum_{a=1}^{r+1} \eta^a h_j^a$ & $ \eta^j$ \\ & \\
$\nu_j^{\infty}$ & $-\frac{M}{2}(r-2j+2)$ \\
& \\
\hline
\end{tabular}
      \captionof{table}{The relevant objects for  $\mg =\mathfrak a_r$. }
      \label{TBB010}
    \end{minipage}
    \vspace{0.05\textwidth}
\end{minipage}

The relevant objects for our analysis are listed in Table~\ref{TBB010}.
The weight diagram, depicted in Figure \ref{Arweight}, of $\bLambda_1$ is very simple and is clearly identical to the Dynkin diagram for the affine algebra. There is a single closed path,
thus the pseudo-differential equation is immediately obtained:
\eq
\psi^{(1)}_{[k]} (z,\bz) = e^{2\pi \mathbbm i k}(-\lambda)^{r+1}\, \mathcal I_1 \mathcal 
I_2 \cdots \mathcal I_{r+1} \; p(z,s) \; \;\psi^{(1)}_{[k]} (z,\bz)~,
\en
leading to the following standard differential equation
\eq
\left[\Delta[\boeta] - e^{2\pi \mathbbm i k}(-\lambda)^{r+1} \,
p(z,s)\right] \,\psi^{(1)}_{[k]}(z,\bz) = 0~,
\label{eq:ardiffeq}
\en
where we introduced the short-hand notation $\Delta[\mathbf{f}] := 
\Delta[f_{r+1}]\Delta[f_{r}] \cdots \Delta[f_1]$ for any vector $\mathbf{f} = 
(f_1,f_2,\ldots,f_{r+1})$ and the operator $\Delta$ was defined in (\ref{eq:diffop}).
\subsection{The $\rho\rightarrow 0$ limit}
\label{subsec:rho0limitar}
Let us analyse the behaviour of the differential equation (\ref{eq:ardiffeq}) in 
the  $z\rightarrow 0$ limit. Heuristically, we expect the possible behaviours around the origin to be independent of the parameter $k$, thus we omit its dependence. Using the asymptotic (\ref{eq:eta0asymptotic}), the 
differential operators simplify to
\eq
\Delta[\eta^j] \equiv (\partial + \partial\eta^j) 
\underset{\rho\rightarrow0}{\sim}  \partial - \frac{\nu^0_j}{z} \; .
\en

We seek solutions that behave as powers in the vicinity of the apex, that is 
$\psi^{(1)} = (\lambda^{\frac{1}{M+1}}z)^{\gamma} + \cdots$; the choice of the constant $\lambda^{\frac{1}{M+1}}$ in front of $z$ comes from the comparison with the so-called \emph{conformal limit} \cite{Dore_Fald_Negr_Tate_12}. In this way the action of the differential 
operator $\Delta$ becomes
\eq
\Delta[\eta^j] \psi^{(1)}(z,\bz) \underset{\rho\rightarrow0}{\sim} 
(\gamma - \nu^0_j) (\lambda^{\frac{1}{M+1}}z)^{\gamma-1} \; ,
\en
and the differential equation reduces to an algebraic one
\begin{align}
&\left(\Delta[\boeta] - e^{2\pi \mathbbm i k}(-\lambda)^{r+1}\, p\right) 
\psi^{(1)}(z,\bz) = 0 \; \Rightarrow \nonumber
\\	\nonumber
\\
 &\Rightarrow \; (\lambda^{\frac{1}{M+1}}z)^{\gamma -r-1} \prod_{j=1}^{r+1} \left(\gamma - \nu^0_j -(j -1)\right) \, 
 \underset{\rho\rightarrow0}{\sim}\, (\lambda^{\frac{1}{M+1}}z)^{\gamma} e^{2\pi \mathbbm i k}(-\lambda)^{r+1}\, s^{M(r+1)} \; ,	\nonumber
\end{align}
which, in turn, becomes an index equation for the exponent $\gamma$; this does not depend on the parameter $k$, as expected:
\eq
\prod_{j=1}^{r+1}\left(\gamma - \nu^0_j -(j-1)\right) = 0 \quad \Rightarrow 
\quad \gamma_l = \nu^0_l + (l-1) \; .
\en
We have thus $r+1$ different leading behaviours in the limit $\rho \rightarrow 0$, that we denote with the symbols $\chi_l$:
\eq
\left(\Delta[\boeta] - e^{2\pi \mathbbm i k}(-\lambda)^{r+1}\, p\right) 
\chi_l (z,\bz) = 0 \; ,\qquad \chi_l(z,\bz) \underset{\rho\rightarrow 
0}{\sim} (\lambda^{\frac{1}{M+1}}z)^{\nu^0_l+(l-1)}~.
\en
These functions are linearly independent, as is easy to show by evaluating their Wronskian in 
the limit $\rho\rightarrow 0$, thus the set
\eq
\mathcal B^{0} = \left\{\chi_l\right\}_{l=1}^{r+1}~,
\en
is a basis in the space of solutions to equation (\ref{eq:ardiffeq}).
\subsection{The $\rho \rightarrow \infty$ limit}
For the large-$\rho$ limit it is useful to consider both the holomorphic and the 
anti-holomorphic equations
\eq
\bar \Delta[-\eta^{r+1}]\bar \Delta[-\eta^1] \ldots \bar \Delta[-\eta^{r}]{\bar\psi}^{(1),r+1}_{[k]} = 
e^{-2\pi \mathbbm i k}\bar p \left(-\lambda^{-1} \right)^{r+1}{\bar\psi}^{(1),r+1}_{[k]} \; ,
\label{eq:arbardiffeq}
\en
at the same time, 
where $\bar\Delta[-\eta^{r+1}]{\bar\psi}^{(1)}_{[k]} = -e^{-2\pi \mathbbm i k}
\lambda^{-1}\bar p {\bar\psi}^{(1),r+1}_{[k]}$.

When considering large values of $\rho$, the differential operators $\Delta$ can be approximated with pure derivatives:
\eq
\Delta[ \eta^j] = (\partial + \partial\eta^j ) \underset{\rho\rightarrow\infty}{\sim}
(\partial - \frac{ \nu_j^\infty}{z}) \underset{\rho\rightarrow\infty}{\sim} \partial~, 
\en
and similarly for the anti-holomorphic ones. This simplification brings us immediately to the following coupled equations
\begin{align}
(-1)^{r+1} \partial^{r+1} \psi^{(1)}_{[k]} &\underset{\rho\rightarrow\infty}{\sim} e^{2\pi \mathbbm i k}\left(e^{\theta} z^M \right)^{r+1} \psi^{(1)}_{[k]} \; ,	\nonumber
\\\label{eq:arasymptinfequation}
\\
(-1)^{r+1}\bar\partial^{r+1} {\bar\psi}^{(1),r+1}_{[k]} &\underset{\rho\rightarrow\infty}{\sim} e^{-2\pi \mathbbm i k}\left( e^{-\theta} \bar z^M \right)^{r+1} 
{\bar \psi}^{(1),r+1}_{[k]} \; ,	\nonumber
\end{align}
where we introduced the variable $\theta = \ln\lambda$.

By applying standard WKB analysis 
\cite{Bend_Orsz_99} we find that the equations (\ref{eq:ardiffeq}), (\ref{eq:arbardiffeq}) admit, for $k=0$, solutions with the following asymptotic behaviours
\begin{align}
&\psi^{(1)} \underset{\rho\rightarrow \infty}{\sim} c\,\left(e^{\theta} z^{M}\right)^{-\frac{r}{2}} e^{- \frac{z^{M+1}}{M+1} 
e^{\theta} + f(\bar z)} \; , 	\nonumber
\\
\\
&{\bar \psi}^{(1)} \underset{\rho\rightarrow \infty}{\sim} \bar c\,\left(e^{\theta} 
\bar z^{M}\right)^{\frac{r}{2}} e^{-\frac{\bar z^{M+1}}{M+1} e^{-\theta} + g(z)} \; ,	\nonumber
\end{align}
if $z=\rho e^{i\phi}$ and $\bar z=\rho e^{-i\phi}$ lie in the following wedge of the $(\rho,\phi)$ complex plane
\eq
-\frac{r+2}{r+1}\frac{\pi}{M+1} <\phi < \frac{r+2}{r+1}\frac{\pi}{M+1}~.
\en
Using the compatibility condition (\ref{eq:compcondonfunctions}) we easily find the expressions 
for $f(\bar z)$, $g(z)$ and $\bar c$, leaving us with a single arbitrary multiplicative constant, $c$ which we set to $1$:
\eq
\psi^{(1)} \underset{\rho\rightarrow \infty}{\sim} \left(-e^{\theta}
z^{M}\right)^{-\frac{r}{2}} e^{-2 \frac{\rho^{M+1}}{M+1} \cosh\big(\theta + \mathbbm i(M+1)\phi\big)} \; .
\label{eq:bigrhoexpansioncomplete}
\en
Since $\psi^{(1)}_{[k]}$ is obtained from $\psi^{(1)}_{[0]}$ through a Symanzik rescaling

\eq
	\psi^{(1)}_{[k]}(z,\bz ;\lambda) = \omega^{k\frac{r}{2}}\,\psi^{(1)}(\omega^k z,\omega^{-k} z;\omega^{-k} \lambda) \; ,
\en
with $\omega = e^{\frac{2\pi \mathbbm i}{(r+1) M}}$, if we define the \emph{Stokes sectors} as
\eq
\mathscr S_k \; : \quad \left\{(\rho,\phi) \; \big/ \;
\Big\vert\phi - \frac{2\pi k}{(r+1)(M+1)}\Big\vert < \frac{\pi}{(r+1)(M+1)}\right\} \; ;
\en
then we can argue that
\begin{Conj}
For any $M>1/2$, the function $\psi^{(1)}_{[k]}(z,\bz;\theta)$ is, for any value of $k$, the unique solution to the differential equations (\ref{eq:ardiffeq}) and (\ref{eq:arbardiffeq}) having the following properties:
\begin{itemize}
\item $\psi^{(1)}_{[k]}(z,\bz;\theta)$ is an entire function of 
$(z,\bz;\theta)$ living, given the branch point at 
$(z,\bz)=(0,0)$ in $p(z)$ and $\bar p(\bz)$, on a suitable cover of the punctured complex plane;
\item $\psi^{(1)}_{[k]}(z,\bz;\theta)$ admit the asymptotic representation

\eq
	\psi^{(1)}_{[k]} \sim \left(e^{\theta + 2\pi \mathbbm i\frac{k}{r+1}}z^M\right)^{-\frac{r}{2}} e^{-2 \frac{\rho^{M+1}}{M+1}\cosh\left[\theta + \mathbbm i(M+1)(\phi+2\pi\frac{k}{(r+1)(M+1)})\right]}
\label{eq:bigrhoexpansioncomplete2}
\en
as $\rho\rightarrow\infty$ in the sector 
$$ 
(z,\bz)\in \bigcup_{j=k}^{r+1+k} \mathscr S_{j-\frac{r+1}{2}}~.
$$
\end{itemize}
\end{Conj}
Let us introduce the following sets of solutions
\eq
\mathcal B^{\infty}_k = \left\{\psi^{(1)}_{[k+\ell]}\right\}_{\ell=1}^{r+1}~,
\label{eq:inftybasis}
\en
we see that in the wedge $\mathscr S_{k+\frac{r+1}{2}}\cup \mathscr S_{k+1+\frac{r+1}{2}}$, 
all the functions in this set can be asymptotically expanded using (\ref{eq:bigrhoexpansioncomplete2}). 
This fact allows us to calculate in the $\rho\rightarrow\infty$ limit their $(r+1)$-Wronskian 
$W^{(r+1)}[\psi^{(1)}_{[k+1]},\ldots,\psi^{(1)}_{[k+r+1]}]$, which turns out to be a constant for any value of $k$ 
(see Section \ref{sec:wronsky}). This means that $\mathcal B^{\infty}_k$ is a basis of the space of 
solutions to (\ref{eq:ardiffeq}), (\ref{eq:arbardiffeq}).

With this basis it is now possible to build other functions by taking Wronskians of its elements. 
Recalling what has been said in Chapter \ref{chap:repthe}, this corresponds to building solutions to differential equations 
associated with the other nodes of the Dynkin diagram.
It is not difficult to evaluate the $r$-Wronskian $W^{(r)}[\psi^{(1)}_{[k+1]},\ldots,\psi^{(1)}_{[k+r]}]$ which, 
in the algebraic picture, should correspond to the $r$-th node in the Dynkin diagram. 
This is done again in the Section \ref{sec:wronsky} and we see that
\eq
\psi^{(r)}_{[k+\frac{r+1}{2}]} = W^{(r)}[\psi^{(1)}_{[k+1]},\ldots,\psi^{(1)}_{[k+r]}] 
\underset{\rho\rightarrow\infty}{\sim} \mathbbm i^{\frac{r}{2}(r-1)} r^{\frac{r-1}{2}} 
\psi^{(1)}_{[k+\frac{r+1}{2}]} \; .
\nonumber
\en
This result is expected: since $\bLambda_{r}$ and $\bLambda_1$ are adjoint one of the other, 
$\psi^{(r)}_{[k+\frac{r+1}{2}]}$ must satisfy the differential equation adjoint to (\ref{eq:ardiffeq}) which, in the 
$\rho\rightarrow\infty$ limit corresponds to (\ref{eq:arasymptinfequation}) with $k$ shifted a half-integer. Inspecting the form of the $\ell$-wronskian of consecutive elements of the basis $\mathcal B_k^{\infty}$ is possible to show that it indeed has the same asymptotics 
as $\psi^{(\ell)}_{[k+\frac{\ell-1}{2}]}$ and satisfies the pseudo-differential equation corresponding to 
the representation $\bLambda_\ell$; however the latter equations  are in general  complicated  and difficult to write explicitly. 
We will give an example below for the simple  $\mathfrak a_3$-related case.

\subsection{The spectral determinants and the Bethe Ansatz equations}
\label{subsec:arbetheansatz}

Now we have got all the elements needed to extract the Bethe Ansatz Equations; let us briefly recall them:

\begin{itemize}
	\item the $\psi$-system (\ref{eq:genpsisystem}):
		\eq
			W^{(2)}[\psi^{(a)}_{[k]} , \psi^{(a)}_{[k']}] = \prod_b \left(\psi^{(b)}_{[k'']}\right)^{A_{ab}}~;
			\label{eq:genpsisystem2}
		\en
	\item the definition of the multi-vector's first component in terms of the basic functions $\psi^{(1)}_{[k]}$, a consequence of (\ref{eq:multivectordefinition}):
		\eq
			\psi^{(a)}_{[k]} = W^{(a)}[\psi_{[-\frac{a-1}{2}]},\psi_{[-\frac{a-3}{2}]},\ldots ,\psi_{[\frac{a-1}{2}]}] ~ ,
			\label{eq:firstcompmultivect}
		\en
	with $\psi_{[k]}(z,\bz ;\lambda) \equiv \psi^{(1)}_{[k]}(z,\bz ;\lambda)$;
	\item the existence of two distinct bases of the space of solutions to the equation (\ref{eq:ardiffeq}):
		\eq
			\mathcal B^{0} = \left\{\chi_l\right\}_{l=1}^{r+1} \; , \quad \chi_l(z,\bz) \underset{\rho\rightarrow 0}{\sim} (\lambda^{\frac{1}{M+1}}z)^{\gamma_l} \; , \quad \gamma_l \doteq \nu^0_l + l-1 \; ,
			\label{eq:rho0basis}
		\en
		
		\eq
			\mathcal B^{\infty}_k = \left\{\psi_{[k+\ell]}\right\}_{\ell=1}^{r+1} \; , \quad \psi_{[k]}(z,\bz ;\lambda) = \omega^{k\frac{r}{2}}\psi(\omega^k z,\omega^{-k} \bz ;\omega^{-k}\lambda) \; ,
			\label{eq:rhoinfbasis}
		\en
	where the functions $\psi_{[k]}$ are defined by their asymptotic behaviour (\ref{eq:bigrhoexpansioncomplete}).
\end{itemize}
Note that for the algebra $\mathfrak a_r$, the relation (\ref{eq:firstcompmultivect}) is valid for any value of $a=1,\ldots,r$ and we set $\psi^{(0)}_{[k]} = \psi^{(r+1)}_{[k]} = 1$. We also impose an ordering $\gamma_i < \gamma_j \; , \ \forall i<j$ on the exponents of (\ref{eq:rho0basis}), which is consistent with the constraints (\ref{eq:nu0constraint}). Let us also recall that, since $\sum_{j=1}^{r+1} \eta_j =0$, we also have $\sum_{j=1}^{r+1} \gamma_j = r\frac{r+1}{2}$.

The first step towards the Bethe Ansatz equations consists in expanding the elements of the basis $\mathcal B^{\infty}_k$ in terms of those of the basis $\mathcal B^0$; for $k=0$ we write

\eq
	\psi(z,\bz ;\lambda) = \sum_{l=1}^{r+1} Q^l(\lambda)\chi_l(z,\bz) \; ,
\en
where we introduced the \emph{connection coefficients} $Q^l(\lambda)$, functions of $\lambda$ only. These functions can be written as

\eq
	Q^{l_\alpha}(\lambda) = \frac{W[\psi(z,\bz ;\lambda),\chi_{l_1}(z,\bz),\ldots ,\chi_{l_r}(z,\bz)]}{W[\chi_{l_\alpha}(z,\bz),\chi_{l_1}(z,\bz),\ldots ,\chi_{l_r}(z,\bz)]} \; , \quad l_1<\ldots <l_r \; ,
\label{eq:conncoeff}
\en
for any $l_\alpha \neq l_i \, , i=1,\ldots ,r$. Recalling what we said in subsection \ref{subsec:stokesmultrel}, we see that they are spectral determinants for the differential equations (\ref{eq:ardiffeq}, \ref{eq:arbardiffeq}). For generic values of $k$, we exploit the relation (\ref{eq:rhoinfbasis}) and the fact that the functions $\chi_l(z,\bz)$ transform trivially under the rotation of the arguments (consequence of their behaviour around the origin):

\eq
	\chi_l(\omega^k z, \omega^{-k} \bz) = \omega^{k\gamma_l\frac{M}{M+1}}\chi_l(z,\bz) \; .
\en
We easily see that

\eq
	\psi_{[k]}(z,\bz ;\lambda) = \sum_{l=1}^{r+1} \omega^{(\gamma_l\frac{M}{M+1} + \frac{r}{2})k}\, Q^l(\omega^{-k}\lambda)\chi_l(z,\bz) \; .
	\label{eq:psikinchi}
\en

Next we use the relation (\ref{eq:firstcompmultivect}) to express all the functions $\psi^{(a)}_{[k]}$ in terms of the basis $\mathcal B^0$ and of the connection coefficients. In order to do so, we need the following two simple properties of the determinants:

\eq
	\det\Big(a_{i,j} + b_{i,j}\delta_{j,k}\Big)_{i,j=1}^n = \det\Big(a_{i,j}\Big)_{i,j=1}^n + \det\Big(a_{i,j} + (b_{i,j}-a_{i,j})\delta_{j,k}\Big)_{i,j=1}^n \; ,
\en

\eq
	\det\Big(c^{\delta_{j,k}}\,a_{i,j}\Big)_{i,j=1}^n = c\det\Big(a_{i,j}\Big)_{i,j=1}^n \; ,
\en
valid for any $n\in \mathbbm N$ and $k=1,\ldots,n$. Plugging (\ref{eq:psikinchi}) into (\ref{eq:firstcompmultivect}) we obtain

\begin{align}
	\psi^{(a)}_{[k]}(z,\bz ;\lambda) &= \sum_{\mathbf l}\omega^{k \left(a\frac{r}{2} +\frac{M}{M+1}\sum_{j=1}^a \gamma_{l_j}\right)}\left(\prod_{j=1}^a \omega^{\frac{M}{M+1}\gamma_{l_j}(j-\frac{a+1}{2})}Q^{l_j}(\omega^{-k}\,\omega^{\frac{a+1}{2}-j}\lambda)\right) \times \nonumber
	\\
	\\
	&\times\, W^{(a)}[\chi_{l_1}(z,\bz),\chi_{l_2}(z,\bz),\ldots ,\chi_{l_a}(z,\bz)] \; ,	\nonumber
\end{align}
which we can rewrite, by exploiting the antisymmetry of the Wronskian, as follows

\eq
	\psi^{(a)}_{[k]}(z,\bz ;\lambda) =\omega^{k a \frac{r}{2}} \sideset{}{'}\sum_{\mathbf l} \widetilde{\omega}^{k\sum_{j=1}^a \gamma_{l_j}}Q_{[k]}^{\{l_1,\ldots ,l_a\}}(\lambda)\,W^{(a)}[\chi_{l_1}(z,\bz),\ldots ,\chi_{l_a}(z,\bz)] \; .
	\label{eq:higherpsiinchi}
\en
In the above two formulae we defined $\widetilde{\omega} \doteq \omega^{\frac{M}{M+1}} \equiv e^{\frac{2\pi \mathbbm i}{(r+1)(M+1)}}$; the symbol $\displaystyle{\sum_{\mathbf l}}$ denotes a sum over all the configurations of the indices $l_1,\ldots ,l_a =1,\ldots ,r+1$, while in $\displaystyle{\sideset{}{'}\sum_{\mathbf l}}$ the additional constraint $0\leq l_1 < \ldots < l_a$ is imposed. We also have introduced the composite functions
\eq
Q^{\{l_1,\ldots ,l_a\}}(\lambda) = \det\left\vert Q^{l_k}_{[j-\frac{a+1}{2}]}(\lambda)\right\vert_{j,k=1}^a\;,
\label{eq:compconncoeff}
\en
and

\eq
	Q^{\{l_1,\ldots ,l_a\}}_{[k]}(\lambda) =  Q^{\{l_1,\ldots ,l_a\}}(\omega^{-k}\lambda) \; .
\en
Now, it is easy to see that for $\rho\rightarrow 0$, the Wronskians in (\ref{eq:higherpsiinchi}) behave as $z^{\sum_{j=1}^a \gamma_{l_j} -a\frac{a-1}{2}}$ which implies, given the ordering of the exponents, that we can order the addends (\ref{eq:higherpsiinchi}) in the following way:

\begin{align}
	\omega^{-k a \frac{r}{2}}&\psi^{(a)}_{[k]}(z,\bz ;\lambda) = \widetilde{\omega}^{k \alpha_a}\mathcal Q^{(a)}_{[k]}(\lambda)\, W^{(a)}[\chi_{l_1}(z,\bz),\ldots ,\chi_{l_a}(z,\bz)] +	\nonumber
	\\ \label{eq:higherpsiinchiexpansion}
	\\
	&+ \widetilde{\omega}^{k(\alpha_{a+1}-\gamma_a)} \bar{\mathcal Q}^{(a)}_{[k]}(\lambda)\, W^{(a)}[\chi_{l_1}(z,\bz),\ldots ,\chi_{l_{a-1}}(z,\bz),\chi_{l_{a+1}}(z,\bz)] + \ldots \ ,	\nonumber
\end{align}
where $\alpha_a \doteq \sum_{j=1}^a \gamma_j$ and

\eq
	\mathcal Q^{(a)}_{[k]}(\lambda) \doteq Q^{\{1,\ldots ,a\}}_{[k]}(\lambda) \; , \qquad \bar{\mathcal Q}^{(a)}_{[k]}(\lambda) \doteq Q^{\{1,\ldots ,a-1,a+1\}}_{[k]}(\lambda) \; .
\en

The last step consists in plugging (\ref{eq:higherpsiinchiexpansion}) into the $\psi$-system (\ref{eq:genpsisystem2}) and identifying the terms with the same power for $\rho\rightarrow 0$. We will need another easily proven functional relation of the Wronskians:

\eq
	W[W^{(m)},\hat W^{(m)}] = W^{(m-1)}W^{(m+1)}\; ,
\en
where $W^{(m)} \doteq W^{(m)}[f_1,\ldots ,f_m]$ and $\hat W^{(m)}\doteq W^{(m)}[f_1,\ldots ,f_{m-1},f_{m+1}]$, for any set of functions $\{f_j\}_{j=1}^{m+1}$.

We end up with the following relation

\eq
 \mathcal Q^{(a+1)}(\lambda)\mathcal Q^{(a-1)}(\lambda) = \widetilde{\omega}^{\frac{\gamma_{a+1}-\gamma_a}{2}}\mathcal Q^{(a)}_{[-\frac{1}{2}]}(\lambda)\bar{\mathcal Q}^{(a)}_{[\frac{1}{2}]}(\lambda) -  \widetilde{\omega}^{-\frac{\gamma_{a+1}-\gamma_a}{2}}\bar{\mathcal Q}^{(a)}_{[-\frac{1}{2}]}(\lambda)\mathcal Q^{(a)}_{[\frac{1}{2}]}(\lambda) \; ,
\en
Finally let us suppose that $\{\lambda_i^{(a)}\}_i$ are the zeroes of the function $\mathcal Q^{(a)}(\lambda)$, which means that $\{\omega^{\pm\frac{1}{2}}\lambda_i^{(a)}\}_i$ are zeroes of, respectively, $\mathcal Q^{(a)}_{\frac{1}{2}}(\lambda)$ and $\mathcal Q^{(a)}_{-\frac{1}{2}}(\lambda)$, and evaluate the above relation at the points $\omega^{\pm\frac{1}{2}}\lambda_i^{(a)}$ for some $i$:

\begin{align}
	\mathcal Q^{(a+1)}(\omega^{-\frac{1}{2}}\lambda_i^{(a)})\mathcal Q^{(a-1)}(\omega^{-\frac{1}{2}}\lambda_i^{(a)}) &= -\widetilde{\omega}^{-\frac{\gamma_{a+1}-\gamma_a}{2}}\bar{\mathcal Q}^{(a)}_{[-\frac{1}{2}]}(\omega^{-\frac{1}{2}}\lambda_i^{(a)})\mathcal Q^{(a)}_{[\frac{1}{2}]}(\omega^{-\frac{1}{2}}\lambda_i^{(a)}) \; ,	\nonumber
	\\
	\\
	\mathcal Q^{(a+1)}(\omega^{\frac{1}{2}}\lambda_i^{(a)})\mathcal Q^{(a-1)}(\omega^{\frac{1}{2}}\lambda_i^{(a)}) &= \widetilde{\omega}^{\frac{\gamma_{a+1}-\gamma_a}{2}}\mathcal Q^{(a)}_{[-\frac{1}{2}]}(\omega^{\frac{1}{2}}\lambda_i^{(a)})\bar{\mathcal Q}^{(a)}_{[\frac{1}{2}]}(\omega^{\frac{1}{2}}\lambda_i^{(a)}) \; .	\nonumber
\end{align}
The ratio of these two equations can be recast in the following nice and compact form:

\eq
	\prod_{b=1}^{r} \Omega^{B_{a,b}\Gamma_b}\frac{\mathcal Q^{(b)}_{[B_{a,b}]}(\lambda_i^{(a)})}{\mathcal Q^{(b)}_{[-B_{a,b}]}(\lambda_i^{(a)})} = -1 \; ,
\label{eq:BAEar}
\en
where we introduced $\Omega \doteq \widetilde{\omega}^{-(r+1)}=e^{2\pi \mathbbm i\frac{M}{M+1}}$ the matrix $B=\frac{1}{2}C$, with $C$ being the Cartan matrix of the algebra $\mathfrak a_r$, $\Gamma_a \doteq \frac{2}{M(r+1)}\left(\sum_{j=1}^a \gamma_j -a \frac{r}{2}\right)$ and we used the relation
\eq
	\frac{1}{M(r+1)} (\gamma_{a+1}-\gamma_a) = -\sum_{b=1}^r B_{a,b} \Gamma_b \; .
\en
The equations (\ref{eq:BAEar}) are the Bethe Ansatz Equations for the algebra $\mathfrak a_r$ and the parameters correspond to those of \cite{Dore_Dunn_Maso_Suzu_Tate_07,Adam_Dunn_13}. The difference here lies in the analytic properties of the functions $\mathcal Q^{(b)}$ which, in our case, present essential singularities at both the origin and the point at infinity. The reason for this is that $\mathcal Q^{(b)}$ are spectral determinants for the pair of equations (\ref{eq:ardiffeq}, \ref{eq:arbardiffeq}) which, in the massive case, are not decoupled; thus the $Q$-functions depend on both $\lambda$ and $\lambda^{-1}$.

Let us spend a few words on the interpretation of the connection coefficients we introduced above. By using the definitions (\ref{eq:conncoeff}, \ref{eq:compconncoeff}) and going through some rather boring algebra, we obtain the following neat expression for the composite connection coefficients in terms of the functions $\psi$ and $\chi$:
\eq
Q^{\{l_1,\ldots ,l_a\}}(\lambda) = \frac{W[\psi_{[-\frac{a-1}{2}]},\psi_{[1-\frac{a-1}{2}]},\ldots ,\psi_{[\frac{a-1}{2}]},\chi_{l_{a+1}},\ldots ,\chi_{l_{r+1}}]}{W[\chi_{l_1},\ldots ,\chi_{l_{r+1}}]}\;,
\en
where $l_i\neq l_j\, ,\ \forall i\neq j$, $l_1 < \ldots < l_a$ and $l_{a+1} < \ldots < l_{r+1}$. From this expression we see that $Q^{\{l_1,\ldots ,l_a\}}$ vanishes if and only if $\lambda$ is such that
\eq
\exists j\in[1,a]\subset \mathbb N\;,\ \exists \alpha\in[a+1,r+1]\subset\mathbb N\;\Big\backslash\; \psi_{[j-\frac{a+1}{2}]} \underset{\rho\rightarrow 0}{\sim} \chi_{\ell_j}\;,\ \ell_j = l_\alpha\;.
\en
In other words, $Q^{\{l_1,\ldots ,l_a\}}(\lambda)$ vanishes at the eigenvalues $\{\lambda_j^{\{l_1,\ldots ,l_a\}}\}_j$ of an appropriate central boundary problem associated with the differential equations (\ref{eq:ardiffeq}, \ref{eq:arbardiffeq}). We can thus deduce that, up to a
factor of an entire function with no zeros, $Q^{\{l_1,\ldots ,l_a\}}(\lambda)$ is the \emph{spectral determinant} associated with said boundary problem.

\section{The $\mathfrak d_r$ case}
\label{sec:drcase}
\setcounter{equation}{0}

Let us now consider the algebra $\hat{\mathfrak d}_r$, we will be sketchier as everything that has been said for the $\hat{\mathfrak a}_{r}$ algebra applies with some minor modifications.
The Dynkin diagram and Cartan matrix of the affine algebra $\hat{\mathfrak d}_r$ are shown in Figure \ref{fig:dralgebra}.
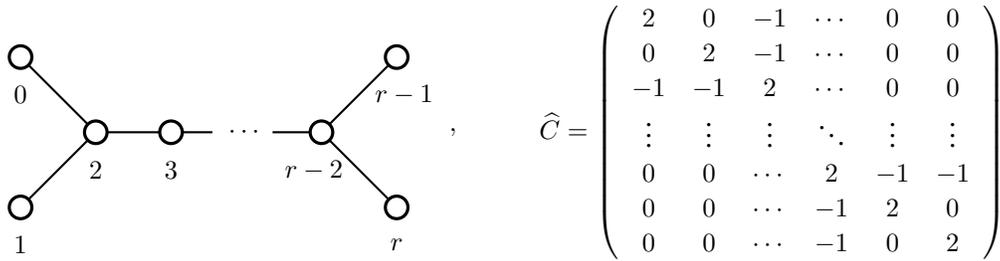
\begin{figure}[h!]
\begin{center}
\begin{tikzpicture}

\node[dynkin node] at (0,1){};
\node[] at (0,0.5){$0$};
\node[dynkin node] at (0,-1){};
\node[] at (0,-1.5){$1$};
\draw[scale=1,domain=0:0.8,smooth,variable=\x, color=black,thick,-] plot ({0.1+\x},{0.9-\x});
\draw[scale=1,domain=0:0.8,smooth,variable=\x, color=black,thick,-] plot ({0.1+\x},{-0.9+\x});
\node[dynkin node] at (1,0){};
\node[] at (1,-0.5){$2$};
\draw[scale=1,domain=1.15:1.85,smooth,variable=\x, color=black,thick,-] plot ({\x},{0});
\node[dynkin node] at (2,0){};
\node[] at (2,-0.5){$3$};
\draw[scale=1,domain=2.15:2.55,smooth,variable=\x, color=black,thick,-] plot ({\x},{0});
\node[] at (3,0) {$\cdots$};
\draw[scale=1,domain=3.35:3.85,smooth,variable=\x, color=black,thick,-] plot ({\x},{0});
\node[dynkin node] at (4,0){};
\node[] at (3.9,-0.5){$r-2$};
\node[dynkin node] at (5,1){};
\node[] at (5.1,0.5){$r-1$};
\node[dynkin node] at (5,-1){};
\node[] at (5,-1.5){$r$};
\draw[scale=1,domain=0:0.8,smooth,variable=\x, color=black,thick,-] plot ({4.1+\x},{0.1+\x});
\draw[scale=1,domain=0:0.8,smooth,variable=\x, color=black,thick,-] plot ({4.1+\x},{-0.1-\x});

\node[] at (5.75,0) {$,$};
\node[] at (10,0) {$\hat C = \left( \begin{array}{cccccc}2&0&-1&\cdots&0&0  \\0&2&-1&\cdots&0&0  \\
	                                            -1&-1&2&\cdots&0&0  \\
				\vdots&\vdots&\vdots&\ddots&\vdots&\vdots\\ 
				                      0&0&\cdots&2&-1&-1 \\0&0&\cdots&-1&2&0 \\
				                      0&0&\cdots&-1&0&2 
				                      \end{array} \right)$};
\end{tikzpicture}
\end{center}
\caption{Dynkin diagram and Cartan matrix for the affine algebra $\hat{\mathfrak d}_r$}
\label{fig:dralgebra}
\end{figure}

The simple roots can be represented in an $r$-dimensional space as
\eq
\left\{\begin{array}{l}
\alpha_i = \epsilon_i - \epsilon_{i+1}, \qquad (0 < i < r) \\
\alpha_r = \epsilon_{r-1} + \epsilon_r, \\
\alpha_0 = - \epsilon_1 - \epsilon_2.
\end{array}\right.
\en
Also in this case $\vert \boldsymbol{\alpha}_i \vert^2 = 2, \; \forall i$, 
the Dynkin indices are $n_i = 2, \; 1< i< r-1$ and $n_i \in \{1, \; i=0,1,r-1,r\}$.

Again, let us consider the lowest-dimensional fundamental representation $\bLambda_1$. 
This representation is $2r$-dimensional; the Cartan subalgebra and the step operators are representable as
\begin{align}
&[\mathscr R^{2 r}(h_i)]_{jk} = \delta_{j,k} (\delta_{i,j} - \delta_{2 r -(i-1),j}), \nonumber \\
&[\mathscr R^{2 r}(e_i)]_{jk} = \left\lbrace
\begin{array}{lcl}
(\delta_{i,j} + \delta_{i,2r - j})\delta_{j+1,k}, &   (0 < i < r), & \\
(\delta_{r+i-1,j} + \delta_{r+1,j})\delta_{j+2,k}, &   (i=r, \, i=0) &
\end{array}
\right. \\
&[\mathscr R^{2 r}(f_i)] = [\mathscr R^{2 r}(e_i)]^{T}. \nonumber
\end{align}
The relevant objects for our analysis are in Table~\ref{tab31} and the weight diagram is depicted in Figure \ref{Drweight}.

\begin{minipage}{\textwidth}
  \begin{minipage}[b]{0.35\textwidth}
  \vspace{0.25\textwidth}
  \centering
    \includegraphics[width=3.8cm]{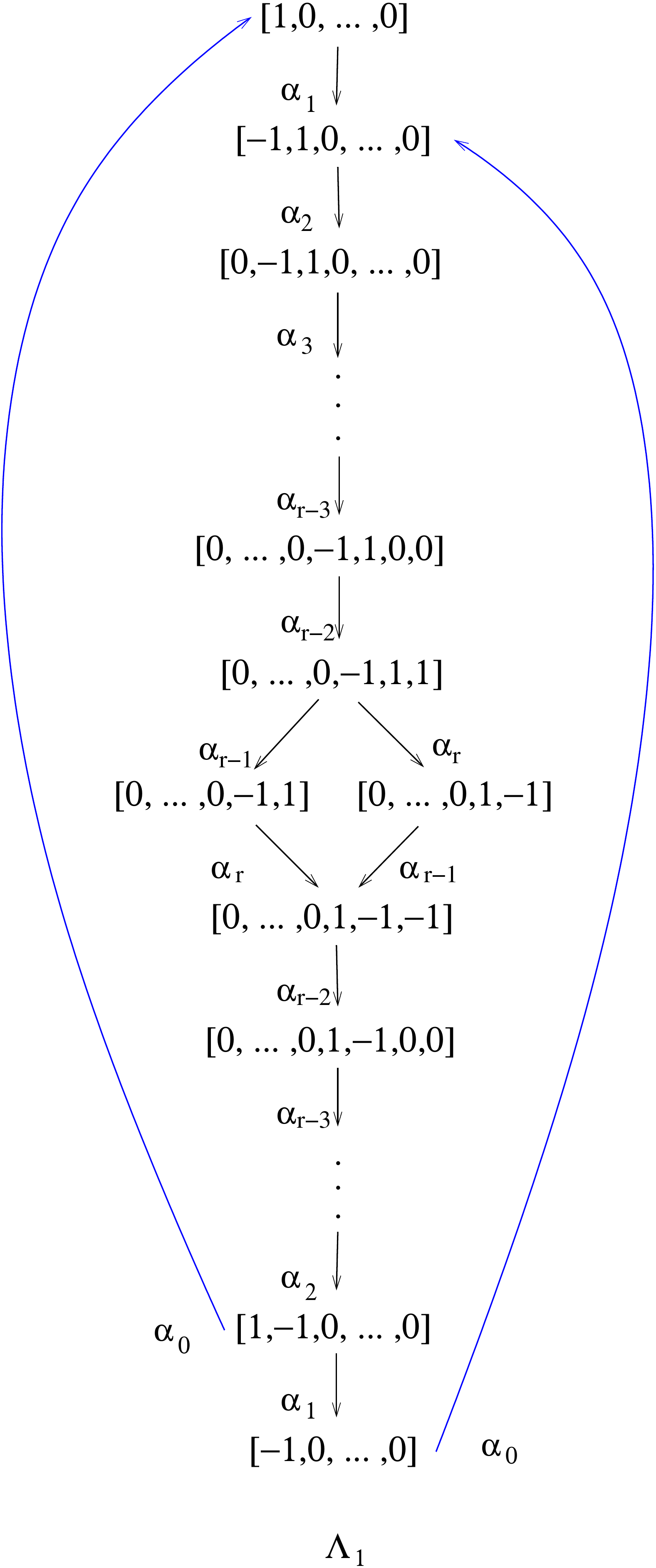}
    \captionof{figure}{The $\hat{\mathfrak d}_r$ weight diagram.}
    \label{Drweight}
  \end{minipage}
\hspace{0.05\textwidth}
  \begin{minipage}[b]{0.4\textwidth}
   \begin{center}
\begin{tabular}{|c | c|}
\hline
Definition & Value in $\bLambda_1$ \\
\hline & \\
$\mu_j^2 = \frac{n_i}{2}\vert\balpha_i\vert^2$ & $\left\{\begin{array}{l} 1, 
\quad (j=0,1,r-1,r)\\ 
2, \quad (1<j<r-1)
\end{array}\right.$ \\ & \\
$\kappa_j = \sum_{a=1}^r \eta^a h_j^a$ & $\left\{\begin{array}{l} \eta^j, 
\qquad \qquad (1\leq j\leq r)  \\ 
-\eta^{2r-(j-1)}, \quad (r+1\leq j\leq 2r)
\end{array}\right.$ \\ & \\
$\nu_j^{\infty}$ & $ (j-r) M$ \\ &\\
\hline
\end{tabular}
 \captionof{table}{The relevant object for  $\mg =\mathfrak d_r$. }
      \label{tab31}
\end{center}
    \end{minipage}
    \vspace{0.05\textwidth}
\end{minipage}

The matrix $\tilde {\mathcal E}$ has the following form:
\eq
\tilde {\mathcal E} =	\left(
\begin{array}{c c c c c c c c c c c}
0      & 1           & 0         & 0           & \cdots     & \cdots & \cdots & \cdots       & \cdots & 0       & 0      \\
0      & 0           & \sqrt{2}  & 0           & \cdots     & \cdots & \cdots & \cdots       & \cdots & 0       & 0      \\
0      & 0           & 0         & \sqrt{2}    & \cdots     & \cdots & \cdots & \cdots       & \cdots & 0       & 0      \\
\vdots & \vdots      & \vdots    & \vdots      & \ddots     & \cdots & \cdots & \cdots       & \cdots & \vdots  & \vdots \\
\cdots & \cdots      & 0         & \cdots      & \cdots     &  1     & 1      & 0            & \cdots & 0       & 0      \\
\cdots & \cdots      & 0         & \cdots      & \cdots     & 0      & 0      &  1  & \cdots & 0       & 0      \\
\cdots & \cdots      & 0         & \cdots      & \cdots     & 0      & 0	&  1 & \cdots	 & 0       & 0      \\
\vdots & \vdots      & \vdots    & \cdots      & \cdots     & \cdots & \cdots & \cdots       & \ddots & \vdots  & \vdots  \\
0      & 0           & 0         & \cdots      & \cdots     & \cdots & \cdots & \cdots       &\cdots  & \sqrt{2} & 0 \\
  p(z,s) & 0   & 0         & \cdots      & \cdots     & \cdots & \cdots & \cdots      &\cdots  & 0        & 1 \\
0      &  p(z,s) & 0      & \cdots      & \cdots     & \cdots & \cdots & \cdots      &\cdots  & 0       & 0	
\end{array}
\right)
\en

Following the same procedure as in the preceding section, we arrive at the following pseudo-differential equation
\bea
\psi^{(1),2}_{[k]} &=& \frac{e^{2\pi \mathbbm i k}}{4}\left(-\lambda \sqrt{2}  \right)^{2r-2} \left( \prod_{j=2}^{r-1} \Delta^{-1}[\kappa_j] \right)\left( \Delta^{-1}[\kappa_r] + \Delta^{-1}[\kappa_{r+1}] \right) \times  \nn \\
&\times& 
\left( \prod_{j=2}^{r-1} \Delta^{-1}[\kappa_{r+j}] \right) \left( \Delta^{-1}[\kappa_{2r}] p + p \Delta^{-1}[\kappa_{1}] \right) \,\psi^{(1),2}_{[k]}\;.
\label{eq:drpathequation}
\eea
The following identities are easily shown to be true, considering that $\kappa_{2r+1-j} = - \kappa_{j}$
\eq
\left( \Delta^{-1}[\kappa_r] + \Delta^{-1}[\kappa_{r+1}] \right) = 2\Delta^{-1}[\kappa_r] \partial\Delta^{-1}[\kappa_{r+1}],
\en

\eq
\left( \Delta^{-1}[\kappa_{2r}] p + p \Delta^{-1}[\kappa_{1}] \right) = \Delta^{-1}[\kappa_{2r}]\Big(2 p \partial + (\partial p)\Big)\Delta^{-1}[\kappa_1] = 
2 \Delta^{-1}[\kappa_{2r}]\sqrt{p}\; \partial \sqrt{p}\;\Delta^{-1}[\kappa_1], \;
\en
where we have also used that $(\partial p)f  = 2\sqrt{p}(\partial \sqrt{p})f = 2\sqrt{p}\partial(\sqrt{p}f )-2p(\partial f)$.

Finally, introducing the vector $\boldsymbol{\eta}^{\dagger} = (-\eta^r,\cdots,-\eta^1)$ and rewriting $\psi^{(1),2}_{[k]}$ in terms of $\psi^{(1)}_{[k]}$:
\eq
\Delta[ \boldsymbol{\eta}^{\dagger}] \partial^{-1} \Delta[ \boldsymbol{\eta}]
\,\psi^{(1)}_{[k]} = e^{2\pi i k}\left(\lambda \sqrt{2}  \right)^{2r-2} 
\sqrt{p} \partial \sqrt{p} \;\psi^{(1)}_{[k]}.
\label{eq:dreqation}
\en
The pseudo-differential equation (\ref{eq:dreqation}) has an interesting property. Let us write the right-hand side of (\ref{eq:dreqation}) in the following form
\eq
\Delta[ \boldsymbol{\eta}^{\dagger}] \partial^{-1} \Delta[ \boldsymbol{\eta}]
\,\psi^{(1)}_{[k]} = \mathcal O(\partial-\partial\eta^r)\partial^{-1}(\partial + \partial\eta^r)\Phi,
\en
where the differential operator $\mathcal O$ and the vector $\Phi$ do not depend on $\eta^r$. It is easy to see that this object is invariant for $\eta^r \longleftrightarrow -\eta^r$:
\eq
\mathcal O(\partial-\partial\eta^r)\partial^{-1}(\partial + \partial\eta^r)\Phi = \mathcal O(\partial^2-\eta^r\partial^{-1}\eta^r)\Phi = \mathcal O(\partial+\partial\eta^r)\partial^{-1}(\partial - \partial\eta^r)\Phi \; . \nonumber
\en
This invariance reflects the $\mathbbm Z_2$ symmetry of the Dynkin diagram of $\mathfrak d_r$ which exchanges the $(r-1)$-th and the $r$-th nodes. In fact by inspecting the $\Psi$-system (\ref{eq:multivectordefinition}) for the $\mathfrak d_r$ algebra:
\begin{align}
&\Psi^{(n)} = \bigwedge_{\ell =1}^n\left(\Psi^{(1)}_{[\ell-\frac{n+1}{2}]}\right) \,; \qquad \forall n\neq r-1,r,	\nonumber
\\
&\Psi^{(r-1)}\otimes\Psi^{(r)} = \bigwedge_{\ell=1}^{r-1} \left(\Psi^{(1)}_{[\ell-\frac{r}{2}]}\right), \label{eq:Drpsi}
\\
&\Psi^{(r-1)}_{[-\frac{1}{2}]}\otimes\Psi^{(r-1)}_{[\frac{1}{2}]} + \Psi^{(r)}_{[-\frac{1}{2}]}\otimes\Psi^{(r)}_{[\frac{1}{2}]} = \bigwedge_{\ell=1}^{r} \left(\Psi^{(1)}_{[\ell-\frac{r+1}{2}]}\right),	\nonumber
\end{align}
we see that the symmetry $\eta^r\longleftrightarrow -\eta^r$  leaves invariant the equations associated to the first $r-2$ nodes, while it might exchange the ones associated to the $(r-1)$-th and $r$-th nodes.

Following the steps outlined in the $\mathfrak a_{r}$ case, we take the $\rho \rightarrow 0$ limit, ask for the solutions to behave as powers near the origin and obtain the index equation:
\eq
\frac{\prod_{j=1}^r [\gamma +\nu^0_j -(j-1)][\gamma -\nu^0_j +(j-1) -2r+2]}{[\gamma -2r+2][\gamma -r+1]}=0.
\label{eq:drindicialeq}
\en
Thus, provided the quantities $\nu^0_j$ are such that no term in the numerator 
simplifies with the denominator, we have $2r$ possible behaviours around the origin
\eq
\chi_l \underset{\rho\rightarrow0}{\sim} z^{\gamma_l} \ , 
\quad \gamma_l = \left\{\begin{array}{l}l-1-\nu^0_l \; , \qquad \qquad (1\leq l \leq r) \\ l-1+ \nu^0_{2r-l},  \quad\, \qquad (r\leq l \leq 2r-1)
\end{array}\right.
\en
Again we can easily see the presence of the above-mentioned $\mathbbm Z_2$ symmetry by noticing that the index equation (\ref{eq:drindicialeq}) is invariant for under
the inversion $\eta^r \longleftrightarrow -\eta^r$.
%
For the large $\rho$ behaviour too we follow the steps underlined in   Section \ref{sec:arcase} and we easily arrive to the asymptotic form of the differential equation
\eq
\partial^{2r-2} \,\psi^{(1)}_{[k]} \underset{\rho\rightarrow\infty}{\sim} e^{2\pi \mathbbm i k}\left(\lambda \sqrt{2}  z^M\right)^{2r-2} \,\psi^{(1)}_{[k]}.
\en
The form of this asymptotic relation is exactly the same as in the $\mathfrak a_{r}$ case, so we can apply what we said then without any modification. The following set
\eq
\mathcal B_k^{\infty} = \left\{\psi^{(1)}_{[k+\ell]}\right\}_{\ell = 1}^{2r-2},
\en
where the functions behave as
\eq
\psi^{(1)}_{[k]} \underset{\rho\rightarrow\infty}{\sim} \left( \sqrt{2}  e^{\theta+2\pi \mathbbm i\frac{k}{2r-2}} z^M\right)^{-\frac{2r-3}{2}}\; e^{-2 \sqrt{2}  \frac{\rho^{M+1}}{M+1} \cosh\left(\theta +\mathbbm i(M+1)\left(\phi + \frac{2\pi k}{(2r-2)(M+1)}\right)\right)},
\en
in the wedge
\eq
(z,\bz)\in \bigcup_{j=k}^{2r-2+k} \mathscr S_{j-r+1}, 
\en
with 
\eq
\mathscr S_k \; : \quad \left\{(\rho,\phi) \big/ \Big\vert\phi - \frac{2\pi k}{(2r-2)(M+1)}\Big\vert < \frac{\pi}{(2r-2)(M+1)}\right\},	
\en
is a basis for the space of solutions. Thus one can build new functions out of this basis, by taking wronskians
\eq
\psi^{(n)}_{[k+\frac{n+1}{2}]} = W^{(n)}[\,\psi^{(1)}_{[k+1]},\ldots,\,\psi^{(1)}_{[k+n]}],
\label{eq:drhigherfunctions}
\en
and show that, for $n<r-1$ they satisfy the pseudo-differential equations associated with the representation $\bLambda_n$. 
The functions $\psi^{(r-1)}_{[k]}$ and $\psi^{(r)}_{[k]}$ associated with the $(r-1)$-th 
and $r$-th nodes of the Dynkin diagram, however, cannot be obtained simply through (\ref{eq:drhigherfunctions}). From the $\Psi$-system (\ref{eq:Drpsi}) we see that the following two functions
\eq
\xi^{(r-1)}_{[k+\frac{r}{2}]} = W^{(r-1)}[\,\psi^{(1)}_{[k+1]},\ldots,\,\psi^{(1)}_{[k+r-1]}],~~
\xi^{(r)}_{[k+\frac{r+1}{2}]} = W^{(r)}[\,\psi^{(1)}_{[k+1]},\ldots,\,\psi^{(1)}_{[k+r]}],
\en
correspond to the representations $\bLambda_{r-1}\otimes\bLambda_r$ and $\bLambda_{r-1}\times\bLambda_{r-1}\; \oplus\; \bLambda_{r}\times \bLambda_{r}$, respectively, meaning that
\eq
\xi^{(r-1)}_{[k+\frac{r}{2}]} = \,\psi^{(r-1)}_{[k+\frac{r}{2}]}\;\psi^{(r)}_{[k+\frac{r}{2}]},~~
\xi^{(r)}_{[k+\frac{r+1}{2}]} = \,\psi^{(r-1)}_{[k+\frac{r}{2}]}\;\psi^{(r-1)}_{[k+\frac{r+2}{2}]} + 
\,\psi^{(r)}_{[k+\frac{r}{2}]}\;\psi^{(r)}_{[k+\frac{r+2}{2}]}.	
\label{eq:xitopsi}
\en
Setting $\xi^{(a)} = \psi^{(a)}$ for $a<r-1$, we see that these functions satisfy an $\mathfrak a_{r-1}$ $\psi$-system:
\eq
W^{(2)}[\xi^{(a)}_{[-\frac{1}{2}]},\xi^{(a)}_{[+\frac{1}{2}]}] = \xi^{(a-1)}\,\xi^{(a+1)}, \quad a = 1,2,\ldots , r-1,
\en
with $\xi^{(0)} = 1$. If we impose the ordering  $\gamma_i < \gamma_j, \; \forall i < j$ for the  exponents, then we can follow what we said in the previous Section and write directly the BA equations for the spectral determinants $\hat{\mathcal Q}^{(a)}$ associated to the functions $\xi^{(a)}$:
\eq
\frac{\hat{\mathcal Q}^{(a-1)}_{[-\frac{1}{2}]}(E_i^{(a)},\bnu^0)}{\hat{\mathcal Q}^{(a-1)}_{[\frac{1}{2}]}(E_i^{(a)},\bnu^0)}\; \frac{\hat{\mathcal Q}^{(a)}_{[1]}(E_i^{(a)},\bnu^0)}{\hat{\mathcal Q}^{(a)}_{[-1]}(E_i^{(a)},\bnu^0)}\; \frac{\hat{\mathcal Q}^{(a+1)}_{[-\frac{1}{2}]}(E_i^{(a)},\bnu^0)}{\hat{\mathcal Q}^{(a+1)}_{[\frac{1}{2}]}(E_i^{(a)},\bnu^0)} = -\Omega^{\frac{\gamma_{a+1}-\gamma_a}{2 M(r-1)}},
\label{eq:D_Abethe}
\en
which is valid for any $a<r$.
Now, by performing the following identifications between the functions $\hat{\mathcal Q}$, associated to the $\xi$, and the $\mathcal Q$, associated to the $\psi$
\begin{align}
&\hat{\mathcal Q}^{(a)}(E,\bnu^0) = \mathcal Q^{(a)}(E,\bnu^0), \quad a = 1, 2, \ldots ,r-2,\nonumber
\\
&\hat{\mathcal Q}^{(r-1)}(E,\bnu^0) = \mathcal Q^{(r-1)}(E,\bnu^0)\mathcal Q^{(r)}(E,\bnu^0),
\\
&\hat{\mathcal Q}^{(r)}(E,\bnu^0) = \mathcal Q^{(r-1)}_{[-\frac{1}{2}]}(E,\bnu^0)\mathcal Q^{(r-1)}_{[\frac{1}{2}]}(E,\bnu^0),	\nonumber
\end{align}
which reflect the relations (\ref{eq:xitopsi}), we recover the BAEs for the algebra $\mathfrak d_r$, apart from the $(r-1)$-th equation
\eq
\prod_{b=1}^r \Omega^{B_{a,b}\Gamma_b} \frac{\mathcal Q^{(b)}_{[B_{a,b}]}(E_i^{(a)},\bnu^0)}{\mathcal Q^{(b)}_{[-B_{a,b}]}(E_i^{(a)},\bnu^0)} = -1 \; , \quad \forall a\neq r-1,
\en
where $B = \frac{1}{2} C$, with $C$ th Cartan Matrix of the algebra $\mathfrak d_r$ and $-\sum_{b=1}^r B_{a,b}\Gamma_{b} = \sum_{b=1}^r(\balpha_a)^b \tilde{\gamma}_b$ with $\tilde{\gamma}_a = \gamma_a - (r-1)$:
\begin{align}
&\Gamma_a = \frac{1}{M(r-1)} \big(\sum_{j=1}^a \gamma_j -a(r-1)\big), \qquad \Gamma_{r-1} = \frac{1}{2 M (r-1)} \big(\sum_{j=1}^r\gamma_j-r(r-1)\big), \nonumber \\
&\Gamma_{r} = \frac{1}{2M(r-1)}\big(\sum_{j=1}^{r-1}\gamma_j - \gamma_r - (r-1)(r-2)\big).
\end{align}
In order to recover the missing equation, we exploit the $\mathbbm Z_2$ symmetry $\eta^r \rightarrow -\eta^r$, which sends $\gamma_r \rightarrow 2(r-1) - \gamma_r$ and, consequently, $\Gamma_{r-1} \leftrightarrow \Gamma_r$. Since the equations for the $(r-1)$-th and $r$-th nodes gets swapped by this symmetry, we 
also have
\eq
\mathcal Q^{(r-1)}(E,\{\nu_1^0,\ldots \nu_{r-1}^0,\nu_r^0\}) \rightarrow \mathcal Q^{(r)}(E,\{\nu_1^0,\ldots \nu_{r-1}^0,-\nu_r^0\}),
\en
while the remaining $\mathcal Q$ are left untouched. Therefore, we finally obtain the full set of BA equations for the algebra $D_r$:
\eq
\prod_{b=1}^r \Omega^{B_{a,b}\Gamma_b} \frac{\mathcal Q^{(b)}_{[B_{a,b}]}(E_i^{(a)},\bnu^0)}{\mathcal Q^{(b)}_{[-B_{a,b}]}(E_i^{(a)},\bnu^0)} = -1, \quad \forall a=1,2,\ldots, r.
\en
While it is possible to extend this setup to the Affine Toda Field Theories based on the algbras $\mathfrak e_6$, $\mathfrak e_7$ and $\mathfrak e_8$: although in general the form of the pseudo-differential equation is extremely complicated, the essential structures needed to arrive at the Bethe Ansatz Equations are the same as the $\mathfrak a_r$ case, with some slight modifications. For what concerns the non simply-laced algebras, on the other hand, the situation is a bit more complicated; recently some results were obtained by K. Ito and C. Locke \cite{Ito_Lock_13}. Instead of pushing forward in the algebras of type $\mathfrak e$ we prefer to apply the framework presented above to some simple though interesting case.

\section{A particular case: $\mathfrak a_3 \cong \mathfrak d_3$}
\label{sec:d3}
\setcounter{equation}{0}

\begin{figure}[h!]
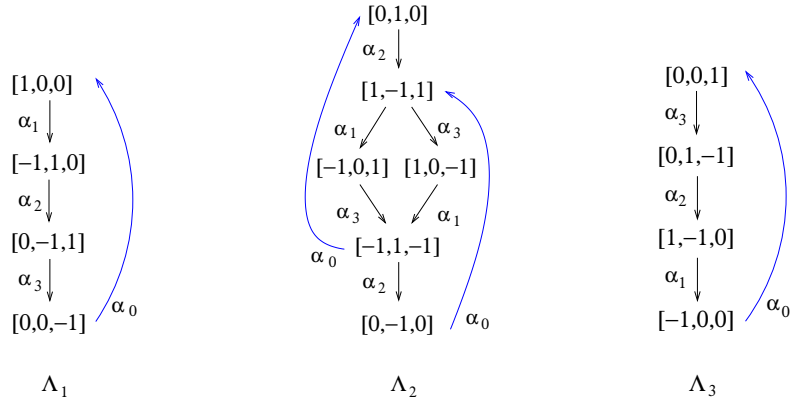

\begin{center}
\includegraphics[width=1.8cm]{A31.pdf} 
\hspace{2.cm}
\includegraphics[width=2.5cm]{A32.pdf} 
\hspace{2.cm}
\includegraphics[width=1.8cm]{A33.pdf}
\end{center}
\caption{The $\mathfrak a_3$ weight systems.}
\label{fig:a3E1_2}
\end{figure}

Let us consider the algebra $\mathfrak a_3$. The weight systems for the fundamental representations are depicted in Figure \ref{fig:a3E1_2}.

For what concerns the fundamental representation $\bLambda_1$, we already have the equation
\eq
\Delta[\eta^4]\Delta[\eta^3]\Delta[\eta^2]\Delta[\eta^1]\,\psi^{(1)}_{[k]} = \left(e^{4 \theta +\pi i 2 k}\right) p(z,s)\,\psi^{(1)}_{[k]} \; ,
\label{eq:a3equation}
\en
and the two basis of the solution space
\eq
\mathcal B^0 = \left\{\chi_l\ \big\vert \ \chi_l \underset{\rho\rightarrow0}{\sim} z^{\gamma_l}\right\}_{l=1}^4,\qquad 
\mathcal B_k^{\infty} = \left\{\,\psi^{(1)}_{[k+\ell]}\right\}_{\ell=1}^k\;,\qquad \gamma_l=\nu_l^0+l-1\;,
\en
where
\eq
\psi^{(1)}_{[k]} \underset{\rho\rightarrow\infty}{\sim} \left(- e^{\theta + \pi \mathbbm i\frac{k}{2}}z^M\right)^{-\frac{3}{2}} 
e^{-2 \frac{\rho^{M+1}}{M+1}\cosh\left(\theta + \mathbbm i(M+1)\big(\phi + \frac{\pi k}{2(M+1)}\big)\right)},
\en
with
$
(z,\bz) \in \bigcup_{j=-2}^{2} \mathscr S_{j-k}.
$
For the representation $\bLambda_3$, the equation has the same  structure as (\ref{eq:a3equation}); what changes are the parameters $\kappa$:
\eq
\kappa_j^{(\bLambda_3)} = - \eta^{5-j}.
\en
As expected,  the resulting equation is the adjoint to (\ref{eq:a3equation}):
\eq
\Delta[-\eta^1]\Delta[-\eta^2]\Delta[-\eta^3]\Delta[- \eta^4]\,\psi^{(3)}_{[k]} = 
\left(e^{\theta +\pi \mathbbm i \frac{k}{2}}\right)^4 p(z,s) \,\psi^{(3)}_{[k]}.
\label{eq:a3adjequation}
\en
Let us now write the equation (\ref{eq:a3equation}) in the following form
\eq
\left(\partial^4 + f_2\partial^2 + f_1\partial + (f_0 -e^{2\pi \mathbbm i k} \lambda^4 p(z,s) )\right)\,\psi^{(1)}_{[k]}=0,
\en
where  $f_i$ are somewhat intricate functions of the fields $\eta^i$. We wish to see what differential function satisfies the following wronskian
\eq
W^{(3)} = W^{(3)}[\,\psi^{(1)}_{[-1]},\,\psi^{(1)}_{[0]},\,\psi^{(1)}_{[1]}].
\en
Direct calculation shows that
\eq
\left(\partial^4 + f_2\partial^2 + \Big(2\big(\partial f_2\big)-f_1\Big)\partial + 
\Big(\big(\partial^2 f_2) - \big(\partial f_1\big) + f_0 - \lambda \Big)\right) W^{(3)} = 0,
\en
and, through some quite boring algebra, one shows that this is indeed the same equation as (\ref{eq:a3adjequation}) 
with $k=0$, given we set $\eta^4 = -\sum_{j=1}^3 \eta^j$.

Now let us work with $\bLambda_2$. 
Looking at its weight system in 
Figure~\ref{fig:a3E1_2} and at the values of $\kappa_j$:
\eq
\kappa_j = \left\{\begin{array}{l r} \tilde \eta^j, & (1\leq j \leq 3), \\ -\tilde \eta^{7-j}, & (4\leq j \leq 6), \end{array}\right.
\en
where we have defined
\eq
\tilde \eta^j = \left\{\begin{array}{l r} \eta^1 + \eta^2, & (j=1), \\ \eta^1 + \eta^3, & (j=2), \\ \eta^2 + \eta^3, & (j=3),
\end{array}\right.
\en
we immediately recognise the structure of $\bLambda_1(\mathfrak d_3)$, which was to be expected, given the isomorphism $\mathfrak a_3\cong \mathfrak d_3$. Thus we have
\eq
\Delta[-\tilde\eta^1]\Delta[-\tilde\eta^2] \Delta[-\tilde\eta^3]\partial^{-1}\Delta[\tilde\eta^3]\Delta[\tilde\eta^2]\Delta[\tilde\eta^1]
\,\psi^{(2)}_{[k]} = \left(e^{4 \theta+2\pi \mathbbm i k }\right)\sqrt{p}\,\partial\,\sqrt{p}\;\psi^{(2)}_{[k]},	
\en
where the only differences with the $r=3$ equation (\ref{eq:dreqation}) can be  reabsorbed trough a redefinition of the fields 
 $\eta^j\rightarrow\tilde\eta^j$ and the mass scale $m=2 \rightarrow m=\sqrt{2}$. 

\section{Another particular case: $\mathfrak d_4$ and triality}
\label{sec:d4}
\setcounter{equation}{0}

The algebra $\mathfrak d_4$ is a peculiar one. In fact its Dynkin diagram possesses a bigger symmetry group than the other algebras, namely the symmetric group $S_3$ (equivalently, the symmetry group Dih$_3$ of an equilateral triangle).
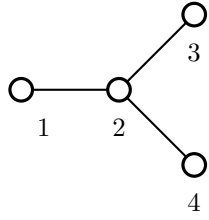
\begin{figure}[h!]
\begin{center}
\begin{tikzpicture}

\node[dynkin node] at (-0.3,0){};
\node[] at (0,-0.5){$1$};
\draw[scale=1,domain=0:1.,smooth,variable=\x, color=black,thick,-] plot ({-0.15+\x},{0});
\node[dynkin node] at (1,0){};
\node[] at (1,-0.5){$2$};
\node[dynkin node] at (2,1){};
\node[] at (2,0.5){$3$};
\node[dynkin node] at (2,-1){};
\node[] at (2,-1.5){$4$};
\draw[scale=1,domain=0:0.8,smooth,variable=\x, color=black,thick,-] plot ({1.1+\x},{0.1+\x});
\draw[scale=1,domain=0:0.8,smooth,variable=\x, color=black,thick,-] plot ({1.1+\x},{-0.1-\x});

\end{tikzpicture}
\end{center}
\caption{Dynkin diagram for the affine algebra $\hat{\mathfrak d}_4$}
\label{fig:dralgebra}
\end{figure}
This symmetry group acts on the representations $\bLambda_1$, $\bLambda_3$ and $\bLambda_4$ permuting them, thus we expect the pseudo-differential equations associated with these to be structurally identical and to be mapped one into the other by the action of $S_3$. This property of the algebra $\mathfrak d_4$ is known as \emph{triality}.
\begin{center}
\begin{figure}[h]
\begin{minipage}[adjusting]{20\linewidth}
\hspace{1.8cm}
\includegraphics[width=3.cm]{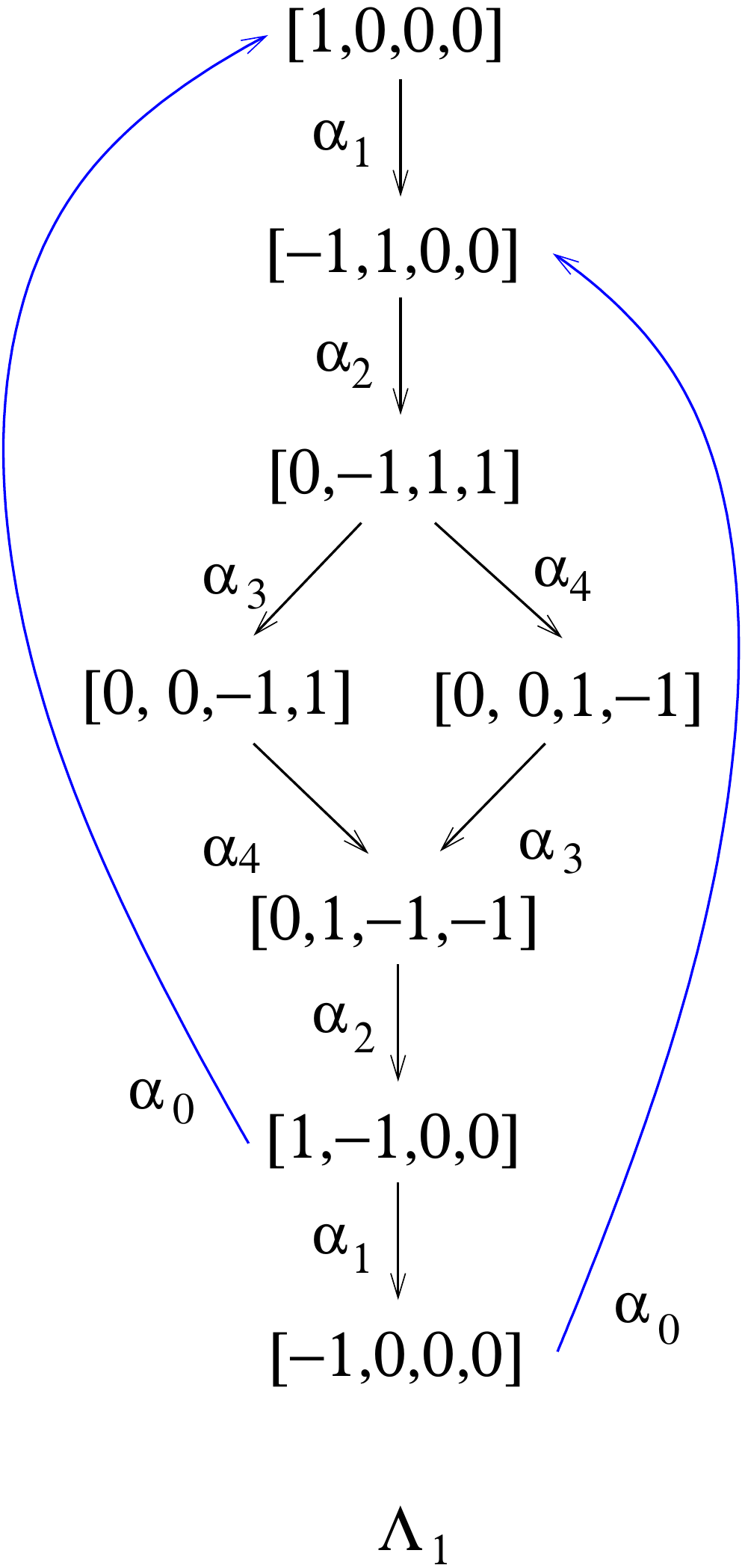} 
\hspace{1.7cm}
\includegraphics[width=3.cm]{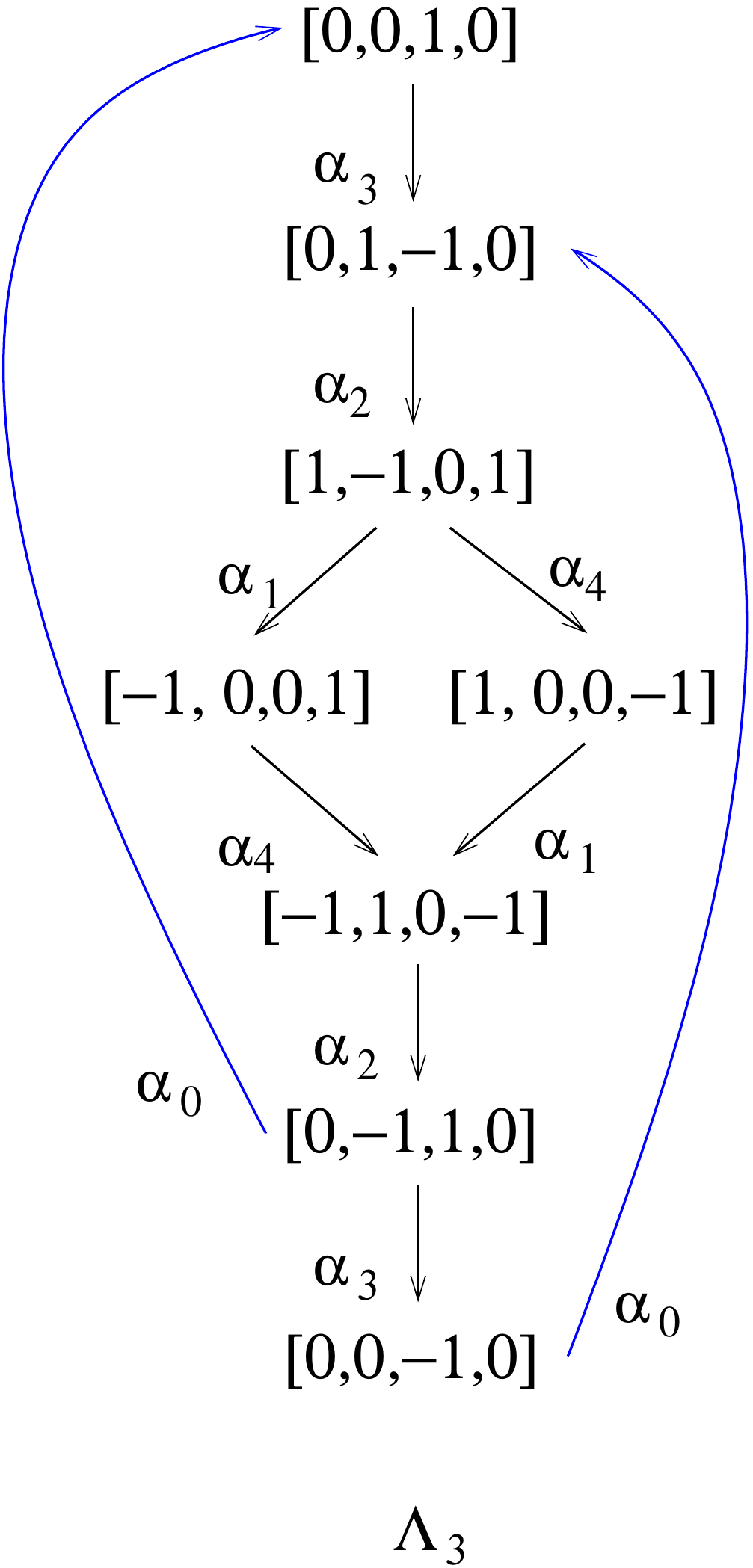} 
\hspace{1.7cm}
\includegraphics[width=3.cm]{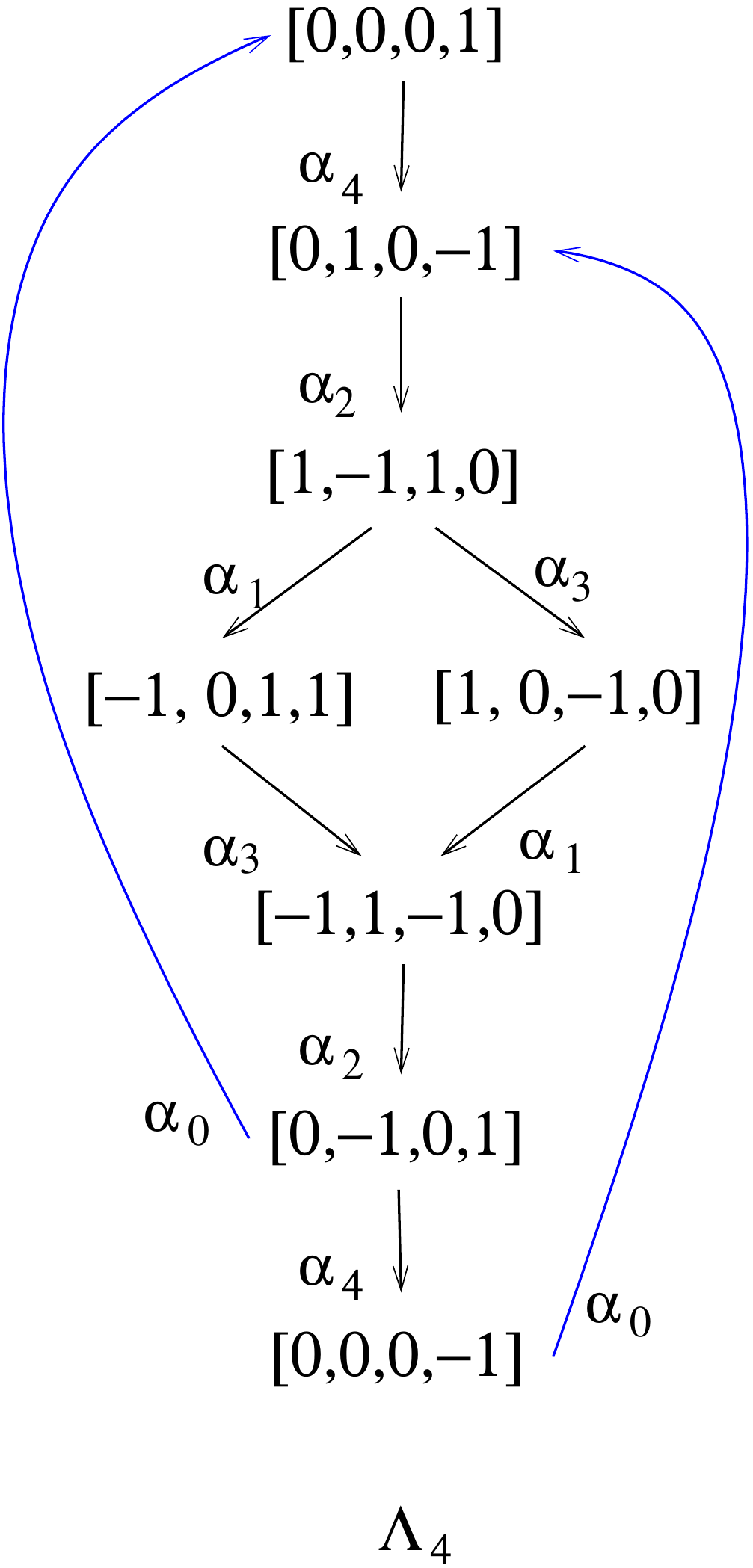}
\end{minipage}
\caption{The $D_4$ weight systems ($\bLambda_1$, $\bLambda_3$, $\bLambda_4$) .}
\label{fig:d4}
\end{figure}
\end{center}
It is indeed sufficient to look at the weight systems in Figure~\ref{fig:d4}, 
to remark that this is the case. The three representations have dimension $8$ and the equations have the form
\eq
\Delta[\boeta^{\dagger}_{(n)}]\partial^{-1}\Delta[\boeta_{(n)}]\,\psi^{(n)}_{[k]} = \left(e^{8 \theta+2\pi \mathbbm i k}\right)\sqrt{p}\,\partial\,\sqrt{p}\,\psi^{(n)}_{[k]}, \quad n=1,3,4,
\label{eq:D4equations}
\en
where the entries vectors $\boeta_{(n)}$ are different in the three cases:
\eq
\boeta_{(1)} = \left(\begin{array}{c}\eta^1\\ \eta^2\\ \eta^3 \\ \eta^4 \end{array}\right),\qquad 
\boeta_{(3)} = \left(\begin{array}{c}-\eta^4+\frac{1}{2}\sum_{i=1}^4\eta^i\\ -\eta^3+\frac{1}{2}\sum_{i=1}^4\eta^i\\ -\eta^2+\frac{1}{2}\sum_{i=1}^4\eta^i 
\\ -\eta^1+\frac{1}{2}\sum_{i=1}^4\eta^i\end{array}\right), \qquad 
\boeta_{(4)} = \left(\begin{array}{c}\frac{1}{2}\sum_{i=1}^4\eta^i\\ -\eta^4-\eta^3+\frac{1}{2}\sum_{i=1}^4\eta^i\\ 
-\eta^2-\eta^4+\frac{1}{2}\sum_{i=1}^4\eta^i \\ -\eta^2-\eta^3+\frac{1}{2}\sum_{i=1}^4\eta^i\end{array}\right)	\nonumber
\en
Let us recall that the following vectors
\eq
\hat\boeta_{(1)} = \left(\begin{array}{c}\eta^1\\ \eta^2\\ \eta^3 \\ -\eta^4 \end{array}\right),\qquad 
\hat\boeta_{(3)} = \left(\begin{array}{c}-\eta^4+\frac{1}{2}\sum_{i=1}^4\eta^i\\ -\eta^3+\frac{1}{2}\sum_{i=1}^4\eta^i\\ -\eta^2+\frac{1}{2}\sum_{i=1}^4\eta^i 
\\ \eta^1-\frac{1}{2}\sum_{i=1}^4\eta^i\end{array}\right), \qquad 
\hat\boeta_{(4)} = \left(\begin{array}{c}\frac{1}{2}\sum_{i=1}^4\eta^i\\ -\eta^4-\eta^3+\frac{1}{2}\sum_{i=1}^4\eta^i\\ 
-\eta^2-\eta^4+\frac{1}{2}\sum_{i=1}^4\eta^i \\ \eta^2+\eta^3-\frac{1}{2}\sum_{i=1}^4\eta^i\end{array}\right)	\nonumber
\en
generate the same equations (\ref{eq:D4equations}).

\begin{figure}[h]
\centering
\includegraphics[width=4cm]{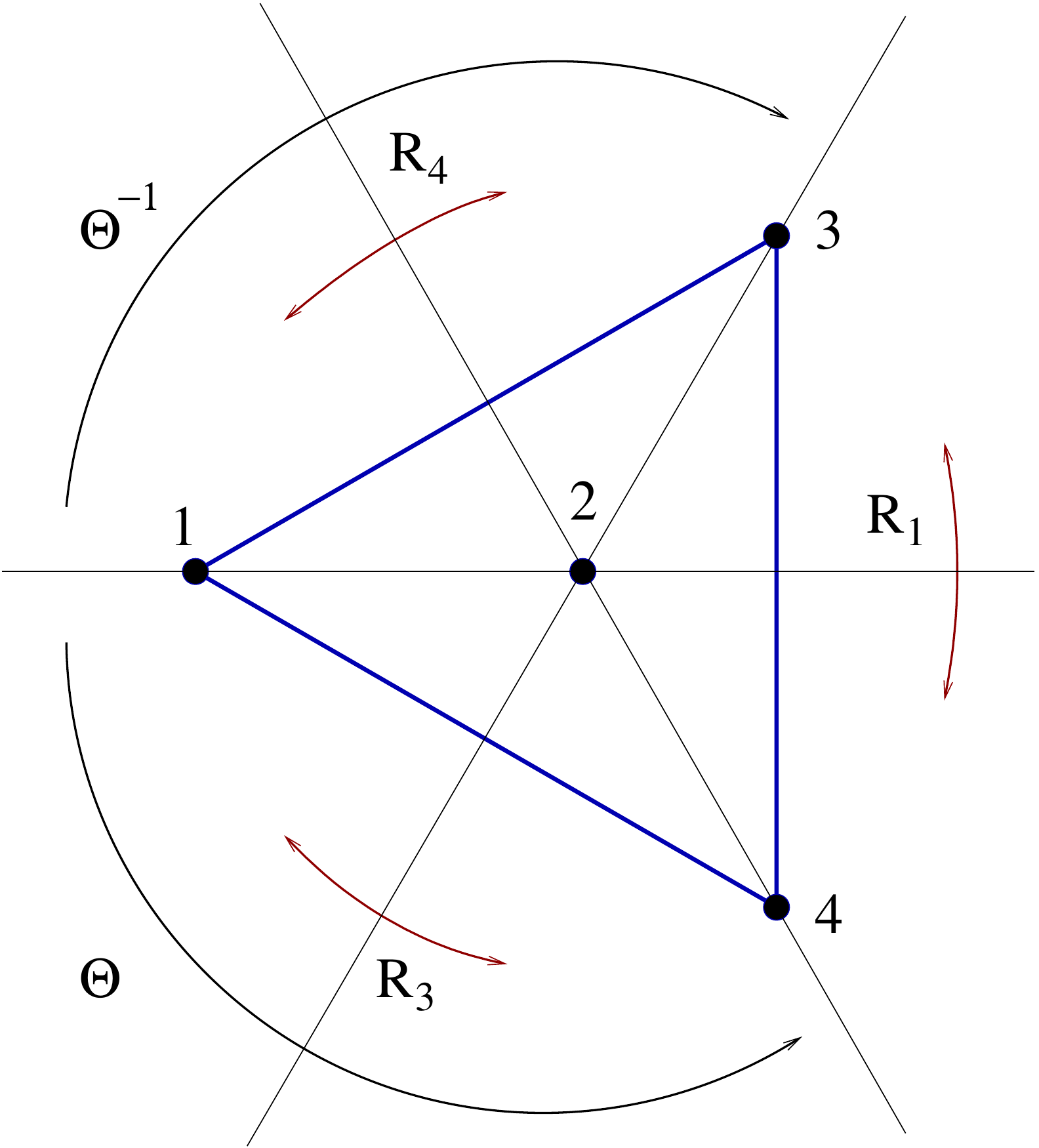}
\caption{The $S_3$ symmetry.}
\label{diag12}
\end{figure}

The action of the group $S_3$ can be represented as the action of $4\times 4$ matrices on the parameter vector:
\eq
\boeta =\left(\begin{array}{c}\eta^1\\ \eta^2\\ \eta^3 \\ \eta^4 \end{array}\right).
\en
It is straightforward to check that the action of the following matrices
\eq
R_3 = \frac{1}{2}\left(\begin{array}{c c c c} 1 & 1 & 1 & 1 \\ 1 & 1 & -1 & -1 \\ 1 & -1 & 1 & -1 \\ 1 & -1 & -1 & 1\end{array}\right), \quad R_4 = \frac{1}{2}\left(\begin{array}{c c c c} 1 & 1 & 1 & -1 \\ 1 & 1 & -1 & 1 \\ 1 & -1 & 1 & 1 \\ -1 & 1 & 1 & 1\end{array}\right).
\en
on the vector $\boeta$ implies
\eq
\begin{array}{c c c c}
& \boeta_{(1)} \rightarrow \boeta_{(4)} & & \boeta_{(1)} \rightarrow \boeta_{(3)}\\
R_3: \quad & \boeta_{(3)} \rightarrow \hat\boeta_{(3)} &,\qquad R_4: \quad & \boeta_{(3)} \rightarrow \boeta_{(1)}\\
& \boeta_{(4)} \rightarrow \boeta_{(1)} & & \boeta_{(4)} \rightarrow \hat\boeta_{(4)}
\end{array}.
\label{eq:r3r4symm}
\en
Direct computation shows that
\eq
R_3^2 = R_4^2 = \mathbb I, \quad (R_3 R_4)^3 = \mathbb I, 
\en
which is the presentation of $S_3$:
\eq
S_3 = \langle s_1,s_2 \vert s_1^2 = s_2^2 = (s_1 s_2)^3 = \mathbb I \rangle.
\en
The remaining elements of the group are the following reflection
\eq
R_1 =R_3R_4R_3= R_4R_3R_4 = \left(\begin{array}{c c c c} 1 & 0 & 0 & 0 \\ 0 & 1 & 0 & 0 \\ 0 & 0 & 1 & 0 \\ 0 & 0 & 0 & -1\end{array}\right),
\en
and the cyclic permutations
\begin{align}
\Theta &= R_4R_3 = \frac{1}{2}\left(\begin{array}{c c c c} 1 & 1 & 1 & -1 \\ 1 & 1 & -1 & 1 \\ 1 & -1 & 1 & 1 \\ 1 & -1 & -1 & -1 \end{array}\right)
\\
\Theta^{-1} &= R_3R_4 = \frac{1}{2}\left(\begin{array}{c c c c} 1 & 1 & 1 & 1 \\ 1 & 1 & -1 & -1 \\ 1 & -1 & -1 & 1 \\ -1 & 1 & 1 & -1 \end{array}\right)
\end{align}
The action of these elements on the parameter vector $\boeta$ results in the following permutations
\eq
\begin{array}{c c c c c c}
& \boeta_{(1)} \rightarrow \hat\boeta_{(1)} & & \boeta_{(1)} \rightarrow \hat\boeta_{(4)} & & \boeta_{(1)}\rightarrow \hat\boeta_{(3)}\\
R_1: \  & \boeta_{(3)} \rightarrow \hat\boeta_{(4)} &,\quad \Theta: \  & \boeta_{(3)} \rightarrow \hat\boeta_{(1)} & ,\quad \Theta^{-1}: \  & \boeta_{(3)}\rightarrow \boeta_{(4)}\\
& \boeta_{(4)} \rightarrow \hat\boeta_{(3)} & & \boeta_{(4)} \rightarrow \boeta_{(3)} & & \boeta_{(4)} \rightarrow \hat\boeta_{(1)}
\end{array}.
\label{eq:r1thetathetam1symm}
\en
The exponents $\Gamma_a$ are
\begin{align}
\Gamma_1 = -\frac{1}{3 M}(\nu_1^0 + 3) \;  &, \quad \Gamma_2 = -\frac{1}{3 M} (\nu_1^0 + \nu_2^0 + 5), 	\nonumber \\
\Gamma_3 = -\frac{1}{6 M}(\nu_1^0 + \nu_2^0 + \nu_3^0 -\nu_4^0 + 6) \; &, \quad \Gamma_4 = -\frac{1}{6 M}(\nu_1^0 + \nu_2^0 + \nu_3^0 +\nu_4^0 + 6),
\end{align}
and it is easy to see that $\Gamma_2$ is invariant under the full $S_3$ group, while the other three exponents are mapped one in the other according to the scheme
given in Figure \ref{diag12}.
%
Given the fact that the differential equations obtained from $\boeta_{(n)}$ and $\hat\boeta_{(n)}$ are the same the relations (\ref{eq:r3r4symm}, \ref{eq:r1thetathetam1symm}), translate in terms of $\mathcal Q$-functions as follows
\begin{align}
&\begin{array}{c c c c c c}
& \mathcal Q^{(1)}\rightarrow \mathcal Q^{(1)} & &\mathcal Q^{(1)} \rightarrow \mathcal Q^{(4)} & & \mathcal Q^{(1)} \rightarrow \mathcal Q^{(3)}\\
R_1: \  & \mathcal Q^{(3)} \rightarrow \mathcal Q^{(4)}&,\quad R_3: \quad & \mathcal Q^{(3)} \rightarrow \mathcal Q^{(3)} &,\quad R_4: \quad & \mathcal Q^{(3)} \rightarrow \mathcal Q^{(1)}\\
& \mathcal Q^{(4)} \rightarrow \mathcal Q^{(3)} & & \mathcal Q^{(4)} \rightarrow \mathcal Q^{(1)} & & \mathcal Q^{(4)} \rightarrow \mathcal Q^{(4)}
\end{array}\;,\nonumber\\\nonumber\\
&\qquad\qquad\qquad\quad\begin{array}{c c c c}
& \mathcal Q^{(1)} \rightarrow \mathcal Q^{(4)} & & \mathcal Q^{(1)}\rightarrow \mathcal Q^{(3)}\\
\Theta\phantom{_1}: \  & \mathcal Q^{(3)} \rightarrow \mathcal Q^{(1)} & ,\quad \Theta^{-1}: \  & \mathcal Q^{(3)}\rightarrow \mathcal Q^{(4)}\\
& \mathcal Q^{(4)} \rightarrow \mathcal Q^{(3)} & & \mathcal Q^{(4)} \rightarrow \mathcal Q^{(1)}
\end{array}.
\label{eq:r3r4symm}
\end{align}

\begin{figure}[h]
\centering
\includegraphics[width=10cm]{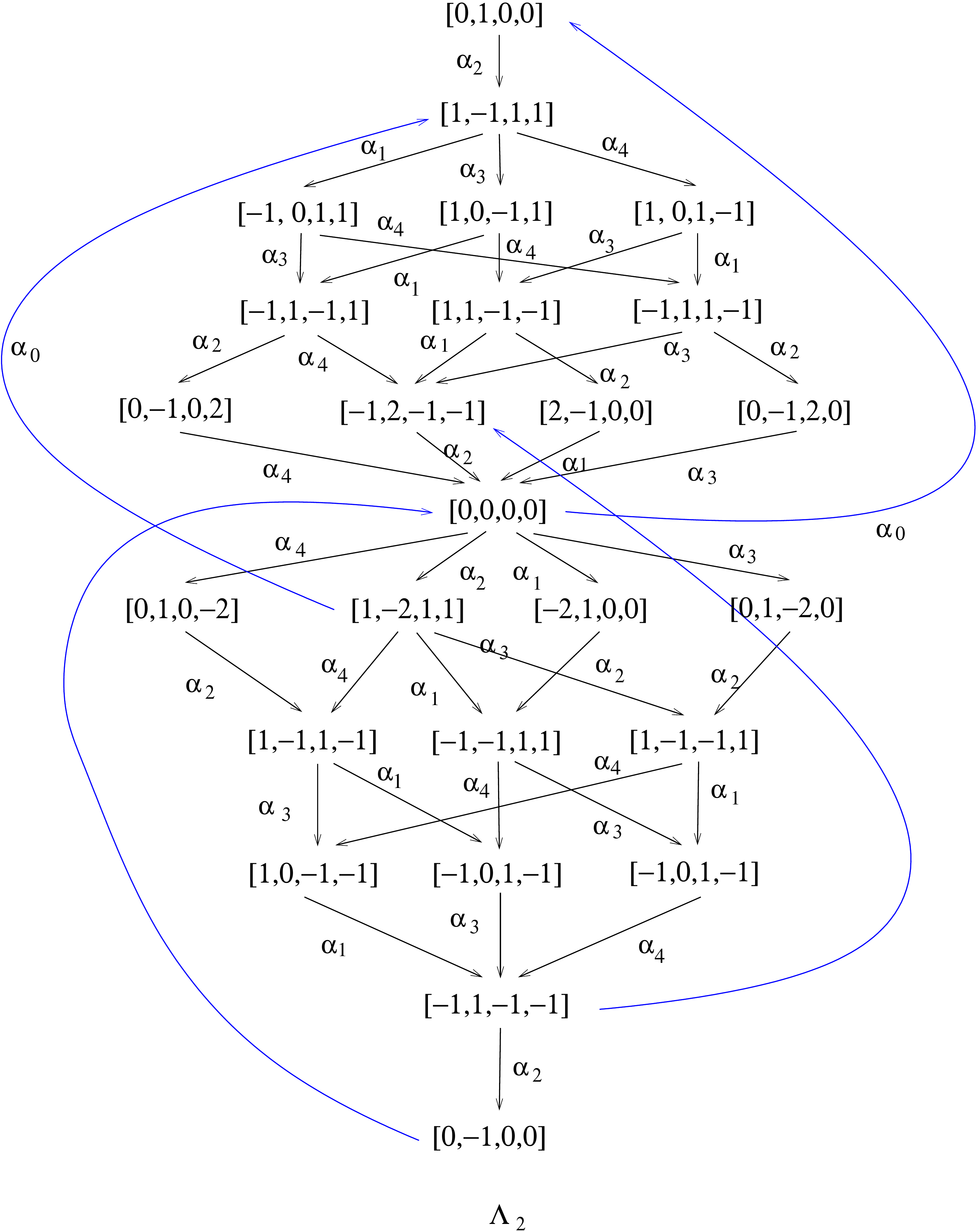}
\caption{The weight system for the representation $\bLambda_2(\mathfrak d_4)$.}
\label{fig:D4_2}
\end{figure}

The study of the representation $\bLambda_2(\mathfrak d_4)$, on the contrary of the simple ones we have seen above, it is extremely complicated, as it result clearly from the weight system in Figure \ref{fig:D4_2}. From this picture we can nonetheless infer that the asymptotic form of the equation will be of the following form:

\eq
	\partial^{11}\psi^{(2)}_{[k]}\underset{\rho\rightarrow\infty}{\sim} 2^2\times 12^2 e^{12\theta+2\pi \mathbbm i k}\sqrt{p}\partial\sqrt{p}\partial^{-3}\sqrt{p}\partial\sqrt{p}\,\psi^{(2)}_{[k]}\;.
\en

The $11$ is found by remarking that the depth of the weight system is $11$. The longest loop touching each level at least once is of length $12$ and there are exactly $12^2$ such loops. Thus the right-hand side has to contain $n$ derivatives and $n+1$ antiderivatives\footnote{The asymptotic equation has to have $n$ derivatives and $m$ anti-derivatives, such that $\vert n-m\vert =12$, the length of the longest loop.}. The right-hand side term is obtained by inspecting how to get from the second node to the next-to-last by moving twice along $\balpha_0$ lines\footnote{Remember that we must go AGAINST the direction of the arrows!}. There are $4$ different ways to go:

\begin{align}
	&[1,-1,1,1]\;{\color{blue}\underset{\balpha_0}{\longrightarrow}}\;[1,-2,1,1]\;\underset{\balpha_2}{\longrightarrow}\;[0,0,0,0]\;{\color{blue}\underset{\balpha_0}{\longrightarrow}}\;[0,-1,0,0]\;\underset{\balpha_2}{\longrightarrow}\;[-1,1,-1,-1] \nonumber \\
	&[1,-1,1,1]\;{\color{blue}\underset{\balpha_0}{\longrightarrow}}\;[1,-2,1,1]\;\underset{\balpha_2}{\longrightarrow}\;[0,0,0,0]\;\underset{\balpha_2}{\longrightarrow}\;[-1,2,-1,-1]\;{\color{blue}\underset{\balpha_0}{\longrightarrow}}\;[-1,1,-1,-1] \nonumber \\
	&[1,-1,1,1]\;\underset{\balpha_2}{\longrightarrow}\;[0,1,0,0]\;{\color{blue}\underset{\balpha_0}{\longrightarrow}}\;[0,0,0,0]\;{\color{blue}\underset{\balpha_0}{\longrightarrow}}\;[0,-1,0,0]\;\underset{\balpha_2}{\longrightarrow}\;[-1,1,-1,-1] \nonumber \\
	&[1,-1,1,1]\;\underset{\balpha_2}{\longrightarrow}\;[0,1,0,0]\;{\color{blue}\underset{\balpha_0}{\longrightarrow}}\;[0,0,0,0]\;\underset{\balpha_2}{\longrightarrow}\;[-1,2,-1,-1]\;{\color{blue}\underset{\balpha_0}{\longrightarrow}}\;[-1,1,-1,-1] \nonumber
\end{align}
producing the following term

\begin{align}
	\partial^{-1}&\left(\partial^{-1}\,p\,\partial^{-2}\,p + p\,\partial^{-3}\,p + \partial^{-1}\,p\,\partial^{-1}\,p\,\partial^{-1} + p\,\partial^{-2}\,p\,\partial^{-1}\right) = \nonumber \\
	&= \partial^{-1}\left(\partial^{-1}\,p+p\,\partial^{-1}\right)\partial^{-1}\left(\partial^{-1}\,p+p\,\partial^{-1}\right) = \partial^{-2}\sqrt{p}\partial\sqrt{p}\partial^{-3}\sqrt{p}\partial\sqrt{p}\;.
\end{align}

\newpage
\section{Some Wronskian identities}
\label{sec:wronsky}
\setcounter{equation}{0}

We are going to calculate, in the $\rho \rightarrow \infty$ limit, the Wronskian of functions belonging to the basis $\mathcal B_k^{\infty}$ for the algebra $\mathfrak a_{r}$. Let us set, for notational convenience,

\eq
	W_k^{(n)} = W^{(n)}[\psi^{(1)}_{[k+1]},\ldots,\psi^{(1)}_{[k+n]}] \; .
\en

Let us begin by writing the asymptotic expansion (\ref{eq:bigrhoexpansioncomplete}) in the following convenient form

\eq
	\psi^{(1)}_{[k]} \underset{\rho\rightarrow\infty}{\sim} (t_k z^{M})^{-\frac{r}{2}} e^{t_k \frac{z^{M+1}}{M+1} + f_k(\bz)} \; ,
\en
with $t_k=-\frac{m}{2} e^{\theta+2\pi \mathbbm i\frac{k}{r+1}}$ and $f_k(\bz) = -\frac{m}{2}e^{-\theta-2\pi \mathbbm i\frac{k}{r+1}}\frac{\bz^{M+1}}{M+1}$ and calculate the asymptotic expressions for its derivatives:

\eq
	\partial^{\ell}\psi^{(1)}_{[k]} \underset{\rho\rightarrow\infty}{\sim} (t_k z^{M})^{-\frac{r}{2}+\ell}\; e^{t_k \frac{z^{M+1}}{M+1} + f_k(\bz)} \; .
\en
Now we express $W_k^{(m)}$ with the help of the $m$-dimensional Levi-Civita totally antisymmetric tensor $\varepsilon_{j_1, \ldots ,j_m}$:

\eq
	W_k^{(n)} \equiv \sum_{j_1,\ldots ,j_n =1}^n \varepsilon_{j_1,\ldots ,j_n} \prod_{\ell =1}^n \partial^{\ell-1}\psi^{(1)}_{[k+j_{\ell}]} \; .
\en
Let us investigate the product of $\psi$ functions

\begin{align}
	&\prod_{\ell =1}^n \partial^{\ell-1}\psi^{(1)}_{[k+j_{\ell}]} \underset{\rho \rightarrow\infty}{\sim}	\nonumber
	\\
	\\
	&\underset{\rho \rightarrow\infty}{\sim} (-\frac{m}{2}e^{\theta}z^M)^{n\frac{n-r-1}{2}}e^{\frac{\pi \mathbbm i}{r+1}\sum_{\ell=1}^n (k+j_\ell)(2\ell -r -2) +\frac{z^{M+1}}{M+1}\sum_{\ell=1}^n t_{k+j_\ell} + \sum_{\ell=1}^n f_{k+j_\ell}(\bz)}	\nonumber \; .
\end{align}

The sums in the exponential are easily calculated, safe for the one involving the product $\ell j_\ell$, once remembering that the Levi-Civita tensor forces each of the indices $j_\ell$ to assume a different value, meaning we can write:

\eq
	\sum_{\ell=1}^n t_{k+j_\ell} = -\frac{m}{2}e^{\theta+2\pi \mathbbm i\frac{k}{r+1}}\sum_{\ell=1}^n e^{2\pi \mathbbm i\frac{\ell}{r+1}} = -\frac{m}{2} \frac{\sin \frac{n\pi}{r+1}}{\sin\frac{\pi}{r+1}} e^{\theta +\frac{\pi \mathbbm i}{r+1}(n+1+2k)} \; ,
\en

\eq
	\sum_{\ell=1}^n f_{k+j_\ell}(\bz) = -\frac{m}{2}\frac{\bz^{M+1}}{M+1}e^{-\theta-2\pi \mathbbm i\frac{k}{r+1}}\sum_{\ell=1}^n e^{-2\pi \mathbbm i\frac{\ell}{r+1}} = -\frac{m}{2}\frac{\bz^{M+1}}{M+1} \frac{\sin \frac{n\pi}{r+1}}{\sin\frac{\pi}{r+1}} e^{-\theta-\frac{\pi \mathbbm i}{r+1}(n+1+2k)} \; ,
\en

\begin{align}
	\sum_{\ell=1}^n (k+j_\ell)(2\ell-r-2) &= \sum_{\ell=1}^n 2\ell j_\ell + \sum_{\ell=1}^n k(2\ell-r-2) - \sum_{\ell=1}^n \ell(r+2) =	\nonumber
	\\
	\\
	& =\sum_{\ell=1}^n 2\ell j_\ell + k n(n-r-1) -n \frac{n+1}{2}(r+2) \; .
\end{align}

Thus we have

\begin{align}
	W_k^{(n)} &\underset{\rho \rightarrow\infty}{\sim} (-\frac{m}{2}e^{\theta}z^M)^{n\frac{n-r-1}{2}}e^{\frac{\pi \mathbbm i}{r+1}\Big(k n(n-r-1) -n \frac{n+1}{2}(r+2)\Big)} \times	\nonumber
	\\
	\\
	&\times e^{ -\frac{m}{2}\frac{\sin\frac{n\pi}{r+1}}{\sin\frac{\pi}{r+1}}\Big(\frac{z^{M+1}}{M+1}e^{\theta + \frac{\pi \mathbbm i}{r+1}(n+1+2k)}+\frac{\bz^{M+1}}{M+1}e^{-\theta - \frac{\pi \mathbbm i}{r+1}(n+1+2k)}\Big)} \sum_{j_1,\ldots ,j_n =1}^n \varepsilon_{j_1,\ldots ,j_n} e^{\frac{2 \pi \mathbbm i}{r+1} \sum_{\ell=1}^n \ell \, j_\ell}	\nonumber\; .
\end{align}

The last factor can be rewritten as

\eq
	\sum_{j_1,\ldots ,j_n =1}^n \varepsilon_{j_1,\ldots ,j_n} e^{\frac{2 \pi \mathbbm i}{r+1} \sum_{\ell=1}^n \ell \, j_\ell} = \sum_{\sigma\in S_n} \vert\sigma\vert e^{\frac{2 \pi \mathbbm i}{r+1} \sum_{\ell=1}^n \ell \, \sigma(\ell)} \; ,
\en
where $S_n$ is the group of permutations of $n$ elements and $\vert\sigma\vert$ is the signature of the permutation. Remembering that each element of $S_n$ can be written as a product of adjacent transpositions $\sigma_{i,i+1}$ and remarking that $\sum_{\ell=1}^n \ell \sigma_{i,i+1}(\ell) = \sum_{\ell=1}^n \ell^2 - 1$, with some work one arrives at the following result

\eq
	\sum_{j_1,\ldots ,j_n =1}^n \varepsilon_{j_1,\ldots ,j_n} e^{\frac{2 \pi \mathbbm i}{r+1} \sum_{\ell=1}^n \ell \, j_\ell} = e^{n \pi \mathbbm i\frac{n+1}{3(r+1)}(2n+1)}\prod_{\ell=1}^{n-1}\left(1-e^{-2 \pi \mathbbm i\frac{\ell}{r+1}}\right)^{n-\ell} \; ,
\en
which brings us to the general formula

\begin{align}
	W_k^{(n)} &\underset{\rho \rightarrow\infty}{\sim} (-\frac{m}{2}e^{\theta}z^M)^{n\frac{n-r-1}{2}}e^{\frac{\pi \mathbbm i}{r+1}\Big(k n(n-r-1) +n \frac{n+1}{6}(4n-3r-4)\Big)} \prod_{\ell=1}^{n-1}\left(1-e^{-2 \pi \mathbbm i\frac{\ell}{r+1}}\right)^{n-\ell} \times	\nonumber
	\\	\label{eq:wronskybeat}
	\\
	&\times e^{ -\frac{m}{2}\frac{\sin\frac{n\pi}{r+1}}{\sin\frac{\pi}{r+1}}\Big(\frac{z^{M+1}}{M+1}e^{\theta + \frac{\pi \mathbbm i}{r+1}(n+1+2k)}+\frac{\bz^{M+1}}{M+1}e^{-\theta - \frac{\pi \mathbbm i}{r+1}(n+1+2k)}\Big)} \nonumber \; .
\end{align}

Let us analyse the particular cases $n=r+1$ and $n=r$.

\subsection{The $n=r+1$ case}
We have

\eq
	W_k^{(r+1)} \underset{\rho \rightarrow \infty}{\sim} e^{\pi \mathbbm i r\frac{r+2}{6}} \prod_{\ell =1}^{r} (1-e^{-2 \pi \mathbbm i \frac{\ell}{r+1}})^{r +1-\ell} \; .
\en

This last product is easily calculated as follows

\begin{align}
	\prod_{\ell =1}^{r} (1-e^{-2 \pi \mathbbm i \frac{\ell}{r+1}})^{r +1-\ell} = e^{-\pi \mathbbm i r \frac{r+2}{6}} \prod_{j=1}^{r}\left(2 \mathbbm i \sin\left( \pi \frac{j}{r+1} \right) \right)^{r+1-l} \; .
\end{align}

Given the symmetry for $j \rightarrow r+1-j$ of the sine function under the product and the fact that $\sin(\frac{\pi}{2}) = 1$, we can write

\begin{align}
	W_k^{(r+1)} &\underset{\rho \rightarrow \infty}{\sim} (2\mathbbm i)^{r\frac{r+1}{2}} \prod_{j=1}^{r}\left(\sin\left(\pi\frac{j}{r+1}\right)\right)^{\frac{r+1}{2}} \; .
\end{align}

Finally, using the known identity $\prod_{k=1}^{n-1} \sin(\pi \frac{k}{n}) = n 2^{1-n}$ we find

\begin{align}
	W_k^{(r+1)} &\underset{\rho \rightarrow \infty}{\sim} \mathbbm i^{\frac{1}{2}r(r+1)} (r+1)^{\frac{r+1}{2}} \; .
\label{eq:rwronsky}
\end{align}

\subsection{The $n=r$ case}
Here we start from

\begin{align}
	W_k^{(r)} &\underset{\rho \rightarrow\infty}{\sim} (-\frac{m}{2}e^{\theta}z^M e^{2\pi \mathbbm i\frac{k}{r+1}})^{-\frac{r}{2}}e^{\pi \mathbbm i\frac{r}{6}(r-4)} \prod_{\ell=1}^{r-1}\left(1-e^{-2 \pi \mathbbm i\frac{\ell}{r+1}}\right)^{r-\ell} \times	\nonumber
	\\
	\\
	&\times e^{ \frac{m}{2}\Big(\frac{z^{M+1}}{M+1}e^{\theta + 2\pi \mathbbm i \frac{k}{r+1}}+\frac{\bz^{M+1}}{M+1}e^{-\theta - 2\pi \mathbbm i \frac{k}{r+1}}\Big)} \nonumber \; .
\end{align}

Again the product is easily computed

\eq
	\prod_{\ell=1}^{r-1}\left(1-e^{-2 \pi \mathbbm i\frac{\ell}{r+1}}\right)^{r-\ell} = (2 \mathbbm i)^{r\frac{r-1}{2}} e^{-\pi \mathbbm i r\frac{r-1}{6}} \left(\prod_{\ell=1}^{r}\sin\frac{\pi\ell}{r+1}\right)^{\frac{r-1}{2}} = \Big(\mathbbm i^{r}\,(r+1)\Big)^{\frac{r-1}{2}}e^{-\pi \mathbbm i r\frac{r-1}{6}}	\nonumber \; .
\en	

Thus we can write

\begin{align}
	W_k^{(r)} &\underset{\rho \rightarrow\infty}{\sim} (-\frac{m}{2}e^{\theta}z^M e^{2\pi \mathbbm i\frac{k}{r+1}})^{-\frac{r}{2}}e^{-\pi \mathbbm i\frac{r}{2}} \mathbbm i^{r\frac{r-1}{2}} (r+1)^{\frac{r-1}{2}}  e^{ \frac{m}{2}\Big(\frac{z^{M+1}}{M+1}e^{\theta + 2\pi \mathbbm i \frac{k}{r+1}}+\frac{\bz^{M+1}}{M+1}e^{-\theta - 2\pi \mathbbm i \frac{k}{r+1}}\Big)} \nonumber \; .
\end{align}

Recalling the form of the asymptotic for the function $^{(1)}\psi_{[k]}$, we finally obtain

\eq
	W_k^{(r)} \underset{\rho \rightarrow\infty}{\sim} \mathbbm i^{r\frac{r-1}{2}}\,(r+1)^{\frac{r-1}{2}} \psi^{(1)}_{[k+\frac{r+1}{2}]} \; .
\label{eq:rm1wronsky}
\en

\vspace{1cm}
These calculations were used in the preceding chapter to prove the completeness of the basis $\mathcal B_{\infty}$ (\ref{eq:inftybasis}) and the fact that the function $\psi^{(r)}$ indeed satisfies the asymptotic expansion expected from the solutions of the differential equation adjoint to (\ref{eq:ardiffeq}). By using the general formula (\ref{eq:wronskybeat}), it is possible to obtain the asymptotics of the functions $\psi^{(\ell)} \; , \  \ell =2,3,\ldots , r-1$ which is much more difficult to extract from the differential equations, given the complexity of the weight diagrams of the fundamental representations $\bLambda_{\ell} \; , \ \ell = 2,3,\ldots ,r-1$.

\part{Thermodynamic Bethe Ansatz, Reflection Relations and Fermionic Basis}
\label{part:Smirnovpart}

\mychapter{4}{Introduction}
\setcounter{equation}{0}
\makeatletter
\long\def\theequation{\ifnum \c@chapter > \z@ \fi \@arabic \c@equation}
\makeatother

This part is dedicated to the study of the particularly simple Toda field theory 
associated with the affine algebra $\hat{\mathfrak a}_1$; this 
theory is known in the literature as \emph{sin(h)-Gordon} model, where ``sin(h)" might stand for \emph{sine} as well as
for \emph{sinh}. In fact depending on the values taken by the coupling constant $b^2$ in (II.\ref{eq:Todalagrangian}) one obtains
two different theories (\textbf{in this section we change the notation, writing $b$ for the coupling constant in (II.\ref{eq:Todalagrangian}),
while we set $\beta= \mathbbm i b$}):

\begin{itemize}
 \item $0 < b^2 < \infty$: this choice corresponds to the \emph{sinh-Gordon} model which actually obeys the duality $b \rightarrow 1/b$,
 thus one can restrict considerations to the region $0< b^2 \leq 1$;
 \item $ -\infty < b^2 < 0$: this choice corresponds to the \emph{sine-Gordon} model; it is acually better to restrict ourselves to the interval
 $-1 \leq b^2 <0$ as for $b^2<-1$ problems arise in the usual treatment of the model\footnote{This is actually due to the fact that, considering the model
 as a perturbation of a $c=1$ CFT, the perturbing operator becomes irrelevant for $b^2<-1$.}.
\end{itemize}

Even though they are described by the same Euclidean action which, in complex coordinates, reads

\eq
  \mathcal A = \int \left\lbrace \frac{1}{4\pi} \partial \eta(z,\bz) \bp \eta(z,\bz) + \frac{2\bmu^2}{\sin(\pi b^2)} \cosh[b\eta(z,\bz)] \right\rbrace
  \frac{dz \wedge d\bz}{2} \; ,
\label{eq:singaction}
\en
these two models behave in quite different fashions and, contrary to what one could na\"{\i}vely argue, it is not always possible to recover 
properties of one model by a simple analytic continuation in the coupling constant from the other one. Nonetheless it is convenient to start from the 
common action (\ref{eq:singaction}) and leave $b^2$ unspecified for the time being.

It is worth to notice now that in this part we will adopt slightly changed normalisation of the dimensional constant and definition of complex coordinates, 
as is clear from (\ref{eq:singaction}); this is done in order to keep formulae closer to those in \cite{Negr_Smir_13_2}. Actually, as discussed in 
the articles cited, the choice of the normalisation for the dimensional constant, aside from being extremely convenient for the subsequent 
discussion, encloses serious physical reasons. Firstly it automatically takes in account the change of sign in the potential energy when passing from 
sinh- to sine-Gordon and encodes also the pole at $b=\mathbbm i$ of this last\footnote{Due, as we said, to the fact that the perturbing operator becomes irrelevant 
for $b^2<-1$. Note that there are poles also for $b\in\mathbbm Z$ which look natural
once one considers the physical scale of the model, namely the mass of the particle \cite{AlZa_06}.}; more importantly, this normalisation lets 
the mass $m$ of both the sinh-Gordon particle and that of the sine-Gordon lowest breather be expressed by an universal formula:

\eq
   \bmu \Gamma(1+b^2) = \left[ \frac{m}{4\sqrt{\pi}} \Gamma\left(\frac{1}{2(1+b^2)}\right)\Gamma\left(1+\frac{b^2}{2(1+b^2)}\right) \right]^{1+b^2} \; .	\label{eq:massformula}
\en
\\

The focus of this part will be on the Quantum version of the sin(h)-Gordon model and, in particular, on the problem of computing its one-point functions. 
These are fundamental data in any quantum field theory; indeed, when applying the Operator Product Expansion (OPE) to extract the ultraviolet asymptotics 
for correlation functions, the required objects are the coefficients of the OPE and the one-point functions. While the former are purely ultraviolet data 
and are, in principle, governed by the perturbation theory of the corresponding ultraviolet CFT, the latter depend essentially on the infrared environment 
of the model and cannot be extracted by means of CFT perturbations: one has to find some other way to calculate them.

The sin(h)-Gordon model is the 
perfect playground in which to experiment new analytical methods aimed at extracting data, in fact it is the most simple example of massive integrable field 
theory and is complicated enough to show interesting structures; moreover this model has been the subject of intensive study during the last 30 years and 
nowadays most of its features are known. In the following we will present a powerful method, recently developed in joint works by H. Boos, M. Jimbo, T. Miwa, 
F. Smirnov and Y. Takeyama \cite{Boos_Jimb_Miwa_Smir_Take_07,Boos_Jimb_Miwa_Smir_Take_09,Jimb_Miwa_Smir_09,Boos_Jimb_Miwa_Smir_10,Jimb_Miwa_Smir_11_1,Jimb_Miwa_Smir_11_2}, 
which allow the computation of any one-point function of sine-Gordon model\footnote{The method was originally developed on the XXZ model and later 
extended to the sine-Gordon model by means of an appropriate scaling limit.} exploiting the existence of a particular fermionic basis in the space of 
states of the theory.

In the first chapter we will review the construction of the fermionic basis in the sin(h)-Gordon model; we will not delve too much in details, given the 
complexity and length of the matter, and refer the interested reader to the above-cited articles. Then, in the second chapter  we will present an interesting application of the fermionic basis to the reflection relations introduced by V. Fateev, D. Fradkin, S. Lukyanov, A. Zamolodchikov and Al. Zamolodchikov in \cite{Fate_Frad_Luky_AZam_AlZa_99}. Finally in the third chapter we expose some comparison with known results, both analytic and numerical.

\chapter{Fermions in the sin(h)-Gordon Model}
\label{chap:fermions}
\markboth{Chapter 1 - Fermions in the sin(h)-Gordon Model}{}
\numberwithin{equation}{chapter}

The starting point of our analysis is the remark that the action (\ref{eq:singaction}) of the sin(h)-Gordon model allows for two different interpretations; in 
fact we might consider it as a perturbation of a $c=1$ CFT, that is a free boson

\eq
  \mathcal A = \int \left\lbrace \left[ \frac{1}{4\pi} \partial \eta(z,\bz) \bp \eta(z,\bz) \right] + \frac{2\bmu^2}{\sin(\pi b^2)} 
  \cos[b\eta(z,\bz)] \right\rbrace  \frac{dz \wedge d\bz}{2} \; ,
  \label{eq:sinegordonactioninheisenbergform}
\en
as well as a perturbation of the complex Liouville model

\begin{align}
  \mathcal A = \int \left\lbrace \left[ \frac{1}{4\pi} \partial \eta(z,\bz) \bp \eta(z,\bz) + \frac{\bmu^2}{\sin(\pi b^2)} e^{b\eta(z,\bz)} 
  \right]\right.& \nonumber
  \\ \label{eq:sinegordonactioninliouvilleform}
  \\ \left. + \frac{\bmu^2}{\sin(\pi b^2)} e^{-b\eta(z,\bz)} \right\rbrace&  \frac{dz \wedge d\bz}{2} \; , \nonumber
\end{align}
where the term in square brackets corresponds to the action of the said complex Liouville model, conventionally identified with the minimal model of CFT 
having central charge

\eq
  c = 1 + 6 Q^2 \quad ; \quad Q = b^{-1} + b\; , 
\en
and the perturbing field is $\Phi_{1,3}(z,\bz) = e^{-b\eta(z,\bz)}$. We let this model live on an infinite cylinder of circumference $2 \pi R$ which 
corresponds to considering the model at finite temperature; the generatrix and the directrix of the cylinder will be called space and Matsubara direction 
respectively.

As stated above, our goal is to find a way to calculate one-point functions; in particular we will consider expectation values of exponential fields parametrised 
as $\Phi_{a}(z,\bz) = e^{a \eta(z,\bz)}$, which we consider as \emph{primary fields} for the actions (\ref{eq:sinegordonactioninheisenbergform},
\ref{eq:sinegordonactioninliouvilleform}), and of their descendants. Let us precise what we mean by ``descendants''. When considering 
(\ref{eq:sinegordonactioninheisenbergform}), the descendants of the primary field $\Phi_a(0) = e^{a\eta(0)}$ are naturally written as normal-ordered products 
of $e^{a\eta(0)}$ with a polynomial in derivatives of $\eta(0)$; we force these polynomials to be of even order, we will explain why later.
Given the symmetry of the action under the reflection $\phi\rightarrow-\phi$, the one-point functions of these fields, that we call 
\emph{Heisenberg descendants}, have to be symmetric with respect to

\eq
  \sigma_1 \; : \; a\rightarrow -a \; .
\en

Now, turning to the action (\ref{eq:sinegordonactioninliouvilleform}), we introduce the components of the energy-momentum tensor\footnote{We shall not make 
use of the trace component of this energy-momentum tensor, which is clearly proportional to $e^{-b\eta(z,\bz)}$.}

\begin{align}
  T(z,\bz) = T_{z,z}(z,\bz) &= - \frac{1}{4}\left[\partial\eta(z,\bz)\right]^2 + \frac{Q}{2}\partial^2\eta(z,\bz) \; ,	\nonumber
  \\ \label{eq:enmomtensor}
  \\ \overline T(z,\bz) = T_{\bz,\bz}(z,\bz) &=- \frac{1}{4}\left[\overline\partial\eta(z,\bz)\right]^2 + \frac{Q}{2}\overline\partial^2\eta(z,\bz) \; .	\nonumber
\end{align}
In this case, the natural form assumed by the descendants is that of normal ordered products of $e^{a\eta(0)}$ with polynomials in the derivatives of $T$ and 
$\overline T$; we will consider only derivatives of even order, again an explanation will be given in a while. It is natural to assume that the one-point functions 
of these fields, which we name \emph{Virasoro descendants}, inherit the symmetry of the Liouville model:

\eq
  \sigma_2 \; : \; a \rightarrow Q-a \; .
\en

For arbitrary values of $a$ (in absence of resonances) it is supposed that the space of the local fields in the perturbed model is the same as in the 
corresponding CFT. This means that, for sin(h)-Gordon model, we have two equivalent descriptions of the space of local fields; writing one description as a 
function of the other give rise to the reflection relations \cite{Fate_Luky_AZam_AlZa_97,Fate_Luky_AZam_AlZa_98,Fate_Frad_Luky_AZam_AlZa_99}, which will be 
discussed later in chap.\ref{chap:refrel}.

In the following we will use various parameters which are interrelated and it is best to give a brief summary of them now:

\begin{center}
\begin{tabular}{| c | c | c | c |}
\hline	Coupling &		Conformal dimension &			Central charge &	Other \\ \hline &&& \\
	$b$ ; $\beta = \mathbbm i b$ &	$a$ ; $\alpha = \frac{2}{Q}a$ &		$c = 1 + 6 Q^2$ & 	$d_{\alpha} = (b^2-b^{-2})(1-\alpha)$ \\
	 &			&					&			$= \frac{1}{6}\sqrt{(c-25)(1+24\Delta_{\alpha}-c)}$\\
	 $\nu = 1+ b^2$ &	&					&			 \\
	 &			$\Delta = a(Q-a)$ & 			&			$x=\frac{2 a - b}{2Q} = \frac{\alpha}{2} + \frac{1-\nu}{2\nu}$\\
	$Q=b+b^{-1}$ &		$=\frac{Q^2}{4}\alpha(2-\alpha)$&	&			\\ &&&\\ \hline
\end{tabular}
\end{center}

\section{Why fermions?}
\label{sec:whyferms}
\numberwithin{equation}{section}
\setcounter{equation}{0}

Computing one-point functions in a perturbed CFT is not easy; the main reason is that, until some years ago, it was unknown how to build an efficient basis in 
the space of states providing a reasonably simple way to calculate expectation values. 

Let us explain this point more clearly: since, as we hinted above, although conformal 
invariance is broken, the local fields in sin(h)-Gordon model are in one-to-one correspondence with local CFT fields, which are organised according to the 
corresponding Virasoro algebra in the usual way, we can identify the space of descendants of the exponential field $\Phi_a(0)$ in sin(h)-Gordon model with the 
tensor product of Verma modules $\mathcal V_a \otimes \overline{\mathcal V}_a$. Thus we can consider operators acting on the space of operators, rather than on the 
space of states, and, by what has just been said above, it is possible to identify these operators, acting on the sin(h)-Gordon fields, as ones acting on the 
corresponding Verma module. This seems to suggest that one point functions would be functionals on the tensor product $\mathcal V_a \otimes \overline{\mathcal V}_a$, 
however things are not this simple: in fact we are not taking into account the integrable structure of the model.

It is known that sin(h)-Gordon model (and, more generally, any integrable $2D$ QFT) possesses an 
infinite quantity of local integrals of motion $I_{2k-1}$ and $\overline{I}_{2k-1}$, including, in particular, the Hamiltonian $H=I_1 + \overline{I_1}$ and the momentum 
operator $P=I_1 - \overline{I}_1$. If we let the compact direction of the cylinder be the time, then the action of these integrals of motion, which we denote as 
$\mathbf{i}_{2k-1},\overline{\mathbf{i}}_{2k-1}$, is represented as the difference of local densities integrated over $\mathbbm R+\mathbbm i0$ and $\mathbbm R-\mathbbm i0$, where 
$\mathbbm R$ is the generatrix of the cylinder (here the spatial dimension). It is clear then that, by moving the contours of integration along the compact 
dimension, the result is zero: \emph{all the one point functions of descendants built out of integrals of motions vanish}. As a consequence, the correct space on which to define the one-point function as a linear 
functional is $\mathcal V^{\textrm{quo}}_a \otimes \overline{\mathcal V}^{\textrm{quo}}_a$: the tensor product of the quotient spaces

\eq
  \mathcal V^{\textrm{quo}}_a = \mathcal V_a \left/ \sum_{k=1}^{\infty} \mathbf{i}_{2k+1} \mathcal V_a \right. \; ,\qquad \overline{\mathcal V}^{\textrm{quo}}_a
  = \overline{\mathcal V}_a \left/ \sum_{k=1}^{\infty} \overline{\mathbf{i}}_{2k+1} \overline{\mathcal V}_a \right. \; .
\en
Notice that $\mathcal V^{\textrm{quo}}_a \otimes \overline{\mathcal V}^{\textrm{quo}}_a$ has non-trivial subspaces of even dimension only and this is the reason 
why we will not consider odd order polynomials in Heisenberg descendants and odd order derivatives in Virasoro descendants.

Now it is becoming clearer what the main issue is: the basis derived from CFT, which is composed of primary fields $\Phi_a(z,\bz)$ and their ``conformal'' 
descendants, is a basis for the full Verma module $\mathcal V_a \otimes \overline{\mathcal V}_a$ and, in order to obtain a basis of the quotient space, one has to 
factor out all the null vectors arising from the action of the integrals of motion. While this is, in principle, a viable strategy, the form of the null vectors quickly becomes extremely complicated, making this construction practically impossible: one would prefer to have a basis 
defined intrinsically in the quotient space, automatically defining states modulo the null vectors. A basis of this kind was actually discovered some years 
ago in the six-vertex model \cite{Boos_Jimb_Miwa_Smir_Take_07,Boos_Jimb_Miwa_Smir_Take_09,Jimb_Miwa_Smir_09} and immediately extended to CFT 
\cite{Boos_Jimb_Miwa_Smir_10}, sine-Gordon \cite{Jimb_Miwa_Smir_10,Jimb_Miwa_Smir_11_1,Jimb_Miwa_Smir_11_2} and sinh-Gordon models \cite{Negr_Smir_13_2}. Very recently a step towards the construction of the fermionic basis for the spin-1 XXZ model has been performed in \cite{Jimb_Miwa_Smir_14}

This basis is built out of primary fields $\Phi_a$ and creation operators acting on them, much like the usual conformal basis; the peculiar fact is that 
these creation operators are \emph{fermions}. There are two of them for each chirality: $\bbeta_{2j-1}^{\ast}$, $\bgamma_{2j-1}^{\ast}$, 
$\overline{\bbeta}_{2j-1}^{\ast}$ and $\overline{\bgamma}_{2j-1}^{\ast}$. In the above-cited articles, it is demonstrated how it is possible to define in a 
mathematical rigorous fashion these fermions in six-vertex, pure CFT and sine-Gordon models; in particular for this last one, the quotient space 
$\mathcal V^{\textrm{quo}}_a \otimes \overline{\mathcal V}^{\textrm{quo}}_a$ was shown to allow the following basis

\eq
  \bbeta_{I^+}^{\ast}\overline{\bbeta}_{\overline I^+}^{\ast}\overline{\bgamma}_{\overline I^-}^{\ast}\bgamma_{I^-}^{\ast}\Phi_a(0) \; , \quad \mathfrak C(I^+)=\mathfrak C(I^-) 
  \; , \quad \mathfrak C(\overline I^+)=\mathfrak C(\overline I^-)\; ,
\en
where $I^{\pm} = \{2i_1^{\pm}-1,\ldots,2i_n^{\pm}-1\}$ and $\overline I^{\pm} = \{2\overline i_1^{\pm}-1,\ldots,2\overline i_{\overline n}^{\pm}-1\}$, the symbol $\mathfrak C(I)$ stands 
for the \emph{cardinality} of the set $I$ and we use the multi-index notation

\eq
  \bbeta_{I^+}^{\ast} = \bbeta^{\ast}_{2i_1^+-1}\cdots \bbeta^{\ast}_{2i_1^+-1} \quad ; \quad \vert I^+ \vert = \sum_{p=1}^n (2 i^+_p-1) \; .
\en
It is understood that, if $I=\{i_1,\ldots,i_n\}$, we use the notation $a I+b = \{ai_1+b,\ldots,ai_n+b\}\ \forall a,b\in \mathbbm Z$.

Although the rigorous construction of this whole setup as presented in \cite{Boos_Jimb_Miwa_Smir_10,Jimb_Miwa_Smir_10,Jimb_Miwa_Smir_11_1,Jimb_Miwa_Smir_11_2} might seem quite 
cumbersome and hard to understand, the power of the fermionic basis reveals itself in the very simple form assumed by the one-point functions:

\eq
  \frac{\langle\bbeta_{I^+}^{\ast}\overline{\bbeta}_{\overline I^+}^{\ast}\overline{\bgamma}_{\overline I^-}^{\ast}\bgamma_{I^-}^{\ast}\Phi_a(0)\rangle_R}
  {\langle \Phi_a(0)\rangle_R} = \mathcal D(I^+ \cup (-\overline I^+)\vert I^- \cup (-\overline I^-) \vert \alpha) \; ,
\label{eq:onepointfunction}
\en
where, for two sets $A=\{a_j\}_{j=1}^n$ and $B=\{b_j\}_{j=1}^n$, we introduce the function

\eq
  \mathcal D(A\vert B\vert \alpha) \doteq \left(\prod_{\ell=1}^n \frac{\textrm{sgn}(a_\ell)\textrm{sgn}(b_\ell)}{\pi} \right) 
  \det \left[\Theta(\mathbbm ia_j,\mathbbm ib_k\vert \alpha)-\pi \delta_{a_j,-b_k} \textrm{sgn}(a_j) t_{a_j}(\alpha)\right]_{j,k=1}^n
\en
and the functions $\Theta(A\vert B\vert \alpha)$ and $t_\ell(\alpha)$ will be defined later on.

A very important fact about these fermions is that one can use them not only to build descendants of primary fields, but also to shift them in $a$. 
In fact, if we give up the conditions $\mathfrak C(I^+)=\mathfrak C(I^-)$, $\mathfrak C(\overline I^+)=\mathfrak C(\overline I^-)$ in favour of the less constraining 
$\mathfrak C(I^+) + \mathfrak C(\overline I^+)=\mathfrak C(I^-) + \mathfrak C(\overline I^-)$ and set $m = \mathfrak C(I^+)-\mathfrak C(I^-)$ (we assume, for 
definiteness, that $m>0$), then one can show \cite{Jimb_Miwa_Smir_11_1} the following

\begin{align}
  &\bbeta_{I^+}^{\ast}\overline{\bbeta}_{\overline I^+}^{\ast}\overline{\bgamma}_{\overline I^-}^{\ast}\bgamma_{I^-}^{\ast}\Phi_{a-m b}(0)	\nonumber
  \\
  \label{eq:shiftformula}
  \\
  &\cong \frac{C_m(a)}{\prod_{j=1}^m t_{2j-1}(a)}\bbeta_{I^+ +2m}^{\ast}\overline{\bbeta}_{\overline I^+ -2m}^{\ast}\overline{\bgamma}_{\overline I^- +2m}^{\ast}
  \bgamma_{I^- -2m}^{\ast}\bbeta_{I_{\textrm{odd}}(m)}^{\ast}\overline{\bgamma}_{I_{\textrm{odd}}(m)}^{\ast}\Phi_{a}(0) \; ,	\nonumber
\end{align}
where $I_{\textrm{odd}(m)} = \{1,3,5,\ldots,2m-1\}$ and we use the symbol $\cong$ to denote identification in the weak sense (that is, under expectation value). 
The constants $C_m(a)$ and $t_{\ell}(a)$ will be given later; the operators with negative indices are to be understood as annihilation operators: 
$\bbeta_{-(2j-1)}^{\ast} = \bgamma_{2j-1}$, $\bgamma_{-(2j-1)}^{\ast} = \bbeta_{2j-1}$ and the same thing for the second chirality. These satisfy the following
relations (the parentheses $[\cdot,\cdot]_+$ stands for the anticommutator)

\begin{align}
  [\bbeta_i,\bbeta_j^{\ast}]_+ = -t_i(a)\delta_{i,j}& \; ,\qquad [\bgamma_i,\bgamma_j^{\ast}]_+ = t_i(-a)\delta_{i,j}	\nonumber
  \\
  \\
  [\overline{\bbeta}_i,\overline{\bbeta}_j^{\ast}]_+ = t_i(-a)\delta_{i,j}& \; ,\qquad [\overline{\bgamma}_i,\overline{\bgamma}_j^{\ast}]_+ = -t_i(a)\delta_{i,j} \; ,	\nonumber
\end{align}
with all the other possible parentheses vanishing.

These relations between the descendants of $\Phi_a(0)$ and those of $\Phi_{a-mb}(0)$ have a relevant consequence on the function $\Theta$ named above; in fact, 
since we can obtain $\Phi_{a-mb}(0)$ either in one go from $\Phi_a(0)$ or in $m$ steps passing through $\Phi_{a-kb} \ k=1,2,\ldots,m-1$, we obtain some quite 
restrictive functional equation for $\Theta(\ell,m\vert\alpha)$ which, joined with some symmetry relations and the request that (\ref{eq:onepointfunction}) 
reproduces, for appropriate values of $a$, one-point functions for components of the energy-momentum tensor (which can be computed from general arguments), 
allowed the authors of \cite{Negr_Smir_13_2} to conjecture the form of the function $\Theta(\ell,m\vert\alpha)$ for the sinh-Gordon model\footnote{In fact 
the lattice regularisation procedure used in \cite{Boos_Jimb_Miwa_Smir_10,Jimb_Miwa_Smir_11_1} for sine-Gordon model is unavailable for the sinh-Gordon one, 
thus it was necessary to rely on conjectures and consistency relations.}.

Let us now review some basic fact about the fermionic basis.

\section{The fermionic basis}
\label{sec:fermbasis}
\setcounter{equation}{0}

We will start on the stable ground granted by the Liouville model: here the fermionic basis can be defined \cite{Negr_Smir_13_1,Negr_Smir_13_2} as an intrinsic property, that is it
allows a pure CFT definition.

The defining property\footnote{It has to be underlined that, for sine-Gordon model, these property and the following ones derive directly from the definition on the lattice of the fermionic basis and, thus, are not ``defining'' properties of the fermions, but rather a consequence of their structure. On the other hand, since we lack a proper lattice construction, for the sinh-Gordon model these properties are assumed as defining and are taken as the starting point for the construction of the fermionic basis. The symmetries listed here are a translation of the reflection relations (see chapter \ref{chap:refrel}) which both sine-Gordon and sinh-Gordon expectation values are believed to obey \cite{Fate_Frad_Luky_AZam_AlZa_99}.} of the fermions relies on their behaviour under the symmetries $\sigma_1$ and $\sigma_2$ introduced above:

\begin{align}
  &\phantom{\sigma_1 \; : \; } \bgamma^{\textrm{CFT}\,\ast}_{2m-1} \rightarrow u(a)\bbeta^{\textrm{CFT}\,\ast}_{2m-1} 
  \phantom{\qquad , \qquad \sigma_2 \; : \qquad\; } \bgamma^{\textrm{CFT}\,\ast}_{2m-1} \rightarrow \bbeta^{\textrm{CFT}\,\ast}_{2m-1}	\nonumber
  \\
  &\sigma_1 \; : \;  \phantom{\bbeta^{\textrm{CFT}\,\ast}_{2m-1} \rightarrow u^{-1}(-a)\bbeta^{\textrm{CFT}\,\ast}_{2m-1}} \qquad , \qquad
  \sigma_2 \; : \;  
  \\
  &\phantom{\sigma_1 \; : \; } \bbeta^{\textrm{CFT}\,\ast}_{2m-1} \rightarrow u^{-1}(-a)\bgamma^{\textrm{CFT}\,\ast}_{2m-1}
  \phantom{\qquad , \qquad \sigma_2 \; : \;\; }\bbeta^{\textrm{CFT}\,\ast}_{2m-1} \rightarrow \bgamma^{\textrm{CFT}\,\ast}_{2m-1}	\nonumber
\end{align}
where

\eq
  u(a) = \frac{-2a+b(2m-1)}{2a+b^{-1}(2m-1)} = \frac{-\nu \alpha +(2m-1)(\nu -1)}{\nu \alpha + (2m-1)}
\en
and for the second chirality we only have to change $a$ in $-a$ in the above function. There is an additional symmetry which was considered in 
\cite{Negr_Smir_13_1}, that is the duality $b \rightarrow b^{-1}$, under which our fermions simply exchange

\begin{align}
  &\phantom{\textrm{duality} \quad\; } \bbeta^{\textrm{CFT}\,\ast}_{2m-1} \rightarrow \bgamma^{\textrm{CFT}\,\ast}_{2m-1}	\nonumber
  \\
  &\textrm{duality} \; :	\label{eq:duality}
  \\
  &\phantom{\textrm{duality} \quad\; } \bgamma^{\textrm{CFT}\,\ast}_{2m-1} \rightarrow \bbeta^{\textrm{CFT}\,\ast}_{2m-1}	\nonumber
\end{align}
The normalisation of the fermions is such that when expressing the fermionic descendants in terms of the Virasoro ones we have

\eq
  \bbeta^{\textrm{CFT}\,\ast}_{I^+}\bgamma^{\textrm{CFT}\,\ast}_{I^-}\Phi_a = C_{I^+,I^-} \left\{\mathbf{l}^n_{-2}+\cdots\right\}\Phi_a
  \; ,\qquad \mathfrak C(I^+) = \mathfrak C(I^-) = n \; ,
\en
with $\mathbf{l}_{n}$ being the components of the Laurent expansion of the components $T(z,\bz)$ and $\bar T(z,\bz)$ of Liouville energy-momentum tensor and $C_{I^+,I^-}$ the determinant of the Cauchy matrix $\{ 1/(i^+_j + i^-_k -1) \}_{j,k = 1}^n$.

These CFT fermions are related to the ones introduced above simply by multiplication for a constant:

\begin{align}
  & \bbeta^{\ast}_{2m-1} = D_{2m-1}(a) \bbeta^{\textrm{CFT}\,\ast}_{2m-1} \qquad , \qquad 
  \bgamma^{\ast}_{2m-1} = D_{2m-1}(Q-a) \bgamma^{\textrm{CFT}\,\ast}_{2m-1} \; ,	\nonumber
  \\	\label{eq:cfttosinhfermions}
  \\ & \overline \bgamma^{\ast}_{2m-1} = D_{2m-1}(a) \overline\bgamma^{\textrm{CFT}\,\ast}_{2m-1} \qquad , \qquad 
  \overline \bbeta^{\ast}_{2m-1} = D_{2m-1}(Q-a) \overline\bbeta^{\textrm{CFT}\,\ast}_{2m-1} \; ,	\nonumber
\end{align}
where

\eq
  D_{2m-1}(a) = \frac{1}{2\pi \mathbbm i} \left(\frac{\bmu\Gamma(1+b^2)}{b^{1+b^2}}\right)^{-\frac{2m-1}{1+b^2}}\frac{\Gamma\left(\frac{a}{Q}+ \frac{2m-1}{2 b Q}\right)
  \Gamma\left(\frac{Q-a}{Q}+b\frac{2m-1}{2Q}\right)}{(m-1)!} \; .
\en
Note that this definition for the constants $D_{2m-1}$ differs from the one used in \cite{Boos_Jimb_Miwa_Smir_10,Jimb_Miwa_Smir_10,Jimb_Miwa_Smir_11_1} by the factor

\eq
(-1)^m\sqrt{\frac{1+b^2}{\mathbbm i}}\frac{\bmu^{-\frac{2m-1}{1+b^2}}}{2\sin\left[\pi\left(\frac{a}{Q}-b\frac{2m-1}{2Q}\right)\right]} \; ;
\en
the reason for this choice is twofold: on the one side, the presence of $\bmu^{-\frac{2m-1}{1+b^2}}$ makes the fermions dimensionless while, on the other, the Q-periodic $\sin\left[\pi\left(\frac{a}{Q}-b\frac{2m-1}{2Q}\right)\right]$ lets to the non CFT fermions inherit the duality (\ref{eq:duality}). Of course this last holds iff the following term is ``self-dual''

\eq
  \frac{[\bmu\Gamma(1+b^2)]^{\frac{1}{1+b^2}}}{b}\; ,
\en
but this follows automatically when expressing $\bmu$ in terms of the sinh-Gordon particle mass, which is explicitly self-dual:

\eq
  \bmu \Gamma(1+b^2) = \left[ \frac{m}{4\sqrt{\pi}} \Gamma\left(\frac{1}{2(1+b^2)}\right)\Gamma\left(1+\frac{b^2}{2(1+b^2)}\right) \right]^{1+b^2} \; .
\en

The constants $t_{\ell}(a)$ and $C_m(a)$ introduced in (\ref{eq:shiftformula}) are defined as follows

\begin{align}
  t^{sG}_{\ell}(a) \doteq \tan^{-1}\left[\frac{\pi}{2}\left(2\frac{a}{Q}+\frac{\ell}{b Q}\right)\right]	\nonumber
  \\
  \\
  t^{shG}_{\ell}(a) \doteq -\frac{1}{2}\sin^{-1}\left[\frac{\pi}{2}\left(2 \frac{a}{Q}+\frac{\ell}{b Q}\right)\right]	\nonumber
\end{align}

\begin{align}
  &C_m(a) \doteq \prod_{j=0}^{m-1} C_1(a-2bj) \; ,	\nonumber
  \\ \label{eq:C-constants}
  \\
  &C_1(a) \doteq [\bmu\Gamma(1+b^2)]^{4 x} \frac{\gamma(x)\gamma\left(\frac{1}{2}-x\right)}{2bQ\gamma(2 b x Q)} \; ,	\nonumber
\end{align}
where $2 Q x = 2 a -b$ and we denote $\gamma(x) = \Gamma(x)/\Gamma(1-x)$.

The time has now come when we have to specify the range of the coupling $b$ or, in other words, to decide whether we are considering sine- or sinh-Gordon. The reason is that, as we hinted at above, even though the two models spawn from the same action and can be regarded, with some na\"{\i}vety, as analytic continuation of one another, truth is they present quite different properties and behaviours. These differences reflect into the strategies that were chosen in order to arrive at the fermionic basis for the two models.

Sine-Gordon model, in the Euclidean field theory formalism, allow for a lattice regularisation in the form of eight-vertex model. The authors of \cite{Jimb_Miwa_Smir_11_1} used this fact and studied the scaling limit of the inhomogeneous six-vertex model \cite{Jimb_Miwa_Smir_09} (which is an Euclidean version of the construction of \cite{Dest_Vega_95}); they introduced the Matsubara transfer matrices $T$ and $Q$ as traces of monodromy matrices associated, respectively, to the two-dimensional and q-oscillator representation of the algebra $U_q(\hat{\mathfrak{sl}}_2)$ (exactly as it is done for the continuous chiral CFT with $c<1$ in \cite{Bazh_Luky_AZam_96,Bazh_Luky_AZam_97,Bazh_Luky_AZam_99}) and they arrived directly at the Destri-DeVega equation \cite{Dest_Vega_95}, circumnavigating the system of TBA equations which, for sine-Gordon, is rather nasty.

For sinh-Gordon model, however, things are not so simple. In fact the eight-vertex model should be replaced, in this case, with a much more complicated model, in which Boltzmann weights are defined in terms of the universal $R$-matrix of the tensor product of two infinite-dimensional representations of $U_q(\hat{\mathfrak{sl}}_2)$ (which do not possess highest weights) \cite{Byts_Tesc_06,Tesc_07}. The status of the phase transition for this model has not been, as far as we know, clarified. On the other hand, in sinh-Gordon model, the $S$-matrix is extremely simple (in fact, there is a single particle in the spectrum) and, as a consequence, the TBA system is composed of a single equation; this makes much more easier to start from TBA and proceed to the construction of the fermionic basis.

\section{One-point functions in sine-Gordon model}
\label{sec:onepointsine}
\setcounter{equation}{0}

We give here a brief review of the results of \cite{Jimb_Miwa_Smir_11_1} that will be useful later.

Let us start from the Destri-DeVega equation for the ground state in Matsubara direction, whose derivation can be found in \cite{Bazh_Luky_AZam_97}:

\eq
  \frac{1}{\mathbbm i}\log \mathfrak a(\theta) = 2\pi m R \sinh\theta - 2 \Im \int\limits_{-\infty}^\infty R(\theta-\theta')\log(1+\mathfrak a(\theta'-\mathbbm i 0))d\theta' \; ,
\en
where the function $R(\theta)$ is essentially the logarithmic derivative of sine-Gordon soliton-soliton scattering amplitude and can be defined as a Fourier transform (remember that $\nu = 1+ b^2 = 1- \beta^2$)

\eq
  R(\theta) = \int\limits_{-\infty}^\infty e^{\mathbbm i \theta k}\hat R(k)\frac{d k}{2\pi} \; ,\qquad \hat R(k) = \frac{\sinh\left(\pi k \frac{2\nu - 1}{2\nu}\right)}{2\sinh\left(\pi k \frac{1-\nu}{2\nu}\right)\cosh\left(\pi\frac{k}{2}\right)} \; .
\label{eq:sinekernel}
\en

The function $\mathfrak a(\theta)$ is the ratio of ground-state eigenvalues of the $\mathbf{Q}$-operator, which is the trace of quantum monodromy matrices in the $q$-oscillator representation of the algebra $U_q(\hat{\mathfrak{sl}}_2)$\footnote{For the sake of precision it would be better to say that $\mathfrak a(\theta)$ is defined as the scaling limit of the ratio of $Q$-function of the six-vertex model: this is the way it was introduced in \cite{Boos_Jimb_Miwa_Smir_10}.}:

\eq
  \mathfrak a(\lambda) = \frac{Q(\lambda q)}{Q(\lambda q^{-1})} \; ,\qquad \mathbf Q(\lambda)\vert 0\rangle = Q(\lambda)\vert 0\rangle \; ,\qquad \lambda = e^{\nu \theta} \; , \quad q = e^{-\mathbbm i\pi b^2} \; ,
\en
with $\vert 0 \rangle$ representing the ground state. The function $Q(\lambda)$ is invariant for $\lambda \rightarrow \lambda^{-1}$, which implies

\eq
  \mathfrak a(\theta) = \frac{1}{\mathfrak a(-\theta)} \; .
  \label{eq:athetasymmetry}
\en

Following an idea proposed in \cite{Boos_Gohm_Klum_Suzu_07}, we introduce a deformed version of the kernel (\ref{eq:sinekernel}):

\begin{align}
  &R(\theta,\alpha) = \int\limits_{-\infty}^\infty e^{\mathbbm i \theta k}\hat R(k,\alpha)\frac{d k}{2\pi} \; ,\qquad \hat R(k,\alpha) = \frac{\sinh\left(\pi k \frac{2\nu - 1}{2\nu}-\mathbbm i\pi\frac{\alpha}{2}\right)}{2\sinh\left(\pi k \frac{1-\nu}{2\nu}+\mathbbm i\pi\frac{\alpha}{2}\right)\cosh\left(\pi\frac{k}{2}\right)} \; ,	\nonumber
  \\ \label{eq:sinedeformedkernel}
  \\
  & R(\theta) \equiv R(\theta,0) \; ,	\nonumber
\end{align}
where $Q \alpha = 2 a$. It is important to remark that this deformed kernel enjoys the following symmetries:

\eq
  \hat R(-k,\alpha) = \hat R(k,-\alpha) \; ,\qquad \hat R(k,\alpha+2) = \hat R(k,\alpha) \; ,\qquad \hat R(k,\alpha +2\frac{1-\nu}{\nu}) = \hat R(k+2\mathbbm i,\alpha) \; .	\nonumber
\en

Now, by introducing a deformed convolution $\ast$, defined as

\begin{align}
  &[f\ast \, g](\theta,\theta')  = \int\limits_{-\infty}^\infty f(\theta,\varphi)g(\varphi,\theta')dm(\varphi) \ ,	\nonumber
  \\
  \\&dm(\varphi) = 2\Re \left(\frac{1}{1+\mathfrak a(\varphi-\mathbbm i0)}\right)d\varphi \; ,	\nonumber
\end{align}
we can build the following ``dressed'' resolvent

\eq
  R_{\textrm{dress}}(\theta,\theta'\vert\alpha) + [R\ast \, R_{\textrm{dress}}](\theta,\theta'\vert\alpha) = R(\theta,\theta'\vert\alpha) \; ,
\label{eq:dressresolvent}
\en
where we agree that $R(\theta,\theta'\vert\alpha) \equiv R(\theta-\theta',\alpha)$.

Finally we define the function $\Theta^{\textrm{sG}}_R(l,m\vert \alpha)$ through the equation

\eq
  R_{\textrm{dress}}(\theta,\theta'\vert\alpha) - R(\theta-\theta',\alpha) = \int\limits_{-\infty}^\infty \int\limits_{-\infty}^\infty \frac{dl}{2\pi}\frac{dm}{2\pi} \hat R(l,\alpha)\Theta^{\textrm{sG}}_R(l,m\vert\alpha)\hat R(m,-\alpha) e^{\mathbbm il\theta+\mathbbm im\theta'} \; ,	\nonumber
\en
which is readily rewritten, using (\ref{eq:dressresolvent}) into

\eq
  \Theta^{\textrm{sG}}_R(l,m\vert\alpha) + G(l+m) + \int\limits_{-\infty}^\infty G(l-k) \hat R(k,\alpha) \Theta^{\textrm{sG}}_R(k,m\vert\alpha) \frac{dk}{2\pi} = 0 \; ,
\en
with $G(k)$ being the k-\emph{moment} of the measure $dm(\theta)$:

\eq
  G(k) = \int\limits_{-\infty}^\infty e^{-\mathbbm i k \theta}dm(\theta) \; .
\en
Another way to express the function $\Theta$ is the following\footnote{Here the convolutions are to be understood as follows: $f\ast \, g = \int\limits_{-\infty}^\infty f(\theta)g(\theta)dm(\theta)$ and $f\ast \, g\ast \, h = \int\limits_{-\infty}^\infty dm(\theta) \int\limits_{-\infty}^\infty dm(\theta') f(\theta)g(\theta,\theta') h(\theta')$.}

\eq
  \Theta^{\textrm{sG}}_R(\mathbbm i l, \mathbbm i m\vert\alpha) = e_{l}\ast \, R_{\textrm{dress}}^{(\alpha)}\ast \, e_{m} - e_{l}\ast \,  e_{m} \; ,
  \label{}
\label{eq:sinthetaerepresentation}
\en
where $e_{\gamma}(\theta) = e^{\gamma \theta} \ , \ \forall \gamma\in\mathbbm C$.

Given the symmetry (\ref{eq:athetasymmetry}), the function $G(k)$ is even; using this fact together with the symmetries of the deformed kernel $\hat R$, we can show that

\eq
  \Theta^{\textrm{sG}}_R(l,m\vert2-\alpha) = \Theta^{\textrm{sG}}_R(-l,-m\vert\alpha)
  \label{eq:thetasymmetry1}
\en

\eq
  \Theta^{\textrm{sG}}_R(l,m\vert-\alpha) = \Theta^{\textrm{sG}}_R(m,l\vert\alpha)
  \label{eq:thetasymmetry2}
\en
and

\begin{align}
  \Theta^{\textrm{sG}}_R(l,m\vert\alpha + 2 \frac{1-\nu}{\nu})-&\Theta^{\textrm{sG}}_R(l+2 \mathbbm i,m-2 \mathbbm i\vert\alpha) =
  \\
  &=-\frac{\Theta^{\textrm{sG}}_R(l+2\mathbbm i,-\mathbbm i\vert\alpha)\Theta^{\textrm{sG}}_R(\mathbbm i,m-2\mathbbm i\vert\alpha)}{\Theta^{\textrm{sG}}_R(\mathbbm i,-\mathbbm i\vert\alpha)-\pi t^{\textrm{sG}}_1(\frac{Q}{2}\alpha)} \; . \nonumber
\end{align}
This last symmetry is easily extended to negative shift $-2\frac{1-\nu}{\nu}$ by changing $\alpha$ in $-\alpha$ and applying the symmetry (\ref{eq:thetasymmetry2}). Directly correlated to this symmetry of the function $\Theta$ is the shift formula (\ref{eq:shiftformula}): it is not difficult to see that one implies the other.

As we anticipated the ratio of one-point functions of any fermionic descendant with the one point-function of the corresponding primary field $\Phi_\alpha(0)$ (here we simplify the notation, writing $\Phi_\alpha(0) \equiv \Phi_{a(\alpha)}(0)$ with $a(\alpha) = \frac{Q}{2}\alpha$) can be written in a determinant form:

\eq
  \frac{\langle\bbeta_{I^+}^{\ast}\overline{\bbeta}_{\overline I^+}^{\ast}\overline{\bgamma}_{\overline I^-}^{\ast}\bgamma_{I^-}^{\ast}\Phi_a(0)\rangle_R}
  {\langle \Phi_a(0)\rangle_R} = \mathcal D(I^+ \cup (-\overline I^+)\vert I^- \cup (-\overline I^-) \vert \alpha) \; ,
\label{eq:onepointfunctionagain}
\en
where, for two sets $A=\{a_j\}_{j=1}^n$ and $B=\{b_j\}_{j=1}^n$, we have

\eq
  \mathcal D(A\vert B\vert \alpha) \doteq \left(\prod_{\ell=1}^n \frac{\textrm{sgn}(a_\ell)\textrm{sgn}(b_\ell)}{\pi} \right) 
  \det \left[\Theta^{\textrm{sG}}_R(\mathbbm ia_j,\mathbbm ib_k\vert \alpha)-\pi \delta_{a_j,-b_k} \textrm{sgn}(a_j) t^{\textrm{sG}}_{a_j}(\alpha)\right]_{j,k=1}^n
\en

This result was obtained in \cite{Jimb_Miwa_Smir_11_1}, where the formula for the one-point function was compared successfully against known results. Of particular importance was the check of the agreement with Zamolodchikov formula (here $\Theta$ stands for the trace of the energy-momentum tensor (\ref{eq:enmomtensor}), not the function defined above!)

\eq
  \langle T\overline T\rangle = \langle T\rangle\langle\overline T\rangle - \langle\Theta\rangle^2 \; ,
\en
which was proven to hold for any 2D Euclidean QFT on a cylinder \cite{AZam_04}. We report this comparison in ch.\ref{chap:knownresults}.

\section{One-point functions in sinh-Gordon model}
\label{sec:onepointsinh}
\setcounter{equation}{0}

Let us now move to the case $0< b^2\leq 1$, which corresponds to the sinh-Gordon model. As said above, since the TBA equation for this model is extremely simple, it is easier to chose them as a starting point and proceed to the construction of the function $\Theta(l,m\vert\alpha)$ relying on consistency equations.
The TBA for sinh-Gordon model consists of a single equation\footnote{Note that the \emph{pseudo-energy} $\epsilon(\theta)$ is minus the logarithm of the function $\mathfrak a(\theta)$ corresponding to the one we used for the sine-Gordon model.}

\eq
  \epsilon(\theta) = 2 \pi m R\cosh\theta - \int\limits_{-\infty}^\infty \Phi(\theta-\theta') \log\left(1+ e^{-\epsilon(\theta')}\right) d\theta' \; ,
  \label{eq:destridevegaequation}
\en
with

\begin{align}
  \Phi(\theta) &= \frac{1}{2\pi\cosh\left(\theta + \pi \mathbbm i \frac{\nu-2}{2\nu}\right)} + \frac{1}{2\pi\cosh\left(\theta - \pi \mathbbm i \frac{\nu-2}{2\nu}\right)} = \int\limits_{-\infty}^\infty e^{\mathbbm ik\theta}\hat{\Phi}(k) \frac{dk}{2\pi} \; ,	\nonumber
  \\
  \\
  \hat{\Phi}(k) &= \frac{\cosh\left(\pi\frac{\nu-2}{2\nu}k\right)}{\cosh\left(\pi\frac{k}{2}\right)} \; .	\nonumber
\end{align}

Starting from this basic equation, one can build all the Matsubara data; namely define

\eq
  \log Q(\theta) = -\frac{\pi m R \cosh\theta}{\sin\frac{\pi}{\nu}} + \int\limits_{-\infty}^{\infty} \frac{\log\left(1+e^{-\epsilon(\theta')}\right)}{\cosh(\theta-\theta')}\frac{d\theta'}{2\pi} \; ,
\en
where we have chosen the first term on the right-hand side for consistency. It is straightforward to check that

\eq
  e^{-\epsilon(\theta)} = Q\left(\theta + \pi \mathbbm i\frac{\nu-2}{2\nu}\right) Q\left(\theta - \pi \mathbbm i\frac{\nu-2}{2\nu}\right) \; ,
\en
from which, recalling the Dirac delta representation $\cosh(\theta+\mathbbm i\frac{\pi}{2})+\cosh(\theta-\mathbbm i\frac{\pi}{2}) = 2\pi \delta(\theta)$, one derive the bilinear equation\footnote{Actually, one should be careful and correctly define the analyticity conditions for the function $Q(\theta)$; a discussion can be found in \cite{AlZa_06}.}

\eq
  Q\left(\theta + \frac{\pi \mathbbm i}{2}\right)Q\left(\theta - \frac{\pi \mathbbm i}{2}\right) -Q\left(\theta + \pi \mathbbm i\frac{\nu -2}{2\nu}\right)Q\left(\theta - \pi \mathbbm i\frac{\nu -2}{2\nu}\right) =1 \; .
  \label{eq:quantumwronskian}
\en

Introducing $\zeta = e^{\nu\theta}$, it is not difficult to see how (\ref{eq:quantumwronskian}) implies that the function $T(\zeta)$, defined from the equation

\eq
  T(\zeta)Q(\theta) = Q\left(\theta +\pi \mathbbm i \frac{\nu-1}{\nu}\right) + Q\left(\theta -\pi \mathbbm i \frac{\nu-1}{\nu}\right) \; ,
\en
is a single-valued function of $\zeta^2$, with essential singularities at $\zeta = 0$ and $\zeta = \infty$. This equation is a second order finite difference equation for the function $Q(\theta)$ and thus admit two different solutions: $Q(\theta)$ and $Q(\theta+\mathbbm i\frac{\pi}{\nu})$, the equation (\ref{eq:quantumwronskian}) being their quantum Wronskian.

It is important to stress that the equations for the functions $Q(\theta)$ and $T(\theta)$ given here are to be considered as \emph{definitions}, thus one should check that they are reasonable. A verification of the correctness of these definition was carried out in \cite{Luky_01}, where the behaviour of $T(\zeta)$ in the ultraviolet region $R\rightarrow 0$ is investigated numerically, showing how the asymptotics of $T(\zeta)$ for $\zeta \rightarrow 0$ and for $\zeta \rightarrow \infty$ correctly reproduce the eigenvalues of CFT integrals of motion and, moreover, that their normalisation is the same as in the sine-Gordon case \cite{Bazh_Luky_AZam_99}; this is an extremely convincing argument.

Now, having the TBA equation (\ref{eq:destridevegaequation}) at our disposal, we proceed in the same exact way we did in the sine-Gordon case, that is, we introduce a deformed kernel $\Phi_{\alpha}(\theta)$. The only difference is that, while in sine-Gordon case the form of the deformed kernel was already known, here we have to make a guess based on the symmetries it has to respect. Namely, we want the Fourier image $\hat{\Phi}(k,\alpha)$ of the deformed kernel to satisfy $\hat\Phi(k,0) = \hat\Phi(k)$, obviously, and the following symmetries

\eq
  \hat{\Phi}(k,\alpha+2) = \hat{\Phi}(k,\alpha) \; ,\qquad \hat{\Phi}(k,-\alpha) = \hat{\Phi}(-k,\alpha) \; ,\qquad \hat{\Phi}(k,\alpha + 2\frac{1-\nu}{\nu}) = \hat{\Phi}(k+2\mathbbm i,\alpha)\; .	\nonumber
\en

It is not hard to find that the kernel we are looking for has the following form:

\begin{align}
  &\Phi_{\alpha}(\theta) = \frac{e^{\mathbbm i\pi\alpha}}{2\pi\cosh\left(\theta +\pi \mathbbm i\frac{\nu-2}{2\nu}\right)}+\frac{e^{-\mathbbm i\pi\alpha}}{2\pi\cosh\left(\theta -\pi \mathbbm i\frac{\nu-2}{2\nu}\right)} = \int\limits_{-\infty}^\infty e^{\mathbbm ik\theta}\hat{\Phi}(k,\alpha)\frac{dk}{2\pi} \; ,	\nonumber
  \\
  \\
  &\hat{\Phi}(k,\alpha) = \frac{\cosh\left(\pi\frac{\nu-2}{2\nu}k - \pi \mathbbm i \alpha\right)}{\cosh\left(\pi\frac{k}{2}\right)} \; .	\nonumber
\end{align}
It is interesting to notice that, contrary to the function $\hat R(k,\alpha)$ of the sine-Gordon model, the deformed kernel $\hat\Phi$ does not have poles in the $k$-plane whose position depends on $\alpha$. This simplification in the kernel structure is directly correlated to the fact that sinh-Gordon one-point functions, as functions of $\alpha$, have much simpler analytical properties than those of sine-Gordon.

Let us proceed by defining the dressed resolvent, which satisfies to the equation 

\eq
  R_{\textrm{dress}}(\theta,\theta'\vert\alpha) - \left[\Phi\ast \, R_{\textrm{dress}}\right](\theta,\theta'\vert\alpha) = \Phi(\theta,\theta'\vert\alpha) \; ,
\en
where $\Phi(\theta,\theta'\vert\alpha) \equiv \Phi_\alpha(\theta-\theta')$ and the $\ast$ denotes, as in the sine-Gordon case, a deformed convolution

\eq
  [f\ast \, g](\theta,\theta') = \int\limits_{-\infty}^\infty f(\theta,\phi)g(\phi,\theta')dm(\phi) \; ,\qquad dm(\phi) = \frac{d\phi}{1+e^{\epsilon(\phi)}} \; .
\en

Now, using the dressed resolvent, we build the function $\Theta^{\textrm{shG}}_R$:

\eq
  R_{\textrm{dress}}(\theta,\theta'\vert\alpha) - \Phi_\alpha(\theta-\theta') = \int\limits_{-\infty}^\infty\int\limits_{-\infty}^\infty \frac{dl}{2\pi}\frac{dm}{2\pi} \hat{\Phi}(l,\alpha)\Theta^{\textrm{shG}}_R(l,m\vert\alpha)\hat \Phi(m,-\alpha)e^{\mathbbm il\theta+\mathbbm im\theta'} \; ;
\en
straightforward calculations show that the function $\Theta^{\textrm{shG}}_R$ satisfies the following equation

\eq
  \Theta^{\textrm{shG}}_R(l,m\vert\alpha) - G(l+m) - \int\limits_{-\infty}^\infty G(l-k)\hat \Phi(k,\alpha) \Theta^{\textrm{shG}}_R(k,m\vert\alpha) \frac{dk}{2\pi} = 0 \; ,
\en
with the function $G(k)$ being, here too, the $k$-moment of the measure $dm(\theta)$

\eq
  G(k) = \int\limits_{-\infty}^\infty e^{-\mathbbm ik\theta} \frac{d\theta}{1+e^{\epsilon(\theta)}} \; .
\en

A useful way to express the function $\Theta^{\textrm{shG}}_R$ is the following

\eq
  \Theta^{\textrm{shG}}_R(\mathbbm il,\mathbbm im\vert\alpha) = e_l\ast \, e_m + e_l\ast \, R_{\textrm{dress}}^{(\alpha)} \ast \, e_m \; .
\label{eq:sinhthetaerepresentation}
\en

Since, for the ground state, the function $\epsilon(\theta)$ is even, one easily derives the following symmetries:

\eq
  \Theta^{\textrm{shG}}_R(l,m\vert - \alpha) = \Theta^{\textrm{shG}}_R(m,l\vert\alpha) \; ,\qquad \Theta^{\textrm{shG}}_R(l,m\vert\alpha +2) = \Theta^{\textrm{shG}}_R(l,m\vert\alpha) \; ,
\en

\begin{align}
  \Theta^{\textrm{shG}}_R(l,m\vert \alpha + 2\frac{1-\nu}{\nu}) &- \Theta^{\textrm{shG}}_R(l+2\mathbbm i,m-2\mathbbm i\vert\alpha) =	\nonumber
  \\ \label{eq:thetashift}
  \\
  & = - \frac{\Theta^{\textrm{shG}}_R(l+2\mathbbm i,-\mathbbm i\vert\alpha)\Theta^{\textrm{shG}}_R(\mathbbm i,m-2\mathbbm i\vert\alpha)}{\Theta^{\textrm{shG}}_R(\mathbbm i,-\mathbbm i\vert\alpha) - \pi t^{\textrm{shG}}_1(\frac{Q}{2}\alpha)} \; .	\nonumber
\end{align}
Again the shift by $-2\frac{1-\nu}{\nu}$ is defined changing $\alpha$ in $-\alpha$ and making use of the preceding properties.

We are now ready to express the main conjecture of this part:

\begin{Mconj}
  We conjecture that, similarly in sine-Gordon model, the one-point functions in the fermionic basis are expressed in terms of a determinant
  
  \eq
    \frac{\langle\bbeta^\ast_{I^+}\overline\bbeta^\ast_{\overline I^+}\overline\bgamma^\ast_{\overline I^-}\bgamma^\ast_{I^-}\Phi_\alpha(0)\rangle_R}{\langle\Phi_\alpha(0)\rangle_R} = \mathcal D\left(I^+ \cup(-\overline I^+) \vert I^-\cup(-\overline I^-)\vert\alpha\right) \; ,
    \label{eq:onepointfunctionsinh}
  \en
  where, for two sets $A=\{a_j\}_{j=1}^n$ and $B=\{b_j\}_{j=1}^n$, we have

\eq
  \mathcal D(A\vert B\vert \alpha) \doteq \left(\prod_{\ell=1}^n \frac{\textrm{sgn}(a_\ell)\textrm{sgn}(b_\ell)}{\pi} \right) 
  \det \left[\Theta^{\textrm{shG}}_R(\mathbbm ia_j,\mathbbm ib_k\vert \alpha)-\pi \delta_{a_j,-b_k} \textrm{sgn}(a_j) t^{\textrm{shG}}_{a_j}(\alpha)\right]_{j,k=1}^n
\en
\end{Mconj}
Notice how, since $\Theta^{\textrm{shG}}_R(l,m\vert\alpha)\underset{R\rightarrow\infty}{\rightarrow}0$ and $\Theta^{\textrm{sG}}_R(l,m\vert\alpha)\underset{R\rightarrow\infty}{\rightarrow}0$, in the infinite volume limit $R\rightarrow\infty$ the formulae for the one-point functions in sinh-Gordon coincide with the analytic continuation with respect to $b$ of the corresponding in sine-Gordon model.

The necessity of the symmetry $\hat\Phi(k,\alpha + 2\frac{1-\nu}{\nu}) = \hat\Phi(k+2 \mathbbm i,\alpha)$, from which the relation (\ref{eq:thetashift}) descends, reveals itself when looking at the shift formulae (\ref{eq:shiftformula}); exactly as it happens in sine-Gordon, those formulae impose certain consistency equations, all of which can be derived, using some combinatorics, from (\ref{eq:thetashift}). For example, the shift formulae (\ref{eq:shiftformula}) imply the following

\eq
  \bbeta^\ast_1\bgamma^\ast_1\Phi_{\alpha+2\frac{1-\nu}{\nu}} = \frac{C_1(\alpha)}{t^{\textrm{shG}}_1(\alpha)} \bbeta^\ast_3\bbeta_1\bbeta^\ast_1\overline\bgamma^\ast_1\Phi_\alpha = -C_1(\alpha)\bbeta^\ast_3\overline\bgamma^\ast_1\Phi_\alpha \; ,
\en
which, in turn, means

\eq
  \frac{\langle\bbeta^\ast_1\bgamma^\ast_1\Phi_{\alpha+2\frac{1-\nu}{\nu}}\rangle_R}{\langle\Phi_{\alpha}\rangle_R} = \frac{\langle\bbeta^\ast_1\bgamma^\ast_1\Phi_{\alpha+2\frac{1-\nu}{\nu}}\rangle_R}{\langle\Phi_{\alpha+2\frac{1-\nu}{\nu}}\rangle_R} \   \frac{\langle\Phi_{\alpha+2\frac{1-\nu}{\nu}}\rangle_R}{\langle\Phi_{\alpha}\rangle_R} = -C_1(\alpha) \frac{\langle\bbeta^\ast_3\overline\bgamma^\ast_1\Phi_{\alpha}\rangle_R}{\langle\Phi_{\alpha}\rangle_R} \; .	\nonumber
\en
Using again the shift formulae we find

\eq
  \frac{\langle\Phi_{\alpha+2\frac{1-\nu}{\nu}}\rangle_R}{\langle\Phi_{\alpha}\rangle_R} = \frac{C_1(\alpha)}{t^{\textrm{shG}}_1(\alpha)} \frac{\langle\bbeta^\ast_1\overline\bgamma^\ast_1\Phi_{\alpha}\rangle_R}{\langle\Phi_{\alpha}\rangle_R} \; ,
  \label{eq:firstshiftexample1}
\en
which brings us to

\eq
  \frac{\langle\bbeta^\ast_1\bgamma^\ast_1\Phi_{\alpha+2\frac{1-\nu}{\nu}}\rangle_R}{\langle\Phi_{\alpha+2\frac{1-\nu}{\nu}}\rangle_R} \   \frac{\langle\bbeta^\ast_1\overline\bgamma^\ast_1\Phi_{\alpha}\rangle_R}{\langle\Phi_{\alpha}\rangle_R} = -t^{\textrm{shG}}_1(\alpha) \frac{\langle\bbeta^\ast_3\overline\bgamma^\ast_1\Phi_{\alpha}\rangle_R}{\langle\Phi_{\alpha}\rangle_R} \; .
  \label{eq:shiftexample1}
\en
Now, if we take the equation (\ref{eq:thetashift}) for $l=m=\mathbbm i$ and use the one-point function equation (\ref{eq:onepointfunctionsinh}), we obtain

\eq
  \pi \frac{\langle\bbeta^\ast_1\bgamma^\ast_1\Phi_{\alpha+2\frac{1-\nu}{\nu}}\rangle_R}{\langle\Phi_{\alpha+2\frac{1-\nu}{\nu}}\rangle_R} = - \pi \frac{\langle\bbeta^\ast_3\overline\bgamma^\ast_1\Phi_{\alpha}\rangle_R}{\langle\Phi_{\alpha}\rangle_R} \left[1-\left(\frac{\langle\bbeta^\ast_1\overline\bgamma^\ast_1\Phi_{\alpha}\rangle_R}{\langle\Phi_{\alpha}\rangle_R}- t^{\textrm{shG}}_1(\alpha)\right)\frac{\langle\Phi_{\alpha}\rangle_R}{\langle\bbeta^\ast_1\overline\bgamma^\ast_1\Phi_{\alpha}\rangle_R}\right] \; ,	\nonumber
\en
which is clearly equivalent to (\ref{eq:shiftexample1}).\\

\chapter{Reflection Relations}
\label{chap:refrel}
\markboth{Chapter 2 - Reflection Relations}{}
\numberwithin{equation}{chapter}

In this chapter we will exploit the fermionic basis approach to one-point functions in order to solve the so called \emph{reflection relations}. These equations, introduced in \cite{Fate_Frad_Luky_AZam_AlZa_99}, arise from the fact that local fields, and thus one-point functions, have to transform in some definite way under the symmetries $\sigma_1,\sigma_2$ introduced in Chap.\ref{chap:fermions}. Let us be clearer: since, for arbitrary values of $a$ (that is, in absence of resonances), the space of local fields is believed to be the same as in the corresponding unperturbed CFT, one can exploit the two possible descriptions of sin(h)-Gordon action in order to describe the descendants in two different bases, namely the Heisenberg and the Virasoro ones. In sine-Gordon this is possible thanks to the Feigin-Fuchs bosonisation of the Virasoro algebra, while in sin(h)-Gordon one argues, following \cite{Seib_90,AZam_AlZa_96}, that both descriptions are available in the domain where Liouville zero-mode $\eta_0$ tends to $-\infty$. So, starting from Virasoro descendants in a definite level $\ell$, for which the symmetry $\sigma_2$ is automatically satisfied, one rewrites them into Heisenberg ones, obtaining ``change of basis'' matrices $U^{(\ell)}(a),\overline U^{(\ell)}(a)$, one for each chirality\footnote{As explained later, one has to factor out the action of the integrals of motion.}. Now, combining all the expectation values of Virasoro descendants in the level $\ell$ into a vector $\mathbf v^{(\ell)}(a)$, one arrives at the following equations (we omit the superscript specifying the level $(\ell)$)

\eq
  \bv(Q-a) = \bv(a) \; , \qquad \bv(a+Q) = \left( S(a) \otimes \overline S(a) \right) \bv(a) \; ,
  \label{eq:genericrefrel}
\en
where

\eq
  S(a) = U(-a)U^{-1}(a) \; , \qquad \overline S(a) = \overline U(-a)\overline U^{-1}(a) \; .
\en
This Riemann-Hilbert problem was called reflection relations in \cite{Fate_Frad_Luky_AZam_AlZa_99}; the reason for the notation adopted is the analogy with scattering theory, with $S(a)$ being the counterpart of the $S$-matrix and $U(a)$ the counterpart of the wave operator.

The question of analyticity of expectation values in the infinite volume and that of the applicability of the reflections $\sigma_1$, $\sigma_2$ to the perturbed model give rise to some subtle issues which we will not present here; they are discussed in details in \cite{Fate_Luky_AZam_AlZa_98}.

While this seems to be a powerful method for obtaining the one-point functions, the reflection relations (\ref{eq:genericrefrel}) are extremely difficult to solve if no hint about the structure of solutions is given. In fact, in the paper \cite{Fate_Frad_Luky_AZam_AlZa_99} only the case $\ell = 2$, which consist of a single non-trivial descendant, was solved; it was also shown how to fix, by minimality assumptions, the CDD factors which usually arise in bootstrap procedures\footnote{The acronym ``CDD'' stands for Castillejo, Dalitz, Dyson \cite{Cast_Dali_Dyso_56} and denotes multiplicative factors which contain no physical information and automatically satisfy the bootstrap requirements.} and how to solve the problem of overall normalisation by requiring the cancellation of resonances. Here, following \cite{Negr_Smir_13_1}, we will show how the fermionic basis automatically grants us solutions to the reflection relations, basically making them diagonal.

Let us now define more carefully the ``change of basis'' matrix $U(a)$ which, from now on, we will call \emph{reflection matrix}; we will focus on the holomorphic sector, since the formulae and considerations are identical for the anti-holomorphic sector.

\section{The reflection matrix}
\label{sec:refmatr}
\numberwithin{equation}{section}
\setcounter{equation}{0}

As we have seen above, the natural form of descendant fields for the $c=1$ CFT is that of normal ordered product of the primary field $e^{a\eta(0)}$ with a polynomial in the derivatives of the field $\eta(0)$, that is $P^{(\ell)}\left(\lbrace\partial^k\eta(0)\rbrace_k\right) e^{a\eta(0)}$, where $\ell$ denotes the order of the polynomial $P$ which we assume to be even.

In a similar fashion, the natural descendants in Liouville CFT are normal ordered products of the primary field $e^{a\eta(0)}$ with a polynomial in the derivatives of the energy-momentum tensor (\ref{eq:enmomtensor}), that is $L^{(\ell)}\left(\lbrace\partial^{2k} T(0)\rbrace_k\right) e^{a\eta(0)}$, where, here, $\ell$ denotes the order of the polynomial $L$ and we assumed that only derivatives of even order appear.

In order to compute the normal ordered expressions, it is more convenient to work in operator formalism. We know that, in the CFT limit, the field $\eta(z,\bz)$ splits into chiral parts $\eta(z,\bz) = \phi(z) + \phi(\bz)$ and we can decompose the field $\phi(z)$ into modes:

\eq
  \phi(z) = \phi_0 - 2\mathbbm i \pi_0 \log(z) + \mathbbm i \sum_{k\in\mathbbm Z_0}\frac{a_k}{k} z^{-k} \; ,
\en
where $\mathbbm Z_0 \equiv \mathbbm Z\backslash 0$, the Heisenberg operators satisfy

\eq
  [a_j,a_k] = 2 j \delta_{j,-k}
\en
and the $\phi_0$ is the zero-mode, canonically conjugated to $\pi_0$

\eq
  \pi_0 = - \mathbbm i \frac{\partial}{\partial\phi_0}
\en

The primary field $e^{a\phi(0)}$ corresponds to the highest vector of the Heisenberg algebra $\vert a;0\rangle$:

\eq
  e^{a\phi(0)} \Longleftrightarrow \vert a;0\rangle \; , \quad \pi_0\vert a;0\rangle = -\mathbbm i a\vert a;0\rangle \; , \quad a_k\vert a;0\rangle = 0 \; , \; \forall k>0\; ,
\en
and, though the second chirality mostly remains a spectator, we will use the following notation

\eq
  \Phi_a = \vert a;0\rangle \otimes \overline{\vert a;0\rangle} \; .
\en
It is straightforward to check that the correspondence with local fields is:

\eq
  P^{(\ell)}(\{\partial^k \eta(0)\}) e^{a\eta(0)} \Longleftrightarrow  P^{(\ell)}(\{\mathbbm i(k-1)!a_{-k}\})\Phi_a \; .
\en

Concerning the Virasoro descendants, we have to introduce the Virasoro generators $\{\mathbf l_n\}_{n\in\mathbbm Z}$ which are the components of the Laurent expansion of the energy-momentum tensor:

\eq
  T(z) = \sum_{n\in\mathbbm Z} z^{-n-2}\, \mathbf{l}_{n} \  \Longrightarrow \  \mathbf l_n = \frac{1}{2\pi \mathbbm i}\oint_0 z^{n+1}\, T(z) dz \; .
\en
By using the expression (\ref{eq:enmomtensor}) for $T(z)$ joint with the mode expansion for the field $\eta(z,\bz)$, we find

\begin{align}
  \mathbf l_n &= \frac{1}{4} \sum_{j\neq 0,n} a_ja_{n-j} + \left(\mathbbm i(n+1)\frac{Q}{2} + \pi_0\right) a_n \; , \quad \forall n \neq 0 \; ,	\nonumber
  \\
  \\
  \mathbf l_0 &= \frac{1}{2} \sum_{j=1}^\infty a_{-j}a_j + \pi_0\left(\pi_0 + \mathbbm i Q\right) \; .	\nonumber
\end{align}
It is rather easy to see that these operators satisfy the Virasoro algebra with central charge $c= 1 + 6 Q^2$:

\eq
  [\mathbf l_m,\mathbf l_n] = (m-n)\mathbf l_{n+m} + c\;\frac{m^3 -m}{12}\, \delta_{m,-n}
\en
and $\mathbf l_0\Phi_a = \Delta \Phi_a$, with $\Delta = a(Q-a)$.

The mode expansion of the energy-momentum tensor shows us directly the correspondence between fields and operators:

\eq
	L^{(\ell)}\Big(\{\partial^{2k}T(0)\}\Big)e^{a\eta(0)} \Longleftrightarrow L^{(\ell)}\Big(\{(2k)!\;\mathbf l_{-2k-2}\}\Big) \Phi_a \; .
\en

Now, let us come to the integrability; as we have seen before, both the Heisenberg and the Virasoro descendants are basis for the whole Verma module $\mathcal V_a$, however, an infinite series of integrals of motion $\mathbf i_{2k-1}$ which survive after the perturbation exists \cite{AZam_87}. These can be represented in terms of the Virasoro generators:

\begin{align}
	& \mathbf i_1 = \mathbf l_{-1} \; , \qquad \mathbf i_3 = 2 \sum_{k=-1}^\infty \mathbf l_{-3-k}\mathbf l_k\; ,	\nonumber
	\\	\nonumber
	\\
	& \mathbf i_3 = 3\left(\sum_{k=-1}^\infty\sum_{l=-1}^\infty\mathbf l_{-5-k-l}\mathbf l_l\mathbf l_k + \sum_{k=-\infty}^{-2}\sum_{l=-\infty}^{-2} \mathbf l_l\mathbf l_k\mathbf l_{-5-k-l}\right) +
	\\	\nonumber
	\\
	& \phantom{\mathbf i_3 = 3\left(\sum_{k=-1}^\infty\sum_{l=-1}^\infty\mathbf l_{-5-k-l}\mathbf l_l\mathbf l_k\right.} + \frac{c+2}{6}\sum_{k=-1}^\infty (k+2)(k+3)\mathbf l_{-5-k}\mathbf l_k \; , \qquad \textrm{etc...} \; ,	\nonumber
\end{align}
and descendants created by the action of these operators must not be taken into account in our discussion since their one-point functions automatically vanish, as we have remarked above. Thus we are interested in the quotient space

\eq
	\mathcal V_a^{\textrm{quo}} = \mathcal V_a / \sum_{k=1}^\infty \mathbf i_{2k-1} \mathcal V_a \; ,
\en
whose nontrivial subspaces $\mathcal V_{a,2k}^{\textrm{quo}}$ are of even degree only, having dimension

\eq
	\textrm{dim}\left(\mathcal V_{a,2k}^{\textrm{quo}}\right) = p(k) \; ,
\en
where $p(k)$ is the number of partition of the number $k$. We shall denote the equality in the quotient space $\mathcal V_a^{\textrm{quo}}$ with the symbol ``$\equiv$''.

In the subspace $\mathcal V_{a,2k}^{\textrm{quo}}$ we choose the two following basis:

\begin{itemize}
	\item $\left\{v_i^{(2k)}\right\}_{i=1}^{p(k)}$, created by the lexicographically ordered action of Virasoro generators with even indices;
	\item $\left\{h_i^{(2k)}\right\}_{i=1}^{p(k)}$, created by the action of an even number of Heisenberg generators.
\end{itemize}
Moreover, we introduce the operators $\mathbf v_i^{(2k)}\, ,\, \mathbf h_i^{(2k)}$ such that

\eq
	v_i^{(2k)} = \mathbf v_i^{(2k)}\Phi_a \; , \qquad h_i^{(2k)} = \mathbf h_i^{(2k)}\Phi_a \; .
\en
The two basis are related by the matrix $U^{(2k)}(a)$:

\eq
	v_i^{(2k)} \equiv \sum_{j=1}^{p(k)} U^{(2k)}_{i,j}(a) h_j^{(2k)} \; ,
\en
which we will refer to as \emph{reflection matrix}.

Now let us give some examples of reflection matrices in order to fix the ideas.

\subsection{Level 2}

This case is quite trivial, since the dimension of the subspace $\mathcal V_{a,2}^{\textrm{quo}}$ equals one. We set

\eq
	\mathbf v_1^{(2)} = \mathbf l_{-2} \; , \qquad \mathbf h_1^{(2)} = (a_{-1})^2 \; ,
\en
and immediately obtain

\eq
	v_1^{(2)} \equiv \frac{1}{4} (b+2a)(b^{-1}+2a) h_1^{(2)} \; .
\en
The zeroes $a_{\pm 1}=-b^{\pm 1}/2$ correspond to singular vectors on level $2$, a fact that does not surprise us.

It is worth remarking that we left out the vector $(\mathbf l_{-1})^2\Phi_a$ which is a descendant of a local integral of motion.

\subsection{Level 4}

This case displays a more interesting structure. Let us set

\eq
	\mathbf v_1^{(4)} = (\mathbf l_{-2})^2 \; , \qquad \mathbf v_2^{(4)} = \mathbf l_{-4} \; , \qquad \mathbf h_1^{(4)} = (a_{-1})^4 \; , \qquad \mathbf h_2^{(4)} = (a_{-2})^2 \; .
\en
The descendants of integrals of motion we need to factor out are those obtained from $\mathbf i_{-1}$, since the only field on level $4$ we can build out of $\mathbf i_3$ is $\mathbf i_3\mathbf i_1 \Phi_a$. After some computations one finds

\eq
	U^{(4)} =-\frac{1}{144} \left(\begin{array}{c c}
		4 a^2 (4 a^2 -3) + 4 a Q (2 a^2 - 3) - 9 & 12 ( 16 a^2 + 3 + 14 a Q + 3 Q^2) \\
		4 a^2 (2 a^2 + 3 a Q + 3) & 12 a (2 a + 3 Q)
	\end{array}\right) \; ,
\en
whose determinant

\eq
	\det \left(U^{(4)}\right) = C^{(4)} \cdot a (2a+b)(2a+b^{-1})(2a+3b)(2a+3b^{-1})(2a+b+b^{-1}) \; ,
\en
where $C^{(4)}$ is an irrelevant numerical multiplier, turns out to be rather simple and instructive. Let us denote $a_{\pm k} = - b^{\pm 1}\, k/2\; , \forall k\in \mathbbm N_{0}$, $a_0 = 0$ and $\tilde a_{l,m} = a_{l} + a_{-m}$; then, the zeroes of the determinant are $a_{\pm 1}$, $a_{\pm 3}$, $\tilde a_{1,1}$ and $a_0$. The first two are obviously associated with the level $2$ null vectors, while $a_{\pm 3}$ and $\tilde a_{1,1}$ correspond to singular vectors of level $4$. The appearance of the zero $a_0$, however, is somewhat peculiar; it clearly descends from the level $1$ null vector $\mathbf l_{-1}\vert 0 \rangle$. This might seem strange, nonetheless the reason is quite simple and contained in the following chain of equalities:

\eq
	2\mathbf l_{-4}\vert 0 \rangle \equiv 2\mathbf l_{-4}\vert 0 \rangle - \mathbf l_{-1}\mathbf l_{-3}\vert 0 \rangle = - \mathbf l_{-3}\mathbf l_{-1}\vert 0 \rangle = 0 \; .
\en

Going higher and higher in the levels one remarks that, while the zeroes corresponding to even level singular vectors appear with regularity, those related to odd levels null vectors follow quite complicated patterns which we will not analyse here, since their structure is not relevant for our goals.

\subsection{Level 6}

As we climb the levels the formulae become more and more heavy, although interesting and peculiar features start to emerge, for this reason we shall display only the most important ones. The Virasoro basis is defined in this case as follows:

\eq
	\mathbf v_1^{(6)} = (\mathbf l_{-2})^3 \; , \qquad \mathbf v_2^{(6)} = \mathbf l_{-4}\mathbf l_{-2} \; , \qquad \mathbf v_3^{(6)} = \mathbf l_{-6} \; ;
\en
this should clarify our lexicographical ordering rule.

We will keep on providing the Heisenberg basis, since it is not clear at first sight when two vectors are linearly dependant, modulo integrals of motion; for this case we set

\eq
	\mathbf h_1^{(6)} = (a_{-1})^6 \; , \qquad \mathbf h_2^{(6)} = (a_{-1})^2(a_{-2})^2 \; , \qquad \mathbf h_3^{(6)} = (a_{-3})^2 \; .
\en

As before, we have to factor out all the descendants of $\mathbf i_1$ and, this time, one descendant of $\mathbf i_3$ has to be taken in account:

\eq
	(\mathbf i_3)^2 \Phi_a \; .
\en
The other possible descendant, $\mathbf i_3\mathbf i_1\mathbf l_{-2}\Phi_a$ is already counted as a descendant of $\mathbf i_1$, exactly as the only possible descendant of $\mathbf i_5$.

The matrix elements of $U^{(6)}$ are rather complicated, however its determinant is given by the following simple formula

\eq
	\det\left(U^{(6)}\right) = C^{(6)} \cdot N^{(6)}(a,b) \cdot \frac{\Delta+2}{3 a^2 - 10 Q^2 - 5}
\en
where $C^{(6)}$ is again a nonrelevant numerical multiplicand and $N^{(6)}(a,b)$ is the null vector contribution factor

\begin{align}
	N^{(6)}(a,b) &= a\,(2a+b)^2(2a+b^{-1})^2(2a+3b)(2a+3b^{-1})(2a+5b)(2a+5b^{-1}) \, \times \nonumber
	\\
	\\
	&\times (2a + b + b^{-1})(2a + 2 b + b^{-1})(2a + b + 2 b^{-1}) \; , \nonumber
\end{align}
whose zeroes hold no surprises. The remaining multiplier, by the way, seems rather odd; its appearance is related to the possibility of decomposing the reflection matrix as follows

\eq
	U^{(6)} = U_0^{(6)} + \frac{1}{3a^2-10Q^2-5}U_1^{(6)} \; , \qquad \left(U^{(6)}\right)^{-1} = \frac{1}{\Delta +2}U^{(6)}_3 + U^{(6)}_4 \; ,
\en
where $U^{(6)}_4$ is regular at $\Delta=-2$, $U^{(6)}_1$ and $U^{(6)}_3$ are of rank $1$ and $U^{(6)}_0$ is linear in $a$. The co-image of the matrix $U^{(6)}_3$ will reveal its importance later and is spanned by

\eq
	\mathbf w^{(6)} = \mathbf l_{-4}\mathbf l_{-2} + \frac{c-16}{2}\mathbf l_{-6} \; .
\label{eq:wvector}
\en

\subsection{Level 8}

The Virasoro basis follows the usual rule, while the Heisenberg basis is set as follows

\begin{align}
	\mathbf h^{(8)}_1 = (a_{-1})^8 \; &, \quad \mathbf h^{(8)}_2 = (a_{-1})^4(a_{-2})^2 \; , \quad \mathbf h^{(8)}_3 = (a_{-2})^4 \; ,	\nonumber
	\\
	\\
	& \mathbf h^{(8)}_4 = (a_{-4})^2 \; , \quad \mathbf h^{(8)}_5 = a_{-2}a_{-6} \; .	\nonumber
\end{align}
The vectors to be factored out are the descendants of $\mathbf i_1$ and 

\eq
	\mathbf i_3 \mathbf l_{-5}\Phi_a \; , \qquad \mathbf i_{3}\mathbf l_{-3}\mathbf l_{-2}\Phi_a \; .
\en

The determinant of the level $8$ reflection matrix is

\eq
	\det\left( U^{(8)}\right) = C^{(8)} \cdot N^{(8)}(a,b)\cdot \frac{(\Delta + 11)(\Delta +4)}{a^2\left(206 a^4 - (1076Q^2+991)a^2 + 21(30 Q^4 + 19 Q^2-76)\right)}
\en
where the null vector contribution is

\begin{align}
	N^{(8)}(a,b) &= a^2(2a+b)^3(2a+b^{-1})^3(2a+3b)^2(2a+3b^{-1})^2(2a+5b)(2a+5b^{-1})\, \times	\nonumber
	\\	\nonumber
	\\
	&\times (2a+7b)(2a+7b^{-1})(2a+ b+b^{-1})^2(2a+2b+b^{-1}) \times
	\\	\nonumber
	\\
	&\times (2a+b+2b^{-1})(2a + 3b+b^{-1})(2a+b+3b^{-1})(a+b)(a+b^{-1}) \; ;	\nonumber
\end{align}
here the the last two factors show that the null vectors on level $3$ contributed for the first time.

For this case too the reflection matrix can be decomposed as follows

\eq
	U^{(8)} = U^{(8)}_0 + \frac{1}{a^2} U^{(8)}_2 + \frac{1}{206 a^4 - (1076Q^2+991)a^2 + 21(30 Q^4 + 19 Q^2-76)}U^{(8)}_3 \; ,
\en
where the ranks of $U^{(8)}_2$ and $U^{(8)}_3$ are respectively $1$ and $2$. More important is the decomposition of the inverse reflection matrix

\eq
	\left(U^{(8)}\right)^{-1} = \frac{1}{\Delta + 4}U^{(8)}_4 + \frac{1}{\Delta + 11}U^{(8)}_5 + U^{(8)}_6 \; ,
\en
where the ranks of the matrices $U^{(8)}_4$ and $U^{(8)}_5$ are both equal to $1$ and the co-images are spanned by the vectors

\begin{align}
	& \mathbf w^{(8)}_4 = -28 \mathbf l_{-4}(\mathbf l_{-2})^2 + 3(c-36) (\mathbf l_{-4})^2 - 2 (5c-12) \mathbf l_{-6}\mathbf l_{-2} + \nonumber
	\\	\nonumber
	\\
	&\phantom{\mathbf w^{(8)}_4} + (4128 -325 c + 5 c^2) \mathbf l_{-8} \; ,
	\\	\nonumber
	\\
	&\mathbf w^{(8)}_{11} = 3(\mathbf l_{-4})^2 + 4 \mathbf l_{-6}\mathbf l_{-2} + (5c-89)	\mathbf l_{-8} \; .	\nonumber
\end{align}

\section{Solving the reflection relations with the fermionic basis}
\label{eq:solvrefferm}
\setcounter{equation}{0}

As we have seen in Sec.\ref{sec:fermbasis}, the CFT fermions $\bbeta^{\textrm{CFT}\ast}_{2m-1},\bgamma^{\textrm{CFT}\ast}_{2m-1}$ transform in a simple way under the reflections $\sigma_1,\sigma_2$; moreover, since by definition these constitute a fermionic basis for the Liouville Verma modules, it is possible to express them in terms of the Virasoro generators in the following way:

\eq
  \bbeta^{\textrm{CFT}\ast}_{I^+}\bgamma^{\textrm{CFT}\ast}_{I^-}\Phi_a = C_{I^+,I^-}\;\bigg[P^{\textrm{even}}_{I^+,I^-}\Big(\{\mathbf l_{-2k}\};c,\Delta\Big) + d_a \; P^{\textrm{odd}}_{I^+,I^-}\Big(\{\mathbf l_{-2k}\};c,\Delta\Big)\bigg] \Phi_a \; ,
\label{eq:fermionicinvirasoro}
\en
where $P^{\textrm{even}}_{I^+,I^-}\Big(\{\mathbf l_{-2k}\};c,\Delta\Big)$ and $P^{\textrm{odd}}_{I^+,I^-}\Big(\{\mathbf l_{-2k}\};c,\Delta\Big)$ are polynomials in Virasoro generators, with coefficients depending rationally on $c$ and $\Delta$; let us recall the definition of $d_a$:

\eq
  d_a = \frac{1}{6}\sqrt{(c-25)(24\Delta +1 -c)} = (b-b^{-1})(Q-2a) \; .
\en
The superscript ``even'' and ``odd'' of the polynomials refer to the behaviour under exchange of the multi-indices $I^+$ and $I^-$, namely

\eq
  P^{\textrm{even}}_{I^+,I^-} = P^{\textrm{even}}_{I^-,I^+}\; ,\qquad P^{\textrm{odd}}_{I^+,I^-} = - P^{\textrm{odd}}_{I^-,I^+}\; ,
\en
in particular

\eq
  P^{\textrm{odd}}_{I^+,I^-} = 0 \; , \ \ \textrm{if} \  I^+=I^- \; .
\en

Now, if the CFT fermions have to satisfy both reflections $\sigma_1$ and $\sigma_2$, we have to assume that they can be expressed also in terms of the Heisenberg basis, through the following formula

\begin{align}
  \bbeta^{\textrm{CFT}\ast}_{I^+}&\bgamma^{\textrm{CFT}\ast}_{I^-}\Phi_a = C_{I^+,I^-} \mathcal U_{I^+,I^-}(a,b) \times	\nonumber
  \\	\label{eq:fermionicinheisenberg}
  \\
  &\times \bigg[Q^{\textrm{even}}_{I^+,I^-}\Big(\{a_{-k}\};a^2,Q^2\Big) + g \; Q^{\textrm{odd}}_{I^+,I^-}\Big(\{a_{-k}\};a^2,Q^2\Big)\bigg] \Phi_a \; ,	\nonumber
\end{align}
with $Q^{\textrm{even}}_{I^+,I^-}\Big(\{a_{-k}\};a^2,Q^2\Big)$ and $Q^{\textrm{odd}}_{I^+,I^-}\Big(\{a_{-k}\};a^2,Q^2\Big)$ being polynomials in Heisenberg generators, with coefficients depending rationally on $a^2$ and $Q^2$. The factor $\mathcal U_{I^+,I^-}(a,b)$ takes into account the nontrivial transformation of the CFT fermions under the symmetry $\sigma_1$ and it is quite straightforward to show that

\eq
  \mathcal U_{I^+,I^-}(a,b) = \prod_{2i^+-1\in I^+}\left(2a + (2 i^+-1)b^{-1}\right)\prod_{2i^--1\in I^-}\left(2a+(2i^--1)b\right) \; .
\en
We also introduced the parameter $g$:

\eq
  g= a(b-b^{-1}) \; .
\en
Let us notice that, since the the duality $b\rightarrow b^{-1}$ exchanges the multi-indices $I^+$ and $I^-$, the coefficients $d_a$ and $g$ grant the two formulae invariance under this transformation.

The assertion that we can express the fermionic basis by means of the formula (\ref{eq:fermionicinheisenberg}) is very serious and critical for our construction. Even though we don't possess a general theorem confirming this statement in general, we checked it up to level $8$.

\subsection{Level 2}

We have

\eq
  P^{\textrm{even}}_{\{1\},\{1\}} = \mathbf l_{-2} \; , \qquad P^{\textrm{odd}}_{\{1\},\{1\}} = 0 \; ,
\en
thus, by operating with $U^{(2)}$ we immediately obtain

\eq
  Q^{\textrm{even}}_{\{1\},\{1\}} = \frac{1}{4} \left(a_{-1}\right)^2 \; , \qquad Q^{\textrm{odd}}_{\{1\},\{1\}} = 0 \; .
\en
From now on we will not write anymore the polynomials which vanish by definition.

\subsection{Level 4}

Using the formulae from \cite{Boos_Jimb_Miwa_Smir_10}, we have

\eq
  P^{\textrm{even}}_{\{1\},\{3\}} = \left(\mathbf l_{-2}\right)^2 + \frac{2c-32}{9}\mathbf l_{-4} \; , \qquad P^{\textrm{odd}}_{\{1\},\{3\}} = \frac{2}{3}\mathbf l_{-4} \; .
\en
Using the reflection matrix $U^{(4)}$ to rewrite the Virasoro basis in term of the Heisenberg one and performing a bit of algebra on the resulting expression for $P^{\textrm{even}}_{\{1\},\{3\}}+d\,P^{\textrm{odd}}_{\{1\},\{3\}}$, one finds that the multiplier $(2a+b^{-1})(2a+3b)$ indeed factorises, leaving us with the following polynomials

\begin{align}
  Q^{\textrm{even}}_{\{1\},\{3\}} &= -\frac{1}{144}\bigg\{\Big[4a^2\big(Q^2-2\big)-3\Big]\big(a_{-1}\big)^4 + 12 \big(Q^2+1\big)\big(a_{-2}\big)^2\bigg\} \; ,	\nonumber
  \\
  \\
  Q^{\textrm{odd}}_{\{1\},\{3\}} &= \frac{1}{216}\bigg\{\big(4a^2-3\big)\big(a_{-1}\big)^4 + 12 \big(a_{-2}\big)^2\bigg\}\; .	\nonumber
\end{align}

\subsection{Level 6}

The polynomials $P^{\textrm{even}}$ and $P^{\textrm{odd}}$ for this level were also computed in \cite{Boos_Jimb_Miwa_Smir_10}, where the appearance of the denominator $\Delta+2$ was left unexplained; now we know this is a consequence of the reflection matrix structure. It is clear that the residues of all the polynomials at the point $\Delta = -2$ have to be proportional to the vector $\mathbf w^{(6)}$ introduced in (\ref{eq:wvector}), which allow us to simplify the formulae to

\begin{align}
  P^{\textrm{even}}_{\{3\},\{3\}} &= \left(\mathbf l_{-2}\right)^3 + \frac{2}{3}(c-19)\mathbf l_{-4}\mathbf l_{-2} + \frac{8(c-28)\Delta + 5 c^2 - 173 c + 1524}{30}\mathbf l_{-6} + \nonumber
  \\	\nonumber
  \\ &- \frac{5c - 158}{6(\Delta +2)}\mathbf w^{(6)} \; ,	\nonumber
  \\	\nonumber
  \\
  P^{\textrm{even}}_{\{1\},\{5\}} &= \left(\mathbf l_{-2}\right)^3 + \frac{2}{3}(c-10)\mathbf l_{-4}\mathbf l_{-2} +
  \\	\nonumber
  \\ & + \frac{8(c-28)\Delta + 3 c^2 - 59 c + 140}{15}\mathbf l_{-6}- \frac{c - 14}{\Delta +2}\mathbf w^{(6)} \; ,	\nonumber
  \\	\nonumber
  \\
  P^{\textrm{odd}}_{\{1\},\{5\}} &= 2 \mathbf l_{-4}\mathbf l_{-2} + \frac{4}{5}(c-13)\mathbf l_{-6} -\frac{4}{\Delta +2}\mathbf w^{(6)} \; .	\nonumber
\end{align}

Now, using the reflection matrix, we see that again the factorisation takes place leaving us with formulae for the polynomials $Q$; these are rather nasty and we will not show them here, but will collect them in the Appendix to this chapter.

\subsection{Level 8}

This case is particularly interesting, since at this level a state containing $4$ fermions appear for the first time: we will display this state only, although we have checked the correctness of our formulae for all the other vectors in this level. The computation of $P^{\textrm{even}}_{\{1,3\},\{1,3\}}$ was carried out by H. Boos in \cite{Boos_11}, following the procedure explained in \cite{Boos_Jimb_Miwa_Smir_10}. This was indeed hard work, since he had to include the descendants in the Matsubara direction in the discussion, but in the end he managed to find the polynomials $P$ for all the level $8$ states; he observed the appearance of two denominators, $\Delta +4$ and $\Delta + 11$, which does not surprise us. Here too, the residues of corresponding to $\Delta =-4$ and $\Delta = -11$ are proportional to the vectors $\mathbf w^{(8)}_4$ and $\mathbf w^{(8)}_{11}$, respectively. In particular, we have

\begin{align}
  &P^{\textrm{even}}_{\{1,3\},\{1,3\}} = \left(\mathbf l_{-2}\right)^4 + 4\;\frac{c-22}{3}\mathbf l_{-4}\left(\mathbf l_{-2}\right)^2 - \frac{c^2 - 34 c - 333 + 8(c-25)\Delta}{9}\left(\mathbf l_{-4}\right)^2 +	\nonumber
  \\	\nonumber
  \\
  & + 2\;\frac{5 c^2 - 193 c + 1544 + 8 (c-28)\Delta}{15}\mathbf l_{-6}\mathbf l_{-2} - 4\;\frac{11c - 71 + 24 \Delta}{3}\mathbf l_{-8} +
  \\	\nonumber
  \\
  & + \frac{5 c-122}{42(\Delta+4)}\mathbf w^{(8)}_4 - \frac{5 c^2-526 c+8648}{42(\Delta + 11)}\mathbf w^{(8)}_{11} \; .	\nonumber
\end{align}

Again, when applying the reflection matrix, the magic happens and the term $(2a+b)(2a+b^{-1})(2a+3b)(2a+3b^{-1})$ is factorised, leaving the function $Q^{\textrm{even}}_{\{1,3\},\{1,3\}}$, even in $a$, which we present in the Appendix.\\

Let us briefly recollect the results of this section. Suppose there is a linear functional $f$ on $\mathcal V^{\textrm{quo}}_a$, such that the vectors

\eq
  V_i(a) = f\left(\mathbf v_i \Phi_a\right) \; , \qquad H_i (a) = f\left(\mathbf h_i\Phi_a\right)
\en
satisfy the following reflections

\eq
  V(Q-a) = V(a) \; , \qquad H(-a) = H(a) \; ,
\label{eq:reflrelformula}
\en
together with the basis transformation

\eq
  V(a) = U(a)H(a) \; .
\en
Then it is straightforward to see that the vector $V(a)$ has to satisfy a nontrivial Riemann-Hilbert problem:

\eq
  V(Q+a) = S(a)V(a) \; , \qquad S(a) = U(-a)U^{-1}(a) \; .
\label{eq:riemannhilbert}
\en

Now, let us introduce the following vector

\eq
  W_{I^+,I^-}(a) = f\left(\bbeta^{\textrm{CFT}\ast}_{I^+}\bgamma^{\textrm{CFT}\ast}_{I^-}\Phi_a\right) \; ;
\en
what we have shown in this section is that $W$ trivially satisfies the reflection relations

\eq
  W(Q-a) = J W(a) \; , \qquad W(-a) = J W(a) \; ,
\label{eq:fermriemannhilbert}
\en
where the matrix $J$ simply exchanges the multi-indices $J: \ (I^+,I^-)\rightarrow (I^-,I^+)$. In short, using the fermionic basis reduces the original nontrivial Riemann-Hilbert problem to a trivial one. Similarly, any solution to (\ref{eq:fermriemannhilbert}) provides a solution to (\ref{eq:riemannhilbert}, \ref{eq:reflrelformula}) up to a quasi-constant coefficient (that is a scalar function $g$ satisfying both reflections $g(Q-a)=g(a)$ and $g(-a)=g(a)$).
The reflection relations do not tell us how the two chiralities should be glued together for the one-point functions in infinite volume, the correct way can be found by looking at the formulae in \cite{Jimb_Miwa_Smir_10}. Going to the second chirality one simply exchange $a\rightarrow Q-a$, then define $\overline{W}(a)$ in the same way as we did for $W(a)$; then the one point functions correspond to the choice $W(a) \times \overline{W}(a)$, that is, to the limit $R\rightarrow \infty$ of the one-point function (\ref{eq:onepointfunction})\footnote{For $R\rightarrow\infty$, the function $\Theta(l,m\vert\alpha)$ vanishes identically.}:

\eq
  \frac{\left\langle\bbeta^\ast_{I^+}\overline\bbeta^\ast_{\overline I^+}\overline\bgamma^\ast_{\overline I^-}\bgamma^\ast_{I^-}\Phi_a(0)\right\rangle_\infty}{\left\langle\Phi_a(0)\right\rangle_\infty} = \delta_{I^+,\overline I^-}\delta_{I^-,\overline I^+} \prod_{2i^+-1\in I^+} t_{2i^+-1}(a) \prod_{2i^--1\in I^-}\left(-t_{2i^--1}(Q-a)\right)\; .	\nonumber
\en

\section{Fermionic basis from reflection relations}
\label{sec:fermfromrefl}
\setcounter{equation}{0}

The procedure described in \cite{Boos_Jimb_Miwa_Smir_10}, although allowing in theory to extract the polynomials $P$ for whatever level one desires, becomes extremely complicated already after level $8$. The reflection relations and the privileged r\^ole played by the fermionic basis provide a much simpler way to obtain the change of basis (\ref{eq:fermionicinvirasoro}): in this section we present this method.

First, one chooses a level $2k$, builds the reflection matrix $U^{(2k)}(a)$ and compute its determinant which will have the following structure

\eq
  \det \left(U^{(2k)} (a)\right) = C^{(2k)} \cdot N^{(2k)}(a,b) \cdot \frac{D_V^{(2k)}(\Delta,c)}{D_H^{(2k)}(a^2,Q^2)} \; ,
\en
where, as usual, $C^{(2k)}$ is a numeric factor, $N^{(2k)}(a,b)$ contains the null-vector contributions and the functions $D_V^{(2k)}(\Delta,c)$, $D_H^{(2k)}(a^2,Q^2)$, which we call respectively \emph{Virasoro denominator} and \emph{Heisenberg denominator}, are the ``curious'' multipliers which start to appear from $2k=6$; e.g.

\eq
  D_V^{(6)}(\Delta,c) = \Delta + 2 \; , \qquad D_H^{(6)}(a^2,Q^2) = 3a^2 - 10 Q^2 -5 \; .	\nonumber
\en

Now, we look for polynomials in the Virasoro basis $\{\mathbf v_i\}$ (we follow the lexicographical order introduced above) of the following form

\begin{align}
  P^{\textrm{even}}_{I^+,I^-} &= \mathbf v_1 + \frac{1}{D^{(2k)}_V(\Delta,c)} \sum_{i=2}^{p(k)} X_{I^+,I^-}^i (\Delta,c) \mathbf v_i \; ,	\nonumber
  \\
  \\
  P^{\textrm{even}}_{I^+,I^-} &= \frac{1}{D^{(2k)}_V(\Delta,c)} \sum_{i=2}^{p(k)} Y_{I^+,I^-}^i (\Delta,c) \mathbf v_i \; ,	\nonumber
\end{align}
where $X_{I^+,I^-}^i (\Delta,c)$ and $Y_{I^+,I^-}^i (\Delta,c)$ are polynomials in $\Delta$ of an unspecified degree $D$. The coefficient of this polynomials are to be considered as unknowns, thus there are $2\big(p(k)-1\big)\big(D+1\big)$ of them.

If we introduce the two following polynomials

\begin{align}
  T^+_{I^+,I^-} (a) = \frac{\mathcal U_{I^+,I^-}(a,b) + \mathcal U_{I^+,I^-}(a,b^{-1})}{2} \; ,	\nonumber
  \\
  \\
  T^-_{I^+,I^-} (a) = \frac{\mathcal U_{I^+,I^-}(a,b) - \mathcal U_{I^+,I^-}(a,b^{-1})}{2a(b-b^{-1})} \; ,	\nonumber
\end{align}
which are invariant under the duality $b\rightarrow b^{-1}$ and thus depend on $b$ through $Q$, straightforward algebra shows that the equation (\ref{eq:fermionicinheisenberg}) is equivalent to the following two requirements:

\begin{itemize}
 \item The following polynomial is even in $a$, for any $1\leq j\leq p(k)$
	\begin{align}
	  &D_V^{(2k)}(\Delta(-a),c)D_H^{(2k)}(a^2,Q^2) \times	\nonumber
	  \\	\nonumber
	  \\
	  &\times\Bigg\{ T^+_{I^+,I^-}(-a) \Big[D_V^{(2k)}(\Delta,c)U_{1,j}^{(2k)}(a) + \sum_{i=2}^{p(k)}X^i_{I^+,I^-}(\Delta,c)U_{i,j}^{(k)}(a)\Big] +
	  \\	\nonumber
	  &- (Q^2-4)(Q-2a) T^-_{I^+,I^-}(-a) \sum_{i=2}^{p(k)} Y^i_{I^+,I^-}(\Delta,c)U_{i,j}^{(k)}(a) \Bigg\} \; ;	\nonumber
	\end{align}
  \item The following polynomial is odd in $a$, for any $1\leq j\leq p(k)$
	\begin{align}
	  &D_V^{(2k)}(\Delta(-a),c)D_H^{(2k)}(a^2,Q^2) \times	\nonumber
	  \\	\nonumber
	  \\
	  &\times\Bigg\{ T^-_{I^+,I^-}(-a) \Big[D_V^{(2k)}(\Delta,c)U_{1,j}^{(2k)}(a) + \sum_{i=2}^{p(k)}X^i_{I^+,I^-}(\Delta,c)U_{i,j}^{(k)}(a)\Big] +
	  \\	\nonumber
	  &- (Q-2a) T^+_{I^+,I^-}(-a) \sum_{i=2}^{p(k)} Y^i_{I^+,I^-}(\Delta,c)U_{i,j}^{(k)}(a) \Bigg\} \; .	\nonumber
	\end{align}
\end{itemize}
If we call $d_V$ the degree of $D_V^{(2k)}(\Delta,c)$ in $\Delta$ and $d_H$ the degree of $D_H^{(2k)}(a^2,Q^2))U^{(k)}(a)$ in $a$, we immediately see that the number of equations corresponding to the above requirements is (obviously $T^{\pm}$ has degree $2\mathfrak C(I^+)\geq 2$ in $a$)

\eq
  p(k)\big(2 d_V + d_H + 2\mathfrak C(I^+) + 2D+1 \big) > 2\big(p(k)-1\big)\big(D+1\big) \; , \quad \forall D,k>0 \; .
\en
Thus the system is overdetermined, nonetheless a nontrivial solution exists. Let us consider the case $k=10$ as an example.

\subsection{Level 10}

We choose to take the following Heisenberg basis

\begin{align}
  \mathbf h^{(10)}_1 = \big(a_{-1}&\big)^{10} \; , \quad \mathbf h^{(10)}_2 = \big(a_{-1}\big)^2\big(a_{-2}\big)^4 \; , \quad \mathbf h^{(10)}_3 = \big(a_{-2}\big)^2\big(a_{-3}\big)^2 \; , \quad \mathbf h^{(10)}_4 = \big(a_{-1}\big)^5 a_{-5} \; ,	\nonumber
  \\	\nonumber
  \\
  &\mathbf h^{(10)}_5 = \big(a_{-5}\big)^2 \; , \quad \mathbf h^{(10)}_6 = \big(a_{-1}\big)^3 a_{-7} \; , \quad \mathbf h^{(10)}_7 = a_{-1}a_{-9}\; ,
\end{align}
and compute the reflection matrix $U^{(10)}$, finding, in particular

\begin{align}
  D_V^{(10)}(\Delta,c) &= \Big(\Delta + 6\Big) \times 	\nonumber
  \\
  \\
  &\times \Big(3\Delta^4 + (c+149)\Delta^3 + (71c+1447)\Delta^2+ (983 c-2285)\Delta + 2905 c-23794\Big) \nonumber
\end{align}
and

\begin{align}
  D_H^{(10)}(a^2,Q^2) &= a^2 \Bigg\{ 1134 a^{10} - (9810 Q^2 - 11097) a^8 + (27282 Q^4 + 21920 Q^2 + 10657) a^6 +	\nonumber
  \\	\nonumber
  \\
  &- (28326 Q^6 + 67739 Q^4 + 72222 Q^2 + 53317) a^4 +
  \\	\nonumber
  \\
  & + 5(1944 Q^8 + 5562 Q^6 + 2793 Q^4 - 5153 Q^2 - 5701)a^2 + 2025(4Q^6 + 16 Q^4 + 19 Q^2 + 6)\Bigg\} \;.	\nonumber
\end{align}
The contributions to $N^{(10)}(a,b)$ coming from the odd level null vectors are the same as in $N^{(8)}$, while the contributions coming from the singular vectors of even level follow the usual routine.

Now we apply the method described above and find out that there exists indeed nontrivial solutions for all possible cases, with actual degrees $D$ as follows

\begin{center}
\begin{tabular}{c | c | c | c | c |}
  & $\{1\},\{9\}$ & $\{3\},\{7\}$ & $\{5\},\{5\}$ & $\{1,3\},\{1,5\}$ \\ \hline & & & & \\
  even & $7$ & $7$ & $7$ & $6$ \\ & & & & \\
  odd & $6$ & $6$ & $\setminus$ & $6$
\end{tabular}
\end{center}

The general structure is similar to that of $k=6$ and $k=8$: we have a vector $\mathbf w^{(10)}_6$, corresponding to the residue at $\Delta = -6$ and a set of $4$ vectors $\{\mathbf w^{(10)}_{\textrm{deg}_4,\, i}\}_{i=1}^4$ which are associated to the residues at the zeroes of the degree $4$ polynomial in $D_V^{(10)}(\Delta,c)$. The explicit formulae are quite nasty and we will not display them.

\section*{Appendix: Polynomials for $k=6$ and $k=8$}\addcontentsline{toc}{subsection}{Polynomials for $k=6$ and $k=8$}
\label{sec:polynomapp}
This Appendix contains formulae for $Q^{\mathrm{even}}$, $Q^{\mathrm{odd}}$ on levels 6 and 8.
\begin{align}
Q_{\{3\},\{3\}}^{\mathrm{even}}(\{a_{-k}\})&=\frac1{129600}\Bigl\{-\bigl[720 a^4 (3 + 2 Q^2) + 12 a^2 (18 + 341 Q^2 + 70 Q^4) \nn\\&+ 
      5 (771 + 2876 Q^2 + 2768 Q^4 + 560 Q^6)\bigr] a_{-1}^6\nn\\& - 
  1800 \bigl[18 + 32 Q^2 + 7 Q^4 + 12 a^2 (3 + Q^2)\bigr] a_{-1}^2 a_{-2}^2 \nn\\&+ 
  240 \bigl[138 + 293 Q^2 + 94 Q^4 - 12 a^2 (-9 + 2 Q^2)\bigr] a_{-3}^2) \nn\\&
  +50  \frac{(2 Q^2 + 1)^2 (14 Q^2 + 51)}{5 - 
      3 a^2 + 10 Q^2}\mathbf{g}\Bigr\}\,,\nn
\end{align}
\begin{align}
Q_{\{1\},\{5\}}^{\mathrm{even}}(\{a_{-k}\})&=\frac1{129600}\Bigl\{\bigl[-1440 a^4 (-1 + 2 Q^2) + a^2 (5412 + 1944 Q^2 - 4080 Q^4)\nn\\& + 
     5 \bigl[1009 + 3080 Q^2 + 224 Q^4 - 2720 Q^6)\bigr] a_{-1}^6\nn\\& - 
  900 \bigl[-11 - 51 Q^2 + 68 Q^4 + 12 a^2 (1 + 4 Q^2)\bigr] a_{-1}^2 a_{-2}^2\nn\\& + 
  480 \bigl[-8 - 33 Q^2 + 146 Q^4 + 12 a^2 (-9 + 2 Q^2)\bigr] a_{-3}^2\nn\\&+100\frac{(1 + 2 Q^2) (-29 - 60 Q^2 + 68 Q^4)}{5 - 
      3 a^2 + 10 Q^2} \mathbf{g}\Bigr\}\,,\nn
\end{align}
\begin{align}
Q_{\{1\},\{5\}}^{\mathrm{odd}}(\{a_{-k}\})&=\frac 1 {32400} \Bigl\{\bigl[2063 + 360 a^4 + 4216 Q^2 + 1920 Q^4 + 
      30 a^2 (34 + 21 Q^2)\bigr] a_{-1}^6 \nn\\&+ 
   450 \bigl[34 + 12 a^2 + 21 Q^2\bigr] a_{-1}^2 a_{-2}^2 - 
   3840 \bigl[4 + 3 Q^2\bigr] a_{-3}^2
\nn\\&+10\frac{ (1 + 2 Q^2) (67 + 48 Q^2)}{5 - 
      3 a^2 + 10 Q^2}\mathbf{g}\Bigr\}\,,\nn
\end{align}
where
$$\mathbf{g}=2 (5 Q^2 + 4) a_{-1}^6 + 45  a_{-1}^2 a_{-2}^2 - 42  a_{-3}^2$$

\begin{align}
&Q_{\{1,3\},\{1,3\}}^{\mathrm{even}}(\{a_{-k}\})= \frac1{1209600a^2(-21 (76 - 19 Q^2 - 30 Q^4) - (991 + 1076 Q^2) a^2 + 206 a^4)}
\nn\\&\times\Bigl\{
-a^2 \bigl[640 a^{10} (1 + 2 Q^2) - 
   16 a^8 (-27011 + 14098 Q^2 + 160 Q^4)  
   \nn\\&  +315 (1748 - 2969 Q^2 + 1830 Q^4) + 
   a^4 (6252242 - 9978784 Q^2 + 4263704 Q^4 - 2042880 Q^6)  \nn\\& + 
   4 a^6 (533225 - 1096594 Q^2 + 465312 Q^4 + 320 Q^6)   \nn\\&- 
   7 a^2 (-941629 + 942172 Q^2 - 466620 Q^4 + 102600 Q^6)\bigr]a_{-1}^8
  \nn\\& 
   +
   280 a^2 \bigl[-96 a^8 (1 + 2 Q^2) + 4 a^6 (-19249 + 9790 Q^2 + 96 Q^4)   \nn\\&-
    a^4 (279425 - 683886 Q^2 + 306784 Q^4 + 192 Q^6) + 
   315 (-380 - 82 Q^2 + 33 Q^4 + 450 Q^6)   \nn\\&+ 
   10 a^2 (-22552 + 129028 Q^2 - 60003 Q^4 + 32256 Q^6)\bigr]a_{-1}^4a_{-2}^2
   \nn\\&
   -420 a^2 \bigl[32 a^6 (1 + 2 Q^2) - 4 a^4 (-2217 + 5482 Q^2 + 32 Q^4)   \nn\\&+ 
   a^2 (49761 - 42996 Q^2 + 148256 Q^4 + 64 Q^6)   \nn\\&- 
   7 (-23408 - 36599 Q^2 + 22474 Q^4 + 19200 Q^6)\bigr]a_{-2}^4
   \nn\\&
   +
   3360 \bigl[16 a^6 (-8399 + 1742 Q^2) + 
   a^4 (473336 + 1270956 Q^2 - 338552 Q^4)   \nn\\&+ 
   a^2 (625801 - 664342 Q^2 - 3416448 Q^4 + 1059120 Q^6)  \nn\\& - 
   1575 (-76 - 401 Q^2 - 402 Q^4 + 372 Q^6 + 360 Q^8)\bigr]a_{-2}a_{-6}
   \nn\\&
   -5040 \bigl[32 a^6 (-2051 + 533 Q^2) + 
   a^4 (72508 + 457080 Q^2 - 124312 Q^4)  \nn\\& + 
   a^2 (85625 - 171248 Q^2 - 651588 Q^4 + 244656 Q^6)   \nn\\&- 
   315 (-76 - 401 Q^2 - 402 Q^4 + 372 Q^6 + 360 Q^8)\bigr]a_{-4}^2
\Bigr\}\,.\nn
\end{align}

\chapter{Comparison Against Known Results}
\label{chap:knownresults}
\markboth{Chapter 3 - Comparison Against Known Results}{}

In this chapter we will use the fermionic basis formalism to extract one-point function ratios for particular values or limits of the parameters $\alpha$ and $R$; this will allow us to make a comparison between our results and those already known in the literature. While this is obviously a relevant check to be made for the sine-Gordon model, it becomes of the utmost importance when considering the sinh-Gordon model where, as we have said many times now, the fermionic basis does not have mathematically rigorous grounds.

\section{Expectation values of the energy-momentum tensor}
\label{sec:expenmomtens}

From \cite{Luky_AZam_10}, we know that the eigenvalues of the local integrals of motion in the Matsubara direction, for both sine-Gordon and Sinh-Gordon models, can be extracted from the asymptotics of $\log T(\zeta)$, as $\zeta\rightarrow 0$ and $\zeta\rightarrow\infty$. In our notations, these read

\eq
  I_{2j-1} = \mathscr C_{2j-1}^{-1} J_{2j-1} \; , \qquad \overline I_{2j-1} = \mathscr C_{2j-1}^{-1} J_{-(2j-1)} \; ,
\en
where $J_{2j-1}$ are the coefficients in the asymptotic expansions of $\log T(\zeta)$:

\begin{align}
  J^{\textrm{sG}}_{2j-1} & = \frac{\pi m R}{2\sin{\frac{\pi}{\nu}}}\delta_{2j-1,\pm1} + (-1)^{j+1}\frac{2}{\pi} \int\limits_{-\infty}^\infty \Im\left\{e^{(2j-1)(\theta- \mathbbm i0)} \log\Big(1+\mathfrak a(\theta-\mathbbm i0)\Big)\right\}d\theta \; ,	\nonumber
  \\
  \\
  J^{\textrm{shG}}_{2j-1} & = \frac{\pi m R}{2\sin{\frac{\pi}{\nu}}}\delta_{2j-1,\pm1} + \frac{1}{\pi} \int\limits_{-\infty}^\infty e^{(2j-1)\theta} \log\Big(1+e^{-\epsilon(\theta)}\Big)d\theta \; .	\nonumber
\end{align}
The normalisation coefficients $\mathscr C_{2j-1}$ read

\eq
  \mathscr C_{2j-1} = \frac{\sqrt{\nu -1}}{\nu} \  \frac{\Gamma\left(\frac{\nu-1}{2\nu}(2j-1)\right)\Gamma\left(\frac{2j-1}{2\nu}\right)}{2\sqrt{\pi}j!}\ \Big(\frac{\bmu \Gamma(\nu)}{(\nu-1)^{\frac{\nu}{2}}}\Big)^{-\frac{2j-1}{\nu}} \; ,
\en
and, since $\sqrt{\nu -1}/\nu = Q^{-1}$, are invariant under the duality $b\rightarrow b^{-1}$.

The expectation values of the energy-momentum tensor components $T$, $\overline T$ and $\Theta$ can be expressed in terms of the ground state energy $E(R)$ in the Matsubara direction (with $P(R)=0$):

\eq
  \langle T\rangle_R = \langle\overline T\rangle_R = \frac{1}{4}\left(\frac{1}{R} - \frac{d}{dR}\right)E(R) \; ,\quad \langle\Theta\rangle_R = - \frac{1}{4}\left(\frac{1}{R} + \frac{d}{dR}\right)E(R) \; .
\en
The ground state energy $E(R)$ is simply the sum of $I_1$ and $\overline I_1$, we have then, considering that $\mathscr C_1 = 4/m$

\begin{align}
  4 \pi E^{\textrm{sG}}(R) &= \frac{\pi^2 m^2 R}{\sin\frac{\pi}{\nu}} + 2m \int\limits_{-\infty}^\infty \Im \left\{\sinh(\theta-\mathbbm i0)\log\Big(1+\mathfrak a(\theta-\mathbbm i0)\Big)\right\}d\theta \; ,	\nonumber
  \\
  \\
  4 \pi E^{\textrm{shG}}(R) &= \frac{\pi^2 m^2 R}{\sin\frac{\pi}{\nu}} + m \int\limits_{-\infty}^\infty \cosh\theta\log\Big(1+e^{- \epsilon(\theta)} \Big)d\theta \; .	\nonumber
\end{align}

Now, with some straightforward calculation we find the following equalities

\begin{align}
  &\frac{1}{\mathbbm i}\frac{d}{dR}\log\mathfrak a(\theta) = \pi m \left((e_{+1} - e_{-1}) - R_{\textrm{dress}}^{\textrm{sG},(\alpha =0)} \ast \, (e_{+1} - e_{-1}) \right)(\theta) \; ,	\nonumber
  \\
  \\
  &\frac{1}{\mathbbm i R}\frac{d}{d\theta}\log\mathfrak a(\theta) = \pi m \left((e_{+1} + e_{-1}) - R_{\textrm{dress}}^{\textrm{sG},(\alpha =0)} \ast \, (e_{+1} + e_{-1}) \right)(\theta) \; ,	\nonumber
  \\	\nonumber
  \\
  &\frac{d}{dR}\epsilon(\theta) = \pi m \left((e_{+1} + e_{-1}) + R_{\textrm{dress}}^{\textrm{shG},(\alpha =0)} \ast \, (e_{+1} + e_{-1}) \right)(\theta) \; ,	\nonumber
  \\
  \\
  &\frac{1}{R}\frac{d}{d\theta}\epsilon(\theta) = \pi m \left((e_{+1} - e_{-1}) + R_{\textrm{dress}}^{\textrm{shG},(\alpha =0)} \ast \, (e_{+1} - e_{-1}) \right)(\theta) \; ,	\nonumber
\end{align}
which immediately lead us to

\begin{align}
  &\left(\frac{1}{R}-\frac{d}{dR}\right) E^{\textrm{sG}}(R) = m^2 (e_1\ast \, R^{\textrm{sG},(\alpha =0)}_{\textrm{dress}} \ast \, e_1 - e_1\ast \, e_1) \; ,	\nonumber
  \\
  \label{eq:lukcomparison1}
  \\
  &\left(\frac{1}{R}+\frac{d}{dR}\right) E^{\textrm{sG}}(R) = m^2 (e_1\ast \, R^{\textrm{sG},(\alpha =0)}_{\textrm{dress}} \ast \, e_{-1} - e_1\ast \, e_{-1}) + \frac{\pi m^2}{2 \sin\frac{\pi}{\nu}} \; ,	\nonumber
  \\	\nonumber
  \\
  &\left(\frac{1}{R}-\frac{d}{dR}\right) E^{\textrm{shG}}(R) = m^2 (e_1\ast \, R^{\textrm{shG},(\alpha =0)}_{\textrm{dress}} \ast \, e_1 + e_1\ast \, e_1) \; ,	\nonumber
  \\
  \label{eq:lukcomparison2}
  \\
  &\left(\frac{1}{R}+\frac{d}{dR}\right) E^{\textrm{shG}}(R) = m^2 (e_1\ast \, R^{\textrm{shG},(\alpha =0)}_{\textrm{dress}} \ast \, e_{-1} + e_1\ast \, e_{-1}) + \frac{\pi m^2}{2 \sin\frac{\pi}{\nu}} \; .	\nonumber
\end{align}
Thus, recalling the representations (\ref{eq:sinthetaerepresentation}) and (\ref{eq:sinhthetaerepresentation}), we can write

\eq
  \langle T\rangle_R = \langle \overline T\rangle_R = \frac{m^2}{4}\Theta(\mathbbm i,\mathbbm i\vert0) \; , \qquad \langle \Theta\rangle_R = \frac{m^2}{4}\Big[\Theta(\mathbbm i,-\mathbbm i\vert0)-\pi t_1(0)\Big] \; ,
\label{eq:lukcomparisonfin}
\en
for both sine-Gordon and sinh-Gordon models.

Concerning the fermionic basis, the energy-momentum tensor is the only non-chiral level-$2$ Virasoro descendent of the identity operator: $\mathbbm I\equiv \Phi_0(0)$; since we know that $\mathbf l_{-2} = \bbeta^{\textrm{CFT}\ast}_1\bgamma^{\textrm{CFT}\ast}_1$ and $\overline{\mathbf l}_{-2} = \overline\bbeta^{\textrm{CFT}\ast}_1\overline\bgamma^{\textrm{CFT}\ast}_1$, we can immediately calculate its expectation value using the fermionic formula (here we agree that the argument of the function $D_1$ is $\alpha$):

\begin{align}
\langle T\overline T\rangle_R &\equiv \langle \mathbf l_{-2}\overline{\mathbf l}_{-2} \mathbbm I\rangle_R \equiv \frac{\langle \bbeta^\ast_1\overline\bbeta^\ast_1\overline\bgamma^\ast_1\bgamma^\ast_1 \Phi_0(0)\rangle_R}{[D_1(0) D_1(2)]^2} = 	\nonumber
\\	\nonumber
\\
&= \frac{m^4}{16} \left\vert\begin{array}{c c}
				\Theta(\mathbbm i,\mathbbm i\vert0) & \Theta(\mathbbm i,-\mathbbm i\vert0) - \pi t_1(0) \\ & \\
				\Theta(-\mathbbm i,\mathbbm i\vert0)+\pi t_{-1}(0) & \Theta(-\mathbbm i,-\mathbbm i\vert0)
                           \end{array}\right\vert \; =
\\	\nonumber
\\
&= \left\vert\begin{array}{c c}
			\langle T\rangle_R & \langle \Theta \rangle_R \\ & \\
			\langle\Theta\rangle_R & \langle\overline T\rangle_R
             \end{array}\right\vert \; ,	\nonumber
\end{align}
where we used, in the last passage, the identity $\Theta(l,m\vert 0) = \Theta(-l,-m\vert0)$. This last equality corresponds to the identity stated in \cite{AZam_04} by A. Zamolodchikov, the only \emph{a priori} known determinant formula.

Finally, in order to close the circle, let us calculate $\langle T\rangle$, $\langle \overline T\rangle$ and $\langle \Theta\rangle$ directly with the fermionic formula. We know that the chiral components $T$ and $\overline T$ are simply the chiral descendent of $\mathbbm I$, while the trace component $\Theta$ is proportional to the field $e^{-b\eta(0)} = \Phi_{2\frac{1-\nu}{\nu}}(0)$; using again the formula for the one-point functions and the shift formula (\ref{eq:shiftformula}) we obtain

\begin{align}
  &\langle T\rangle_R = \langle \mathbf l_{-2}\mathbbm I\rangle_R = \frac{\langle\bbeta^\ast_1\bgamma^\ast_1\Phi_0(0)\rangle_R}{D_1(0) D_1(2)} = \frac{m^2}{4}\Theta(\mathbbm i,\mathbbm i\vert0) \; ,
  \\	\nonumber
  \\
  &\langle \overline T\rangle_R = \langle \overline{\mathbf l}_{-2}\mathbbm I\rangle_R = \frac{\langle\overline\bbeta^\ast_1\overline\bgamma^\ast_1\Phi_0(0)\rangle_R}{D_1(0) D_1(2)} = \frac{m^2}{4}\Theta(-\mathbbm i,-\mathbbm i\vert0)  \; ,
  \\	\nonumber
  \\
  &\langle \Theta\rangle_R = - 2 \pi \nu \frac{\bmu^2}{\sin\pi\nu} \langle \Phi_{2\frac{1-\nu}{\nu}}(0)\rangle_R = - 2 \pi \nu \frac{\bmu^2}{\sin\pi\nu} \frac{C_1(0)}{t_1(0)}\langle\bbeta^\ast_1\overline\bgamma^\ast_1\Phi_0(0)\rangle_R =	\nonumber
  \\
  \\
  &\phantom{\langle \Theta\rangle_R = - 2 \pi \nu \frac{\bmu^2}{\sin\pi\nu} \langle \Phi_{2\frac{1-\nu}{\nu}}(0)\rangle R } =  \frac{m^2}{4}\Big[\Theta(\mathbbm i,-\mathbbm i\vert0) - \pi t_1(0)\Big] \; ,	\nonumber
\end{align}
where the multiplier $2 \pi \nu$ in the last term takes into account the CFT normalisation of the energy-momentum tensor and the scaling dimension of $\bmu$. These equalities correspond exactly to the identities (\ref{eq:lukcomparisonfin}); this completes our check.

\section{LeClair-Mussardo formula}
\label{sec:leclairmussardo}
\setcounter{equation}{0}

From now on, let us concentrate on the sinh-Gordon model. In this chapter we wish to check the results obtained from the fermionic basis framework against a result found by LeClair and Mussardo in \cite{Lecl_Muss_99}.

Let us introduce the following function

\eq
  F(\alpha) = \frac{\langle\Phi_{\alpha}(0)\rangle_R}{\langle\Phi_{\alpha}(0)\rangle_\infty} \; ,
\en
which is obviously periodic of period $2$: $F(\alpha+2)=F(\alpha)$. Using the formulae (\ref{eq:shiftformula}), (\ref{eq:sinhthetaerepresentation}) and (\ref{eq:onepointfunctionsinh}) we obtain

\eq
  \frac{F(\alpha + 2\frac{1-\nu}{\nu})}{F(\alpha)} = 1 + \frac{2}{\pi}\sin\left[\pi\left(\alpha + \frac{1}{\nu}\right)\right] \big(e_1 \ast \, e_{-1} + e_1\ast \, R_{\textrm{dress}} \ast \, e_{-1}\big) \; .
\label{eq:leclairmussardocheck}
\en

In the paper \cite{Lecl_Muss_99}, LeClair and Mussardo express the large-$R$ expansion of the function $F(\alpha)$ in the following form:

\eq
  F(\alpha) = 1 + \sum_{n=1}^\infty \int \left(\prod_{j=1}^n dm(\theta_j)\right)\prod_{i<j\leq n} \Phi_{\alpha}(\theta_i-\theta_j) F_j(\theta_1,\ldots,\theta_n) \; ,
\label{eq:formfactorexpansion}
\en
giving explicitly the first three terms

\begin{align}
  F_1 &= 2\frac{\sin\frac{\pi}{\nu}}{\pi} [k]^2 \; ,	\nonumber
  \\	\nonumber
  \\
  F_2 &= 2\frac{\sin\frac{\pi}{\nu}}{\pi} [k]^2 \left([k]^2 c_{12} - \frac{[k-1][k+1]}{c_{12}}\right) \; ,
  \\	\nonumber
  \\
  F_3 &= \frac{[k]}{12} \left(A + B(c_{12}^2 + c_{23}^2 + c_{13}^2) + \frac{C}{c_{12}c_{23}c_{13}} + D \frac{c_1^2 + c_2^2 + c_3^2}{c_{12}c_{23}c_{13}}\right) \; ,	\nonumber
\end{align}
where we introduced the notations

\eq
  2 k=\nu\alpha \; , \quad [m] = \frac{\sin\left(m\frac{\pi}{\nu}\right)}{\sin\frac{\pi}{\nu}} \; ,\quad c_{ij} = \cosh(\theta_i-\theta_j) \;, \quad c_i = \cosh(2\theta_i -\theta_j-\theta_k) \; ,	\nonumber
\en
and defined

\begin{align}
  A &= - 28 [k-1][k][k+1] \left([k]^2 + 1\right) +	\nonumber
  \\
  & + 8 \left([k-2][k]^2[k+1]^2 + [k-1]^2[k]^2[k+2]\right) +	\nonumber
  \\
  & - 2 \left([k-2][k-1][k+1]^3 +[k-1]^3[k+1][k+2] - [k-2][k]^3[k+2] - [k]^5\right) \; ,	\nonumber
  \\	\nonumber
  \\
  B &= 8 [k]^5 \; ,	\nonumber
  \\	\nonumber
  \\
  C &= [k]^5 + 5 [k-1][k]^3[k+1] + 2 [k-1]^2[k][k+1]^2 + [k-2][k]^2[k+1]^2 +	\nonumber
  \\
  & + [k-1]^2[k]^2[k+2] - [k-2][k-1][k+1]^3 - [k-1]^3[k+1][k+2] +	\nonumber
  \\
  & - [k-2][k]^2[k+2] - 3 [k-2][k-1][k][k+1][k+2] \; ,	\nonumber
  \\	\nonumber
  \\
  D &= -4 [k-1][k]^3[k+1] \; .	\nonumber
\end{align}

Through some tedious algebra one finds that the ratio $F(\alpha + 2\frac{1-\nu}{\nu})/F(\alpha)$ has the same structure as (\ref{eq:formfactorexpansion}) with the following coefficients

\begin{align}
  F_1 &= \frac{2}{\pi} \sin\left[\pi\left(\alpha +\frac{1}{\nu}\right)\right] \; ,	\nonumber
  \\	\nonumber
  \\
  F_2 &= \frac{2}{\pi} \sin\left[\pi\left(\alpha +\frac{1}{\nu}\right)\right] \left[1 + \frac{1-\cos\left[\pi\left(\alpha + \frac{1}{\nu}\right)\right] \cos\frac{\pi}{\nu}}{\sin^2\frac{\pi}{\nu}} c_{12}\right]t^2_{12} \; ,
  \\	\nonumber
  \\
  F_3 &= \widetilde A + \widetilde B(c_{12}^2 + c_{23}^2 + c_{13}^2) + \frac{\widetilde C}{c_{12}c_{23}c_{13}} + \widetilde D \frac{c_1^2 + c_2^2 + c_3^2}{c_{12}c_{23}c_{13}} \; ,	\nonumber
\end{align}
with

\begin{align}
  \widetilde A &= \frac{[2k+1][2]}{6}\left(3[2k+1][2k-1][2] - [2k+1][2k]([2]^2 + 6) - 2[4k]\right) \; ,	\nonumber
  \\	\nonumber
  \\
  \widetilde B &= \frac{2}{3} [2k+1]^3 \; ,	\nonumber
  \\	\nonumber
  \\
  \widetilde C &= \frac{[2k+1][2]}{24}\left([4k](3[2]^2 - 4)-2 [2k+1][2k]([2]^2-6)\right)\; ,	\nonumber
  \\	\nonumber
  \\
  \widetilde D &= -\frac{1}{6} [2k+1]^2[2k][2] \; .	\nonumber
\end{align}

It is not difficult to check that these coefficients coincide with those one finds by iterating $R_{\textrm{dress}} = \Phi_\alpha + \Phi_\alpha\ast\Phi_\alpha + \cdots$ in the formula (\ref{eq:leclairmussardocheck}), that is

\eq
  \frac{F(\alpha + 2\frac{1-\nu}{\nu})}{F(\alpha)} = 1 + \frac{2}{\pi}\sin\left[\pi\left(\alpha + \frac{1}{\nu}\right)\right] \big(e_1 \ast \, e_{-1} + e_1\ast \, \Phi_\alpha \ast \, e_{-1} + e_1\ast \, \Phi_\alpha \ast \, \Phi_\alpha \ast \, e_{-1} + \cdots\big) \; .	\nonumber
\en

\section{Classical limit}
\label{sec:classical}
\setcounter{equation}{0}

In this section we investigate the behaviour of the function $F(\alpha)$ introduced above in the classical limit and compare it against a result presented by Lukyanov in \cite{Luky_01}.

In sinh-Gordon model the r\^ole of the Planck constant is played by $b^2$, thus the semiclassical regime corresponds to $\nu\gtrsim 1$; in \cite{Luky_01} a formula for the classical approximation of $F(\alpha)$ was found in two different ways, namely by applying the steepest descent method to the integral obtained by separation of variables and by evaluating the classical action on the solution to the sinh-Gordon equation with a puncture. The result reads, in our notations, as follows

\eq
  \log F(\alpha) = \frac{1}{b^2} \int\limits_0^\alpha d\alpha' \int\limits_{-\infty}^\infty \frac{d\theta}{2\pi \mathbbm i} \log\left(\frac{1- e^{-r\cosh\theta -\pi \mathbbm i \alpha'}}{1- e^{-r\cosh\theta +\pi \mathbbm i \alpha'}}\right) + O(b^0) \; ,
\en
with $r=2\pi m R$. From this formula, we immediately obtain

\eq
  \frac{F(\alpha + 2\frac{1-\nu}{\nu})}{F(\alpha)} = \exp \Bigg[-\frac{1}{\pi \mathbbm i}\int\limits_{-\infty}^\infty \log\Bigg(\frac{1- e^{-r\cosh\theta -\pi \mathbbm i \alpha'}}{1- e^{-r\cosh\theta +\pi \mathbbm i \alpha'}}\Bigg)\Bigg] + O(b^0) \; ,
\label{eq:lukyfrac}
\en
while, using (\ref{eq:leclairmussardocheck}) we see that

\eq
  \frac{F(\alpha + 2\frac{1-\nu}{\nu})}{F(\alpha)} = 1 - 2\frac{\sin(\pi\alpha)}{\pi} e_{1}\ast^{\textrm{cl}} E_{-1} + O(b^0) \; ,
\label{eq:fedorfrac}
\en
where we introduced the function $E_{-1}$, which satisfies the equation

\eq
  E_{-1} = e_{-1} + \Phi_{\alpha}^{\textrm{cl}} \ast^{\textrm{cl}}E_{-1} \; .
\label{eq:Em1equation}
\en

These two last formulae contain the classical limits of the kernel $\Phi$ and of the deformed convolution $\ast$; it is straightforward to compute them. In fact

\eq
  \Phi_\alpha^{\textrm{cl}}(\theta) = \frac{e^{-\pi \mathbbm i \alpha}}{2\pi \mathbbm i \sinh(\theta-\mathbbm i0)} - \frac{e^{\pi \mathbbm i \alpha}}{2\pi \mathbbm i \sinh(\theta +\mathbbm i0)} \; ,
\en
which, in particular, implies

\eq
  \Phi_0^{\textrm{cl}}(\theta) = \delta (\theta) \; .
\en
This identity allow us to explicitly solve the DDV equation (\ref{eq:destridevegaequation}), obtaining

\eq
  1+ e^{\epsilon^{\textrm{cl}}(\theta)} = e^{r\cosh\theta} \; ;
\en
thus the classical limit of the deformed convolution is

\eq
  f\ast^{\textrm{cl}}g = \int\limits_{-\infty}^\infty f(\theta)g(\theta) e^{-r\cosh\theta} d\theta \; .
\en

Now, introducing the function

\eq
  G(\theta) = \int\limits_{-\infty}^\infty \frac{E_{-1}(\theta')e^{-r\cosh\theta'}}{2\pi\cosh(\theta-\theta')} d\theta' \; ,
\en
we easily see that

\eq
  e^{-r\cosh\theta}E_{-1}(\theta) = G(\theta + \frac{\pi \mathbbm i}{2}) + G(\theta - \frac{\pi \mathbbm i}{2}) \; ,
\en
and the equation (\ref{eq:Em1equation}) can be recast in the following simple boundary problem for $G(\theta)$

\eq
  G(\theta + \frac{\pi \mathbbm i}{2})\left(1-e^{-\pi \mathbbm i \alpha - r\cosh\theta}\right) + G(\theta - \frac{\pi \mathbbm i}{2})\left(1-e^{\pi \mathbbm i \alpha - r\cosh\theta}\right) = e^{-\theta - r\cosh\theta} \; .
\en

Let us now introduce the following functions:

\eq
  H(\theta) = \frac{1-e^{-r\cosh\theta - \pi \mathbbm i \alpha}}{1-e^{-r\cosh\theta + \pi \mathbbm i \alpha}} -1 \; ,
\en
and

\eq
  X_{\pm}(\theta) = \exp\left(-\frac{1}{2\pi \mathbbm i}\int\limits_{-\infty}^\infty \frac{e^{\theta-\theta'}}{\sinh(\theta-\theta'\pm \mathbbm i0)}\log(1+H(\theta')) d\theta'\right) \; .
\en
With some simple calculations one can see how the equality of (\ref{eq:lukyfrac}) and (\ref{eq:fedorfrac}) is equivalent to the following identity

\begin{align}
  &\exp\left(-\frac{1}{\pi \mathbbm i} \int\limits_{-\infty}^\infty\log(1+H(\theta))d\theta\right) = 1-\frac{1}{2\pi \mathbbm i}\int\limits_{-\infty}^\infty \frac{2+H(\theta)}{1+H(\theta)}H(\theta)d\theta + 	\nonumber
  \\	\label{eq:identitylukyfedor}
  \\
  & + \frac{1}{(2\pi \mathbbm i)^2} \int\limits_{-\infty}^\infty\int\limits_{-\infty}^\infty \frac{H(\theta)H(\theta')}{\sinh(\theta-\theta')}\left(e^{\theta-\theta'}\frac{X_-(\theta)}{X_+(\theta')} - e^{\theta'-\theta}\frac{X_-(\theta')}{X_+(\theta)}\right)d\theta d\theta' \; ,	\nonumber
\end{align}
which is true for any function $H(\theta)$ rapidly decreasing at $\pm\infty$; the proof goes as follows.

\subsection{Proof of (\ref{eq:identitylukyfedor})}

The function $X_+(\theta)$, originally defined on the real axis, can be analytically continued in the strip $0 < \Im \theta < \pi$ and it coincides with $X_-(\theta)$ on the upper boundary of said strip. One easily sees that

\begin{align}
  &X_+(\theta) = (1+H(\theta))X_-(\theta) \; ,	\nonumber
  \\
  \\
  &X_+(-\infty) = 1 \; , \qquad X_+(\infty) = \exp \left(-\frac{1}{\pi \mathbbm i}\int\limits_{-\infty}^\infty \log(1+H(\theta))d\theta\right) \; .	\nonumber
\end{align}

Let us start by considering the two-fold integral in (\ref{eq:identitylukyfedor})

\eq
  I_2 = \frac{1}{(2\pi \mathbbm i)^2} \int\limits_{-\infty}^\infty\int\limits_{-\infty}^\infty \frac{H(\theta)H(\theta')}{\sinh(\theta-\theta')}\left(e^{\theta-\theta'}\frac{X_-(\theta)}{X_+(\theta')} - e^{\theta'-\theta}\frac{X_-(\theta')}{X_+(\theta)}\right)d\theta d\theta' \; ,
\en
and understand, for definiteness the denominator as $\sinh(\theta-\theta'+ \mathbbm i0)$; then, changing integration variables, we have

\eq
  I_2 = \frac{1}{2(\pi \mathbbm i)^2} \int\limits_{-\infty}^\infty\int\limits_{-\infty}^\infty \frac{H(\theta)H(\theta')e^{\theta-\theta'}}{\sinh(\theta-\theta' + \mathbbm i 0)} \frac{X_-(\theta)}{X_+(\theta')} d\theta d\theta' + \frac{1}{2\pi \mathbbm i}\int\limits_{-\infty}^\infty\frac{H(\theta)^2}{1+H(\theta)}d\theta \; .
\en
Now we perform the integration over $\theta$ of the first term

\begin{align}
  &\int\limits_{-\infty}^\infty \frac{H(\theta)X_-(\theta)}{2\pi \mathbbm i\sinh(\theta-\theta' + \mathbbm i 0)} e^{\theta-\theta'} d\theta = \int\limits_{-\infty}^\infty \frac{X_+(\theta) -X_-(\theta)}{2\pi \mathbbm i\sinh(\theta-\theta' + \mathbbm i 0)} e^{\theta-\theta'} d\theta =	\nonumber
  \\	\nonumber
  \\
  &=\int\limits_{-\infty}^\infty \frac{X_+(\theta)}{2\pi \mathbbm i\sinh(\theta-\theta' + \mathbbm i 0)} e^{\theta-\theta'} d\theta - \int\limits_{-\infty}^\infty \frac{X_-(\theta)}{2\pi \mathbbm i\sinh(\theta-\theta' - \mathbbm i 0)} e^{\theta-\theta'} d\theta + X_-(\theta') = 	\nonumber
  \\
  \\
  &= \oint_{\mathcal C}\frac{X_+(\theta)}{2\pi \mathbbm i\sinh(\theta-\theta')} e^{\theta-\theta'} d\theta -\pi \mathbbm i \left(\frac{X_+(\theta)}{2\pi \mathbbm i\sinh(\theta-\theta')} e^{\theta-\theta'}\right)\Bigg\vert_{-\infty}^\infty + X_-(\theta') =	\nonumber
  \\	\nonumber
  \\
  &= -\exp\left(-\frac{1}{\pi \mathbbm i}\int\limits_{-\infty}^\infty \log(1+H(\theta))d\theta\right) + X_-(\theta') \; ,	\nonumber
\end{align}
where the contour $\mathcal C$ is a rectangle having vertices at $\theta_1 = -\infty + \mathbbm i0$, $\theta_2 =\infty + \mathbbm i0$, $\theta_3 =\infty + \mathbbm i \pi - \mathbbm i0$, $\theta_4 =-\infty + \mathbbm i\pi - \mathbbm i0$ and clearly contains no poles.

Thus we have

\eq
  I_2 = -\frac{1}{\pi \mathbbm i}\exp\left(-\frac{1}{\pi \mathbbm i}\int\limits_{-\infty}^\infty \log(1+H(\theta))d\theta\right)\int\limits_{-\infty}^\infty\frac{H(\theta')}{X_+(\theta')}d\theta' + \frac{1}{2\pi \mathbbm i}\int\limits_{-\infty}^\infty\frac{2+H(\theta)}{1+H(\theta)}H(\theta)d\theta \; .	\nonumber
\en
we are left with the computation of the integral over $\theta'$:

\begin{align}
  &\int\limits_{-\infty}^\infty\frac{H(\theta')}{\pi \mathbbm i X_+(\theta')}d\theta' = \frac{1}{\pi \mathbbm i}\int\limits_{-\infty}^\infty\left(\frac{1}{X_-(\theta')}-\frac{1}{X_+(\theta')}\right)d\theta' = 	\nonumber
  \\
  \\
  &= -\oint_{\mathcal C} \frac{1}{\pi \mathbbm i X_+(\theta)}d\theta +\pi \mathbbm i \left(\frac{1}{\pi \mathbbm i X_+(\theta)}\right)\Bigg\vert_{-\infty}^\infty = \exp\left(\frac{1}{\pi \mathbbm i}\int\limits_{-\infty}^\infty \log(1+H(\theta))d\theta\right) -1 \; ,	\nonumber
\end{align}
meaning

\eq
 I_2 = -1 + \frac{1}{\pi \mathbbm i}\exp\left(-\frac{1}{\pi \mathbbm i}\int\limits_{-\infty}^\infty \log(1+H(\theta))d\theta\right) + \frac{1}{2\pi \mathbbm i}\int\limits_{-\infty}^\infty\frac{2+H(\theta)}{1+H(\theta)}H(\theta)d\theta \; ,	\nonumber
\en
which, inserted into (\ref{eq:identitylukyfedor}) completes the proof.

\begin{flushright}
 $\blacksquare$
\end{flushright}

\section{Numerical analysis in the $R\rightarrow 0$ limit}
\label{sec:numanal}
\setcounter{equation}{0}

We now turn to the numerical evaluation of the one-point functions of the sinh-Gordon model in the UV limit $R\rightarrow 0$. We will begin by studying the behaviour of the descendant fields and then move to the primary ones. In order to correctly perform the UV limit we need to rescale the theory on a cylinder of fixed radius $2\pi$, as done in \cite{Luky_01}; this amounts to a renormalisation of the physical mass $m\rightarrow m R$, so that $\bmu \propto R^{1+b^2}$. We take $r=2\pi m R$ as the parameter to be sent to zero.

For the sake of readability, the tables and figures concerned in this section are collected in a dedicated subsection at the bottom.

\subsection{Descendant fields}
\label{subsec:descendant}

We are interested in the UV behaviour of the following class of one-point functions

\eq
	F_{2j-1,2k-1}(\alpha,r) \doteq \frac{\langle\bbeta^\ast_{2j-1}\bgamma^\ast_{2k-1}\Phi_\alpha\rangle_r}{\langle\Phi_\alpha\rangle_r} \; , \qquad j,k \in \mathbbm N \; ,
\en
which can be rewritten using (\ref{eq:cfttosinhfermions}) as

\eq
	F_{2j-1,2k-1}(\alpha,r) = D_{2j-1}(\alpha) D_{2k-1}(2-\alpha) \frac{\langle\bbeta^{\textrm{CFT}\,\ast}_{2j-1} \bgamma^{\textrm{CFT}\,\ast}_{2k-1} \Phi_\alpha(0)\rangle_r}{\langle\Phi_\alpha(0)\rangle_r} \; , \nonumber
\en
In the $r\rightarrow 0$ limit, these functions should behave like CFT ratios of one-point functions. In particular, using the formulae found in the appendix of \cite{Jimb_Miwa_Smir_11_1}, we see that

\eq
	F_{2j-1,2k-1} \underset{r\rightarrow 0}{\sim} -\left(\frac{2 \pi m}{r}\right)^{2j+2k-2} \frac{D_{2j-1}(\alpha)D_{2k-1}(2-\alpha)}{j+k-1}\Omega_{2j-1,2k-1} \; ,
\label{eq:theodesc}
\en
where $\Omega_{2j-1,2k-1}$ are functions of the vacuum eigenvalues $I_{2n-1}$ of the integrals of motion, which can be found, for example, in \cite{Bazh_Luky_AZam_96}. For the cases we are interested in we have

\begin{align}
	&\Omega_{1,1}(\alpha ,r) = I_1(r) - \frac{\Delta_\alpha}{12} \; , \nonumber
	\\
	\\
	&\Omega_{1,3}(\alpha ,r) = I_3(r) - \frac{\Delta_\alpha}{6} I_1(r) + \frac{\Delta_\alpha^2}{144} + \frac{c+5}{1080}\Delta_\alpha - \frac{\Delta_\alpha}{360} d_\alpha \; . \nonumber
\end{align}

The vacuum eigenvalues of the integrals of motion do not depend directly on the radius $r$, but rather on the momentum $P(r)$, which is itself a function of $r$:

\eq
	I_1(r) = P(r)^2 - \frac{1}{24} \; , \qquad I_3(r) = I_1(r)^2 + \frac{1}{6}I_1(r) + \frac{c}{1440} \; .
\en
As explained neatly in \cite{AZam_AlZa_96}, in the limit $r\rightarrow 0$, the main contribution to the one-point functions $\langle e^{a\eta}\rangle$, with $a>0$, comes from the following region in the configuration space

\eq
	\vert b \eta_0 \vert < -\log \frac{\mu^2}{\sin\pi b^2} \; ,
\label{eq:quantisationregion}
\en
where $\eta_0$ is the zero mode of the field $\eta(z,\bz)$; here the interaction term in sinh-Gordon action can be neglected. This means that in this region we can consider $\eta$ as a free field and that the ground state wave functional $\boldsymbol{\Psi}_0[\eta]$ can be approximated by the superposition of two zero-modes plane waves

\eq
	\boldsymbol{\Psi}_0[\eta] \underset{r\rightarrow \infty}{\sim} \left( c_1 e^{\mathbbm i P(r) \eta_0} + c_2 e^{-\mathbbm i P(r) \eta_0}\right) \; ,
\en
where the momentum $P(r)$ is quantised thanks to the presence of the potential walls $b\eta_0 \sim \pm \log \frac{\mu^2}{\sin\pi b^2}$. The quantisation condition reads

\eq
	S(P)^2 = 1 \; \Rightarrow \; \delta(P) = \pi \; , \ S(P)\doteq e^{-\mathbbm i\delta(P)} \; ,
\en
where $S(P)$ is the Liouville reflection amplitude

\eq
	S(P) = -\left(\bmu\frac{\Gamma(1+b^2)}{b^2}\right)^{-4 \mathbbm i \frac{P(r)}{b}}\frac{\Gamma\big(1+2 \mathbbm i P(r) b\big)\Gamma\big(1+2 \mathbbm i P(r) b^{-1}\big)}{\Gamma\big(1-2 \mathbbm i P(r) b\big)\Gamma\big(1-2 \mathbbm i P(r) b^{-1}\big)} \; .
\en
Using (\ref{eq:massformula}) and remembering that we rescaled the mass $m\rightarrow m R$, we easily obtain the quantisation condition for the momentum

\begin{align}
	&2 P(r) Q \log\left[\frac{r}{8 \pi^{\frac{3}{2}} \big(b^2\big)^{\frac{1}{1+b^2}}}\Gamma\Big(\frac{1}{2(1+b^2)}\Big)\Gamma\Big(1+\frac{b^2}{2(1+b^2)}\Big)\right] = \nonumber
	\\	\label{eq:quantcond}
	\\
	&= -\frac{\pi}{2} + \frac{1}{2 \mathbbm i}\log\left[\frac{\Gamma\big(1+2 \mathbbm i P(r) b\big)\Gamma\big(1+2 \mathbbm i P(r) b^{-1}\big)}{\Gamma\big(1-2 \mathbbm i P(r) b\big)\Gamma\big(1-2 \mathbbm i P(r) b^{-1}\big)}\right] \; .	\nonumber
\end{align}

\begin{figure}[h]
\centering
\includegraphics[scale=1]{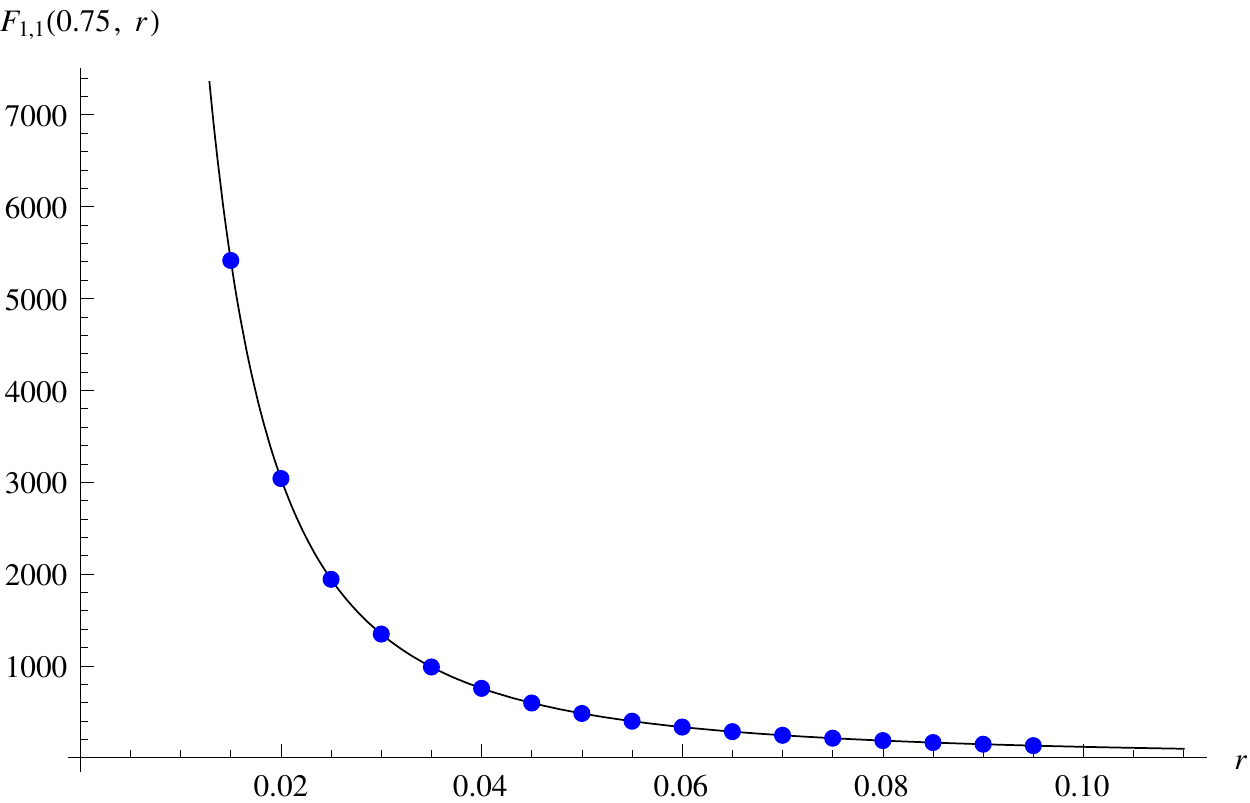}
\caption{Plot of $F_{1,1}(\alpha,r)$ against its theoretical behaviour for $\alpha =0.75$ and $b=0.4$}
\label{fig:f11_a075_b4}
\end{figure}

\begin{figure}[h]
\centering
\includegraphics[scale=1]{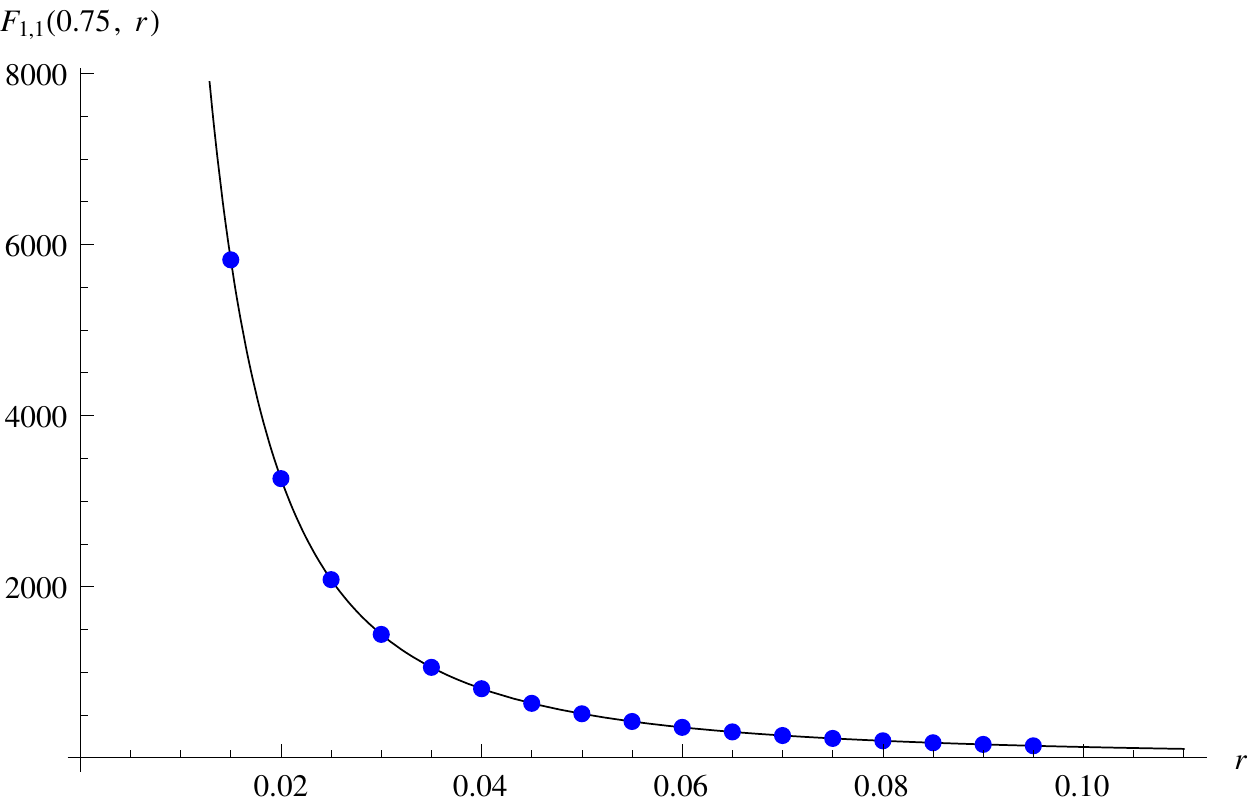}
\caption{Plot of $F_{1,1}(\alpha,r)$ against its theoretical behaviour for $\alpha =0.75$ and $b=0.8$}
\label{fig:f11_a075_b8}
\end{figure}

\begin{figure}[h]
\centering
\includegraphics[scale=1]{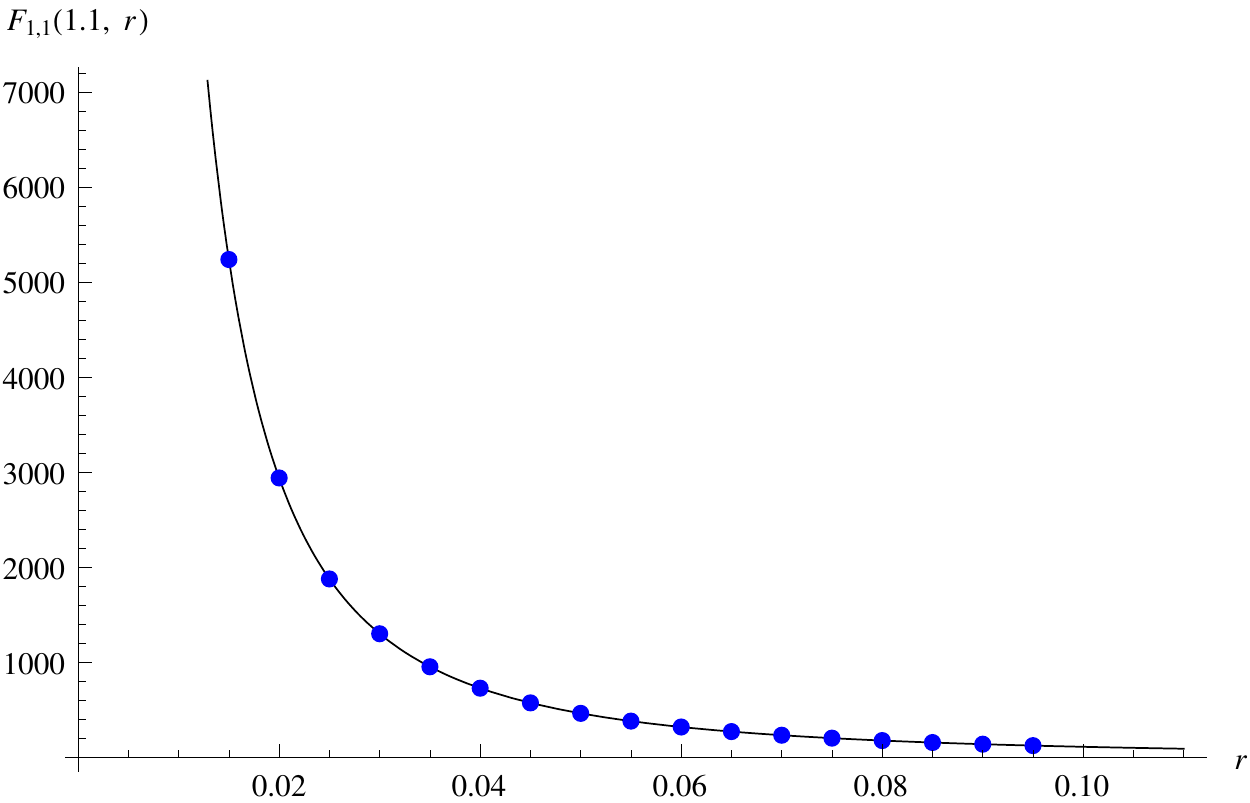}
\caption{Plot of $F_{1,1}(\alpha,r)$ against its theoretical behaviour for $\alpha =1.1$ and $b=0.4$}
\label{fig:f11_a11_b4}
\end{figure}

\begin{figure}[h]
\centering
\includegraphics[scale=1]{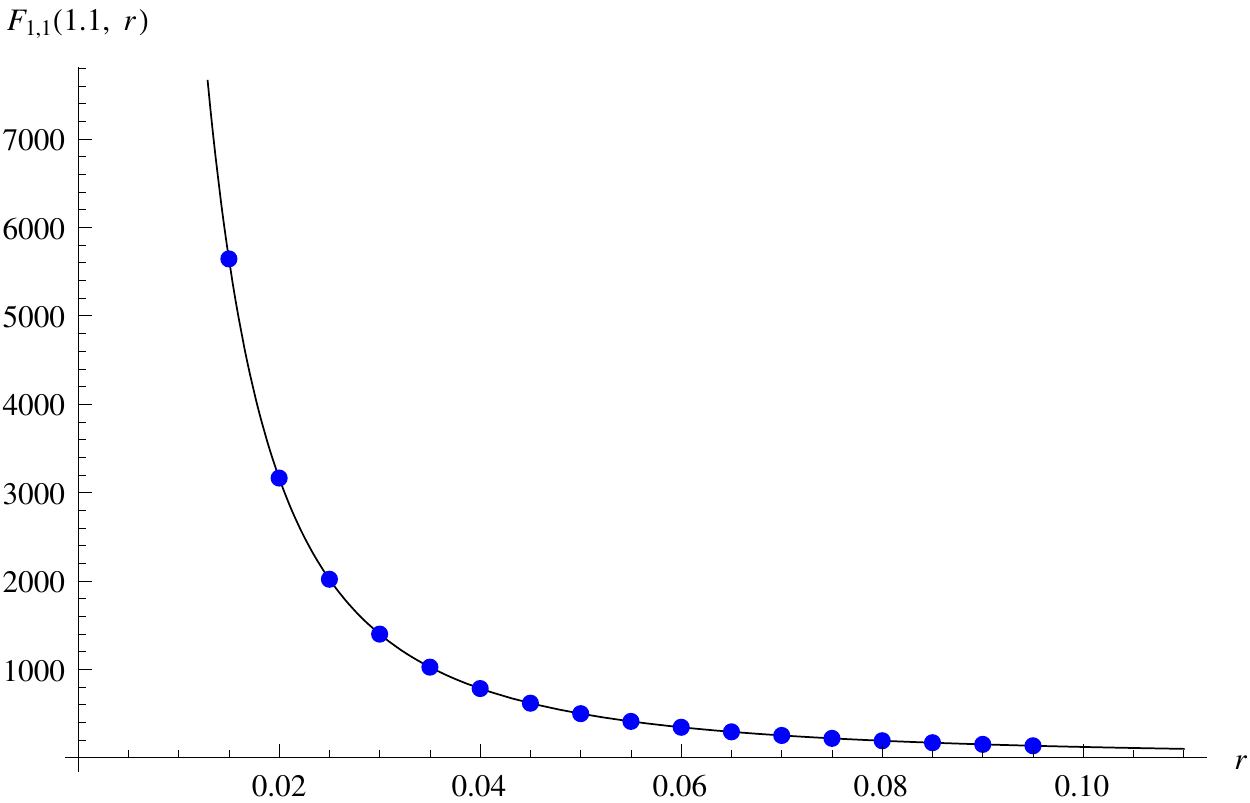}
\caption{Plot of $F_{1,1}(\alpha,r)$ against its theoretical behaviour for $\alpha =1.1$ and $b=0.8$}
\label{fig:f11_a11_b8}
\end{figure}

\begin{figure}[h]
\centering
\includegraphics[scale=1]{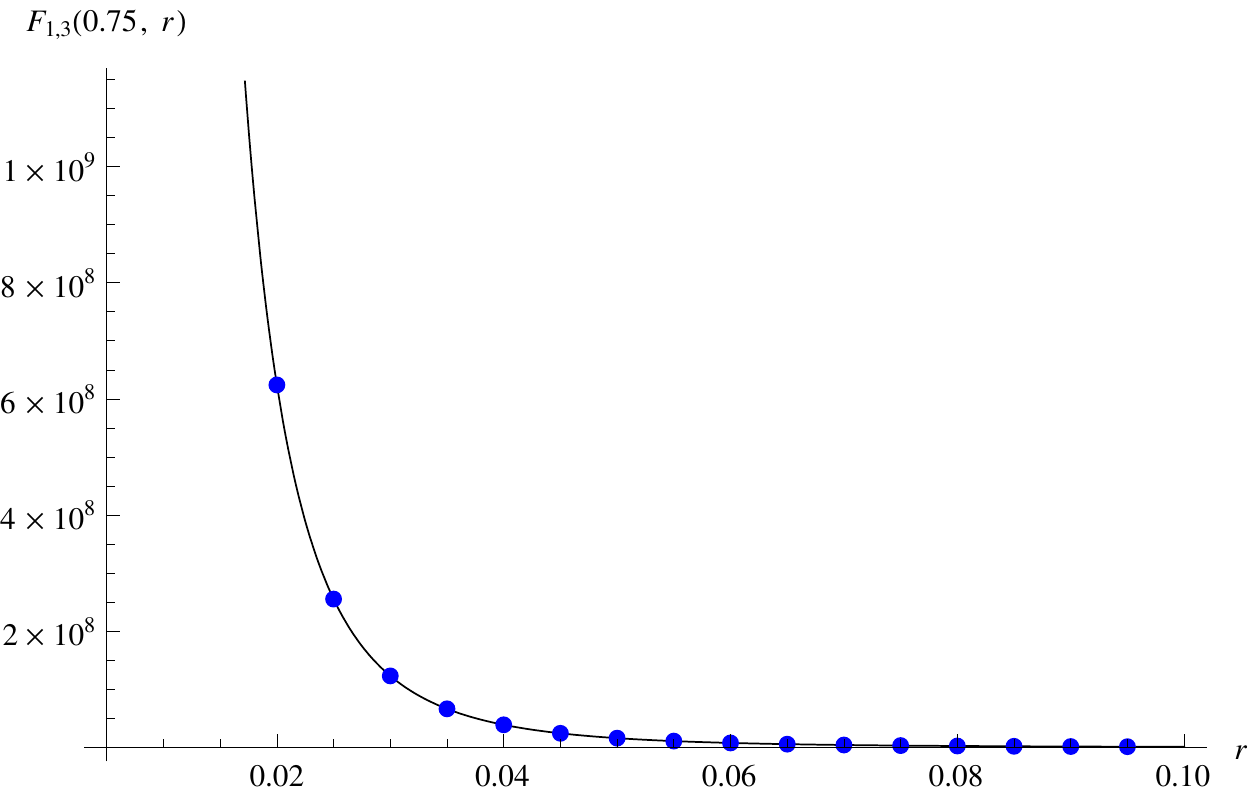}
\caption{Plot of $F_{1,3}(\alpha,r)$ against its theoretical behaviour for $\alpha =0.75$ and $b=0.4$}
\label{fig:f13_a075_b4}
\end{figure}

\begin{figure}[h]
\centering
\includegraphics[scale=1]{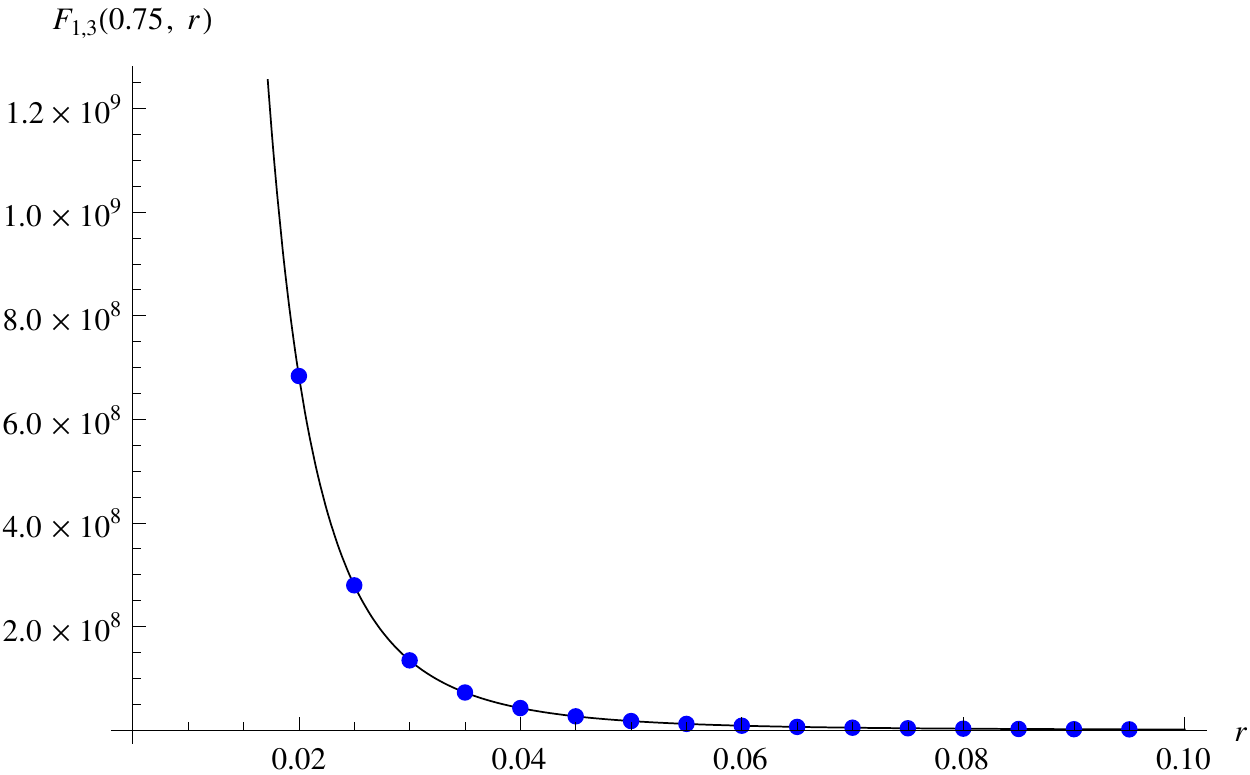}
\caption{Plot of $F_{1,3}(\alpha,r)$ against its theoretical behaviour for $\alpha =0.75$ and $b=0.8$}
\label{fig:f13_a075_b8}
\end{figure}

\begin{figure}[h]
\centering
\includegraphics[scale=1]{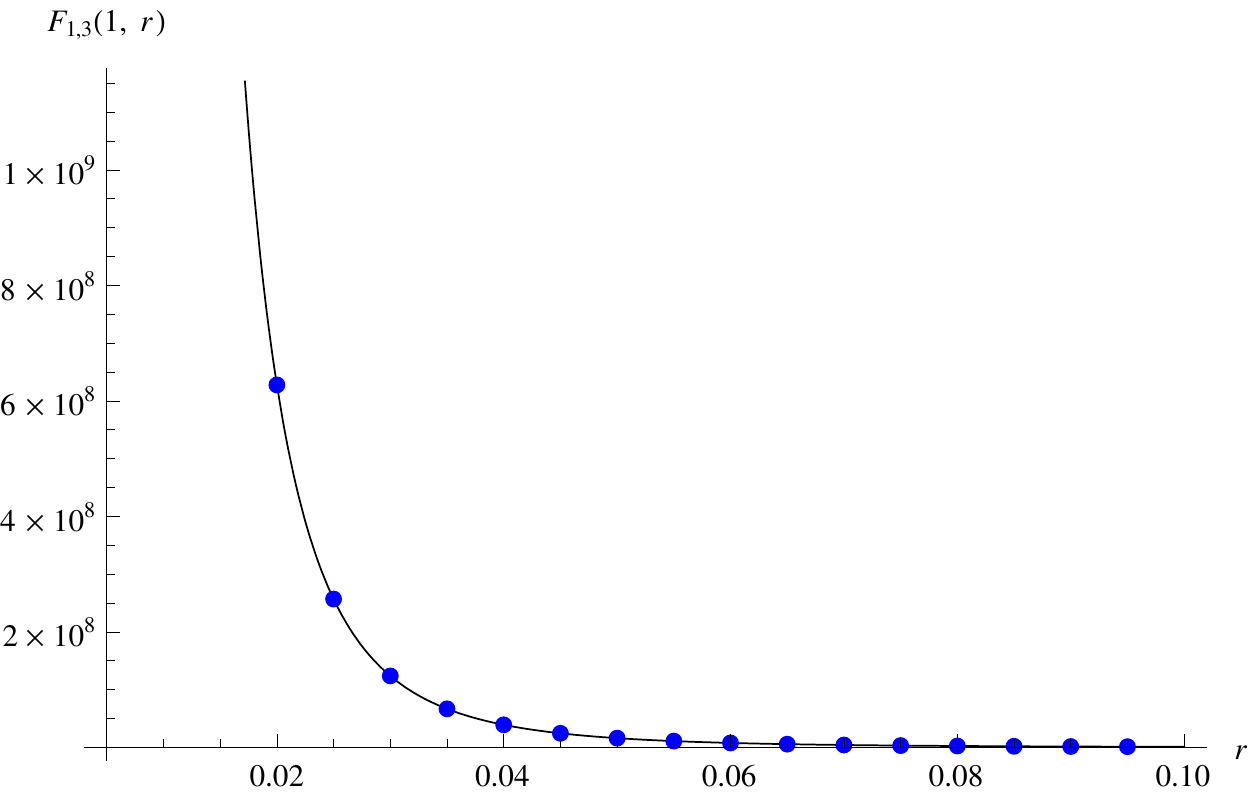}
\caption{Plot of $F_{1,3}(\alpha,r)$ against its theoretical behaviour for $\alpha =1$ and $b=0.4$}
\label{fig:f13_a11_b4}
\end{figure}

\begin{figure}[h]
\centering
\includegraphics[scale=1]{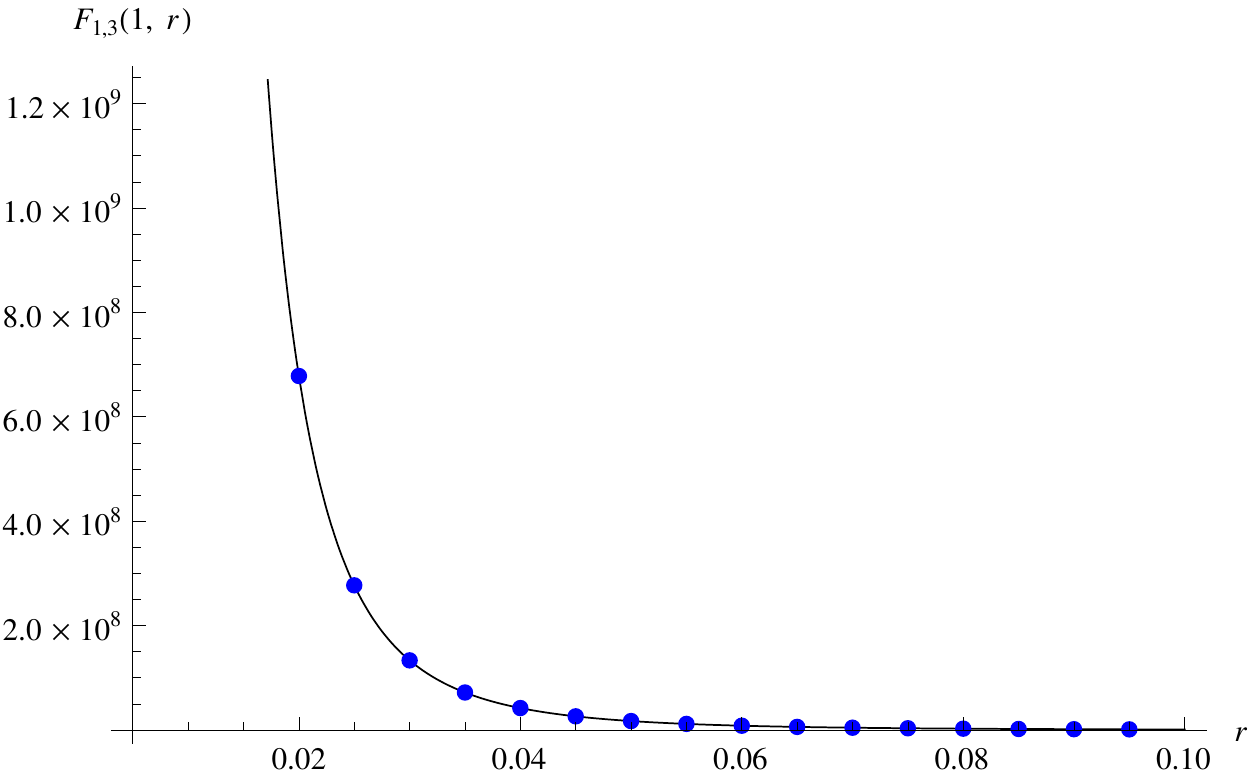}
\caption{Plot of $F_{1,3}(\alpha,r)$ against its theoretical behaviour for $\alpha =1$ and $b=0.8$}
\label{fig:f13_a11_b8}
\end{figure}

We have considered the following two ratios of expectation values

\eq
	F_{1,1}(\alpha,r) \doteq \frac{\langle\bbeta^\ast_1\bgamma^\ast_1\Phi_\alpha(0)\rangle_r}{\langle\Phi_\alpha(0)\rangle_r} \; , \qquad	F_{1,3}(\alpha,r) \doteq \frac{\langle\bbeta^\ast_1\bgamma^\ast_3\Phi_\alpha(0)\rangle_r}{\langle\Phi_\alpha(0)\rangle_r} \; ,
\en
and evaluated numerically the corresponding functions $\Theta_r^{\textrm{shG}}(i,i\vert\alpha)$ and $\Theta_r^{\textrm{shG}}(i,3i\vert\alpha)$ for values of $\alpha$ ranging from $0.75$ up to $1.5$, with $b\in [0.4,1.0]$ and $r\in [0.005,0.95]$. Figures \ref{fig:f11_a075_b4}-\ref{fig:f13_a11_b8} show some of these numerical estimates plotted against the curve (\ref{eq:theodesc}); the agreement of the data with the theoretical prevision is really good for the whole range of $r$ considered. The tables Table \ref{tab:f11relerr} and Table \ref{tab:f13relerr}, displaying the values of the relative error $\sigma$

\eq
	\sigma_{2j-1,2k-1} \doteq \left\vert 1-\frac{F_{2j-1,2k-1}(\alpha,r)}{F_{2j-1,2k-1}^{\textrm{CFT}}(\alpha,r)}\right\vert
\en
with

\eq
 F_{2j-1,2k-1}^{\textrm{CFT}}(\alpha ,r) = -\left(\frac{2\pi m}{r}\right)^{2j+2k-2} \frac{D_{2j-1}(\alpha)D_{2k-1}(2-\alpha)}{j+k-1}\Omega_{2j-1,2k-1} \; ,
\en
are a remarkable evidence in support of the conjecture introduced in Sec.\ref{sec:onepointsinh}.

\subsection{Primary fields}
\label{subsec:primary}

Let us now consider the following ratio of primary fields' expectation values

\eq
	\mathcal F(\alpha,r) \doteq \frac{\langle\Phi_{\alpha -2\frac{b^2}{b^2+1}}\rangle^{\textrm{shG}}_r}{\langle\Phi_\alpha\rangle^{\textrm{shG}}_r}\; .
\en
Using the shift formula (\ref{eq:shiftformula}) and the determinant one (\ref{eq:onepointfunctionsinh}) we can write

\eq
	\mathcal F(\alpha,r) = \frac{C_1(\alpha)}{t_1(\alpha)} \frac{\langle\bbeta^\ast_1\bar\bgamma^\ast_1\Phi_{\alpha}\rangle^{\textrm{shG}}_r}{\langle\Phi_\alpha\rangle^{\textrm{shG}}_r} = - \frac{C_1(\alpha)}{\pi t_1(\alpha)}\Big[\Theta(\mathbbm i,-\mathbbm i\vert\alpha) - \pi t_1(\alpha)\Big] \; .
\label{eq:finthetaterms}
\en

On the other hand, from \cite{Luky_01} we know that we can approximate the behaviour of the expectation value of a primary field $\Phi_\alpha$ in the region (\ref{eq:quantisationregion}) with that of a three-point function of Liouville CFT:

\eq
	\langle\Phi_\alpha\rangle^{\textrm{shG}}_r \underset{r\rightarrow 0}{\sim} \mathcal N(r,b) \langle 0\vert e^{a(-P)\eta(-\infty)} \Phi_\alpha e^{a(P)\eta(\infty)}\vert 0\rangle_r^{\textrm{Liou}} \; ,	\label{eq:hypothesis}
\en
where the function $\mathcal N(r,b)$ is a normalisation constant and

\eq
	a(P) \doteq \frac{Q}{2} + \mathbbm i P(r) \; \Rightarrow \qquad \Delta_{a(P)} = \frac{Q^2}{4} - P(r)^2
\en
with $P(r)$ satisfying the quantisation condition (\ref{eq:quantcond}).

The form of Liouville three-point function was found in \cite{Dorn_Otto_94,AZam_AlZa_96} and reads

\eq
	\langle 0\vert e^{a(-P)\eta(-\infty)} \Phi_\alpha e^{a(P)\eta(\infty)}\vert 0\rangle_r^{\textrm{Liou}} = \left(\bmu \frac{\Gamma(1+b^2)}{b^{1+b^2}}\right)^{-Q\frac{\alpha}{b}} \Upsilon_0 \frac{\Upsilon(2a)\Upsilon(Q-2 \mathbbm i P)\Upsilon(Q+2 \mathbbm i P)}{\Upsilon(a)^2\Upsilon(a-2 \mathbbm i P)\Upsilon(a+2 \mathbbm i P)} \; ,
\label{eq:liouvillethreepoint}
\en
where the function $\Upsilon(x)$ is defined by the equations

\eq
	\frac{\Upsilon(x+b)}{\Upsilon(x)} = \gamma(b\,x) b^{1-2bx} \; , \quad \frac{\Upsilon(x+b^{-1})}{\Upsilon(x)} = \gamma\left(\frac{x}{b}\right) b^{-1+2\frac{x}{b}} \; , \quad \Upsilon_0 \doteq \frac{d \Upsilon}{dx}\Big\vert_{x=0} \; .	\nonumber
\en
The general form of the normalisation $\mathcal N(r,b)$ is not known, but this is irrelevant to our needs, since we are considering the ratio of two one-point functions.

With some simple calculations one finds

\begin{align}
	\mathcal F(\alpha,r) &\underset{r\rightarrow 0}{\sim} \mathcal F^{\textrm{CFT}}(\alpha,r) = \left[\frac{r}{8\pi^{\frac{3}{2}}}\Gamma\left(\frac{1}{2(1+b^2)}\right)\Gamma\left(1+\frac{b^2}{2(1+b^2)}\right)\right]^2 \times	\nonumber
	\\	\label{eq:fcftbehaviour}
	\\
	&\times \frac{\gamma\big(b(a-b)\big)^2}{\gamma\big(b(2a-b)\big)\gamma\big(2b(a-b)\big)} \gamma\big(b(a-b+2 \mathbbm i P)\big)\gamma\big(b(a-b-2 \mathbbm i P)\big) \; .	\nonumber
\end{align}

\begin{figure}[h]
\centering
\includegraphics[scale=1]{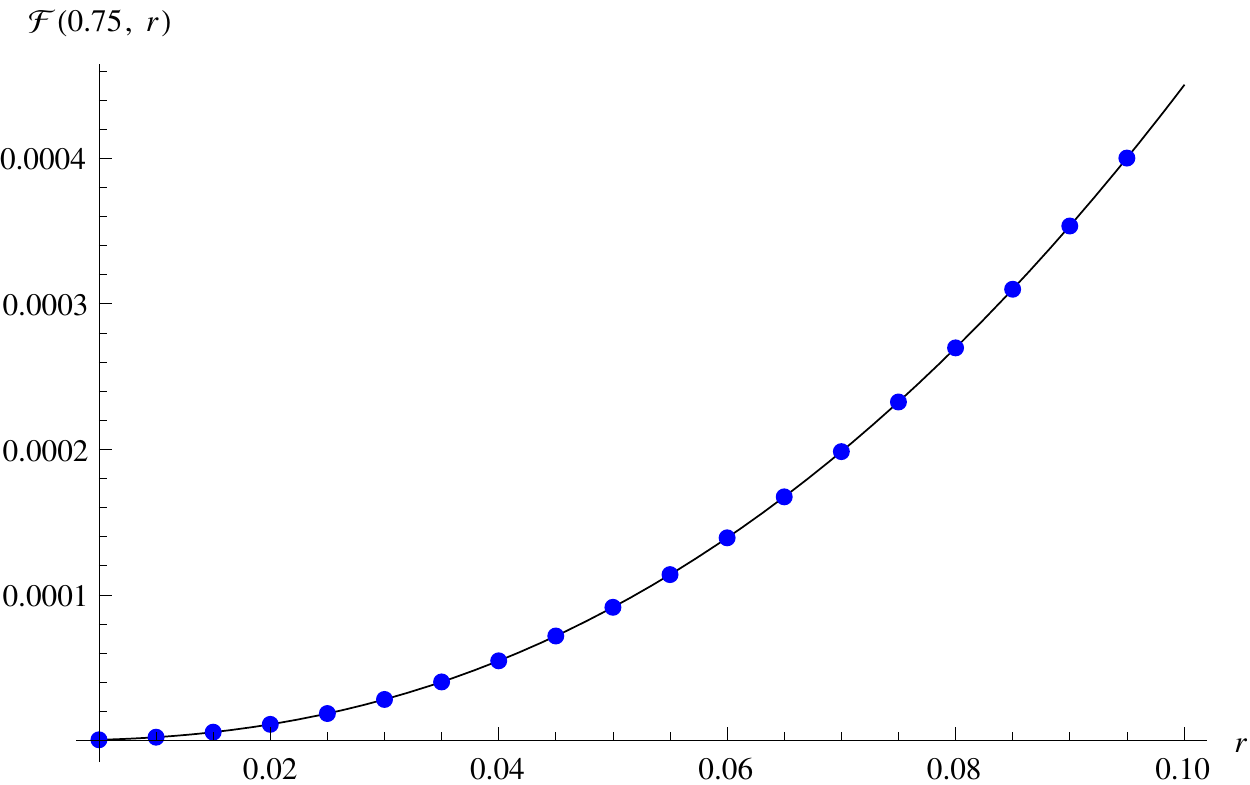}
\caption{Plot of $\mathcal F(\alpha,r)$ against its theoretical behaviour for $\alpha =0.75$ and $b=0.4$}
\label{fig:eff_a075_b4}
\end{figure}

\begin{figure}[h]
\centering
\includegraphics[scale=1]{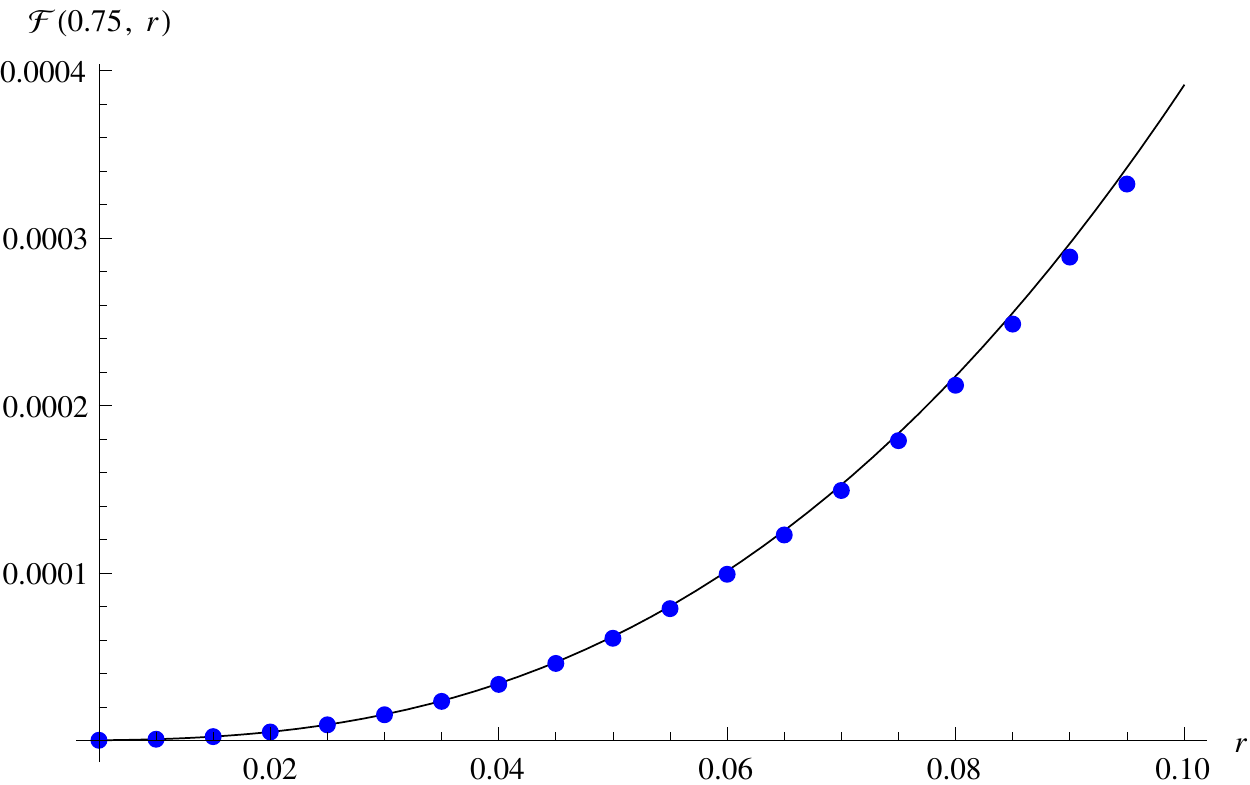}
\caption{Plot of $\mathcal F(\alpha,r)$ against its theoretical behaviour for $\alpha =0.75$ and $b=0.7$}
\label{fig:eff_a075_b7}
\end{figure}

\begin{figure}[h]
\centering
\includegraphics[scale=1]{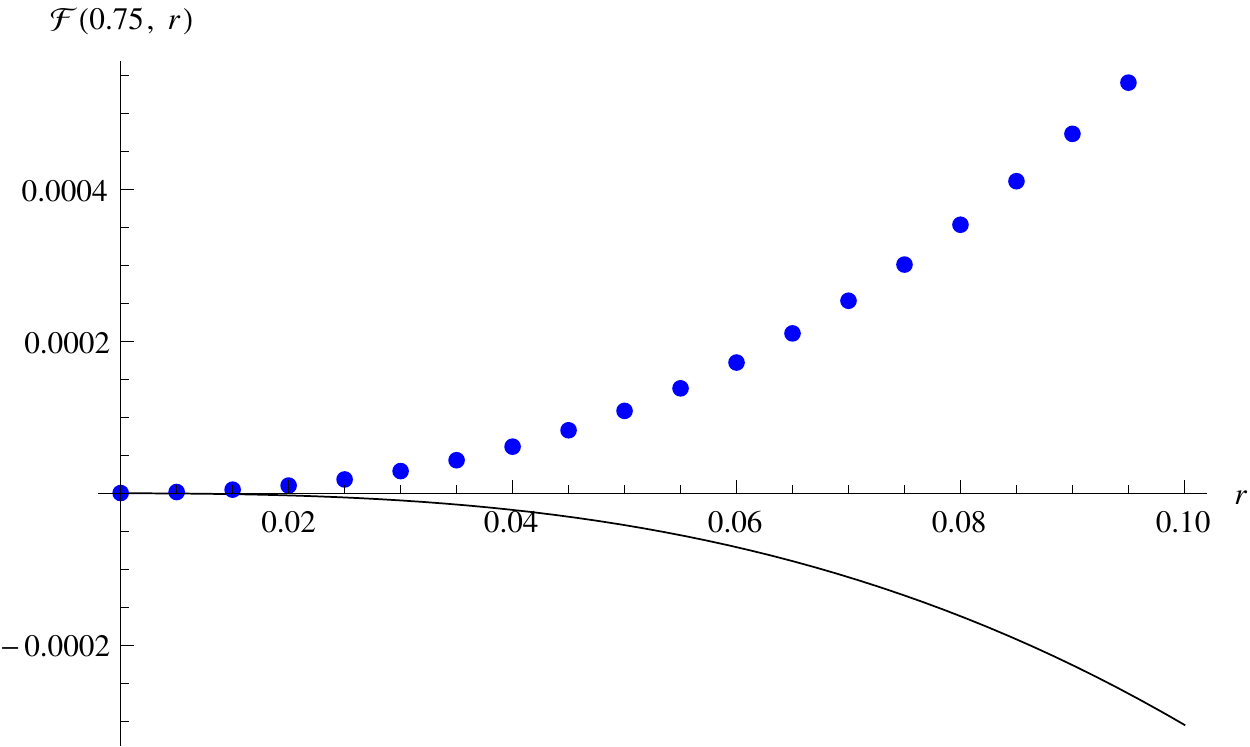}
\caption{Plot of $\mathcal F(\alpha,r)$ against its theoretical behaviour for $\alpha =0.75$ and $b=0.8$}
\label{fig:eff_a075_b8}
\end{figure}

\begin{figure}[h]
\centering
\includegraphics[scale=1]{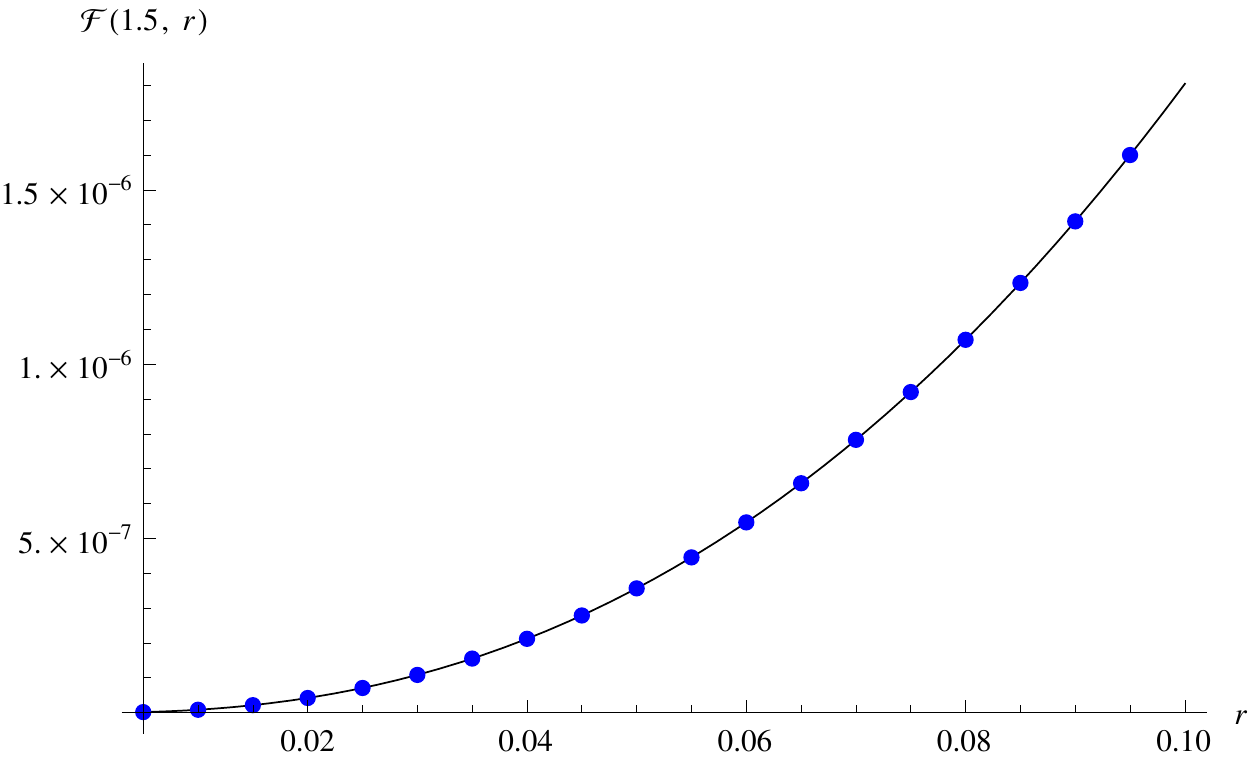}
\caption{Plot of $\mathcal F(\alpha,r)$ against its theoretical behaviour for $\alpha =1.5$ and $b=0.4$}
\label{fig:eff_a15_b4}
\end{figure}

\begin{figure}[h]
\centering
\includegraphics[scale=1]{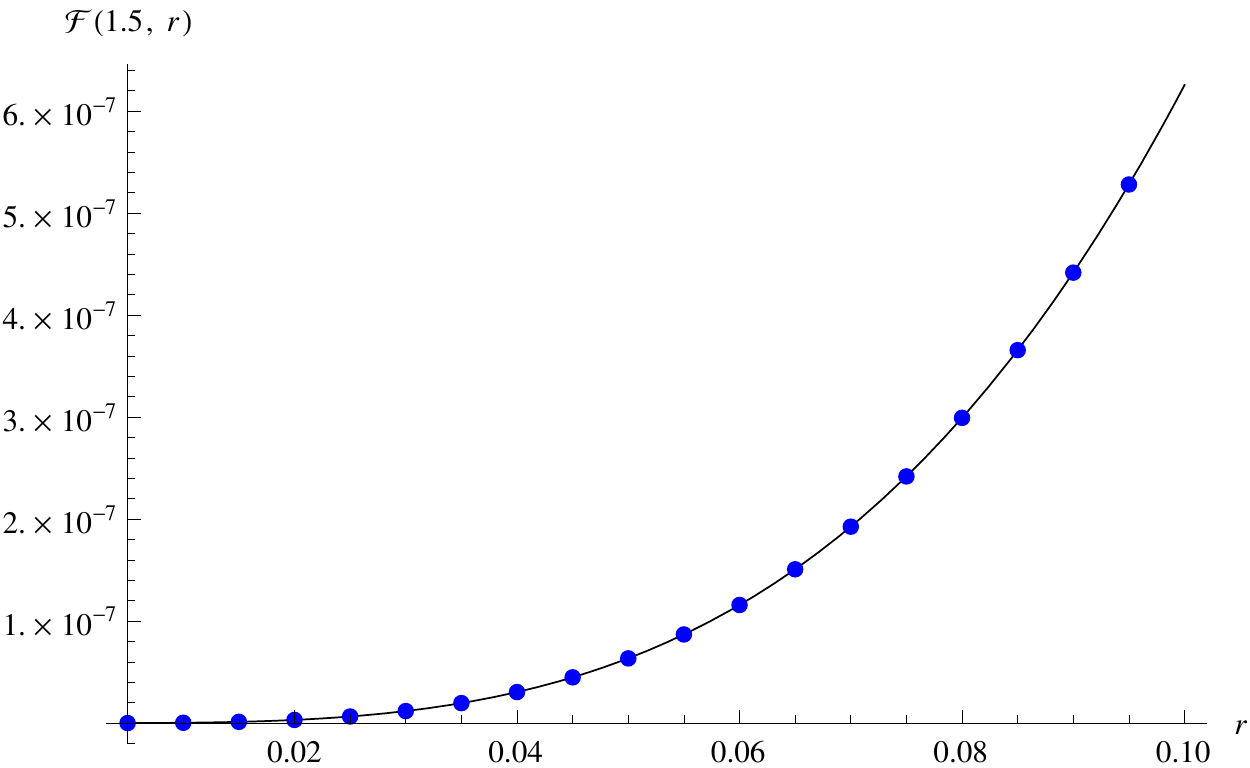}
\caption{Plot of $\mathcal F(\alpha,r)$ against its theoretical behaviour for $\alpha =1.5$ and $b=0.8$}
\label{fig:eff_a15_b8}
\end{figure}

We have evaluated numerically the function $\Theta(\mathbbm i,-\mathbbm i\vert\alpha)$ and used it to extract the value of $\mathcal F(\alpha,r)$ by means of the formula (\ref{eq:finthetaterms}). We then compared the data we obtained with the theoretical CFT behaviour (\ref{eq:fcftbehaviour}). Figures \ref{fig:eff_a075_b4}-\ref{fig:eff_a15_b8} show the result, while collected in table \ref{tab:effrelerr} are the values of the relative error $\varsigma$

\eq
	\varsigma \doteq \left\vert 1 - \frac{\mathcal F(\alpha,r)}{\mathcal F^{\textrm{CFT}}(\alpha,r)}\right\vert \; .
\en

The agreement between the data and the CFT behaviour is incredibly good until $b\gtrsim 0.7$, when $\alpha=0.75$, as is clearly visible from figures \ref{fig:eff_a075_b7} and \ref{fig:eff_a075_b8}. The reason for this discrepancy is that the hypothesis (\ref{eq:hypothesis}) we made above holds true only if the conformal dimension of the involved field is positive. This means that we have to restrict our analysis to the region of parameter space $(\alpha ,b)$ where both fields appearing in $\mathcal F(\alpha , r)$ have a positive dimension, which means

\eq
	\left\lbrace \begin{array}{l}
					0 < \alpha < 2 \\
					0 < \alpha -2 \frac{b^2}{1+b^2} < 2
				\end{array}\right.
	\quad \Rightarrow \quad 2 \frac{b^2}{1+b^2} < \alpha < 2 \; .
\en
Outside this \emph{natural region}, that is for $b\geq\sqrt{\frac{\alpha}{2-\alpha}}$, the sinh-Gordon model no more approaches na\"\i vely the Liouville CFT in its UV limit: there are contributions not taken into account which become important. This fact can be nicely visualised by sending $\alpha\rightarrow 0$; in this case, the expectation value of the field $\Phi_{-2 b/Q}(0)$ can be calculated explicitly in terms of the ground-state energy $E(R) \underset{R\rightarrow 0}{\sim} -\frac{\pi}{6R} c_{\textrm{eff}}(R)$, where \cite{Luky_01}:

\eq
	c_{\textrm{eff}}(R) \underset{R\rightarrow 0}{\sim} 1-\frac{24 \pi}{\left(\delta_1 - 4Q\log\frac{R}{2\pi}\right)^2} \; ,
\en
and $\delta_1$ is a constant whose form is irrelevant for our argument. Making use of formulae (\ref{eq:onepointfunctionsinh}),(\ref{eq:firstshiftexample1}) and (\ref{eq:lukcomparison2}) we obtain

\eq
	\langle e^{-b\eta(0)}\rangle = -\frac{C_1(0)}{\pi m^2 t_1(0)}\left(\frac{1}{R} + \frac{d}{dR}\right)E(R) \underset{R\rightarrow 0}{\sim} \frac{\pi c_1(0,b)}{2m^2Q^2}\frac{R^{-2(b^2+1)}}{(-\log R)^3} \; ,
\en
where we used (\ref{eq:C-constants}), setting $C_1(a)/t_1(a)\underset{R\rightarrow 0}{\sim}c_1(\alpha,b)R^{2 b(Q\alpha-b)}$, and $c_1(a,b)$ is a function of $\alpha$ and $b$ only. On the other hand, the formula (\ref{eq:liouvillethreepoint}) gives us a completely different answer:

\eq
	\langle e^{-b\eta(0)}\rangle = \frac{\langle\Phi_{\frac{Q}{2}-P}(-\infty) \vert \Phi_{-2 \frac{b}{Q}}(0)\vert \Phi_{\frac{Q}{2}+P}(\infty)\rangle}{\langle\Phi_{\frac{Q}{2}-P}(-\infty)\vert \Phi_{\frac{Q}{2}+P}(\infty)\rangle} \underset{R\rightarrow 0}{\sim} k(\alpha ,b)R^{2(1+b^2)} \; .
\en

Returning to our numerical analysis, we see that, when $\alpha=0.75$, the \emph{critical} value of $b$ is $b^{\textrm{crit}} = \sqrt{\frac{0.75}{2-0.75}} = \sqrt{3/5} \sim 0.774$, which explains why figure \ref{fig:eff_a075_b7} still shows a good agreement for very small values of $r$, while in figure \ref{fig:eff_a075_b8} we see that the data and the CFT curve behave in radically different ways.
\newpage
\subsection*{Figures and Tables}
Here we collect the plots and tables we addressed to in the previous subsection.

\begin{table}[h]
\centering
\begin{tabular}{l | c | c |c| c | c |}
\cline{2-6}
 & \multicolumn{5}{c|}{$\sigma_{1,1}$}\\
 \cline{2-6}
 & \multicolumn{2}{c|}{$\alpha = 0.75$} & & \multicolumn{2}{c|}{$\alpha = 1.1$}\\
 \hline
 \multicolumn{1}{|c|}{$r$} & $b=0.4$ & $b=0.8$ & & $b=0.4$ & $b=0.8$\\
 \hline
 \multicolumn{1}{|l|}{0.005} & $1.5 \times 10^{-4}$  & $2.0 \times 10^{-5}$ & &  $2.4 \times 10^{-4}$  & $6.0 \times 10^{-5}$ \\
 \multicolumn{1}{|l|}{0.01}  & $5.5 \times 10^{-5}$  & $3.3 \times 10^{-6}$ & &  $1.1 \times 10^{-4}$  & $1.5 \times 10^{-5}$ \\
 \multicolumn{1}{|l|}{0.015} & $2.3 \times 10^{-5}$  & $1.2 \times 10^{-6}$ & &  $6.1 \times 10^{-5}$  & $8.1 \times 10^{-6}$ \\
 \multicolumn{1}{|l|}{0.02}  & $1.3 \times 10^{-5}$  & $1.8 \times 10^{-6}$ & &  $3.7 \times 10^{-5}$  & $4.6 \times 10^{-6}$ \\
 \multicolumn{1}{|l|}{0.025} & $7.5 \times 10^{-6}$  & $3.0 \times 10^{-7}$ & &  $2.2 \times 10^{-5}$  & $2.3 \times 10^{-6}$ \\
 \multicolumn{1}{|l|}{0.03}  & $5.1 \times 10^{-6}$  & $7.3 \times 10^{-7}$ & &  $1.6 \times 10^{-5}$  & $2.8 \times 10^{-6}$ \\
 \multicolumn{1}{|l|}{0.035} & $1.7 \times 10^{-6}$  & $1.1 \times 10^{-6}$ & &  $1.1 \times 10^{-5}$  & $1.1 \times 10^{-6}$ \\
 \multicolumn{1}{|l|}{0.04}  & $1.4 \times 10^{-6}$  & $1.1 \times 10^{-6}$ & &  $7.0 \times 10^{-6}$  & $3.1 \times 10^{-7}$ \\
 \multicolumn{1}{|l|}{0.045} & $1.4 \times 10^{-6}$  & $1.1 \times 10^{-6}$ & &  $6.7 \times 10^{-6}$  & $2.2 \times 10^{-6}$ \\
 \multicolumn{1}{|l|}{0.05}  & $1.3 \times 10^{-6}$  & $1.3 \times 10^{-6}$ & &  $2.4 \times 10^{-6}$  & $2.5 \times 10^{-6}$ \\
 \multicolumn{1}{|l|}{0.055} & $4.2 \times 10^{-6}$  & $3.3 \times 10^{-6}$ & &  $7.1 \times 10^{-6}$  & $8.0 \times 10^{-7}$ \\
 \multicolumn{1}{|l|}{0.06}  & $1.1 \times 10^{-6}$  & $2.5 \times 10^{-6}$ & &  $2.2 \times 10^{-6}$  & $3.2 \times 10^{-7}$ \\
 \multicolumn{1}{|l|}{0.065} & $4.0 \times 10^{-7}$  & $2.4 \times 10^{-7}$ & &  $2.4 \times 10^{-6}$  & $4.4 \times 10^{-7}$ \\
 \multicolumn{1}{|l|}{0.07}  & $2.7 \times 10^{-7}$  & $2.9 \times 10^{-7}$ & &  $2.2 \times 10^{-6}$  & $1.7 \times 10^{-7}$ \\
 \multicolumn{1}{|l|}{0.075} & $3.1 \times 10^{-7}$  & $2.8 \times 10^{-7}$ & &  $1.3 \times 10^{-6}$  & $1.2 \times 10^{-6}$ \\
 \multicolumn{1}{|l|}{0.08}  & $1.3 \times 10^{-7}$  & $1.0 \times 10^{-7}$ & &  $1.0 \times 10^{-6}$  & $4.3 \times 10^{-8}$ \\
 \multicolumn{1}{|l|}{0.085} & $5.4 \times 10^{-7}$  & $1.3 \times 10^{-7}$ & &  $3.1 \times 10^{-8}$  & $5.4 \times 10^{-7}$ \\
 \multicolumn{1}{|l|}{0.09}  & $2.8 \times 10^{-8}$  & $2.2 \times 10^{-7}$ & &  $1.4 \times 10^{-6}$  & $2.8 \times 10^{-6}$ \\
 \multicolumn{1}{|l|}{0.095} & $1.2 \times 10^{-6}$  & $2.3 \times 10^{-7}$ & &  $2.6 \times 10^{-6}$  & $6.8 \times 10^{-11}$ \\
 \hline
\end{tabular}
\caption{Values of the relative error for $F_{1,1}(\alpha,r)$.}
\label{tab:f11relerr}
\end{table}

\begin{table}[h]
\centering
\begin{tabular}{l | c | c |c| c | c |}
\cline{2-6}
 & \multicolumn{5}{c|}{$\sigma_{1,3}$}\\
 \cline{2-6}
 & \multicolumn{2}{c|}{$\alpha = 0.75$} & & \multicolumn{2}{c|}{$\alpha = 1$}\\
 \hline
 \multicolumn{1}{|c|}{$r$} & $b=0.4$ & $b=0.8$ & & $b=0.4$ & $b=0.8$\\
 \hline
 \multicolumn{1}{|l|}{0.005} & $2.4 \times 10^{-4}$  & $6.2 \times 10^{-5}$ & &  $1.4 \times 10^{-4}$  & $1.8 \times 10^{-5}$ \\
 \multicolumn{1}{|l|}{0.01}  & $1.0 \times 10^{-4}$  & $1.6 \times 10^{-5}$ & &  $5.2 \times 10^{-5}$  & $5.0 \times 10^{-6}$ \\
 \multicolumn{1}{|l|}{0.015} & $6.2 \times 10^{-5}$  & $9.3 \times 10^{-6}$ & &  $2.5 \times 10^{-5}$  & $3.1 \times 10^{-6}$ \\
 \multicolumn{1}{|l|}{0.02}  & $3.7 \times 10^{-5}$  & $4.1 \times 10^{-6}$ & &  $1.2 \times 10^{-5}$  & $1.1 \times 10^{-6}$ \\
 \multicolumn{1}{|l|}{0.025} & $2.1 \times 10^{-5}$  & $8.9 \times 10^{-7}$ & &  $7.0 \times 10^{-6}$  & $5.0 \times 10^{-7}$ \\
 \multicolumn{1}{|l|}{0.03}  & $1.8 \times 10^{-5}$  & $8.1 \times 10^{-7}$ & &  $4.2 \times 10^{-6}$  & $8.8 \times 10^{-7}$ \\
 \multicolumn{1}{|l|}{0.035} & $1.0 \times 10^{-5}$  & $8.5 \times 10^{-7}$ & &  $1.7 \times 10^{-6}$  & $9.9 \times 10^{-7}$ \\
 \multicolumn{1}{|l|}{0.04}  & $7.4 \times 10^{-6}$  & $1.4 \times 10^{-6}$ & &  $1.0 \times 10^{-6}$  & $6.7 \times 10^{-7}$ \\
 \multicolumn{1}{|l|}{0.045} & $5.9 \times 10^{-6}$  & $1.3 \times 10^{-6}$ & &  $4.1 \times 10^{-7}$  & $2.1 \times 10^{-6}$ \\
 \multicolumn{1}{|l|}{0.05}  & $3.3 \times 10^{-6}$  & $2.2 \times 10^{-6}$ & &  $3.4 \times 10^{-7}$  & $1.0 \times 10^{-6}$ \\
 \multicolumn{1}{|l|}{0.055} & $2.4 \times 10^{-6}$  & $1.2 \times 10^{-6}$ & &  $1.3 \times 10^{-6}$  & $2.2 \times 10^{-6}$ \\
 \multicolumn{1}{|l|}{0.06}  & $2.2 \times 10^{-6}$  & $4.8 \times 10^{-7}$ & &  $7.1 \times 10^{-7}$  & $8.0 \times 10^{-7}$ \\
 \multicolumn{1}{|l|}{0.065} & $1.6 \times 10^{-6}$  & $7.0 \times 10^{-7}$ & &  $3.9 \times 10^{-7}$  & $4.2 \times 10^{-8}$ \\
 \multicolumn{1}{|l|}{0.07}  & $1.3 \times 10^{-8}$  & $4.9 \times 10^{-7}$ & &  $3.0 \times 10^{-7}$  & $4.7 \times 10^{-7}$ \\
 \multicolumn{1}{|l|}{0.075} & $3.8 \times 10^{-7}$  & $1.1 \times 10^{-6}$ & &  $7.6 \times 10^{-9}$  & $5.8 \times 10^{-7}$ \\
 \multicolumn{1}{|l|}{0.08}  & $8.0 \times 10^{-7}$  & $1.2 \times 10^{-6}$ & &  $9.2 \times 10^{-8}$  & $3.6 \times 10^{-7}$ \\
 \multicolumn{1}{|l|}{0.085} & $2.8 \times 10^{-6}$  & $4.2 \times 10^{-7}$ & &  $4.5 \times 10^{-7}$  & $6.1 \times 10^{-7}$ \\
 \multicolumn{1}{|l|}{0.09}  & $1.6 \times 10^{-6}$  & $2.8 \times 10^{-6}$ & &  $7.0 \times 10^{-7}$  & $4.0 \times 10^{-7}$ \\
 \multicolumn{1}{|l|}{0.095} & $3.0 \times 10^{-6}$  & $1.0 \times 10^{-6}$ & &  $7.1 \times 10^{-7}$  & $6.1 \times 10^{-7}$ \\
 \hline
\end{tabular}
\caption{Values of the relative error for $F_{1,3}(\alpha,r)$.}
\label{tab:f13relerr}
\end{table}

\begin{table}[h]
\centering
\begin{tabular}{l | c | c | c |c| c | c |}
\cline{2-7}
 & \multicolumn{6}{c|}{$\varsigma$}\\
 \cline{2-7}
 & \multicolumn{3}{c|}{$\alpha = 0.75$} & & \multicolumn{2}{c|}{$\alpha = 1.5$}\\
 \hline
 \multicolumn{1}{|c|}{$r$} & $b=0.4$ & $b=0.7$ & $b=0.8$ & & $b=0.4$ & $b=0.8$\\
 \hline
 \multicolumn{1}{|l|}{0.005} & $6.2 \times 10^{-3}$ & $1.1 \times 10^{-3}$  & $1.2$ & &  $1.1 \times 10^{-2}$  & $8.2 \times 10^{-4}$ \\
 \multicolumn{1}{|l|}{0.01}  & $2.7 \times 10^{-3}$ & $1.7 \times 10^{-3}$  & $1.2$ & &  $3.6 \times 10^{-3}$  & $2.5 \times 10^{-4}$ \\
 \multicolumn{1}{|l|}{0.015} & $1.4 \times 10^{-3}$ & $5.9 \times 10^{-3}$  & $1.3$ & &  $1.7 \times 10^{-3}$  & $1.1 \times 10^{-4}$ \\
 \multicolumn{1}{|l|}{0.02}  & $7.8 \times 10^{-4}$ & $7.3 \times 10^{-3}$  & $1.3$ & &  $9.6 \times 10^{-4}$  & $5.4 \times 10^{-5}$ \\
 \multicolumn{1}{|l|}{0.025} & $5.7 \times 10^{-4}$ & $9.4 \times 10^{-3}$  & $1.3$ & &  $5.8 \times 10^{-4}$  & $2.7 \times 10^{-5}$ \\
 \multicolumn{1}{|l|}{0.03}  & $3.1 \times 10^{-4}$ & $1.1 \times 10^{-2}$  & $1.3$ & &  $3.7 \times 10^{-4}$  & $1.6 \times 10^{-5}$ \\
 \multicolumn{1}{|l|}{0.035} & $2.2 \times 10^{-4}$ & $1.2 \times 10^{-2}$  & $1.3$ & &  $2.4 \times 10^{-4}$  & $1.0 \times 10^{-5}$ \\
 \multicolumn{1}{|l|}{0.04}  & $1.7 \times 10^{-4}$ & $1.4 \times 10^{-2}$  & $1.4$ & &  $1.7 \times 10^{-4}$  & $5.3 \times 10^{-6}$ \\
 \multicolumn{1}{|l|}{0.045} & $1.6 \times 10^{-4}$ & $1.6 \times 10^{-2}$  & $1.4$ & &  $1.2 \times 10^{-4}$  & $1.2 \times 10^{-6}$ \\
 \multicolumn{1}{|l|}{0.05}  & $3.6 \times 10^{-5}$ & $1.7 \times 10^{-2}$  & $1.4$ & &  $8.5 \times 10^{-5}$  & $5.9 \times 10^{-6}$ \\
 \multicolumn{1}{|l|}{0.055} & $9.5 \times 10^{-5}$ & $1.9 \times 10^{-2}$  & $1.4$ & &  $5.9 \times 10^{-5}$  & $1.1 \times 10^{-6}$ \\
 \multicolumn{1}{|l|}{0.06}  & $3.6 \times 10^{-5}$ & $2.0 \times 10^{-2}$  & $1.4$ & &  $4.4 \times 10^{-5}$  & $6.6 \times 10^{-7}$ \\
 \multicolumn{1}{|l|}{0.065} & $7.2 \times 10^{-5}$ & $2.1 \times 10^{-2}$  & $1.4$ & &  $3.5 \times 10^{-5}$  & $5.6 \times 10^{-7}$ \\
 \multicolumn{1}{|l|}{0.07}  & $5.5 \times 10^{-5}$ & $2.3 \times 10^{-2}$  & $1.4$ & &  $2.6 \times 10^{-5}$  & $9.8 \times 10^{-7}$ \\
 \multicolumn{1}{|l|}{0.075} & $2.8 \times 10^{-5}$ & $2.4 \times 10^{-2}$  & $1.4$ & &  $1.8 \times 10^{-5}$  & $3.3 \times 10^{-7}$ \\
 \multicolumn{1}{|l|}{0.08}  & $3.1 \times 10^{-5}$ & $2.5 \times 10^{-2}$  & $1.5$ & &  $1.3 \times 10^{-5}$  & $9.2 \times 10^{-8}$ \\
 \multicolumn{1}{|l|}{0.085} & $3.1 \times 10^{-5}$ & $2.7 \times 10^{-2}$  & $1.5$ & &  $9.3 \times 10^{-6}$  & $8.2 \times 10^{-8}$ \\
 \multicolumn{1}{|l|}{0.09}  & $7.5 \times 10^{-6}$ & $2.8 \times 10^{-2}$  & $1.5$ & &  $9.1 \times 10^{-6}$  & $2.0 \times 10^{-7}$ \\
 \multicolumn{1}{|l|}{0.095} & $3.7 \times 10^{-6}$ & $2.9 \times 10^{-2}$  & $1.5$ & &  $4.8 \times 10^{-6}$  & $3.9 \times 10^{-7}$ \\
 \hline
\end{tabular}
\caption{Values of the relative error for $\mathcal F(\alpha,r)$.}
\label{tab:effrelerr}
\end{table}

\part*{Conclusions and perspectives}\addcontentsline{toc}{part}{Conclusions and perspectives}
\label{part:concl_persp}
\markright{Conclusions and Perspectives}{}
\markboth{Conclusions and Perspectives}{}
\markright{Conclusions and Perspectives}{}
\setcounter{equation}{0}
\makeatletter
\long\def\theequation{\ifnum \c@chapter > \z@ \thechapter --\fi \@alph \c@equation}
\makeatother
\mysection{0}{The ODE/IM correspondence}

In the first part of this thesis we have presented the ODE/IM correspondence starting with a general overview of the subject: its realisation on the simple case of the six-vertex model. We have then proceeded to the study of the correspondence in the affine Toda field theories related to the simply-laced affine algebras $\hat{\mathfrak a}_r$ and $\hat{\mathfrak d}_r$. We managed to successfully establish the link between the linear problem associated to a classical simply-laced affine Toda field theory and the Bethe ansatz equations (BAEs) related to its massive quantum version. As a byproduct of this analysis we gave an interesting interpretation of the $\psi$-system in terms of the structure of the underlying Lie algebra's representations. This point of view on the $\psi$-system appears to be a fruitful perspective with which to delve deeper in the analysis of the structure of the ODE/IM correspondence.

These results are being collected in a paper for future publication \cite{Negr_Tate_14}, while an article, specialising on the case of the $\hat{\mathfrak a}_2^{(2)}$ affine Toda Field theory (that is, the Tzitz\'{e}ica-Bullough-Dodd model), was already published \cite{Dore_Fald_Negr_Tate_12}.

\mysection{0}{The fermionic basis}

In the second part of this thesis we have presented an application of the method of fermionic basis to the sin(h)-Gordon models. After an overview on the construction of the fermionic basis, we have investigated its relations with the reflection relations of \cite{Fate_Frad_Luky_AZam_AlZa_99}: our results show that the latter are trivially solved when the space of states of the model is described in the fermionic basis. Moreover, this fact provides an interesting interpretation to the fermions: they are particular linear combinations of the Virasoro generators, whose action on the highest-weight vector $\Phi_a(0)$ produce a complete basis of the quotient space $\mathcal V^{\textrm{quo}}_a \otimes \bar{\mathcal V^{\textrm{quo}}_a}$; the peculiarity of this basis is its invariance under the symmetries $\sigma_1$ and $\sigma_2$ (and also under the duality $b\rightarrow b^{-1}$). This point of view gives the fermionic basis a more ``physical" interpretation with respect to the formal mathematical introduction of \cite{Jimb_Miwa_Smir_11_1} and, in addition, allows to extend its application to the sinh-Gordon model for which a rigorous definition of the fermions was not available, given the complicated nature of the lattice regularisation; the only ``weak spot" in this interpretation is the fact that the formula (\ref{eq:onepointfunctionsinh}) has to be introduced as a conjecture. In the chapter \ref{chap:knownresults} we presented a series of tests for the validity of this conjecture, by comparing its predictions against known results. Both the analytical and the numerical studies were found to be in perfect accordance with the literature on the subject and we believe they make the conjecture strongly reliable. Nonetheless it would be desirable to introduce the formula (\ref{eq:onepointfunctionsinh}) in a more rigorous fashion; a possible way to arrive at this might require to give a ``physical" interpretation to the function $\Theta(k,l\vert\alpha)$: below we sketch a tentative approach to this question.

The results of this second section were collected and published in three separate articles \cite{Negr_Smir_13_1,Negr_Smir_13_2,Negr_14}

\mysection{0}{The generalised Gibbs' ensemble and the Yang-Yang action: hints for a connection}

The function $\Theta(k,l\vert\alpha)$ has been defined in section \ref{sec:onepointsinh} starting from the TBA equation (\ref{eq:destridevegaequation}); let us generalise the TBA by considering, instead of the usual partition function of the Gibbs' ensemble

\eq
	Z(R) = \textrm{Tr} \left(e^{-2\pi R \mathcal H} \right)\; ,
\en
where $\mathcal H$ is the Hamiltonian of the system, the partition function of the \emph{generalised Gibbs' ensemble} \cite{Dunj_Olsh_Rigo_Yuro_07}:

\eq
	Z(\{g\}) = \textrm{Tr}\left[\exp\left(-\sum\limits_{\ell=-\infty}^\infty g_{2\ell-1}\mathcal H_{2\ell -1}\right) \right]\; ,
\en
where, for $j\geq 1$, the operators $\mathcal H_{2j-1}$ and $\mathcal H_{-2j+1} \equiv \bar{\mathcal H}_{2j-1}$ are the local integrals of motion in the space direction; we use the calligraphic letters in order to distinguish them from the integrals of motion in the Matsubara direction, introduced in \ref{sec:expenmomtens}. The ``chemical potentials" are chosen to be positive $g_{2j-1} >0$ and the usual partition function for the Gibbs' ensemble is recovered by specialising

\eq
	\mathscr G\; : \quad g_{2j-1} = 0 \; , \quad \vert 2j-1\vert >1 \; , \qquad g_{\pm 1} = 2\pi R \; .
\label{eq:usualGibbs}
\en
Following the standard procedure, shown for example in \cite{Yang_Yang_69,AlZa_90}, one introduces the pseudoenergy $\epsilon(\theta)$ and minimises the free energy, obtaining

\eq
	\epsilon(\theta) = \sum\limits_{j=-\infty}^\infty g_{2j-1}e^{(2j-1)\theta} - \int\limits_{-\infty}^\infty \Phi(\theta-\theta') \log\left(1+ e^{-\epsilon(\theta')}\right) d\theta' \; .
\en
If one assumes that the series $\sum_{j=-\infty}^\infty g_{2j-1}z^{2j-1}$ has an infinite radius of convergence, then this TBA equation is perfectly well-defined. We can further define the $Q$-function

\eq
	\log Q(\theta) = -\sum\limits_{j=-\infty}^\infty \frac{g_{2j-1}}{2\cos\left(\pi(2j-1)\frac{\nu-2}{2\nu}\right)}e^{(2j-1)\theta} + \int\limits_{-\infty}^\infty \frac{d\theta'}{2\pi}\frac{\log\left(1+e^{-\epsilon(\theta')}\right)}{\cosh(\theta-\theta')} \; ,
\en
although its meaning is not clear; it might be possible to follow the lead of \cite{Klum_Saka_02} and define it passing by the lattice regularisation \cite{Byts_Tesc_06,Tesc_07} on a cylinder and put inhomogeneities into the transfer matrix in the Matsubara direction. If such a construction was available, then one could generalise the formulae for the eigenvalues of the integrals of motion in the Matsubara direction:

\eq
	J_{2j-1} = \frac{\pi g_{-(2j-1)}}{2\sin \left( \pi\frac{\vert 2j-1\vert}{\nu} \right)} - \int\limits_{-\infty}^\infty \log\left(1+e^{-\epsilon(\theta)}\right)e^{(2j-1)\theta}d\theta \; , \quad \forall j\in \mathbbm Z \; .
\label{eq:generalisedintegral}
\en

However, as we have already observed, the lattice regularisation for the sinh-Gordon model is a problematic subject and we cannot rely on this hypothetical construction; for this reason we consider (\ref{eq:generalisedintegral}) as a formal definition.

It is not hard to see that

\eq
	\frac{\partial\epsilon(\theta)}{\partial g_{2j-1}} = \left[e_{2j-1} + R_{\textrm{dress}}^{(0)}\ast e_{2j-1}\right](\theta) \; ,
\en
where $R_{\textrm{dress}}^{(\alpha)}$ was introduced in section \ref{sec:onepointsinh}, along with the definition of the deformed convolution $\ast$. With this we can easily evaluate the derivatives of the integrals of motion with respect to the ``chemical potentials":

\eq
	\frac{\partial J_{2k-1}}{\partial g_{2j-1}} = \frac{\pi \delta_{j,-k}}{2\sin \left( \pi\frac{\vert 2j-1\vert}{\nu} \right)} + e_{2j-1}\ast e_{2k-1} + e_{2j-1}\ast R_{\textrm{dress}}^{(0)}\ast e_{2k-1} \; .
\en
The fact that the resolvent $R_{\textrm{dress}}$ is symmetric ensures us that the formula above is invariant under the exchange $k\leftrightarrow j$, which in turn means that there must exist a potential $Y(\{g\})$ such that

\eq
	J_{2j-1} = \frac{\partial Y(\{g\})}{\partial g_{2j-1}} \; .
\en
We refer to $Y(\{g\})$ as the \emph{on-shell Yang-Yang action} since it corresponds to the same object introduced with this name in \cite{Luky_11} in the case of the usual Gibbs' ensemble. Using the formulae (\ref{eq:sinhthetaerepresentation}) and (\ref{eq:onepointfunctionsinh}) we derive

\eq
	\Theta\left(\mathbbm i a,\mathbbm i b\vert 0\right) - \delta_{a,-b}\, \textrm{sgn}(a)\pi t_{a}(0) = \frac{\partial^2 Y(\{g\})}{\partial g_a\partial g_b}\bigg\vert_{\mathscr G} \; , \quad \forall a,b \in 2\mathbbm Z +1 \; ,
\en
where $\Big\vert_{\mathscr G}$ denotes the specialisation (\ref{eq:usualGibbs}) to the usual Gibbs ensemble. Finally we see that the one-point functions of all the fermionic descendants of the identity $\Phi_0\equiv \mathbbm I$ are expressed in terms of Hessians of the on-shell Yang-Yang action:

\begin{align}
	\langle\bbeta^\ast_{I^+}\bar{\bbeta}^\ast_{\bar{I^+}}\bar{\bgamma}^\ast_{\bar{I^-}}\bgamma^\ast_{I^-}\mathbbm I\rangle = \left(\prod_{\ell =1}^{\mathfrak C(A)} \frac{\textrm{sgn}(a_\ell)\textrm{sgn}(b_\ell)}{\pi}\right) \det\left( \frac{\partial^2 Y(\{g\})}{\partial g_{a_j}\partial g_{b_k}}\bigg\vert_{\mathscr G} \right)_{j,k=1}^{\mathfrak C(A)}\nonumber
	\\	\label{eq:onepointYang}
	\\	
	a_{\ell}\in A\equiv I^+\cup(-\bar{I}^+) \; , \quad b_{\ell}\in B\equiv I^-\cup (-\bar{I}^-) \; , \quad \mathfrak C(A) = \mathfrak C(B) \; .\nonumber
\end{align}

This nice formula is valid only for $\alpha = 0$, however, due to (\ref{eq:shiftformula}), the fermionic descendants of $\mathbbm I$ include the Virasoro descendants of all the fields $\Phi_{2 n \frac{1-\nu}{\nu}}\; , \  \forall n\in\mathbbm Z$. Moreover the one-point functions at radius $R$, normalised to the one-point functions at $R=\infty$, are $2$-periodic in $\alpha$. These two facts joined tell us that the formula (\ref{eq:onepointYang}) can be extended to values $\alpha = 2 p + 2 q \frac{1-\nu}{\nu}\; , \ \forall p,q\in\mathbbm Z$ which, for irrational values of $\nu$ are dense in $\mathbbm R$; generic $\alpha$ appears then by continuity. The consistency of this argument is guaranteed by the formula (\ref{eq:thetashift}).

The connection with the ODE/IM can be made by following the recent work of Lukyanov \cite{Luky_11}, where he showed how, for the usual Gibbs ensemble, the on-shell Yang-Yang action essentially coincides with the regularised action of the classical Euclidean sinh-Gordon model with special boundary conditions. It would be extremely interesting to extend the construction of Lukyanov to the case of the generalised Gibbs ensemble as this would not only give an interpretation of the one-point functions in the quantum model in terms of classical data, but could also provide a hint on how to generalise the whole setup of the fermionic basis to different models.\\

\mysection{0}{Further perspectives}

Aside from the analysis of the idea sketched above, there are various other possible directions of future investigation.

First of all we wish to recall that, in some aspects, the ODE/IM correspondence is not fully understood; in particular the relation between classical and quantum data is still mysterious. Some hypothesis on this relation have been proposed, see for example \cite{Feig_Fren_07}, and it would be interesting to gain some more insight on the deep structure of the ODE/IM correspondence; this direction might also provide an answer to the question whether the correspondence is limited to integrals of motion of the quantum integrable models or encompasses bigger structures and put them in relation with the classical world. We believe that the interpretation we gave of the $\psi$-system can be of help in this task, especially when declined on the non simply-laced algebras for which the Langlands dual algebra is different from the starting one. Another possible future investigation concerns the generalisation of the ODE/IM correspondence to other models; of particular interest on this regard is the emergence of determinant relations, very similar to the $\psi$-system, in the context of Heisenberg spin chains \cite{Grom_Kaza_Leur_Voli_14}, where they are called \emph{$QQ$-system}. It would be nice to investigate the possible connections between these relations and those that arise in the ODE/IM correspondence.

Concerning the fermionic basis formalism, the primary interest of future studies points towards the possible generalisations. In particular we would like to understand if the fermionic basis is a peculiar feature of the sin(h)-Gordon and XXZ spin-$1/2$ models or if it can be extended to other models. A first step in this direction has been made very recently in \cite{Jimb_Miwa_Smir_14}, where the fermionic creation operators for the XXZ spin-$1$, rather the $\hat{\mathfrak a}_1^{(1)}$ nineteen-vertex model, are constructed. Further generalisation might be obtained by studying the relation of the fermionic basis with the ODE/IM correspondence sketched above.

\bibliographystyle{ieeetr}
\bibliography{bibliography}
\end{document}